\documentclass[10pt,oneside]{article}

\usepackage{amsthm}
\usepackage{amsmath}
\usepackage{natbib}
\usepackage[colorlinks,citecolor=blue,urlcolor=blue,filecolor=blue,backref=page]{hyperref}
\usepackage{xcolor}
\usepackage{graphicx}
\usepackage[normalem]{ulem}

\usepackage{bbm}
\usepackage{bm}
\usepackage{amssymb}
\usepackage{authblk}
\usepackage{mathrsfs}

\def\bSig\mathbf{\Sigma}

\newcommand{\bbS}{\bm{S}}
\newcommand{\bZ}{\bm{Z}}
\newcommand{\bY}{\bm{Y}}
\newcommand{\bC}{\bm{C}}
\newcommand{\bX}{\bm{X}}

\newcommand{\bD}{\bm{D}}
\newcommand{\TY}{\tilde{Y}^{\mathrm{obs}}}
\newcommand{\TS}{\tilde{S}^{\mathrm{obs}}}

\newcommand{\Cobs}{C}
\newcommand{\Yobs}{Y^{\mathrm{obs}}}
\newcommand{\Sobs}{S^{\mathrm{obs}}}

\newcommand{\ACE}{\textup{ACE}}
\newcommand{\DCE}{\textup{DCE}}
\newcommand{\cDCE}{\textup{cDCE}}

\newcommand{\cA}{{\cal{A}}}

\newcommand{\Dr}{{\cal{I}}^{r}}

\newcommand{\bS}{\overline{{\mathbb{S}}}}
\newcommand{\R}{{\mathbb{R}}}
\newcommand{\bE}{{\mathbb{E}}}
\newcommand{\bI}{{\mathbb{I}}}
\newcommand{\ca}{{\mathrm{cand}}}

\newcommand{\btheta}{\bm{\theta}}
\newcommand{\eeta}{\bm{\eta}}
\newcommand{\bzeta}{\bm{\zeta}}
\newcommand{\bmu}{\bm{\mu}}

\newtheorem{assumption}{Assumption}

\title{Assessing causal effects in the presence of treatment switching through principal stratification}
\author[1,2]{Alessandra Mattei\thanks{alessandra.mattei@unifi.it}}
\author[3]{Peng Ding\thanks{pengdingpku@berkeley.edu}} 
\author[1]{Veronica Ballerini\thanks{veronica.ballerini@unifi.it}} 
\author[1,4]{Fabrizia Mealli\thanks{Fabrizia.Mealli@eui.eu}}
\affil[1]{University of Florence}
\affil[2]{Florence Center of Data Science}
\affil[3]{University of California Berkeley}
\affil[4]{European University Institute} 
\date{}

\begin{document}

\maketitle
	





\vspace{-1cm}
\begin{abstract}
Clinical trials often allow patients in the control arm to switch to the treatment arm if their physical conditions are worse than certain tolerance levels. 
For instance, treatment switching arises in the Concorde clinical trial, which aims to assess causal effects on the time-to-disease progression or death of immediate versus deferred treatment with zidovudine among patients with asymptomatic HIV infection. 
The Intention-To-Treat analysis does not measure the effect of the actual receipt of the treatment and ignores the information on treatment switching. 
Other existing methods reconstruct the outcome a patient would have had if they had not switched under strong assumptions. 
Departing from the literature, we re-define the problem of treatment switching using principal stratification and focus on causal effects for patients belonging to subpopulations defined by the switching behavior under control.
We use a Bayesian approach to inference, taking into account that
$(i)$ switching happens in continuous time; 
$(ii)$ switching time is not defined for patients who never switch in a particular experiment; 
and $(iii)$ survival time and switching time are subject to censoring. 
We apply this framework to analyze synthetic data based on the Concorde study. 
Our data analysis reveals that immediate treatment with zidovudine increases survival time for never switcher and that treatment effects are highly heterogeneous across different types of patients defined by the switching behavior.
\end{abstract}
		


\section{Introduction}\label{s:intro}
Treatment switching is a post-randomization event that commonly occurs in clinical trials designed to assess the effect of a treatment on the incidence of a disease. 
There exist various types of treatment switching. 
During the follow-up period, the treatment may cause unwanted side effects for some patients, preventing them from continuing the treatment; such a kind of treatment switching is known as ``treatment discontinuation". 
For instance, in clinical trials where the control group is the standard of care, patients may be allowed to switch from the active to the control treatment if unbearable toxicity occurs under treatment. 
In clinical trials aiming to assess the causal effects of an active versus a placebo treatment plus the existing standard of care, patients may be allowed to discontinue both treatments while remaining on the standard of care. 
In other cases, a sudden disease worsening for some weaker patients forces physicians to allow them to switch to the treatment arm or take a non-trial treatment. 
In this work, we focus on clinical trials where patients in the treatment arm never switch to the control arm, but patients in the control arm can switch to the treatment arm if their physical conditions are worse than certain tolerance levels. 
This type of switching often happens in clinical trials 
for patients suffering from AIDS-related illnesses or particularly painful cancers in advanced stages  \citep[see, e.g.,][]{robins1991correcting, robins1994correcting, white1997impact, white1999randomization, zeng2011estimating, chen2013estimating}. 
Such a type of switching also occurs in the Concorde Trial \citep{Concorde1994}, which we will use as a running example to illustrate the methodological framework we propose to deal with the problem of treatment switching. 
An additional example of treatment switching is the BREAK-3 Trial \citep{hauschild2012dabrafenib}.
The BREAK-3 Trial and the CheckMate 067 phase III trial, which is an example of treatment discontinuation \citep{larkin2015combined}, will serve as additional case studies to describe our methodology at work, even though no data will be analyzed (see details in Section~\ref{sec:examples}).

The Concorde Study is a randomized clinical trial that aims to assess the causal effects of immediate versus deferred treatment with an antiretroviral medication (zidovudine) on time-to-disease progression or death among symptom-free individuals infected with HIV. 
According to the trial protocol, patients assigned to the control group should not receive the active treatment until they progress to AIDS-related complex (ARC) or AIDS. 
However, physicians may judge it unethical to keep patients in the control arm if their physical conditions worsen considerably, e.g., if they experience persistently low CD4 cell counts even before the onset of ARC or symptoms of HIV.

Intention-to-treat (ITT) analysis compares groups formed by randomization regardless of the treatment actually received, ignoring the information on treatment switching in the control group. 
It is valid for measuring the effect of assignment but does not estimate the effect of the actual receipt of the treatment. 
In the Concorde study, an ITT analysis compares outcomes by the assignment to immediate versus deferred treatment with zidovudine, ignoring whether control patients stay on the control arm for the entire follow-up period (or, according to the protocol, up to the onset of ARC or AIDS). 
However, we cannot ignore treatment switching if the focus is on assessing the effects of the treatment itself, that is, of receiving zidovudine immediately versus subsequently after the onset of ARC or AIDS. 

Unfortunately, we cannot adjust for treatment switching by simply conditioning on its observed value because treatment switching is a non-randomized post-assignment variable. 
Imagine that immediate treatment with zidovudine increases every individual's survival, but weaker patients, who are most at risk of death, would switch very early if assigned to control. 
A naive analysis that compares observed immediate versus observed deferred treatment with zidovudine may unfairly conclude that the first has no or little effect on survival.
Web Appendix A reviews various existing methods to evaluate the effect of a treatment accounting for treatment switching. 
They focus on causal effects for the whole population under the assumption that each individual has an outcome that would have happened under assignment to treatment and an outcome that would have happened under assignment to control if that individual had not switched. 
The recent release of an Addendum to the E9 guideline on `Statistical principles in clinical trials' by the ICH (ICH E9(R1) addendum) refers to this approach as a ``hypothetical strategy'' for dealing with inference on treatment effects in the presence of intercurrent events, such as treatment switching \citep{ICH2019}.
In Web Appendix B, we also describe and discuss a semi-competing risks approach to the analysis of randomized studies with survival outcomes suffering from treatment switching.
In the classical competing risks literature, controlled direct effects and total effects are usually the targets of inference. 
Controlled direct effects are hypothetical estimands as those usually considered in the treatment switching literature, comparing the time to the primary event (e.g., disease progression or death) under assignment to treatment versus control after somehow eliminating the competing event (e.g., switching).  
Total effects are also causal effects for the whole population.
They are a type of ITT effect, namely the contrasts of the probabilities of experiencing the primary event before a time $t$. 
As total effects, they do not account for the mechanisms by which the treatment affects the occurrence of the primary event, e.g., through other (secondary) events like tolerance implying treatment switching.

We propose to re-define the problem of treatment switching using principal stratification \citep{FrangakisRubin2002}, which is also recognized in the ICH E9(R1) addendum \citep{ICH2019} as a strategy to deal with intercurrent events.
The novel causal estimands are the principal causal effects (PCEs) for subpopulations defined by the switching behavior under control. 
The key insight underlying our approach is that treatment switching can be viewed as a general form of noncompliance. Principal stratification plays an important role in the analysis of randomized studies with all-or-none noncompliance, where it classifies units into groups defined by compliance status.
These studies usually focus on the causal effects for the principal stratum of compliers \citep{AIR1996}.
In clinical trials with treatment switching, classifying patients into subpopulations defined by the switching behavior is an extension of classifying units based on the compliance status (see Web Appendix C for details on the connection to the noncompliance literature). 
To the best of our knowledge, no published studies before our study first published on ArXiv \citep{mattei2020assessing} used principal stratification to deal with the problem of treatment switching.
Principal stratification has been recently used to define the causal effects of treatment with semi-competing risks \citep{comment2019survivor, xu2022bayesian}, and strong connections exist	between our study and the existing studies.
Nevertheless, some distinguishing features make our contribution unique.
\cite{comment2019survivor} and \cite{xu2022bayesian} focus on assessing the causal effects of treatment on non-terminal time-to-event outcomes and use principal stratification to account for the fact that the non-terminal endpoints are subject to truncation by death, that is, they are not well-defined after death. 
Specifically, their causal estimands are time-varying survivor causal effects for the principal strata of patients who would survive regardless of treatment assignment.  
Our causal estimands are PCEs on a terminal time-to-event outcome (i.e., time-to-disease progression or death), with principal strata defined by the switching behavior considering that the occurrence of the primary terminal endpoint precludes any future non-terminal switching event. 
Because switching is a non-terminal event that does not truncate death, here the causal effects are well-defined for each principal stratum. 
See Web Appendix B for an in-depth discussion of the similarities and differences between our framework and the existing frameworks in the presence of semi-competing risks.

The PCEs are \textit{local} causal effects for patients who are homogeneous with respect to the switching behavior. 
Therefore, the PCEs provide information on treatment effect heterogeneity with respect to the switching behavior.
In the Concorde trial, the principal stratum of non-switchers will be of particular interest. 
Non-switchers are patients who would never switch to the active treatment if assigned to the control. 
They take the treatment and control according to the protocol and thus can provide evidence on the causal effect of treatment versus control.

Treatment switching complicates causal inference. 
First, the switching of patients under control either never happens or happens in continuous time. 
Second, assumptions such as exclusion restrictions \cite[][]{AIR1996}, typically invoked in the noncompliance setting, are untenable in studies with treatment switching. 
Section~\ref{sec:2} will discuss these issues in detail. 
We deal with inferential issues in the analysis of the Concorde trial using a flexible model-based Bayesian approach, which allows us to take into account that $(i)$ switching happens in continuous time, generating a continuum of principal strata, $(ii)$ switching time is not defined for patients who never switch in a particular experiment, and $(iii)$ survival time and switching time are subject to censoring.

\section{The Concorde trial}\label{concorde_data}
The Concorde trial is a double-blind, randomized clinical trial aimed to evaluate the effect of immediate/active versus deferred/control treatment with zidovudine in symptom-free individuals infected with HIV  \citep{Concorde1994}. 
Due to privacy constraints, we use a synthetic dataset produced by \cite{White_et_al:2002}, which closely mimics the Concorde trial.
The data comprise $n=1000$ patients with asymptomatic HIV infection. Half the patients are randomized to immediate zidovudine, and the other half to deferred zidovudine. 
In principle, patients in the deferred arm should not receive zidovudine until they progress to AIDS-related complex (ARC) or AIDS. 
Nevertheless, some patients in the deferred arm are allowed to switch to the active treatment arm, starting zidovudine before the onset of ARC or symptoms of HIV on the basis of persistently low CD4 cell counts. 
The outcome is time-to-disease progression or all causes of death. 
The survival time and the switching time are subject to censoring. 
The trial lasts 3 years, with staggered entry over the first 1.5 years; therefore, the censoring time ranges from 1.5 to 3 years. 
The data do not include any pretreatment covariates.

For each patient $i$, let $Z_i$ denote the treatment assignment: $Z_i=1$ for immediate zidovudine and $Z_i = 0$ for deferred zidovudine. 
Let $\Yobs_i$ and $\Sobs_i$ denote the survival time and switching time under the actual treatment assignment without censoring. 
Let $C_i$ be the censoring time. 
Let $\tilde{Y}^{\mathrm{obs}}_i=\min\{\Yobs_i, \Cobs_i\}$ denote the censored time-to-disease progression or death. 
Because patients cannot switch from the treatment to control, for patients assigned to immediate zidovudine, we set the switching time to be $\tilde{S}^{\mathrm{obs}}_i=\Sobs_i=\bS$, where the symbol ``$\bS$'' is a non-real value. 
Under control, patients could either experience the event of interest (disease progression or death) without switching or switch before progressing or dying; they can switch from the control to the treatment arm only before their time-to-disease progression or death under control. 
Therefore, for patients assigned to  deferred treatment with zidovudine, we observe the censored switching time:
$$
\tilde{S}^{\mathrm{obs}}_i = \begin{cases}
	\Sobs_i  & \hbox{if }  \Sobs_i \in \R_+ \hbox{ and }  \Sobs_i \leq \Cobs_i,  \\
	C_i & \hbox{if }  \Sobs_i \in \R_+ \hbox{ and }  \Sobs_i > \Cobs_i , \\
	C_i & \hbox{if }  \Sobs_i = \bS,
\end{cases}
$$
where we set $ \Sobs_i = \bS$ for control patients who progress the disease/die without switching.

\begin{figure}
	\begin{center}
		\begin{tabular}{c}
			$P\left\{\tilde{Y}^{\mathrm{obs}}_i>y \mid Z_i\right\}$   \vspace{-0.5cm}\\
			\includegraphics[width = 0.5\textwidth]{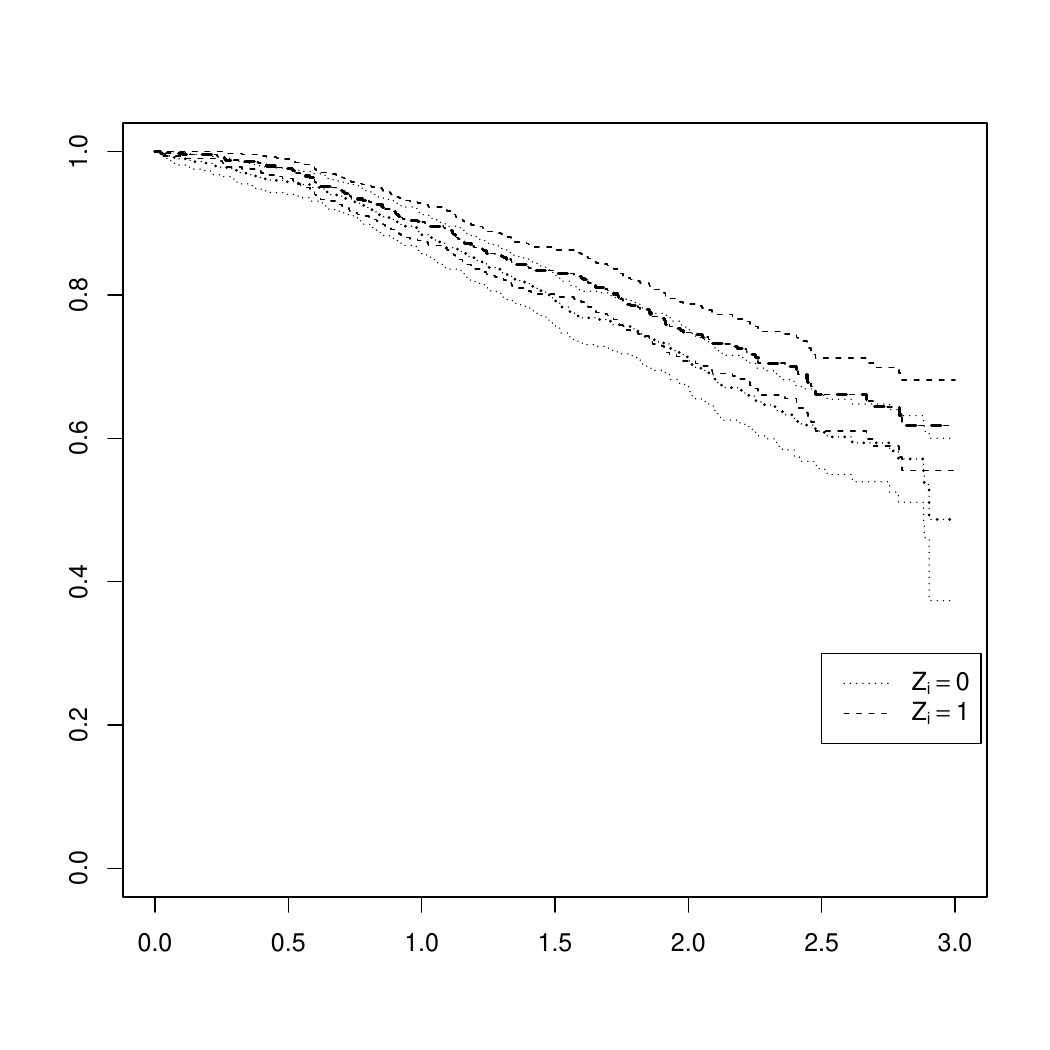}  	
			\vspace{-0.5cm}\\
		\end{tabular}	
	\end{center}
	\caption{ITT analysis:  Kaplan--Meier estimates of the survival functions with 95\% confidence intervals. 
		The solid line corresponds to the control, and the dashed line corresponds to the treatment. } \label{Fig_km_des}
\end{figure}

Table~\ref{tab2} presents some summary statistics.
The upper panel in Table~\ref{tab2} provides some insights that immediate treatment with zidovudine increases survival time. 
However, this simple comparison between survival times under treatment and control cannot even be interpreted as the average causal effect of the assignment due to censoring.
Figure~\ref{Fig_km_des} shows the Kaplan--Meier estimates of the survival functions. The one under treatment dominates the one under control. 
This suggests that being assigned to immediate treatment with zidovudine is beneficial, although the difference between the two survival curves is quite small, and the 95\% confidence intervals overlap. 
The comparison between the survival curves provides a non-parametric estimate of the ITT effect.
Nevertheless, to assess the effect of immediate versus deferred treatment with zidovudine, we cannot ignore information on the switching status. 
\begin{table}
\caption{Synthetic Concorde data: Descriptive statistics}\label{tab2}
$$
\begin{array}{l|ccc}
\hline
\hbox{Variable}& \hbox{All} & Z_i=0&Z_i=1 \\
\multicolumn{1}{r|}{\hbox{}}&(n=1000) & (n=500) & (n=500) \\
\hline
\vspace{-0.3cm}\\
\hbox{Treatment assignment } (Z_i)                                             & 0.5  & 0    & 1 \\
\vspace{-0.3cm}\\
\hbox{Indicator for the switching time being censored}	     & -    &  0.62&-\\
\hbox{or taking on a non-real value}\\
(\mathbb{I}\{\left(S_i^{\mathrm{obs}} \in \mathbb{R}_+  \hbox{ and } S_i^{\mathrm{obs}}  > \Cobs_i \right) \hbox{ or } S_i^{\mathrm{obs}}  =\bS\})  &&&\\
\vspace{-0.3cm}\\
\hbox{Censored switching time } (\tilde{S}^{\mathrm{obs}}_i)                            & -    &  1.55 &-\\
\vspace{-0.3cm}\\
\hbox{Censoring indicator for the survival time }
        & 0.69 & 0.66 & 0.71 \\
        (\mathbb{I}\{Y_i^{\mathrm{obs}} > \Cobs_i \})    &&&\\
\vspace{-0.3cm}\\
\hbox{Censored survival time } (\tilde{Y}^{\mathrm{obs}}_i)      & 1.93 & 1.89 & 1.97 \\
\hline
\end{array}
$$
$$
\begin{array}{l|ccccc}
\hline
& \multicolumn{3}{c}{Z_i=0}\\
\cline{5-6}\vspace{-0.3cm}\\
&  \Yobs_i \leq \Cobs_i   &   &   \tilde{Y}^{\mathrm{obs}} = \Cobs_i 
\\
\hbox{Variable}	&  \tilde{S}_i^{\mathrm{obs}} = \Cobs_i  &  \Sobs_i \leq \Cobs_i &     \tilde{S}^{\mathrm{obs}} = \Cobs_i    \\
\vspace{-0.3cm}\\
\multicolumn{1}{r|}{\hbox{}}& (n=119) & (n=189)  & (n=192) \\
\hline
\vspace{-0.3cm}\\
\hbox{Indicator for the switching time}	 
  & 1 & 0 & - \\
   \hbox{taking on a non-real value }	  (\mathbb{I}\{S_i^{\mathrm{obs}}=\bS\})&&&\\
\vspace{-0.3cm}\\
\hbox{Indicator for the switching time being censored}       & 1 & 0 & 1\\
\hbox{or taking on a non-real value}\\
(\mathbb{I}\{\left(S_i^{\mathrm{obs}} \in \mathbb{R}_+  \hbox{ and } S_i^{\mathrm{obs}}  > \Cobs_i \right) \hbox{ or } S_i^{\mathrm{obs}}  =\bS\})  &&&\\
\vspace{-0.3cm}\\
\hbox{Censored switching time } (\tilde{S}^{\mathrm{obs}}_i)                            & - & 1.24 & 2.11^\ast   \\
\vspace{-0.3cm}\\
\hbox{Censoring indicator for the survival time }
        & 0 & 0.74 & 1 \\
        (\mathbb{I}\{Y_i^{\mathrm{obs}} > \Cobs_i \}) \\
        \vspace{-0.3cm}\\
\hbox{Censored survival time } (\tilde{Y}^{\mathrm{obs}}_i)                             & 1.16 & 2.14 &2.11^\ast \\
\hline
\multicolumn{4}{l}{^\ast \hbox{Average censoring time}}
\end{array}
$$
\end{table}

\section{Treatment switching with censoring} \label{sec:2}
\subsection{Potential outcomes}

The objective is to assess the causal effects of immediate versus deferred treatment with zidovudine on the time-to-event outcome, $Y$ (e.g., survival time or time-to-disease progression). 
We use the potential outcomes to define causal effects and make the stable unit treatment value assumption. 
Let $Y_i(z) \geq 0$ and $C_i(z) \geq 0$ be the potential survival time and censoring time for patient $i$  under treatment assignment $z$ $(z=0,1)$. 
The survival time is subject to censoring. The trial starts and ends at specific calendar times, which determine a fixed duration of the study, $c=3$ years. 
Therefore, the censoring time depends on patients' entry, which is staggered over time. Thus, $C_i(z)\leq c$ represents the duration till the end of the study for patient $i$ given treatment assignment $z$. 
Because the censoring time is determined by the date of entrance in the study (which varies with $i$) and the date the study ended (which is determined a priori and is the same for all $i$), we assume $C_i(0)=C_i(1)=C_i$ for all $i=1 \ldots, n$.

As the trial goes on, keeping the patients in the control arm is unethical if their physical conditions are worse than certain tolerance levels.
Therefore, some patients might switch to the treatment arm even if they had been assigned to the control. 
Some trials permit patients in the treatment arm to switch to control if, e.g., they experience an adverse reaction to the treatment. 
Here, we focus on one-sided switching, as in the Concorde trial, where only patients in the control arm can switch to the treatment.  
Let $S_i(z)$ be the potential switching time of patient $i$ under treatment assignment $z$. 
The value of $S_i(z)$ needs careful discussion. 
First, in the presence of one-sided switching behavior, patient $i$'s switching time is the potential switching time under control $S_i(0)$. Because patients in the treatment arm cannot switch, we define $S_i(1)= \bS$.  
Second, a patient $i$ may not switch from the control to treatment no matter how long the follow-up is, implying $S_i(0)=\bS$. 
Third, a patient $i$ can switch to the treatment arm only before their survival time, implying a natural constraint $S_i(0) < Y_i(0)$. 
The natural constraint implies that for patients who would die under control without switching to the active treatment, the switching time, $S_i(0)$, is censored by death, with the censoring event (death/survival time) defined by the potential outcome under control for the primary endpoint, $Y_i(0)$. 
For this type of patients, the switching time is not only not observed but also undefined; thus, $S_i(0)=\bS$. 
Fourth, the switching time is also subject to censoring.

\subsection{Causal estimands} \label{sec:estimands}
Causal effects are comparisons of the treatment and control potential outcomes for a common set of units. 
The average causal effect of treatment assignment equals 
\begin{equation}\label{eq:ace}
	\ACE= \mathbb{E} \left\{ Y_i(1)\right\}- \mathbb{E}\left\{ Y_i(0)\right\} .
\end{equation}
When assessing whether the treatment can prolong the survival of patients, we are also interested in the distributional causal effect:
\begin{equation}\label{eq:dce}
	\DCE(y)= P\left\{Y_i(1) > y \right\}-P\left\{Y_i(0) > y \right\} ,  \qquad  ( y \in \R_+ ).
\end{equation}
\citet{ju2010criteria} noted that
$
\ACE = \int_0^{+\infty} \DCE(y) ~\mathrm{d}y
$.
Although the average causal effect in \eqref{eq:ace} and the distributional effect in \eqref{eq:dce} measure well-defined ITT causal effects, they ignore the information on treatment switching in the control group.

We adopt principal stratification \citep{FrangakisRubin2002} to define causal estimands \textit{adjusted} for the treatment switching behavior. 
A principal stratification with respect to the switching behavior classifies patients into latent groups named principal strata, defined by the joint potential values of the switching time under control and under treatment, $(S_i(0), S_i(1))$. 
In the presence of one-sided switching from the control to the treatment arm, $S_i(1) = \bS$ for all patients; thus, principal strata are defined by the potential outcome of the treatment switching time under control, $S_i(0)$, only.
\citet{FrangakisRubin2002} pointed out that $S_i(0)$ acts as a pretreatment covariate unaffected by the treatment assignment. The variable $S(0)$ is semi-continuous because switching either does not happen or happens in continuous time. 
Therefore, the basic principal stratification with respect to the treatment switching behavior consists of a continuum of principal strata. 
Each principal stratum comprises patients with the same value of the switching time: $\{i: S_i(0) = s\}$, $s \in \{\bS\} \cup \R_+$. 
Throughout the paper, we refer to patients with $S_i(0)=\bS$ as
\textit{non-switchers}, and to patients with a positive real value $S_i(0) = s$, $s \in \R_+$, as \textit{switchers}. 
Non-switchers are patients who experience disease progression or death without switching if assigned to control. 
Switchers belong to $\cup_{s \in \R_+}\{i: S_i(0) = s\}$, the union of  the basic principal strata $\{i: S_i(0) = s\}$, $s \in \R_+$. 
Hereafter we also refer to $S_i(0)$ as the ``switching status'' of patient $i$: ``non-switcher'' is the switching status of a patient $i$ with $S_i(0) =\bS$, and ``switcher (at some point in time)'' is the switching status of a patient $i$ with $S_i(0) =s$, $s \in \mathbb{R}_+$.

The causal effects within principal strata are called principal causal effects (PCEs). 
For instance, 
\begin{equation}\label{eq:pace}
	\ACE(s)= \mathbb{E}\left\{ Y_i(1)  \mid S_i(0) = s \right\} -\mathbb{E}\left\{  Y_i(0)  \mid S_i(0) = s \right\}  
\end{equation}
is the principal average causal effect for $s \in \{\bS\} \cup \R_+$, and 
\begin{equation}\label{eq:pdce}
	\DCE(y\mid s)= P\left\{Y_i(1) > y \mid S_i(0) = s \right\}-P\left\{Y_i(0) > y  \mid S_i(0) = s \right\} 
\end{equation}
is the principal distributional causal effect,
for $y\in \R_+$ and $s \in \{\bS\} \cup \R_+$.

Because non-switchers would not switch to treatment if assigned to control, for them, the treatment received coincides with the treatment assigned. 
Thus, the PCEs for non-switchers are attributable to treatment received, that is, $\ACE(\bS)$  and $\DCE(y\mid \bS)$
can be interpreted as the effects of the treatment.
They provide information on the causal effects of immediate versus deferred treatment with zidovudine for the subpopulation of patients who would never start zidovudine before the onset of ARC or AIDS if assigned to deferred zidovudine.

\begin{figure}
	\begin{center}
		\includegraphics[width=7cm]{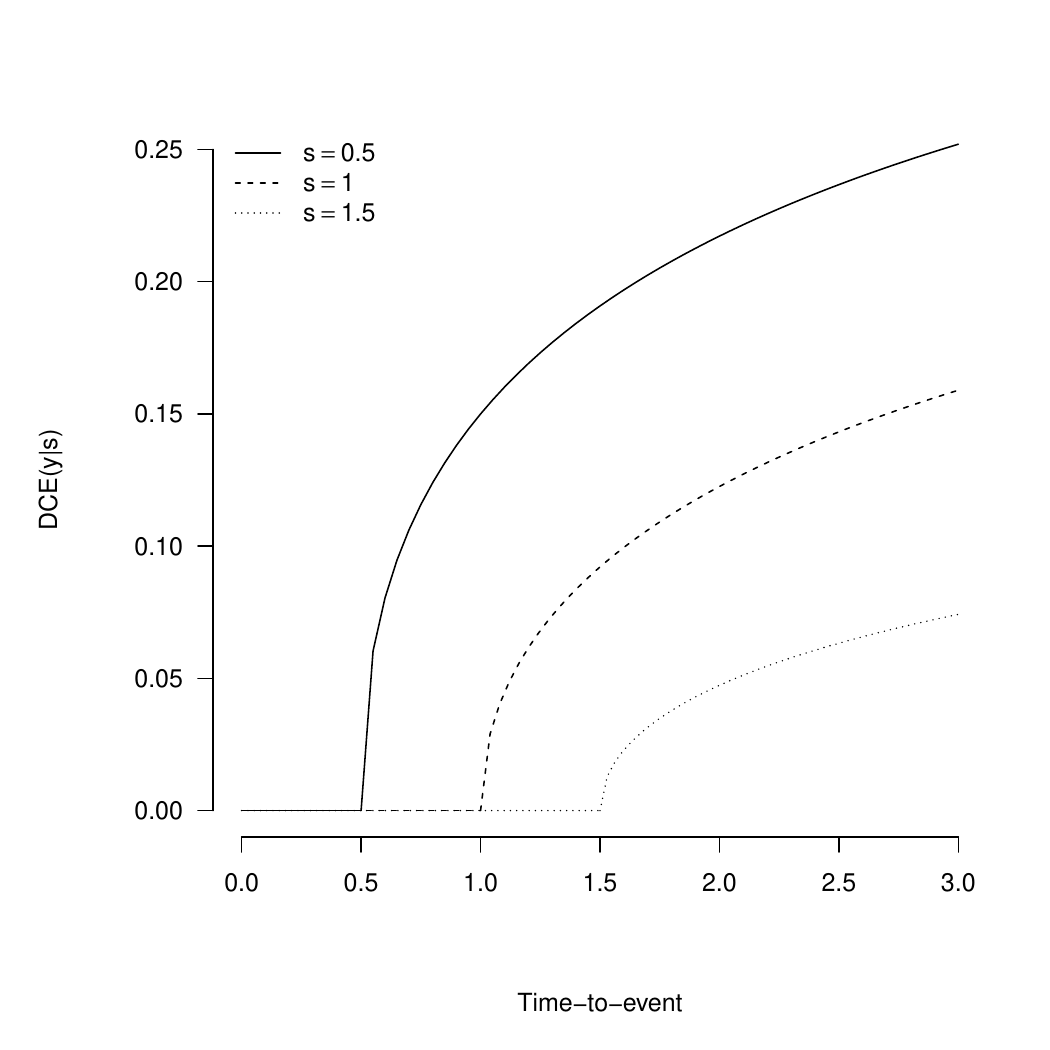}
	\end{center}
	\caption{Examples of principal distributional causal effects under monotonicity $(c=3)$} \label{Fig0}
\end{figure}

The estimands $\ACE(y\mid s)$ and $\DCE(y\mid s)$ for $s \in \R_+$ measure the average causal effect and the distributional causal effect for patients who would switch to the treatment arm at time $s$ had they been assigned to the control arm.
For $y \in \R_+$ and $s \in \R_+$, $\DCE(y\mid s)$ defines a two-dimensional surface on $\R_+ \times \R_+$.
The  natural constraint $S_i(0)< Y_i(0)$ implies that  $P\left\{Y_i(0) > y  \mid S_i(0) = s \right\} = 1$  for $y < s$, and thus the principal distributional causal effect reduces to $\DCE(y\mid s)=P\left\{Y_i(1) > y \mid S_i(0) = s \right\}-1$  for $y < s$.
If we further assume monotonicity of survival time with respect to treatment assignment: 
$Y_i(1) \geq Y_i(0)$, then $Y_i(1)> S_i(0)$
and the principal distributional causal effect   reduces to $\DCE(y\mid s)=0$ for $y< s$.
In this case, for a fixed
value of $S_i(0) = s$, $s \in \R_+$, the principal distributional causal effect curve is non-negative within the interval $[s,c]$ as depicted by Figure~\ref{Fig0}.

Monotonicity states that the treatment does not shorten survival compared to the control. 
In the Concorde trial, monotonicity amounts to assuming that immediate zidovudine does not shorten time-to-disease progression or death compared to deferred zidovudine.
This assumption cannot be directly validated and can be suspicious.
Without monotonicity, the principal distributional causal effect $\DCE(y\mid s)$ is negative (or at most zero) by construction for  $y < s$. 
A  structural negative effect may lead to a misleading interpretation of the effectiveness of the treatment. 
Therefore, it is sensible to consider the conditional principal distributional causal effect for the subpopulation with $Y_i(1)> S_i(0)$: 
\begin{eqnarray}\label{eq:pcdce_s}
	\lefteqn{\cDCE(y   \mid s ) }
	\\&=& P\left\{Y_i(1) > y \mid Y_i(1)> S_i(0), S_i(0) = s \right\}-P\left\{Y_i(0) > y  \mid  Y_i(1)> S_i(0), S_i(0) = s \right\} \nonumber\\
	&=& P\left\{Y_i(1) > y \mid Y_i(1)>  s, S_i(0) = s \right\}-P\left\{Y_i(0) > y  \mid  Y_i(1)> s, S_i(0) = s \right\}, \nonumber
\end{eqnarray}
for $y, s \in \R_+$. 
For $y \leq s$, $\cDCE(y   \mid s )=1-1=0$. 
The estimand $\cDCE(y   \mid s )$, $s \in \R_+$ measures the distributional causal effect on the residual survival time from the switching time for patients who would switch to the treatment arm at time $s$ had they been assigned to the control.

The principal causal effects in \eqref{eq:pace}--\eqref{eq:pcdce_s} are causal effects for groups of patients defined by the detailed value of $S_i(0)$, and thus they are basic PCEs \citep{FrangakisRubin2002}. 
Furthermore, we can define coarsened PCEs
\begin{eqnarray}
	\ACE(\cal{A} ) &=&  \mathbb{E}\left\{ Y_i(1)  \mid S_i(0) \in {\cal{A}}\right\}-\mathbb{E}\left\{  Y_i(0)  \mid S_i(0) \in {\cal{A}} \right\}, \label{eq:ace_A}\\
	\DCE(y   \mid \cal{A} ) &=& P\left\{Y_i(1) > y \mid  S_i(0) \in {\cal{A}} \right\}-P\left\{Y_i(0) > y  \mid S_i(0) \in {\cal{A}} \right\}, \label{eq:dce_A}\\
	\cDCE(y   \mid \cal{A} ) &=& P\left\{Y_i(1) > y \mid Y_i(1)> S_i(0), S_i(0) \in {\cal{A}} \right\} \label{eq:cdce_A}\\&&-P\left\{Y_i(0) > y  \mid Y_i(1)> S_i(0), S_i(0) \in {\cal{A}} \right\}, 
	\nonumber
\end{eqnarray}
where ${\cal{A}}$ is a subset of $\R_+$.
The simplest example is ${\cal{A}}=\R_+$, which implies that the causal effects in \eqref{eq:ace_A}--\eqref{eq:cdce_A} are the causal effects for the coarsened stratum of all switchers.
They are the causal effects for patients who would switch earlier than or at and later than time $s$ if assigned to control for ${\cal{A}}= [0, s]$ and for ${\cal{A}}= (s, +\infty)$, respectively.  
Explicit formulae for these examples are shown in Web Appendix D.

In general, we can discretize the switching time into several disjoint intervals $\R_+ = {\cal{A}}_1 \cup \cdots \cup {\cal{A}}_K$, and define  $\ACE({\cal{A}}_k ) $, $\DCE(y   \mid {\cal{A}}_k ) $ and $\cDCE(y   \mid {\cal{A}}_k ) $ for $k = 1, \ldots , K$. 
When $K = 2$, ${\cal{A}}_1 = [0, s]$ and ${\cal{A}}_2 = (s, +\infty)$, these coarsened PCEs reduce to causal effects for patients who would switch earlier and later than time $s$ for $k=1$ and for $k=2$, respectively. 
However, patients switching at different times have different characteristics, and the basic PCEs conditioning on the potential switching time give detailed information on treatment effect heterogeneity.

Randomized clinical trials on survival outcomes with treatment switching have a similar structure to survival studies with semi-competing risks if we view the switching time and the time-to-disease progression or death as semi-competing events. 
The switching time for patients who would switch is a non-terminal competing event to the event of interest, and the event of interest (i.e., disease progression or death) is a terminal truncating event for the switching time.
Our principal stratification approach offers an innovative approach to deal with semi-competing risks, with distinguishing features that make it crucially different from	existing approaches, including the classical approach based on competing risks models. 
The classical semi-competing risks literature focuses on the causal effects for the whole population, namely, on total effects or (controlled) direct effects \citep{young2020causal}.
Recently, \cite{stensrud2022translating} proposed to target separable effects in the presence of semi-competing risks (see Section~\ref{s:intro} and Web-Appendix B).
Our principal stratification analysis does not target causal effects for the whole population; it is a type of ``sub-group'' analysis with a focus on the PCEs, which are local causal effects for latent subpopulations of units defined by the switching behavior. 
The focus on local causal effects offers methodological and substantive advantages. 
The PCEs are defined without envisaging any hypothetical scenarios or hypothetical decomposition of the treatment. 
The PCEs may be of great interest because they naturally provide information on the heterogeneity of the treatment effect with respect to the switching behavior and on the effect of the treatment for the subpopulation of non-switchers. 

\subsection{Observed data}
The potential outcome for the switching status under control, $S_i(0)$, and the potential outcomes for survival, $Y_i(0)$ and $Y_i(1)$, are well-defined and a-priori observable for all patients in the sense that they could be observed if the patients were assigned to the corresponding treatment level (at least in the absence of censoring). 
A-posteriori, once the treatment has been assigned, for each patient, only the potential outcome corresponding to the treatment actually assigned is observed; the other potential outcome is missing. 
Specifically, for each patient $i$, we observe: $\tilde{Y}^{\mathrm{obs}}_i=\min\{\Yobs_i, \Cobs_i\}$, where $\Yobs_i = Z_i  Y_i(1)  +  (1-Z_i) Y_i(0)$; and 
$\tilde{S}^{\mathrm{obs}}_i=\Sobs_i=S_i(1)=\bS$ under treatment and
$$
\tilde{S}^{\mathrm{obs}}_i = \begin{cases}
	\Sobs_i = S_i(0) & \hbox{if }  \Sobs_i = S_i(0)\in \R_+ \hbox{ and }  \Sobs_i = S_i(0) \leq \Cobs_i, \\
	C_i & \hbox{if }  \Sobs_i = S_i(0) \in \R_+ \hbox{ and }  \Sobs_i = S_i(0) > \Cobs_i, \\
	C_i & \hbox{if }  \Sobs_i = S_i(0) = \bS
\end{cases}  
$$
under control.

In general, we do not observe the principal stratum of a patient for different reasons in the treatment and control arms.
In the treatment arm, we observe no treatment switching,
and the potential outcome $S_i(0)$ is missing. 
Therefore, a patient in the treatment arm may belong to any principal stratum defined by $S_i(0)$, and the treatment group results in an infinite mixture of principal strata.
In the control arm, both the survival time and switching time are subject to censoring. 
We have the following cases.
\begin{description}
    \item[$\mathrm{(a)}$] The patient dies at time $\Yobs_i \leq \Cobs_i$ and does not switch to zidovudine before the onset of ARC or AIDS, i.e., $\tilde{S}^{\mathrm{obs}}_i=\Cobs_i$ and $\tilde{Y}^{\mathrm{obs}}_i=\Yobs_i $. 
    The natural constraint implies $\Sobs_i=S_i(0) = \bS$. 
    Since we observe the time-to-disease progression or death without switching and the switching time is not well-deﬁned after disease progression/death, this patient is a non-switcher belonging to the stratum $ \{i: S_i(0)=\bS\}$.
    \item[$\mathrm{(b)}$] The patient switches to the treatment arm, starting zidovudine before the onset of ARC or AIDS, at time $\Sobs_i$ and dies at time $\Yobs_i$ with $\Sobs_i<\Yobs_i \leq \Cobs_i$, i.e., $\tilde{S}^{\mathrm{obs}}_i=\Sobs_i=S_i(0)= s \in \R_+$, and $\tilde{Y}^{\mathrm{obs}}_i=\Yobs_i=  Y_i(0)$.
    This patient is a switcher belonging to the stratum $ \{i: S_i(0)=s\}$.
    \item[$\mathrm{(c)}$] The patient switches to the treatment arm, starting zidovudine before the onset of ARC or AIDS, at time $\Sobs_i \leq \Cobs_i$ but does not die before the end of the study, i.e.,  $\tilde{S}^{\mathrm{obs}}_i=\Sobs_i=  S_i(0)= s \in \R_+$,  and $\tilde{Y}^{\mathrm{obs}}_i=\Cobs_i < \Yobs_i $.
    This patient is a switcher belonging to the stratum $ \{i: S_i(0)=s\}$. 
    \item[$\mathrm{(d)}$] The patient neither switches to zidovudine before the onset of ARC or AIDS nor dies before the end of the study (with $\Sobs_i \in \{\bS\} \cup (\Cobs_i, +\infty)$ and $\Yobs_i > \Cobs_i$), i.e., $\tilde{S}^{\mathrm{obs}}_i=\tilde{Y}^{\mathrm{obs}}_i=\Cobs_i$. 
    This patient may be a switcher with $\Cobs_i < \Sobs_i=S_i(0) < \Yobs_i=Y_i(0)$, or a non-switcher with $\Sobs_i=S_i(0) =\bS$ and  $\Yobs_i=Y_i(0)> \Cobs_i$. 
    This patient belongs to either the stratum $\{i: S_i(0)=\bS\} $ or the union of strata $ \cup_{s > C_i}\{i: S_i(0)=s\} $.
\end{description}

Cases (a)--(c) have clear values of the switching time and survival time, at least hypothetically, so we directly observe the principal strata for these types of patients. 
Case (d) is less clear due to censoring because the principal stratum membership for patients with $\tilde{S}^{\mathrm{obs}}_i=\tilde{Y}^{\mathrm{obs}}_i= \Cobs_i$ is missing.
Table~\ref{tab1} shows the data pattern and latent principal strata associated with each observed group.
\begin{table}
\caption{Observed data pattern and possible latent principal strata} \label{tab1}
\begin{center} $
\begin{array}{|c|c|c|c|c|}
	\hline
	\vspace{-0.35cm}\\
	Z_i & \tilde{S}^{\mathrm{obs}}_i  &  \tilde{Y}^{\mathrm{obs}}_i  & \hbox{Principal strata} & \hbox{Principal stratum label} \\ 
&&&&	\vspace{-0.35cm}\\
	\hline
&&&&	\vspace{-0.35cm}\\
	0 & \Cobs_i   & \Yobs_i \in [0,\Cobs_i]  &  \{i: S_i(0)=\bS\} &   \hbox{Non-switchers} \\ \hline 
&&&&	\vspace{-0.35cm}\\
	0 & \Sobs_i \leq \Cobs_i     &   \Yobs_i \in (\Sobs_i,\Cobs_i]   & \{i: S_i(0)=\Sobs_i \}&  \hspace{-0.2cm}\hbox{Switchers at time }  \\ 
	&  &  & (\Sobs_i  \in \R_+) &   \hspace{-0.2cm}   \Sobs_i \in \R_+\\ \hline 
&&&&	\vspace{-0.35cm}\\
	0 & \Sobs_i \leq \Cobs_i    &\Cobs_i  &  \{i: S_i(0)=\Sobs_i \}  &   \hbox{Switchers at time }    \\ 
	&  &  & (\Sobs_i  \in \R_+) &     \hspace{-0.2cm} \Sobs_i \in \R_+\\ \hline 
&&&&	\vspace{-0.35cm}\\
	0 & \Cobs_i   & \Cobs_i    &    \left\{i: S_i(0)=\bS  \right\} \hbox{ or } &   \hbox{Non-switchers or }  \\ 
	&    && \left\{i: S_i(0) =s\in (\Cobs_i, + \infty)   \right\} &   \hbox{Switchers at }    \\
	&&&&  \hbox{some time }  s > \Cobs_i\\ \hline 
&&&&	\vspace{-0.35cm}\\
	1 & \bS     &   \Yobs_i \in [0,\Cobs_i]  & \left\{i: S_i(0)=\bS\right\} \hbox{ or} &   \hbox{Non-switchers or} \\  
	& & & \left\{S_i(0) \in \R_+ \right\}&  \hbox{Switchers} \\ \hline 
&&&&	\vspace{-0.35cm}\\
	1 & \bS    & \Cobs_i & \left\{i: S_i(0)=\bS\right\} \hbox{ or}&   \hbox{Non-switchers or}    \\ 
	& &  & \left\{S_i(0) \in \R_+ \right\}&  \hbox{Switchers} \\
	\hline
\end{array}
$
\end{center}
\end{table}

\subsection{Identification issues under randomization}
Although the synthetic data of the Concorde trial do not have any covariates, we discuss a general case with a $K$-dimensional vector of pretreatment variables $X_i$ for each patient.
We consider a completely randomized trial where the following assumption holds by design:
\begin{assumption}  \label{cre}
	$
	P\left\{Z_i \mid S_i(0), Y_i(0),   Y_i(1),  C_i, X_i \right\}=P\left\{Z_i\right\} .
	$ 
\end{assumption}

We assume that the censoring mechanism is independent of both the survival time and the switching time.

\begin{assumption}  \label{icm}
	$
	P\left\{ C_i\mid  S_i(0), Y_i(0), Y_i(1), X_i  \right\}=
	P\left\{ C_i   \right\}.
	$
\end{assumption}

Assumption \ref{icm} implies that the distribution of the censoring times contains no information about the distributions of the potential survival and switching time.
We can also extend the discussion under unconfounded treatment assignment $ P\{Z_i \mid S_i(0), Y_i(0),$  $Y_i(1), C_i, X_i \}=P\left\{Z_i\mid X_i\right\}, $ and ignorability
of the censoring mechanism conditional on observed pretreatment variables $X_i$, $ P\left\{ C_i\mid  S_i(0), Y_i(0), Y_i(1), X_i \right\} = P\left\{ C_i \mid X_i \right\} $. 
The following discussion would be applicable within cells defined by $X_i$.

Randomization helps inference. 
It implies that the distribution of the switching behavior, $S_i(0)$, is the same in the treatment and control arms. 
Moreover, it allows us to express the distributional causal effects of the treatment assignment on the survival time in \eqref{eq:dce} by the distribution of the observed data:
$$
\DCE(y)= P\left\{\Yobs_i>y \mid Z_i=1\right\}-P\left\{\Yobs_i>y \mid Z_i=0\right\}.
$$
Under ignorability of the censoring mechanism, we can estimate the survival functions $P\left\{\Yobs_i>y \mid  Z_i=z\right\}$ for $y\in [0,c]$ by the empirical survival functions under treatment $z$, $z=0,1$. 
Without imposing further assumptions, the data provide no information about the survival functions for $y\in (c,+\infty)$. 
Therefore, the identification of the average causal effect must rely on further (parametric) assumptions on $Y$ because
$
\ACE = \int_{0}^c \DCE(y) \mathrm{d}y+\int_{c}^{+\infty} \DCE(y) \mathrm{d} y
$
depends on the distributional causal effect within both the intervals $[0,c]$ and $(c,+\infty)$. 

Identifying the principal average and distributional causal effects is even more challenging. 
For instance, the distributional effect for the non-switchers, $\DCE(y \mid \bS)$, is, in general, different from the prima facie distributional effect,
$$
\textup{FDCE} (y \mid \bS) = P\left\{\Yobs_i>y \mid Z_i=1\right\}-P\left\{\Yobs_i>y \mid Z_i=0, \Sobs_i=\bS\right\},
$$
i.e., the naive comparison between the patients that do not switch under treatment and control. 
The prima facie effect would differ from $\DCE(y \mid \bS)$, even if no censored cases existed. 
Without censoring, randomization implies $$P\left\{Y_i(0)>y \mid S_i(0)=\bS\right\} =P\left\{\Yobs_i>y \mid Z_i=0, \Sobs_i=\bS\right\},$$ and if we assume that switchers are less healthy people than non-switchers, then $\textup{FDCE}(y \mid \bS)$ is a lower bound for $\DCE(y \mid \bS)$. 
More precisely, if $P\left\{Y_i(1)>y \mid S_i(0)=\bS\right\} \geq P\left\{Y_i(1)>y \mid S_i(0) \in \R_+\right\}$, then
\begin{eqnarray*}
	\lefteqn{ P\left\{\Yobs_i>y \mid Z_i=1\right\}}\\&  =&  P\left\{Y_i(1)>y \!\mid\! S_i(0)=\bS\right\}\! P\left\{ S_i(0)=\bS\right\}\!+\!
	P\left\{Y_i(1)>y \!\mid \!S_i(0)\in \R_+\right\} \! P\left\{ S_i(0)\in \R_+\right\}\\
	& \leq &   P\left\{Y_i(1)>y \mid S_i(0)=\bS\right\},
\end{eqnarray*}
which implies that $\textup{FDCE}(y \mid \bS) \leq \DCE(y \mid \bS)$.

\section{Bayesian Inference}\label{sec:BaysianInference}
In the Concorde trial, inference on the PCEs is particularly challenging due to the nature of the intermediate variable, which is a time-to-event outcome subject to censoring. 
Since we generally do not observe the principal stratum membership, we have to deal with a large amount of missing data, and the PCEs of interest are either not or only partially identified.
We propose to use a flexible Bayesian parametric approach, which is often adopted in principal stratification analysis where inference involves techniques for incomplete data \cite[see, e.g.,][]{MatteiMealli2007, JinRubin2008, JinRubin2009, ZiglerBelin:2012, Schwartz11, Kim_et_al_2017}.
Conceptually, the Bayesian approach does not require full identification. Bayesian inference is based on the posterior distribution of the parameters of interest, which is derived by updating a prior distribution via a likelihood, irrespective of whether the parameters are fully or partially identified, and it is always proper if the prior distribution is proper \cite[e.g.,][]{lindley1972bayesian, ding2018causal}. 
Nevertheless, in finite samples, posterior distributions of partially identified parameters may be \textit{weak identifiable} in the sense that they may have a substantial region of flatness \cite[e.g.,][]{ImbensRubin1997, Gustafson2010, Schwartz11}. 
Another appealing feature of the Bayesian approach is that it allows us to deal with all complications -- missing data, truncation by death, and censoring -- simultaneously in a
natural way. 
Moreover, in Bayesian analysis, inferences are directly interpretable in probabilistic terms. 
The following subsections introduce and discuss a specific parametric model, and Web Appendix E details the description of our Bayesian principal stratification approach. 
Nevertheless, it is worth noting that alternative model specifications, possibly with a different parameterization, can be used, also depending on the specific substantive setting. 
The principal stratification method we propose is general and does not rely on any particular model.

\subsection{Parametric assumptions} \label{sec:par_ass}
We adopt flexible parametric models for the switching status and survival times.
We use the Weibull distribution to model the potential switching time and the potential survival times.
The Weibull model has appealing features: its hazard and survival functions have a simple form, and it is flexible and easy to interpret. 
We can similarly consider alternative survival models, such as Burr models \cite[e.g.,][]{MealliPudney2003} or Bayesian semi-parametric or non-parametric models \cite[e.g.][]{Ibrahim_at_al:2001, Schwartz11, Kim_et_al_2017}. 
The Weibull model has two positive parameters $\alpha$ and $\xi$. 
The parameter $\alpha$ allows for different shapes of the hazard function. 
The hazard function monotonically decreases if $\alpha < 1$, is constant if $\alpha = 1$,
and monotonically increases if $\alpha > 1$. 
We write the Weibull model in terms of the parameterization $(\alpha, \log(\xi) ) $.

First, we model $S_i(0)$. 
We assume that the binary indicator $\mathbb{I}\{S_i(0)= \bS\}$ follows a Bernoulli distribution with probability of success
$$
\pi(x_i) = \dfrac{\exp(\eta_0 + x'_i\eeta)}{1+\exp(\eta_0 + x'_i\eeta)},  \quad
(\eta_0,\eeta) \in \R^{K+1} .
$$
Conditionally on $S_i(0)$ taking on real values, we assume that it follows a Weibull distribution: $S_i(0) \mid S_i(0) \in \R_+, X_i \sim \mathrm{Weibull}\left(\alpha_S, \log(\xi_S) =\beta_S+  X_i'\eeta_S\right)$, $\alpha_S>0, \beta_S \in \R$, $  \eeta_S \in \R^{K}$.

Second, we model $Y_i(0) \mid S_i(0), X_i$. 
Conditionally on $S_i(0)=\bS$, we model $Y_i(0) \mid S_i(0)=\bS, X_i$ as a Weibull distribution with parameters $\left(\bar{\alpha}_Y, \log(\bar{\xi}_0) = \bar{\beta}_Y +  X_i'\bar{\eeta}_Y\right)$, $\bar{\alpha}_Y >0$,  $\bar{\beta}_Y \in \R$ and $ \bar{\eeta}_Y \in \R^{K}$.
Conditionally on $S_i(0) = s \in \R_+$, we model $Y_i(0)$ as a location shifted Weibull distribution: $$
Y_i(0) \mid S_i(0)=s, X_i \sim  s+\mathrm{Weibull}\left(\alpha_Y, \log(\xi_{0}) = \beta_Y +\lambda_0 \log(s)+ X_i'\eeta_Y \right)$$ 
with $\alpha_Y>0$, $\beta_Y, \lambda_0 \in \R$, $\eeta_Y \in \R^K$. 
This location shift parameterization reflects the constraint $Y_i(0) > S_i(0)$ for switchers.

Third, we model $Y_i(1) \mid S_i(0), Y_i(0), X_i$. 
Conditionally on $S_i(0)=\bS$, we model $Y_i(1)$ for non-switchers as a location shifted Weibull distribution: 
 $$Y_i(1) - \kappa Y_i(0) \mid    S_i(0)=\bS, Y_i(0), X_i \sim \mathrm{Weibull}\left(\bar{\nu}_Y, \log(\bar{\xi}_{1}) = \bar{\gamma}_Y + X_i'\bar{\bzeta}\right),$$ with $\kappa \in [0,1]$, $\bar{\nu}_Y>0$, $\bar{\gamma}_Y \in \R$, $\bar{\bzeta} \in \R^K$. 
Conditionally on $S_i(0) = s \in \R_+$, we model $Y_i(1)$ as a location shifted Weibull distribution:   
$$
Y_i(1)- \kappa Y_i(0) \mid   S_i(0)=s,    Y_i(0), X_i \sim  \mathrm{Weibull}\left(\nu_Y, \log(\xi_{1}) = \gamma_Y +\lambda_1 \log(s) + X_i' \bzeta \right)
$$
with  $\kappa \in [0,1]$,  $\nu_Y>0$, $\gamma_Y, \lambda_1 \in \R$, $\bzeta \in \R^K$. 

In Web Appendix F, we explicitly show the probability density functions, the survivor functions, and the hazard functions corresponding to these model assumptions. 
The entire parameter vector is 
$\btheta = $
$\big[(\eta_0, \eeta)$, $ \left(\alpha_S, \beta_S, \eeta_S\right)$,
$\left(\bar{\alpha}_Y, \bar{\beta}_Y, \bar{\eeta}_Y\right)$, $\left(\alpha_Y, \beta_Y, \lambda_0, \eeta_Y \right)$,
$\left(\bar{\nu}_Y, \bar{\gamma}_Y,\bar{\bzeta}_Y\right)$, $\left(\nu_Y, \gamma_Y,\lambda_1,  \bzeta_Y\right)$,  $\kappa\big].$

\subsection{Identification of some model parameters}

The parameters $\lambda_1$ and  $\kappa$ deserve some discussion. 
The parameter $\kappa$ characterizes the dependence between  $Y_i(1)$ and $Y_i(0)$ given $\{  S_i(0), X_i \} $. 
The observed data provide little information on $\kappa$ because we can observe only one of the potential survival times for each patient. 
We can view $\kappa$ as a sensitivity parameter: when $\kappa = 0$, the potential survival times, $Y_i(1)$ and $Y_i(0)$ are conditionally independent; when $\kappa = 1$, monotonicity $Y_i(1) \geq Y_i(0)$ holds, which describes a type of perfect dependence structure. 
In practice, we suggest conducting sensitivity analysis by varying $\kappa $ within the range $[0, 1]$.

The parameter $\lambda_1$ describes the association between $Y_i(1)$ and $S_i(0)$ given $Y_i(0)$ for switchers. 
Because $S_i(0)$ is never observed for treated patients, the observed data provide no direct information about the partial association between $Y_i(1)$ and $S_i(0)$  given $Y_i(0)$. 
This lack of information may affect the causal analysis, leading to imprecise inference on the causal estimands of interest.

We propose to deal with this identifiability issue by introducing parametric assumptions that allow us to leverage better the information we have in the observed data,   borrowing information on  $\lambda_1$ from other parameters and the modeling structure. 
We assume equality of the association parameters $\lambda_0$ and $\lambda_1$: $\lambda \equiv \lambda_0=\lambda_1$, so that a common parameter, $\lambda$, is used to describe the association between  $Y_i(1)$ and $S_i(0)$ given $Y_i(0)$ and between $Y_i(0)$ and $S_i(0)$.
Because $Y_i(0)$ and $S_i(0)$ are jointly observed for some control patients, we have some direct information on the association between $Y_i(0)$ and $S_i(0)$, and thus on the parameter $\lambda$. 
It is worth further highlighting that Bayesian principal stratification analysis does not require the assumption of equality of the association parameters $\lambda_0$  and $\lambda_1$. 
Nevertheless, this parametric assumption may help sharpen inference, leading to more informative and firm causal conclusions, unless results are to some extent sensitive to it.  

Under the parametric assumption that  $\lambda \equiv \lambda_0=\lambda_1$ the entire parameter vector is 
$\btheta = $
$\big[(\eta_0, \eeta)$, $ \left(\alpha_S, \beta_S, \eeta_S\right)$,
$\left(\bar{\alpha}_Y, \bar{\beta}_Y, \bar{\eeta}_Y\right)$, $\left(\alpha_Y, \beta_Y,   \eeta_Y \right)$,
$\left(\bar{\nu}_Y, \bar{\gamma}_Y,\bar{\bzeta}_Y\right)$, $\left(\nu_Y, \gamma_Y,  \bzeta_Y\right)$, $\lambda$, $\kappa\big].$

It is worth noting that some values of the parameters $(\lambda, \kappa)$ correspond to invoking specific structural assumptions.
For instance, under our model specification,  if $\kappa = 1$ and $\lambda <0$, then $Y_i(1)\approx Y_i(0)$ for each patient $i$ with $S_i(0) \in \R_+$ and $S_i(0) \approx 0$.
In fact, if $\kappa = 1$ and $\lambda <0$, then
$$
\lim\limits_{s \rightarrow 0} G_{Y(1)}(y\mid s, y_0, x_i) = \lim\limits_{s \rightarrow 0} P\{Y_i(1)-y_0> y\mid S_i(0)=s, Y_i(0)=y_0, X_i=x_i\} =0
$$ 
for each $y,y_0 \in \R_+$, and thus  $Y_i(1)\approx Y_i(0)$ with probability one.
This is a type of ``exclusion restriction,'' which assumes that the assignment has no or little effect on the survival outcome for switchers if they would immediately switch to the treatment arm had they been assigned to the control arm. 

The association between $Y_i(1)$ and $S_i(0)$ given $Y_i(0)$ for switchers and the dependence between $Y_i(1)$ and $Y_i(0)$ given $S_i(0)$, conditional on covariates, are not identifiable nonparametrically. 
The little information contained in the data on these associations affects inference on the parameters $(\lambda, \kappa)$. 
Under our parametric assumptions, $(\lambda, \kappa)$ enter the observed data likelihood and thus enter the Bayesian posterior inference. 
Therefore, they are at least partially identified and could be parametrically identified depending on the modeling assumptions \citep{Gustafson2010}.
Information on these parameters is implicitly embedded in the model for the joint potential outcomes $Y_i(0)$ and $Y_i(1)$ conditional on $\{ S_i(0), X_i\} $. 
Such a model provides the structure to recover the relationship between the observed and missing potential outcomes.
We factorize the joint conditional distribution of  $P \{ Y_i(0), Y_i(1) \mid S_i(0), X_i \} $
into the product of $P\{  Y_i(0)  \mid S_i(0), X_i \} $ and $P\{  Y_i(1) \mid Y_i(0), S_i(0), X_i \} $.
The model for $P\{  Y_i(0) \mid S_i(0), X_i \}$ characterizes the relationship between the survival outcome and the switching status under control.
The model for $P\{  Y_i(1) \mid Y_i(0), S_i(0), X_i \} $ provides the structure to recover the relationship among the potential survival outcomes and the switching status under control.
The data-augmentation algorithm, detailed in  Web Appendix H for the Concorde study, further provides intuition about how the observed data and model specification together allow for drawing information on the missing potential outcomes and thus on the parameters $(\lambda, \kappa)$. 
We draw the missing switching status for control patients from a distribution that depends on $\tilde{S}_i^{\mathrm{obs}}$ through the distribution of $\tilde{Y}_i^{\mathrm{obs}}$.
Then, we draw the missing switching status and the missing survival time under control for treated patients from a joint distribution that depends on the distribution of $\tilde{Y}_i^{\mathrm{obs}}$.

Given the possible sensitivity of the prior specifications for $(\lambda, \kappa)$, we will conduct various sensitivity checks in the data analysis.

\subsection{Prior distribution, posterior distribution and sensitivity checks} \label{sec:mcmc}
We assume that the parameters are a priori independent. 
We propose to use Normal prior distributions for the parameters of the logistic regression model for the mixing probability, $\pi(X_i)$, Gamma prior distributions for the shape parameters of the Weibull distributions, and Normal prior distributions for the other parameters of the Weibull distributions. 
Finally, we use a Dirac delta prior for the sensitivity parameter $\kappa$ concentrated at a pre-fixed value $\kappa_0 \in [0,1]$, which is essentially the same as fixing $\kappa$ at $\kappa_0 $ a priori. See Web Appendix G for details.

The observed-data likelihood has a complex form involving infinite mixtures because we do not observe the switching status under treatment and only partially observe the switching status under control due to censoring.
Therefore, it is extremely complicated to infer the causal estimands of interest based on the observed-data likelihood directly. 
We use the data augmentation algorithm to derive the complete-data posterior, which is easy to deal with because it does not involve any mixture distributions. 
We can compute the causal estimands as byproducts, and therefore, we can simulate their posterior distributions. 

We investigate the sensitivity of the results with respect to the prior specification for $\lambda$  using both more informative priors (e.g., Normal priors with a smaller variance) as well as a less informative prior (e.g., a uniform prior distribution). 
We also investigate the sensitivity of the results with respect to the parametric assumption of equality of the association parameters $\lambda_0$ and $\lambda_1$.

We conduct the main analysis fixing $\kappa=0$, that is, assuming that $Y_i(1)$ and $Y_i(0)$ are conditionally independent	given $S_i(0)$.
We then assess the sensitivity of the conclusions to different assumptions on $\kappa$ by examining how the posterior distributions of the causal estimands change with respect to different $\kappa_0$ within the range $[0, 1]$.

\section{Bayesian causal inference in the Concorde Trial}\label{concorde}
\subsection{Bayesian ITT analysis}
We first conduct a Bayesian model-based ITT analysis, which compares survival times by assignment, ignoring the switching status (see the estimands in \eqref{eq:ace} and \eqref{eq:dce}). 
This analysis aims to further highlight our substantive contribution to the analysis of clinical trials suffering from treatment switching.

\begin{figure}
	\begin{center}
		\begin{tabular}{c}
			$\DCE(y)=P\left\{Y(1)>y\right\}-P\left\{Y(0)>y\right\}$ \vspace{-0.5cm} \\
			\includegraphics[width = 0.5\textwidth]{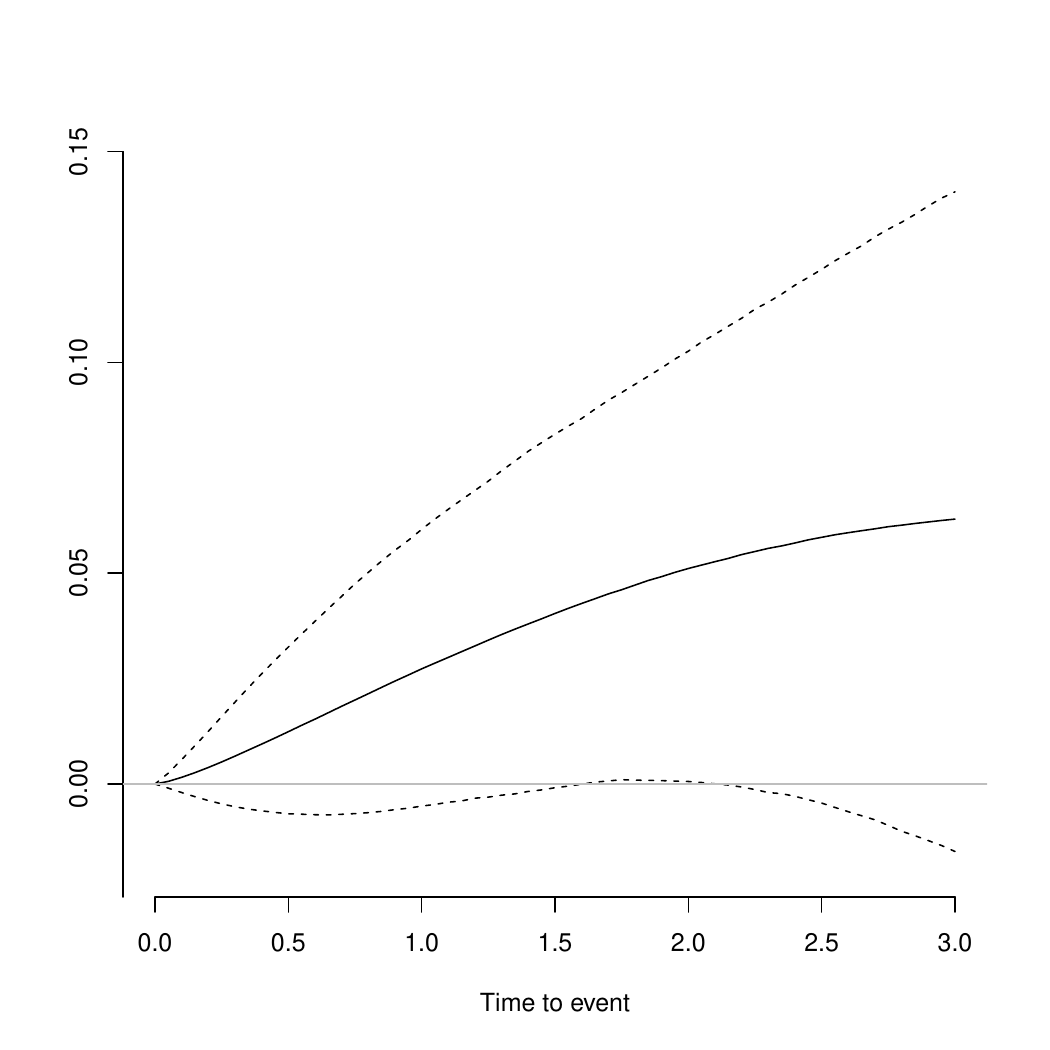} \\
			\\
		\end{tabular}
		
	\end{center}
	\caption{Bayesian ITT analysis using Weibull models. 
 The solid line corresponds to the posterior median, and the dashed lines correspond to the 95\% posterior credible interval.} \label{Fig_itt}
\end{figure}

We assume that $Y_i(0)$ and $Y_i(1)$ marginally follow  Weibull distributions, with parameters $\left( \alpha_Y,\beta_Y\right)$, and $\left( \nu_Y,\gamma_Y\right)$, respectively, where $\alpha_Y, \nu_y >0$, and $ \beta_Y, \gamma_Y \in \R$. 
We conduct Bayesian inference using Gamma prior distributions with shape parameter $0.01$ and scale parameter $100$, and thus with mean $1$ and variance $100$, for $\alpha_Y$ and $\nu_Y$, and Normal prior distributions with zero mean and variance $10\,000$ for $\beta_Y$ and $\gamma_Y$.
The posterior median of the average causal effect is approximately 0.42, with a relatively wide 95\% posterior credible interval, $(-0.51, 1.42)$, which covers zero. 
Although the posterior probability that this effect is positive is relatively high (approximately 0.83), there is little evidence that being assigned to immediate treatment with zidovudine increases the average survival time. 
Similarly, the estimated distributional causal effects are positive and increase monotonically over time. 
Still, there is little difference between survival curves, with the 95\% posterior credible intervals always covering zero except for durations between $1.65$ and $2.10$, where the lower bound of the point-wise credible intervals is very close to zero though.
See Figure~\ref{Fig_itt}  showing the posterior medians and 95\% posterior credible intervals of the distributional causal effects. 
Thus, there is evidence that immediate treatment with zidovudine extends life in individuals infected with HIV, but the estimated effects are small and statistically negligible. 
Nevertheless, it is sensible to expect that the causal effects are heterogeneous across non-switchers and switchers or, more generally, across principal strata, making the ITT analysis an inadequate summary of the evidence in the data for the efficacy of the treatment.

To overcome the limitations of the ITT analysis, we conduct Bayesian inference on the PCEs using the framework and the parametric assumptions introduced in Section~\ref{sec:BaysianInference}.

\subsection{Bayesian principal stratification analysis} \label{sec:BPS}
As a starting point, we assume $\kappa=0$, i.e., $Y_i(0)$ and $Y_i(1)$ are independent given $S_i(0)$. 
We simulate the posterior distributions of the causal estimands of interest using three independent chains from different starting values. 
We run each chain for $125\,000$ iterations, discarding the first $25\,000$ iterations and saving every $20th$ iteration. 
The Markov chains mix well. 
We combine the three chains and use the remaining $15\,000$ iterations to draw inferences. 
See Web Appendices H and I for details on the model and prior specification, the posterior distribution of the model parameters, the  MCMC algorithm, and convergence checks.

\begin{figure}
\begin{center}
\begin{tabular}{c}
$\DCE(y \mid \bS)$
\vspace{-0.75cm}\\
\includegraphics[width=7cm]{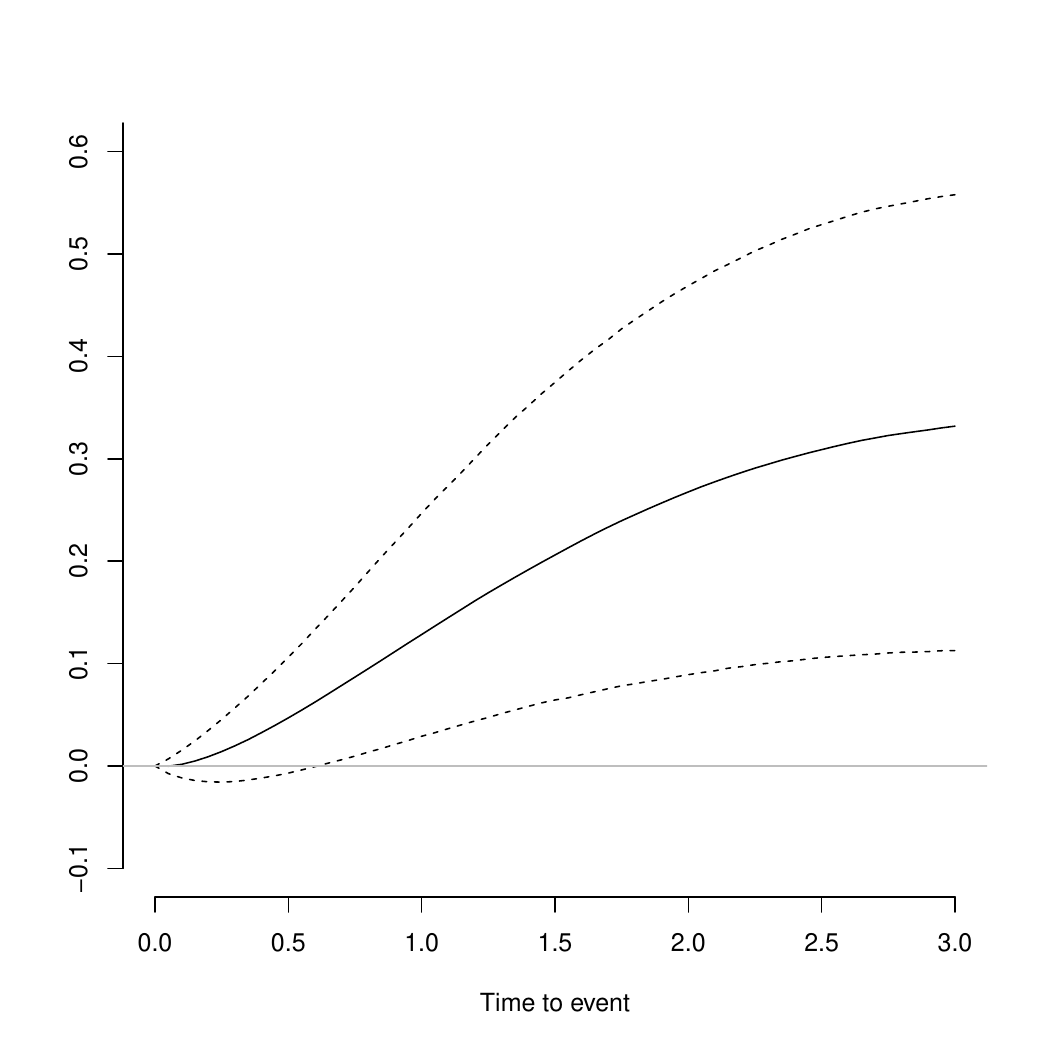}\\
\\
\end{tabular}
\end{center}
\caption{Principal stratification analysis: Posterior median (solid line) and 95\% posterior credible interval (dashed lines) of the distributional causal effects for non-switchers} \label{Fig2} 
\end{figure}

Based on the posterior medians, on average, immediate treatment with zidovudine increases survival time for non-switchers by $2.66$ years, from $2.05$ under deferred treatment with zidovudine to $4.76$ years under immediate treatment with zidovudine. 
The 95\% posterior credible interval $(0.71, 7.73)$ only comprises positive values. 
In Figure~\ref{Fig2}, the posterior medians of the distributional causal effects for non-switchers, $\DCE(y\mid \bS)$, are positive and increase over time from 0 to 0.33 (approximately 4 months). 
The posterior credible intervals include only positive values except for survival times less than $y=0.60$ (about 7 months), where the lower bound is very close to zero though.  
Thus, there is evidence that immediate versus deferred treatment with zidovudine increases survival time for non-switchers.

The interpretation of the results for switchers deserves some care. 
For switchers, $Y_i(1)$ is the survival value if they were assigned and actually exposed to the active treatment. 
The potential outcome under assignment to control, $Y_i(0)$, is the value of survival if switchers were initially assigned to the control treatment, exposed to the control treatment up to the time of switching, e.g., $s$, $s \in \R_+$, and then exposed to the active treatment from the time of switching, $s$, onward.
Therefore, the PCEs for switchers at time $s$ compare the potential outcome that would have happened if they had been initially assigned to treatment and the potential outcome that would have happened if they had been initially assigned to control and received control treatment up to time $s$, and active treatment from $s$ onward.

\begin{figure}
\begin{center}
\begin{tabular}{cc}
$\ACE(s)$ & $\cDCE(y \mid s)$ \\
\includegraphics[width=0.49\textwidth]{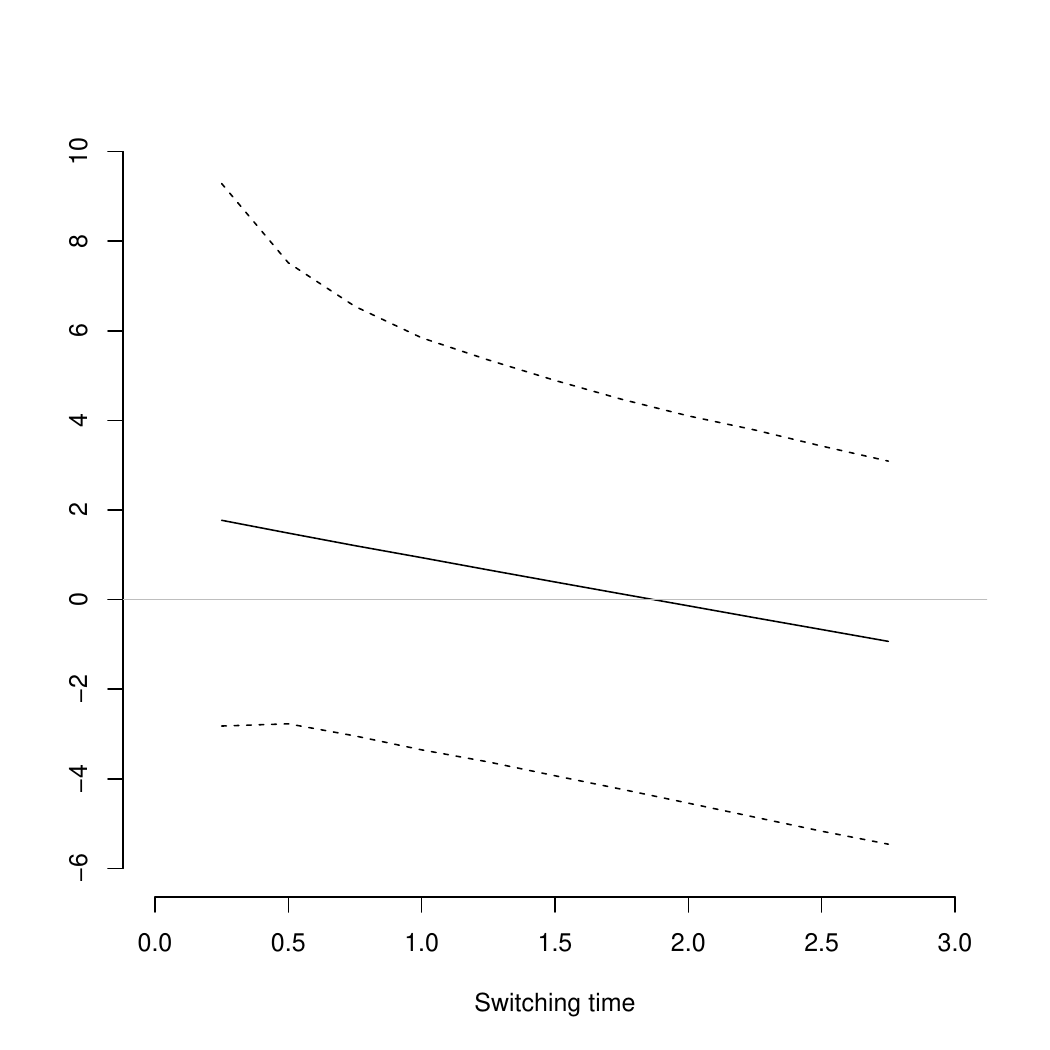} &
\includegraphics[width=0.49\textwidth]{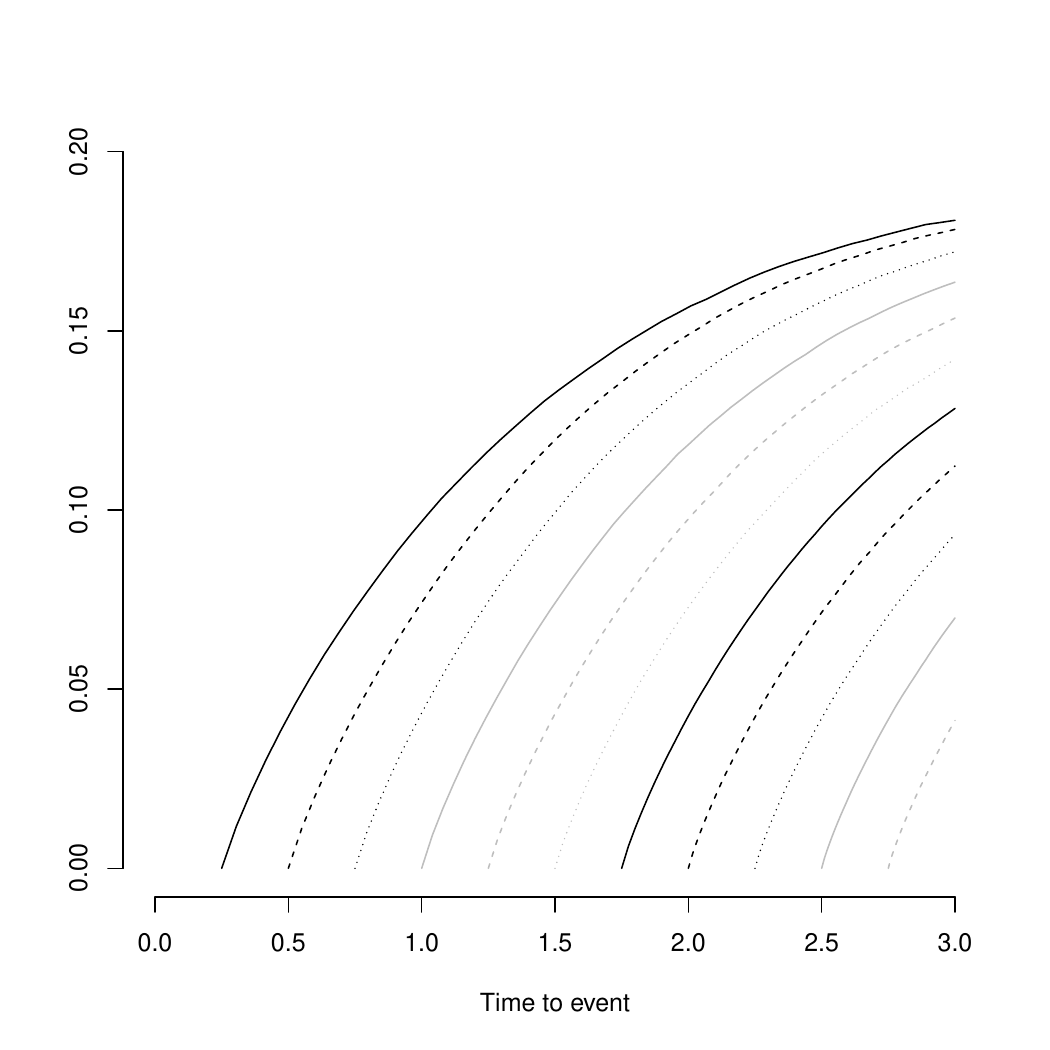} \\
(a)  $\ACE(s)$, $s \in \R_+$ &
(b) $\cDCE(y \mid s)$\\
with 95\% credible interval&
for $s=0.25, 0.50, \ldots, 2.50, 2.75$\\
\\
\end{tabular}
\end{center}
\caption{Posterior medians of the PCEs for switchers} \label{Fig3} 
\end{figure}

The average causal effects for switchers are very small and statistically negligible, irrespective of the time to switching.
See Figure~\ref{Fig3}(a). 
Therefore, the assignment to immediate treatment with zidovudine does not affect the average survival time of patients who would have switched to zidovudine before the onset of ARC or symptoms of HIV had they been assigned to deferred treatment with zidovudine. 
We can interpret these results as evidence that for switchers starting to take the active treatment before the onset of ARC or symptoms of HIV is beneficial, in the sense that their survival is the same as if they had received the active treatment from the time of assignment.  

We focus on the conditional distributional causal effects for switchers and relegate the results on the (unconditional) distributional causal effects to Web Appendix I. 
Figure~\ref{Fig3}(b) shows that the conditional distributional causal effects are always positive and show a trend increase throughout the years irrespective of the time to switching.
Nevertheless, the longer the time to switching, the smaller the effects. 
Therefore, the distributional causal effects for switchers are highly heterogeneous with respect to the switching time. 
This seems plausible scientifically. 
For example, patients switch later because their CD4 cell counts remain sufficiently high for longer. 
Therefore, early switchers comprise sicklier patients, and spending even a short time under control may harm them. 
Under this mechanism, the benefits of immediate versus deferred treatment with zidovudine will be bigger for early switchers, i.e., the conditional distributional causal effects for early switchers will be larger than those for late switchers.
Most of the posterior credible intervals for the conditional	distributional causal effects only include positive values, except for patients who would switch later than 2.75 years had they been assigned to deferred treatment with zidovudine. 
Thus, taking the active treatment from the beginning rather than later increases the survival time for switchers. 
The posterior credible intervals for the conditional distributional causal effects are not shown to make Figure \ref{Fig3}(b) easy to read.
	 
\subsection{Sensitivity Analyses and Model Checking}
Previous results are obtained by fixing $\kappa = 0$ and using a weakly informative prior distribution for  $\lambda$, namely, N$(0,10^4)$.
We assess the sensitivity of the results to $\kappa$, the partial association between $Y_i (1)$ and $Y_i(0)$ given the switching status, $S_i(0)$, and to the prior specification for $\lambda$, by investigating how the posterior distributions of causal estimands change under different $\kappa$ values and different prior specifications for $\lambda$. 
Results appear robust to prior specifications for $\lambda$; different prior distributions for $\lambda$ change the posterior distribution of the causal estimands only slightly.
We find some sensitivity of inferences on the causal estimands to $\kappa$, especially for switchers.

We also investigate the robustness of the results with respect to the parametric assumption imposing prior equality of the association parameters $\lambda_0$ and $\lambda_1$. 
Relaxing the parametric assumption $\lambda_0=\lambda_1$ only slightly changes the results for switchers by leading to posterior distributions of the causal effects for switchers with a larger posterior variability. 
The increased uncertainty in the causal estimands for switchers makes it more difficult to draw firm causal conclusions for them, especially for early switchers.

Our analysis is based on parametric modeling assumptions
and weakly informative priors. 
It is important to conduct model checking. 
We use posterior predictive $p$-values to evaluate parametric assumptions. 
We find no evidence against the model. 
We relegate details on sensitivity analyses and model checks to Web Appendix I for brevity.

\section{Principal stratum strategy at work}\label{sec:examples}
We have used the Concorde trial as an illustrative case study to introduce, describe, and discuss our methodology's key concepts and provide useful insights into the interpretation of the results.
However, the principal stratification approach we propose is general, defining an innovative methodological framework for the analysis of randomized clinical trials with time-to-event primary outcomes suffering from problems of treatment switching or treatment discontinuation, which may lead to important contributions from a substantial perspective. 
To better convey the great power of our methodology in answering substantial questions, in this Section, we briefly revisit two randomized controlled oncology trials involving patients with metastatic melanoma: the BREAK-3 Trial \citep{hauschild2012dabrafenib, latimer2015adjusting} and the CheckMate 067 phase III trial \citep{larkin2015combined}.

The BREAK-3 Trial is a multicenter, open-label, phase 3 randomized controlled clinical trial conducted between December 23, 2010, and September 1, 2011, where patients with previously untreated metastatic melanoma (BRAF V600E mutation-positive melanoma) are randomly assigned to receive either dabrafenib, a selective BRAF inhibitor, or dacarbazine, a chemotherapy medication \citep{hauschild2012dabrafenib}. 
The primary endpoint is progression-free survival. 
In the BREAK-3 study, the trial protocol allows patients to switch from dacarbazine to dabrafenib at disease progression. 
Patients who permanently stop taking dacarbazine for any reason other than the progression of the disease are not eligible for crossover. 
The BREAK-3 Trial has been previously analyzed using an intention-to-treat approach \citep{hauschild2012dabrafenib} and a hypothetical approach   \citep{latimer2015adjusting}. 
Our principal stratum strategy offers an appealing alternative to assess the efficacy of the treatment accounting for the problem of treatment switching. 
The BREAK-3 Trial presents features very similar to the Concorde Trial. 
Both studies are randomized controlled trials with one-sided treatment switching, where patients are allowed to switch from the control to the active treatment during the follow-up period if their physical conditions worsen above certain tolerance levels. 
In both studies, the outcome of interest is a time-to-event outcome, and the time to switch is censored by death, with the censoring event defined by the potential outcome under control for the primary endpoint.
Therefore, the methodological setup we have introduced for the Concorde Trial can be used for the BREAK-3 Trial, with the principal causal effects defined as local causal effects for the subpopulation of non-switchers, patients who would not switch from dacarbazine to dabrafenib if assigned to dacarbazine, and for the subpopulations of switchers, patients who would switch from dacarbazine to dabrafenib if assigned to dacarbazine at some time point (before death).
The average and distributional causal effects for non-switchers provide information on the efficacy of dabrafenib versus dacarbazine. 
The average and distributional causal effects for switchers provide information on the heterogeneity of treatment effects with respect to the switching time. 
Physicians may also be interested in investigating the heterogeneity of the treatment effects across non-switchers and switchers by comparing the principal causal effects for non-switchers and switchers.

The  CheckMate 067 phase III trial is a multicenter, double-blinded, randomized trial where patients with metastatic melanoma are randomly assigned to receive a combination of nivolumab and ipilimumab, ipilimumab monotherapy, or nivolumab monotherapy.
Patients randomized to the combination therapy receive ipilimumab combined with nivolumab once every 3 weeks for four doses, followed by nivolumab alone every 2 weeks. 
Patients randomized to the monotherapy receive either ipilimumab or nivolumab combined with placebo once every 3 weeks for four doses, followed by placebo alone every 2 weeks. 
An outcome of interest is progression-free survival, defined as the time between the date of randomization and the first date of documented progression, as determined by the Investigator, or death due to any cause, whichever occurs first.
Per protocol, patients are treated until progression or unacceptable toxicity and are allowed to discontinue the assigned treatment in the presence of Adverse Events (AEs).
Although patients participating in the  CheckMate 067 study may discontinue both the combination therapy and the monotherapy due to AEs, discontinuation of the combination therapy is particularly interesting \citep{schadendorf2017efficacy}. 
Thus, we revisit the CheckMate 067 study focusing on one-sided discontinuation of the combination therapy.  
We can deal with the problem of treatment discontinuation using the proposed principal stratification framework. 
Let $S_i(z)$ denote the discontinuation time under treatment assignment $z$, with $z=1$ for the combination therapy and $z=0$ for the monotherapy.
Under one-sided discontinuation of the combination therapy, the discontinuation time under monotherapy, $S_i(0)$, is not defined, so we set it to the non-real value $\bS$, and define principal strata by the discontinuation behavior under the active treatment (the combination therapy), $S_i(1)$. 
Specifically, we can cross-classify patients into the following principal strata: 
$(i)$ the principal stratum of patients who would never discontinue the combination therapy if assigned to it;
and $(ii)$ the principal strata of patients who would discontinue the combination therapy at some point in time if assigned to it.
For the former type of patients, the discontinuation time under treatment is not defined: $S_i(1)= \bS$;   
for the latter, $S_i(1) \in \mathbb{R}_+$.
Similar to the Concorde Trial and the BREAK-3 Trial, which suffer from one-sided treatment switching, the key features of the CheckMate 067 phase III trial with one-sided treatment discontinuation are as follows: 
$(a)$ Since discontinuation never happens or happens in continuous time, there is a continuum of principal strata with patients who would discontinue the combination therapy; 
$(b)$ time-to-discontinuation is censored by death, with the censoring event defined by the potential outcome under the combination therapy for the primary endpoint, time-to-disease progression or death; 
$(c)$ both time to discontinuation under treatment and time-to-disease progression or death are subject to censoring due to the end of follow-up.
Patients who would discontinue the combination therapy if assigned to it probably have prognosis factors good enough to experience the progression-free survival event not immediately but are at the same time fragile from suffering treatment discontinuation due to adverse events before disease progression. 
Our principal stratification framework may provide useful information to physicians. 
The principal causal effects for patients who would discontinue the combination therapy if assigned to it can be interpreted as evidence of whether patients who would discontinue the combination therapy due to AE still benefit from it; 
the principal causal effects for patients who would not discontinue the combination therapy can be interpreted as the ``pure effect" of the combination therapy versus the monotherapy because these patients are essentially compliers who would take the treatment assigned and would not discontinue.  
Moreover, comparing the principal causal effects provides information on the heterogeneity of the treatment effects with respect to the discontinuation behavior. 
We can investigate the heterogeneity of the treatment effects between patients who would not discontinue the combination therapy and patients who would discontinue the combination therapy. 
We can also study the differences in treatment effects across subsets of patients who would discontinue the combination therapy.

In some clinical trials, especially oncology trials, the sample size may be relatively small, and thus, the number of patients who switch/discontinue may be minimal. 
The Bayesian principal stratification approach we propose, not relying on asymptotic approximations, is a natural mode of inference in small samples by conveying and correctly quantifying uncertainty due to small sample sizes and a large amount of missingness. 
For instance, in clinical trials suffering from treatment switching and involving a small number of patients, a small proportion of switchers will increase uncertainty in the PCEs for switchers, but inference on the PCEs for the never-switchers will be more precise, and because never-switchers will represent a larger portion of the population, the PCE for them will be even more meaningful and easier to generalize.    

\section{Discussion}
In this work, we have proposed to use principal stratification to assess the causal effects in the Concorde trial with one-sided treatment switching. 
The \textit{principal causal effects} (PCEs) allow for treatment comparisons with proper adjustment for the post-treatment switching behavior. 
The PCEs provide valuable information on treatment effect heterogeneity across different types of patients: non-switchers and switchers, and switchers at different time points.
In particular, the PCEs for non-switchers provide information on the ``pure effect" of the treatment because they are essentially compliers who would take the treatment assigned and would not switch. 

Although we focus on a specific setting -- clinical trials with one-sided switching from the control arm to the treatment arm --  we can extend our approach to general cases. 
For example, we can extend our principal stratification framework to analyze: 
$(a)$ multi-arm clinical trials where patients in one treatment group may switch to another active treatment group; 
$(b)$ clinical trials where the control group is standard of care and patients are allowed to switch from the active treatment to the control treatment if unbearable toxicity occurs \citep[a type of treatment discontinuation; see, e.g.,][]{LipkovichRatitchMallinckrodt2020}; 
$(c)$ clinical trials where treatment switching or treatment discontinuation is two-sided so patients can switch or discontinue both treatments. 
Regarding point $(c)$, consider, for instance, a clinical trial aiming to assess the causal effects of an active versus a placebo oncological treatment on time-to-disease progression or death. 
Suppose all patients are exposed to the existing standard of care and are allowed to discontinue both treatments while remaining on the standard of care; the study suffers from two-sided treatment discontinuation. 
Let $S_i(z)$ denote the potential outcome for the discontinuation behavior under assignment to treatment $z$; $z=0$ for placebo or control, and $z=1$ for treatment.
The joint potential discontinuation behavior under control and under treatment defines the principal stratum to which a patient belongs. 
Specifically, patients can be cross-classified into the following principal strata:
$(i)$ the principal stratum of patients who would discontinue neither the placebo nor the active treatment, irrespective of their assignment. 
For this type of patients, the time to discontinuation is not defined under either treatment status.
Using the notation we introduce in Section~\ref{sec:2}, we can set the discontinuation  time for patients who would not discontinue either treatment to the non-real value $\bS$, so $S_i(0)=S_i(1)=\bS$ for the principal stratum of patients who would discontinue neither the placebo nor the active treatment, irrespective of their assignment;
$(ii)$ the principal strata of patients who would discontinue the placebo treatment but would not discontinue the active treatment. 
For this type of patients  $S_i(0)=\bS$ and $S_i(1) \in \mathbb{R}_+$;
$(iii)$ the principal strata of patients who would discontinue the active treatment but would not discontinue the placebo treatment. 
For this type of patients $S_i(0) \in \mathbb{R}_+$ and $S_i(1)=\bS$;
$(iv)$ the principal strata of patients who would discontinue both the placebo and the active treatment, irrespective of their assignment. 
For this type of patients, $S_i(0) \in \mathbb{R}_+$ and $S_i(1)\in \mathbb{R}_+$. 
The principal average and distributional causal effects can be defined as causal effects for the subpopulations of patients with the same discontinuation behavior under treatment and under control (the principal strata), as we have defined the principal average and distributional causal effects for non-switchers and switchers.
The data structure is similar to that in the Concorde trial. 
For patients assigned to treatment $z$ who do not discontinue before experiencing either disease progression or death, the discontinuation time under treatment $z$ is undefined; the discontinuation time under treatment $z$ is censored by death with the censoring event defined by the potential outcome under treatment $z$ for the primary endpoint.	
Moreover, both the survival and the discontinuation times are subject to censoring due to the end of follow-up.

Background covariate information is valuable in various ways. First, if pretreatment variables enter the treatment assignment mechanism, such as in stratified randomized experiments, analyses must be conditional on them. 
Second, in completely randomized experiments, although pretreatment covariates do not enter the treatment assignment mechanism, they can make parametric assumptions more plausible. 
Moreover, they can improve the prediction of the missing potential outcomes, leading to more precise inferences. 
Third, relevant information could also be obtained in the principal stratification analysis by looking at the distribution of baseline characteristics within each principal stratum. 
Characterizing the latent subgroups of patients in terms of their background characteristics can provide insights into the type of patients for which the treatment is more effective.
Therefore, covariates might help explain the heterogeneity of the effects across principal strata defined by the switching status.

In clinical trials involving duration outcomes, censoring may be due to other events such as dropout and loss to follow-up.
We have assumed the ignorability of the censoring mechanism, which implies that the censoring mechanism is independent of the survival potential outcomes and the switching time. 
A valuable topic for future research is to relax the assumption of an ignorable censoring mechanism addressing the problem of treatment switching with non-ignorable random censoring.
An appealing approach to deal with non-ignorable random censoring is to extend the principal stratification analysis we present here to multiple intermediate variables, the switching status and the censoring time, considering alternative sets of assumptions on the censoring mechanism and investigating the sensibility of the results with respect to them.

\subsection*{Acknowledgments}
The authors thank Kaifeng Lu for the precious comments and suggestions.






\bibliographystyle{ba}
\bibliography{TreatmentSwitchingR1.bib}

\newpage


\appendix
\renewcommand{\thesection}{\Alph{section}}
\setcounter{figure}{0} \renewcommand{\thefigure}{A.\arabic{figure}} 
\setcounter{table}{0} \renewcommand{\thetable}{A.\arabic{table}} 

\begin{center}
    \LARGE{Web appendix for \\ ``Assessing causal effects in the presence of treatment	switching through principal stratification''}
\end{center}


\section{Methods for treatment switching: A review} \label{s:review}

In causal inference, various methods have been proposed to evaluate the effect of a treatment accounting for treatment switching.
To the best of our knowledge, all the existing methods generally focus on causal effects for the whole population, which are defined under the assumption that, for each individual, both the outcome that would have happened under assignment to treatment and the outcome that would have happened under assignment to control exist, if that individual had not switched. 
Unfortunately, for switchers, the outcome that would have happened if they had not switched does not exist conceptually in the data; it is an a-priori counterfactual \cite[][]{FrangakisRubin2002}. 
The data contain no or little information on these a-priori counterfactual outcomes for switchers; thus, assumptions that allow one to extrapolate from the observed data information on them are required.

It is worth noting that the problem of introducing assumptions that allow one to extrapolate from the observed data information on quantities that do not exist in the data for some units also arises in randomized experiments with non-compliance when the focus is on causal effects for the whole population.
In these experiments, the Instrumental Variable (IV) assumptions (exogeneity of the instrument, existence of an association between the instrument and the treatment, exclusion restrictions, and monotonicity) are sufficient to identify average causal effects for the subpopulation of compliers.
Still, they are not sufficient to identify the average effect of the treatment for the whole population; in addition to the IV assumptions (with or without monotonicity), we need to introduce additional assumptions that allow one to infer the overall average effect, i.e., assumptions on a-priori counterfactual outcomes for never-takers and always-takers.
In the literature, alternative sets of assumptions have been considered, including the relatively strong assumption of identical treatment effect for all units and the weaker homogeneity assumption of no additive effect modification across levels of the instrument within the treated and the untreated \cite[][]{Robins:1989, HernanRobins:2006}.

In the treatment switching literature, naive methods include excluding patients who switch, censoring patients who switch, and using the treatment as a time-varying covariate in a regression model. 
See \citet{Morden_et_al2011} for a review.
Excluding switchers results in a comparison of all patients who receive the treatment to patients who are assigned to the control and do not switch.
This analysis compares groups that are not formed by randomization and, therefore, it may produce heavily biased results unless the switching behavior is completely at random.
Censoring the survival time at the switch relies on the assumption that the switching status is ignorable, i.e., the prognosis of patients who switch is equal to that of patients who do not switch. 
This assumption is untenable in studies with non-ignorable switching behavior.
An alternative approach considers the treatment as a  time-varying covariate and includes a time-varying indicator for the treatment received in a (Cox proportional hazards) model. 
It is difficult to interpret the regression coefficients in these models \citep{fisher1999time}, especially their relationships with causal effects of interest. 
Moreover, this model-based approach compares groups that are not formed by treatment assignment, and thus it loses the benefits of randomization and can bias the estimates.

More sophisticated approaches address treatment switching by reconstructing the outcome a patient would have had if they had not switched. 
These are inverse probability of censoring weighting
(IPCW) methods, marginal structural models, and rank-preserving structural failure time (RPSFT) models. 
The IPCW approach censors the switchers at the time point of switching and weights the subjects inversely proportional to their probability to switch \citep{RobinsFinkelstein2000}. 
Marginal structural models impose structure on potential outcomes that would have been observed under different treatment histories \citep{HernanBrumbackRobins2000}. 
A key assumption underlying these approaches is that the switching status is independent of the switch-free outcomes conditional on the observed covariates. 
The plausibility of this assumption rests on the information contained in the covariates. 
It is worth noting that the IPCW approach is generally not applicable if no pretreatment variable is available, as in our synthetic study. 
Moreover, when the covariates are strong predictors of the switching behavior, the estimated switching probability will be close to zero or one for some patients, and the weights can be large. 
As a result, in such settings, IPCW estimators can be sensitive to minor changes in the specification	of the model for the probability of switching.

The RPSFT model relates the observed survival time for each individual to the time-to-event that would have been observed for that individual if s/he had never received the treatment.
The RPSFT model is rank preserving in the sense that, given any two patients, $i$ and $i'$, if patient $i$ survives longer than patient $i'$ under a treatment regime, then $i$ survives longer than  $i'$ under another treatment regime.
This approach, initially proposed by \citet{robins1991correcting} and further developed by \citet{white1999randomization}, explicitly assumes that the time-varying treatment received status is the actual intervention and the random treatment assignment acts as an instrumental variable. 
Since the instrumental variable is binary, \citet{robins1991correcting} and \citet{white1999randomization} require a model linking the potential outcome and the observed outcome by a scalar parameter. 
This scalar parameter is a real value by which the treatment would extend each patient's baseline lifetime, regardless of when the patient eventually switches. 
Therefore, the scalar parameter is the causal effect of interest, which is assumed to be the same for all patients regardless of the switching time. 
The assumption that the treatment effect is constant allows one to extrapolate treatment effects across different subpopulations of patients (i.e., from non-switchers to switchers, irrespective of their switching time). 
Along this line, \citet{walker2004parametric} and \citet{zhang2016correcting} further imposed additional parametric assumptions.

Other researchers have focused on modeling the observed data using parametric or semiparametric approaches \citep{zeng2011estimating, chen2013estimating}. 
However, they usually rely on strong assumptions, like that there exists no relation between a patient's prognosis and switching behavior.
Clearly, the switching status of the patients in the control group contains important post-treatment information, which is useful to characterize treatment effect heterogeneity. \citet{shao2005statistical} realize this problem and propose a model incorporating the ``switching effect.'' However, as pointed by \citet{white2006estimating}, in their likelihood-based inference, \citet{shao2005statistical} again assume independence of the switching time and the survival time.

\section{Competing risks models, Survivor PCEs and PCEs with respect to treatment switching: A comparison} \label{s:competing risks}

Randomized clinical trials on survival outcomes with treatment switching have a similar structure to survival studies with semi-competing risks, where typically two (or multiple) events happen over time, and at least one of the events is an absorbing/terminal event, that is, it prevents from observing the other event(s). 
Therefore, it is worthwhile to clarify similarities and differences between the principal stratification approach we propose and both approaches based on classical semi-competing risks models and other methods based on principal stratification analysis with semi-competing risks.

We first summarize the distinguishing features of the principal stratification framework we propose (see the main text for details).
In our setting, the outcome of primary interest is time-to-disease progression or death, and the switching status is an intercurrent event. 
Two distinguishing features characterize our setting: $(i)$ it is biologically possible that a patient could either progress/die without switching or switch before progressing/dying and $(ii)$ patients can switch from the control to the treatment arm if assigned to control only before their time-to-disease progression or death under control. 
Therefore the switching status and the time-to-disease progression or death can be viewed as semi-competing events: 
the switching time for patients who would switch is a non-terminal competing event to the event of interest; the event of interest (i.e., disease progression or death) is instead a terminal truncating event for the switching time. 
Since the time to switching is not well defined after disease progression or death, the switching time is ``censored by death'' with the censoring event defined by the primary endpoint. 
Thus, assessing the effect of treatment on a terminal event, such as time-to-disease progression or death, in the presence of treatment switching requires accounting for the fact that switching is a non-terminal semi-competing event that is not well-defined after progression or death.

We focus on describing and addressing these complications in our study using a Bayesian approach with the principal stratification framework. 
We first introduce principal causal effects (PCEs) for subpopulations of patients defined by their switching behavior under the control treatment, that is, for non-switchers and switchers. 
It is worth noting that, in our setting, causal effects of treatment on time-to-disease progression or death are well-defined for all types of patients defined by the switching status (non-switchers and switchers at some point in time); only causal effects for non-switchers are interpretable as the ``pure'' effects of the treatment though.
Then, we adopt a Bayesian parametric approach to inference, specifying parametric models for the switching status, the switching time for switchers, and the joint distribution of the potential outcomes for the time-to-event primary endpoint conditional on the switching status.

\subsection*{Principal stratification versus semi-competing risks approaches}

Two key aspects make the principal stratification framework we propose crucially different from classical approaches based on competing risks models: $(i)$ the target causal estimands, and $(ii)$ the use of the observed data to draw inferences on the causal effects of interest.

The causal estimands we focus on, namely PCEs for non-switchers and switchers by time to switching, are particular to our principal stratification approach. 

The classical semi-competing risks literature focuses on statistical estimands generally defined as contrasts of risks, without using a formal framework for characterizing causal effects and their identifying conditions. 
Recently, \cite{young2020causal} clarified that total effects or (controlled) direct effects \citep{robins1992identifiability, Pearl2001} of the treatment on the event of interest are usually the targets in the classical competing risks literature.
Generally, controlled direct effects are related to the marginal cumulative incidence or net risk, and total effects are related to the concept of sub-distribution function, cause-specific cumulative incidence function, or crude risk \citep{geskus2016data}.

(Controlled) direct effects measure treatment effects on the event of interest not mediated through the competing event. 
They are causal effects for the whole population, defined under the assumption that there exists, for each individual, the outcome that would have happened under assignment to a treatment if the competing events had been somehow eliminated, assuming the existence of a-priori counterfactuals. 
In clinical trials with one-sided treatment switching, controlled direct effects compare the outcome that would have happened under assignment to the active treatment and the outcome that would have happened under assignment to control if that individual had not switched.
Controlled direct effects are the hypothetical estimands usually targeted by the literature on treatment switching, where the focus is on causal effects for the whole population in the hypothetical scenario that the competing event (i.e., switching) can be somehow eliminated for all patients (see Web-Appendix \ref{s:review} for a review of the literature on methods for treatment switching). 

Total effects of the treatment on the event of interest are defined as a comparison of the joint distribution of the time to the event of interest (e.g., disease progression or death) and the time to the competing event (e.g., switching time) under treatment versus control. 
Therefore they are a type of ITT effect; thus, they do not account for the mechanisms by which the treatment affects the occurrence of the primary event, e.g., through other (secondary) events like tolerance implying treatment switching.

Recently, \cite{stensrud2022translating} proposed to target separable effects in the presence of semi-competing risks, under the assumption that the treatment can be, at least conceptually, separated into components such that each component affects a different competing event.

In a principal stratification analysis, potential outcomes are defined as a function of the initial treatment assignment only, and no a-priori counterfactual is required. 
In principle, a principal stratification analysis is like a ``sub-group'' analysis, where groups are defined by a latent variable (the principal stratum membership).
The focus is not on the causal effects for the whole population but on the principal causal effects, which are local causal effects for the principal strata.
Although we cannot observe the principal stratum membership for any patient, principal strata exist in the data; we know that each patient belongs to a principal stratum, which can be viewed as an intrinsic latent characteristic of each patient. 
Principal causal effects are sensible and may be of great interest in randomized clinical trials with treatment switching. 
They provide information on the heterogeneity of the treatment effect across principal strata, that is, with respect to the switching behavior, and on the `pure' effect of the treatment for the subpopulation of non-switchers.
In the principal stratification framework, we can also naturally deal with the problem that the switching time under control is not well-defined for patients who would experience disease progression or death under control without switching from the control to the treatment arm.

Another critical difference between a principal stratification analysis and an analysis based on models for semi-competing risks concerns the use of the observed data to draw inferences on the causal effects of interest.

Suppose that the censoring mechanism is ignorable. 
In principal stratification analysis, the observed-data likelihood involves infinite mixtures (see Web-Appendix~\ref{s:bayesian}). 
Models typically used in the classical competing risk literature for analyzing semi-competing data can be divided into two broad classes: models for the distribution of the observable data, which usually target total effects, and models for the distribution of latent failure times, which usually target controlled direct effects \cite[see][for a review]{varadhan2014semicompeting}. 
Models for the distribution of the observable data, which include cause-specific models and sub-distribution functions, only consider the time and type of the first event that occurs to an individual, ignoring the information available after the non-terminal event.
Models for latent failure times attempt to model the joint distribution of the time to the non-terminal event and the time to the terminal event or the marginal distributions of the time to the non-terminal event and the time to the terminal event under the assumption that the time to the non-terminal event without the terminal event is well defined for all subjects.

\subsection*{Principal stratification with semi-competing risks}
Principal stratification analysis has been previously proposed for evaluating causal effects of treatment with semi-competing risks \citep{comment2019survivor, xu2022bayesian}, and there exist strong connections between our study and the existing studies, although distinguishing features make our contribution unique.

\cite{comment2019survivor} and \cite{xu2022bayesian} focus on assessing the causal effects of treatment on non-terminal time-to-event outcomes -- hospital readmission and disease progression, respectively -- accounting for the fact that readmission and disease progression are subject to truncation by death; since patients could die without experiencing hospital readmission/disease progression, assessing the effect on hospital readmission/disease progression requires to take into account that readmission and disease progression are not well defined after death.
In these types of studies, death is not the primary endpoint, but it is a terminal event that precludes the occurrence of the primary non-terminal time-to-event outcome. 

\cite{comment2019survivor} and \cite{xu2022bayesian} propose to handle the problem of truncation by death with principal stratification, defining principal strata by the pair of potential death times, to account for the fact that causal effects on the primary outcome are well defined only for principal strata of patients who would not die regardless of treatment assignment. 
They introduce new survivor causal effects for patients who would survive regardless of treatment assignment that explicitly account for the time-to-event nature of the non-terminal outcome. 
A nice methodological contribution of these papers is that survivor causal effects are defined over time rather than at a single point in time so that they can also investigate how the proportion of patients who would survive regardless of treatment assignment evolves. 
Recently, \cite{nevo2022causal} used the potential outcome approach with principal stratification in time-to-event studies with two semi-competing risks, where the focus is on assessing causal effects on both event times. 
Their key insight is to define principal stratification with respect to the order of the two events under both treatment and control.

In principal stratification analysis, inference is usually conducted by factorizing the joint distribution of all the potential outcomes for the intercurrent outcome and the primary outcome into the product of the marginal distribution of the principal stratum membership (the joint distribution of the potential outcomes for the intercurrent outcome) and the conditional distribution of the potential outcomes for the primary outcome given the principal stratum membership.
We use this approach in our study. 
\cite{xu2022bayesian}  also use this approach, developing a Bayesian non-parametric model under a principal ignorability assumption \citep{jo2009use, DingLu2017, Feller_et_al_2016, mattei2023assessing}. 
\cite{comment2019survivor} propose an alternative factorization of the joint distribution of all the potential outcomes.
This is factorized into the product of the two joint distributions of potential intercurrent outcome and potential main outcome under treatment and under control. 
Inference is then conducted using a Bayesian parametric approach under a conditional independence assumption between potential outcomes under treatment and under control, given covariates and an individual-level latent trait. 
A similar frailty-based approach with parametric assumptions is proposed by \cite{nevo2022causal}.

\section{Connection to the noncompliance literature} \label{s:connections}
Treatment switching is a general form of the noncompliance problem. 
Consider the case where the switching of the patients under the control arm either occurs within a short period or never happens, i.e., $S_i(0) \in  \{\bS\} \cup [0, \epsilon]$, with $\epsilon>0$ being a number smaller than any survival or censoring time. 
Some patients immediately switch to the treatment arm after the treatment assignment. 
In this case, treatment switching is equivalent to the so-called ``all-or-none compliance problem'' \citep{AIR1996, FrangakisRubin1999}. 
Non-switchers, i.e., those units such that $S_i(0) = \bS$, and switchers, for whom $S_i(0) \in  [0, \epsilon]$, correspond to compliers and an always-takers, respectively.
Therefore, $\DCE(y \mid \bS)$ is the distributional effect for non-switchers or compliers, and $\DCE(y \mid [0, \epsilon])$ is the distributional effect for switchers or always-takers.

Since $\epsilon$ is small, it is reasonable to assume that the treatment assignment affects only the outcomes of compliers but not those of always-takers. 
This is the exclusion restriction assumption \citep{AIR1996}, meaning $\DCE(y \mid [0, \epsilon])=0$ for all $y$. 
Therefore, the compliers' distributional effect can be identified by
$$
\DCE(y \mid \bS)= \dfrac{P\left\{\Yobs_i>y \mid Z_i=1\right\}-P\left\{\Yobs_i>y \mid Z_i=0\right\}}{P\left\{\Sobs_i=\bS \mid Z_i=0\right\}},
$$
i.e., the ratio of the distributional effect on the outcome divided by
the proportion of non-switchers.

\subsection*{Connection to partial noncompliance and dose-response relationship}
The switching status is a semi-continuous post-treatment variable, with a binary component that classifies patients into non-switchers and switchers, and a non-negative continuous component that classifies switchers according to their switching time. 
In this subsection, we focus on the switchers, a coarsened principal stratum defined by the union of uncountable sets.

Recently, assessing principal causal effects in the presence of continuous intermediate variables and infinitely many principal strata have received increasing attention \citep{JinRubin2008, BartolucciGrilli2011, MaRoyMarcus2011, Schwartz11, ZiglerBelin:2012, Kim_et_al_2017}.
Interest may lie either in principal causal effects for specific unions of principal strata (such as the average and distributional principal causal effects in~(6)--(8) in the main text) or in entire dose-response functions or surfaces describing how the causal effect on the outcome varies as a function of the basic principal strata membership. 
In our setting, the dose is the time to switching for switchers.
In the main text, the average principal causal effect in (3) defines a dose-response function, and the distributional causal effects in (4) and (5) define dose-response surfaces. 
They describe how causal effects on survival time vary as functions of the dose.
They are similar to the ``causal effect predictiveness surfaces'' in the literature on surrogate endpoints \cite[e.g.,][]{GilbertHudgens2008, ZiglerBelin:2012}.

Our setting is related to randomized experiments with partial compliance \citep{JinRubin2008, MaRoyMarcus2011}. 
In particular, a monotonicity assumption holds by design because no patient in the treatment group can switch to control. 
The switching status, $S_i(0)$, can be viewed as the level (time) of control received by patient $i$ if assigned to control, and (3), (4), and (5) in the main text are causal effects on survival time for patients who would comply with the assignment to the control arm for a specific amount of time, $s$, had they been assigned to the control arm. 
Similarly, the coarsened principal causal effects in (6)--(8) in the main text can be interpreted as causal effects in specific compliance regions \citep{MaRoyMarcus2011}. 
The principal causal effects are generally not identifiable with continuous intermediate variables.
Flexible parametric \cite[e.g.,][]{JinRubin2008, JinRubin2009, MaRoyMarcus2011, ZiglerBelin:2012} and semi-parametric models \cite[e.g.,][]{Schwartz11, BartolucciGrilli2011, Kim_et_al_2017}, possibly coupled with structural assumptions, have been developed to face the identification and estimation issues.

\section{Coarsened principal causal effects}\label{s:cpce}
We define coarsened principal causal effects compared to Equations (6), (7), and (8) in the main text.

The simplest example is ${\cal{A}}=\R_+$ and the causal effects for all switchers are
\begin{eqnarray*}
	\ACE( \R_+ ) &=&  \mathbb{E}\left[Y_i(1)  \mid S_i(0) \in  \R_+\right]-\mathbb{E}\left[ Y_i(0)  \mid S_i(0) \in  \R_+ \right],\\
	\DCE(y   \mid \R_+ ) &=& P\left\{Y_i(1) > y \mid  S_i(0) \in \R_+ \right\}-P\left\{Y_i(0) > y  \mid S_i(0) \in \R_+ \right\},\\
	\cDCE(y   \mid\R_+ ) &=& P\left\{Y_i(1) > y \mid Y_i(1)> S_i(0), S_i(0) \in \R_+ \right\}\\&&- P\left\{Y_i(0) > y  \mid Y_i(1)> S_i(0), S_i(0) \in \R_+ \right\}.
\end{eqnarray*}
If ${\cal{A}}= [0, s]$, then the causal effects for units that switch earlier than or at time $s$ are
\begin{eqnarray*}
	\ACE( [0,s]  ) &=& \mathbb{E}\left[Y_i(1)  \mid S_i(0) \leq s \right]-\mathbb{E}\left[ Y_i(0)  \mid S_i(0) \leq s \right], \label{eq:ace-}\\
	\DCE(y   \mid [0,s] )&=& P\left\{Y_i(1) > y \mid  S_i(0) \leq s \right\}-P\left\{Y_i(0) > y  \mid S_i(0) \leq s \right\}, \label{eq:dce-}\\
	\cDCE(y   \mid [0,s]) &=& P\left\{Y_i(1) > y \mid Y_i(1)> S_i(0), S_i(0) \leq s\right\} \label{eq:cdce-}\\ && - P\left\{Y_i(0) > y  \mid Y_i(1)> S_i(0), S_i(0) \leq s \right\}. \nonumber
\end{eqnarray*}
If ${\cal{A}}= (s, +\infty)$, then the causal effects for units that switch later than time $s$ are
\begin{eqnarray*}
	\ACE( (s, +\infty)  ) &=&  \mathbb{E}\left[Y_i(1)  \mid S_i(0) > s  \right]-\mathbb{E}\left[ Y_i(0)  \mid S_i(0) > s \right], \label{eq:ace+}\\
	\DCE(y   \mid (s, +\infty) ) &=& P\left\{Y_i(1) > y \mid  S_i(0) > s \right\}-P\left\{Y_i(0) > y  \mid S_i(0) > s \right\}, \label{eq:dce+}\\
	\cDCE(y   \mid (s, +\infty)) &=& P\left\{Y_i(1) > y \mid Y_i(1)> S_i(0), S_i(0) > s\right\}\label{eq:cdce+} \\ && -P\left\{Y_i(0) > y  \mid Y_i(1)> S_i(0),  S_i(0) > s \right\}. \nonumber
\end{eqnarray*}

\section{Bayesian Inference} \label{s:bayesian}
Let  $\bZ$, $\bC$, $\bbS(0)$, $\bY(0)$, and $\bY(1)$  be $n$-vectors with $i$th elements equal to $Z_i$,  $C_i$, $S_i(0)$, $Y_i(0)$, and $Y_i(1)$, respectively.
Let $\bX$ be a $n \times K$ matrix with $i$-th row equal to $X_i$.
Under  Assumption 1,  the joint probability (density) function of these random variables is
$$
P\left\{ \bZ, \bC, \bbS(0),  \bY(0),  \bY(1), \bX\right\}
= P\left\{ \bC, \bbS(0),  \bY(0),  \bY(1), \bX \right\} P\left\{\bZ\right\}.
$$
This allows us to ignore the model of $P\left\{\bZ\right\}$.

We assume that $P\left\{\bC, \bbS(0), \bY(0), \bY(1), \bX  \right\}$ is unit-exchangeable. 
By appealing to de Finetti's theorem \citep{de1937prevision}, there exists an unknown parameter vector $\btheta$ with prior distribution $P(\btheta)$ such that
\begin{eqnarray*}
	\lefteqn{
		P\left\{\bC, \bbS(0),  \bY(0),  \bY(1),\bX \right\} }\\
	&=&  \int \prod_{i=1}^{n} P\left\{ C_i,
	S_i(0), Y_i(0), Y_i(1), X_i\mid  \btheta\right\}
	P(\btheta) \mathrm{d}\, \btheta\\
	&=& \int \prod_{i=1}^{n}  P\left\{X_i \mid \btheta \right\}  P\left\{S_i(0) \mid X_i; \btheta\right\}
	P\left\{Y_i(0) \mid S_i(0),   X_i; \btheta\right\}
	\\
	&&~~~~~~~~~~~P\left\{Y_i(1)\mid Y_i(0), S_i(0),  X_i; \btheta\right\}  P\left\{ C_i \mid S_i(0) ,  Y_i(0),  Y_i(1),  X_i; \btheta\right\}
	P(\btheta) \mathrm{d}\, \btheta.
\end{eqnarray*}

We condition on the observed distribution of covariates and assume that the parameters of the
distribution of covariates are a priori independent of the other parameters.
Then we do not need to model $P\left\{X_i \mid \btheta \right\}$. 
Under Assumption~2, 
$$P\left\{ C_i \mid S_i(0) ,  Y_i(0),  Y_i(1),  X_i; \btheta\right\} = P\left\{C_i \mid  \btheta\right\}
$$
Assuming that the parameters of the censoring mechanism are a priori independent of the other parameters, we can then ignore the model of $P\left\{ C_i \mid S_i(0), Y_i(0), Y_i(1), X_i; \btheta\right\} $.
Therefore, Bayesian inference for principal stratification involves two sets of models: one for the principal strata defined by the switching status, $S_i(0)$, given the covariates, $X_i$, and the other for the distribution of potential survival times $Y_i(0)$ and $Y_i(1)$ conditional on the switching status and covariates, $X_i$.

First, we postulate a two-part model for $S_i(0)$. 
Let $\pi(x_i) = P\left\{ S_i(0)=\bS \mid X_i=x_i; \btheta\right\} $ be the probability of being a non-switcher, and let $f_{S(0)}\left(\cdot \mid x_i \right) = f_{S(0)}(\cdot \mid S_i(0) \in \R_+, X_i=x_i; \btheta)$ and $G_{S(0)}(\cdot \mid x_i) = P\left\{S_i(0)> \cdot \mid S_i(0) \in \R_+, X_i=x_i; \btheta\right\}$ denote the probability density function and the survival function of the switching time for switchers (that is, given that $S_i(0)$ does not take on value $\bS$).
Since we focus on time-to-event variables, it is helpful to introduce the notation for hazard functions.
Let $h_{S(0)}\left(\cdot \mid x_i\right) = h_{S(0)}(\cdot \mid S_i(0) \in \R_+, X_i=x_i; \btheta)$ be the hazard function of  $S_i(0)$ for switchers, which satisfies $f_{S(0)}\left(\cdot \mid x_i \right)= h_{S(0)}\left(\cdot \mid x_i\right) \times G_{S(0)}(\cdot \mid x_i)$.
Second, we specify a model for the joint conditional distribution of $Y_i(0)$ and $Y_i(1)$ given $S_i(0)$ and $X_i$ by factorizing it as the product of the conditional distribution of $Y_i(0)$  given $S_i(0)$ and $X_i$, and the conditional distribution of  $Y_i(1)$  given $Y_i(0)$, $S_i(0)$ and $X_i$.
Table~\ref{tab:pdfsf} shows the notation for the probability density functions, hazard functions, and the survival functions of the potential survival times $Y_i(0)$ and $Y_i(1)$.

\begin{table}	
\caption{Probability density functions, hazard functions, and survival functions of the potential survival times conditional on the switching status under non-informative type one censoring\label{tab:pdfsf}.}
	$$
	\begin{array}{lccl}
		\hline
		\vspace{-0.2cm}\\
		\multicolumn{4}{l}{\hbox{Variable}} \\
		& \multicolumn{3}{c}{\hbox{Probability density function, hazard function and survival function}}\\
		\hline
		\vspace{-0.2cm}\\
		\multicolumn{4}{l}{S_i(0) \mid S_i(0) \in \R_+, X_i=x_i}\\
		\vspace{-0.2cm}\\
		& f_{S(0)}\left(\cdot \mid x_i\right) &=& f_{S(0)}(\cdot \mid S_i(0) \in \R_+, X_i=x_i; \btheta)\\
		\vspace{-0.3cm}\\
		& h_{S(0)}\left(\cdot \mid x_i\right) &=& h_{S(0)}(\cdot \mid S_i(0) \in \R_+, X_i=x_i; \btheta)\\
		\vspace{-0.3cm}\\
		& G_{S(0)}(\cdot \mid x_i) &=& P\left\{S_i(0)> \cdot \mid S_i(0) \in \R_+, X_i=x_i; \btheta\right\}\\
		\\
		\multicolumn{4}{l}{Y_i(0) \mid S_i(0) =\bS, X_i=x_i} \\
		\vspace{-0.2cm}\\
		&f_{Y(0)}^{\bS}\left(\cdot \mid x_i\right) &=& f_{Y(0)}(\cdot \mid S_i(0) =\bS, X_i=x_i; \btheta)\\
		\vspace{-0.3cm}\\
		&h_{Y(0)}^{\bS}\left(\cdot \mid x_i \right) &=& h_{Y(0)}(\cdot \mid S_i(0) =\bS, X_i=x_i; \btheta) \\
		&G^{\bS}_{Y(0)}(\cdot\mid x_i)&=& P\left\{Y_i(0)> \cdot \mid S_i(0) =\bS, X_i=x_i; \btheta\right\}\\
		\\
		\multicolumn{4}{l}{Y_i(0) \mid S_i(0) = s, X_i=x_i \quad (s\in \R_+)}\\
		\vspace{-0.2cm}\\
		& f_{Y(0)}\left(\cdot \mid s, x_i\right) &=& f_{Y(0)}(\cdot \mid S_i(0) =s, X_i=x_i; \btheta)\\
		\vspace{-0.3cm}\\
		& h_{Y(0)}\left(\cdot \mid s, x_i\right) &=& h_{Y(0)}(\cdot \mid S_i(0) =s, X_i=x_i; \btheta)\\
		\vspace{-0.3cm}\\
		& G_{Y(0)}(\cdot \mid s, x_i)&=& P\left\{Y_i(0)> \cdot \mid S_i(0) = s, X_i=x_i; \btheta\right\}\\
		\\
		\multicolumn{4}{l}{Y_i(1) \mid S_i(0) =\bS, Y_i(0)=y_0, X_i=x_i  }\\
		\vspace{-0.2cm}\\
		& f_{Y(1)}^{\bS}\left(\cdot\mid y_0, x_i\right) &=& f_{Y(1)}(\cdot \mid S_i(0) =\bS, Y_i(0)=y_0, X_i=x_i; \btheta)\\
		\vspace{-0.3cm}\\
		& h_{Y(1)}^{\bS}\left(\cdot\mid y_0, x_i \right) &=& h_{Y(1)}(\cdot \mid S_i(0) =\bS, Y_i(0)=y_0, X_i=x_i; \btheta)\\
		\vspace{-0.3cm}\\
		& G^{\bS}_{Y(1)}(\cdot\mid y_0, x_i)&=& P\left\{Y_i(1)> \cdot \mid S_i(0) =\bS, Y_i(0)=y_0, X_i=x_i; \btheta\right\}\\
		\\
		\multicolumn{4}{l}{Y_i(1) \mid S_i(0) = s, Y_i(0)=y_0, X_i=x_i \quad (s\in \R_+)}\\
		\vspace{-0.2cm}\\
		& f_{Y(1)}\left(\cdot \mid s, y_0, x_i\right) &=& f_{Y(1)}(\cdot \mid S_i(0) =s, Y_i(0)=y_0, X_i=x_i; \btheta)\\
		\vspace{-0.3cm}\\
		& h_{Y(1)}\left(\cdot \mid s, y_0, x_i\right) &=& h_{Y(1)}(\cdot \mid S_i(0) =s, Y_i(0)=y_0, X_i=x_i; \btheta)\\
		\vspace{-0.3cm}\\
		& G_{Y(1)}(\cdot \mid s, y_0, x_i)&=& P\left\{Y_i(1)> \cdot \mid S_i(0) = s, Y_i(0)=y_0, X_i=x_i; \btheta\right\}\\
		\vspace{-0.3cm}\\
		\hline
	\end{array}
	$$
\end{table}

Let $\bD^{\mathrm{obs}}= \left[ \bZ, \mathbf{C}, \tilde{\mathbf{S}}^{\mathrm{obs}},  \mathbb{I}\{\mathbf{S}^{\mathrm{obs}} \leq \mathbf{C}\},
\tilde{\bY}^{\mathrm{obs}}, \mathbb{I}\{\mathbf{Y}^{\mathrm{obs}} \leq \mathbf{C}\} \right]$   be an
$n \times 6$ matrix, with $i$th row equal to $D_i^{\mathrm{obs}} =[Z_i$, $\Cobs_i$  $\tilde{S}^{\mathrm{obs}}_i$,
$\mathbb{I}\{S^{\mathrm{obs}}_i \leq \Cobs_i\}$,    $\tilde{Y}^{\mathrm{obs}}_i$,
$\mathbb{I}\{Y^{\mathrm{obs}}_i \leq \Cobs_i\}] $. The complete-data contain the observed data $\bX$ and $\bD^{\mathrm{obs}}$, as well as the vector of switching statuses $\bbS^\ast(0)$ with the $i$th element $S_i^\ast(0)= (1-Z_i)[\tilde{S}^{\mathrm{obs}}_i \, \mathbb{I}\{S_i(0) \in \R_+\} + \bS\, \mathbb{I}\{S_i(0) = \bS\}] + Z_i S_i(0)$ and the vector of survival times under control  $\bY^\ast(0)$ with the $i$th element $Y_i^\ast(0)= (1-Z_i)\tilde{Y}^{\mathrm{obs}}_i + Z_i Y_i(0)$. 
We can then write the observed data likelihood 
function in terms of the observed data as:

{\small
	\begin{eqnarray*}\label{eq:obs}
		\lefteqn{ {\mathscr{L}}\left\{\btheta \mid  \bX, \bD^{\mathrm{obs}}\right\}  = \prod_{i: Z_i=0, 
				\mathbb{I}\{S_i^{\mathrm{obs}} \leq \Cobs_i\}=0, \mathbb{I}\{\Yobs_i\leq\Cobs_i\}=1    } \pi(X_i) \,
			f_{Y(0)}^{\bS} \left(\Yobs_i\mid X_i \right)   
		}\\&& \nonumber
		\hspace{-0.6cm} \times\prod_{i: Z_i=0, 	\mathbb{I}\{S_i^{\mathrm{obs}} \leq \Cobs_i\}=1, \mathbb{I}\{\Yobs_i\leq\Cobs_i\}=0} 
		\left[1-\pi(X_i)\right] \, 
		f_{S(0)}\left(\Sobs_i \mid X_i\right) \cdot G_{Y(0)}\left(\Cobs_i \mid \Sobs_i, X_i\right)  \\&& \nonumber
		\hspace{-0.6cm} \times \prod_{i: Z_i=0, 	\mathbb{I}\{S_i^{\mathrm{obs}} \leq \Cobs_i\}=1, \mathbb{I}\{\Yobs_i\leq\Cobs_i\}=1} 
		\left[1-\pi(X_i)\right] \, 
		f_{S(0)}\left(\Sobs_i \mid X_i\right)  f_{Y(0)}\left(\Yobs_i \mid \Sobs_i,X_i\right)  
		\\&& \nonumber
		\hspace{-0.6cm} \times \prod_{i: Z_i=0, 	\mathbb{I}\{S_i^{\mathrm{obs}} \leq \Cobs_i\}=0, \mathbb{I}\{\Yobs_i\leq\Cobs_i\}=0} 
		\pi(X_i) \, G_{Y(0)}^{\bS}\left(\Cobs_i \mid  X_i\right) 
		+ \left[1-\pi(X_i)\right] \, 
		G_{S(0)}\left(\Cobs_i \mid  X_i\right) \cdot 1  \\&& \nonumber
		\hspace{-0.6cm} \times \quad\, \; \prod_{i: Z_i=1, \mathbb{I}\{\Yobs_i\leq\Cobs_i\}=1} \left[ \pi(X_i) \,\int_{\R_+} f_{Y(1)}^{\bS}\left(\Yobs_i \mid Y_i(0)=y_0,  X_i\right)
		f_{Y(0)}^{\bS}\left(y_0\mid  X_i\right) \mathrm{d} y_0 + \right.
		\\&& \nonumber
		\left. \left[1-\pi(X_i)\right] \int_{\R_+}\int_{s}^{+\infty} f_{Y(1)}\left(\Yobs_i \mid  S_i(0)=s, Y_i(0) =y_0, X_i\right) 
		f_{Y(0)}\left(y_0 \mid S_i(0)=s,X_i\right)
		f_{S(0)}\left(s\mid X_i\right)  \mathrm{d}y_0 \, \mathrm{d}s \right]
		\\&& \nonumber
		\hspace{-0.6cm} \times \quad\, \; \prod_{i: Z_i=1,  \mathbb{I}\{\Yobs_i\leq\Cobs_i\}=0}  \left[\pi(X_i) \,\int_{\R_+} G_{Y(1)}^{\bS}\left(\Cobs_i \mid Y_i(0)=y_0, X_i\right) f_{Y(0)}^{\bS}\left(y_0\mid  X_i\right) \mathrm{d} y_0 +\right.
		\\&& \nonumber
		\left.  \left[1-\pi(X_i)\right]   \int_{\R_+}\int_{s}^{+\infty} G_{Y(1)}\left(\Cobs_i \mid  S_i(0)=s, Y_i(0)=y_0, X_i\right) 
		f_{Y(0)}\left(y_0 \mid S_i(0)=s, X_i\right)
		f_{S(0)}\left(s \mid X_i\right)  \mathrm{d}y_0 \, \mathrm{d}s  \right]
	\end{eqnarray*}
}
The posterior distribution of $\btheta$ based on the complete data is
\begin{eqnarray*}\label{eq:complete}
	\lefteqn{P\left\{\btheta \mid \bX, \bD^{\mathrm{obs}}, \bbS^\ast(0), \bY^\ast(0)\right\}  \propto P\left( \btheta \right) }\\&& \nonumber
	\hspace{-0.5cm} \times \; \prod_{i: Z_i=0, S_i^\ast(0)=\bS} \pi(X_i)\,
	f_{Y(0)}^{\bS} \left(\Yobs_i \mid X_i\right)^{\mathbb{I}\{\Yobs_i \leq \Cobs_i\}} 
	G_{Y(0)}^{\bS}\left(\Cobs_i\mid X_i\right)^{\mathbb{I}\{\Yobs_i > \Cobs_i\}} \\&& \nonumber
	\hspace{-0.5cm} \times\prod_{i: Z_i=0, S^\ast_i(0) \in \R_+} 
	\left[1-\pi(X_i)\right] \, G_{S(0)}\left(\Cobs_i\mid X_i\right)^{\mathbb{I}\{\Sobs_i > \Cobs_i\}}\\&&\nonumber
	\quad  
	\left[f_{S(0)}\left(\Sobs_i\mid X_i\right)  f_{Y(0)}\left(\Yobs_i \mid \Sobs_i, X_i\right)^{\mathbb{I}\{\Yobs_i \leq \Cobs_i\}}
	G_{Y(0)}\left(\Cobs_i \mid \Sobs_i, X_i\right)^{\mathbb{I}\{\Yobs_i > \Cobs_i\}} \right]^{\mathbb{I}\{\Sobs_i \leq \Cobs_i\}}  \\&& \nonumber
	\hspace{-0.5cm} \times \prod_{i: Z_i=1, S^\ast_i(0)=\bS}  \pi(X_i) \,f_{Y(0)}^{\bS}\left(Y_i^\ast(0)\mid  X_i\right)\,
	f_{Y(1)}^{\bS}  \left(\Yobs_i \mid Y^\ast_i(0), X_i\right)^{\mathbb{I}\{\Yobs_i \leq \Cobs_i\}} 
	G_{Y(1)}^{\bS}\left(\Cobs_i \mid Y^\ast_i(0), X_i \right)^{\mathbb{I}\{\Yobs_i > \Cobs_i\}} \\&& \nonumber
	\hspace{-0.5cm}\times \!\! \prod_{i: Z_i=1, S^\ast_i(0) \in \R_+}  \left[1-\pi(X_i)\right] \, 
	f_{S(0)}\left(S^\ast_i(0) \mid X_i\right) f_{Y(0)}\left(Y^\ast_i(0) \mid S^\ast_i(0), X_i\right) \\&& \nonumber 
	\hspace{2cm}f_{Y(1)}\left(\Yobs_i \mid S^\ast_i(0), Y^\ast_i(0),X_i\right)^{\mathbb{I}\{\Yobs_i \leq \Cobs_i\}}
	G_{Y(1)}\left(\Cobs_i \mid S^\ast_i(0), Y^\ast_i(0), X_i\right)^{\mathbb{I}\{\Yobs_i > \Cobs_i\}}.
\end{eqnarray*}

\section{Parametric assumptions} \label{s:parametric_ass}
\subsection*{Weibull distribution} 
A Weibull random variable $T$ with parameters $(\alpha, \eta)$ has pdf
$$
f_T(t)=\begin{cases}
	\alpha \eta t^{\alpha-1}  \exp\{- \eta  t^{\alpha}\} & \hbox{for }  t>0, \alpha>0, \eta>0 , \\
	0 & \hbox{otherwise} . 
\end{cases}
$$

The survivor function, the hazard function, and the cumulative hazard function of $T$ are
$$
G_T(t) =  \exp\{- \eta  t^{\alpha}\} ,  \qquad    \qquad
h_T(t) =\alpha \eta t^{\alpha-1} ,  \qquad    \qquad
H_T(t) = \int_{0}^{t} h(u)\, du = \eta  t^{\alpha} . 
$$
Under the parameterization $\beta = \log(\eta)$, we have 
\begin{eqnarray*}
	&	f_T(t) = \alpha   t^{\alpha-1}  \exp\{\beta - e^{\beta}  t^{\alpha}\} , & \\
	&	G_T(t) = \exp\{- e^{\beta}  t^{\alpha}\} ,  \qquad \qquad
	h_T(t; \alpha, \beta) = \alpha t^{\alpha-1} e^{\beta} \qquad \qquad H_T(t)= e^{\beta}  t^{\alpha} . &
\end{eqnarray*}

\subsection*{Sub-model for $\mathbb{I}\{S_i(0)= \bS\}$}

$\mathbb{I}\{S_i(0)= \bS\} \sim \mathrm{Bernoulli}(\pi(x_i))$ with
$$
\pi(x_i) = \dfrac{\exp(\eta_0 + x'_i\eeta)}{1+\exp(\eta_0 + x'_i\eeta)},  \quad
(\eta_0,\eeta) \in \R^{K+1} .
$$

\subsection*{Sub-model for $S_i(0)$ for switchers}
$S_i(0) \mid S_i(0) \in \R_+, X_i \sim \mathrm{Weibull}\left(\alpha_S,\beta_S+  X_i'\eeta_S\right)$, $\alpha_S>0, \beta_S \in \R$, $  \eeta_S \in \R^{K}$:
$$
\begin{array}{c}
	f_{S(0)}(s\mid x_i) = \alpha_S s^{\alpha_S-1}  \exp\{[\beta_S   + x_i'\eeta_S] - e^{\beta_S   + x'_i\eeta_S}  s^{\alpha_S}\},  \\
	h_{S(0)}(s\mid x_i) = \alpha_S s^{\alpha_S-1} \exp\{\beta_S   + x'_i\eeta_S\}, \qquad   G_{S(0)}(s\mid x_i) = \exp\{- e^{\beta_S    + x'_i\eeta_S}  s^{\alpha_S}\} .
\end{array}
$$

\subsection*{Sub-model for $Y_i(0)$ for non-switchers}
$Y_i(0) \mid S_i(0)=\bS, X_i \sim  \mathrm{Weibull}\left(\bar{\alpha}_Y,\bar{\beta}_Y+  X_i'\bar{\eeta}_Y\right)$, $\bar{\alpha}_Y >0$,  $\bar{\beta}_Y \in \R$ and $ \bar{\eeta}_Y \in \R^{K}$:
$$
\begin{array}{c}
	f_{Y(0)}^{\bS}(y\mid x_i) = \bar{\alpha}_Y y^{\bar{\alpha}_Y-1}  \exp\{[\bar{\beta}_Y   + x'_i\bar{\eeta}_Y ]- e^{\bar{\beta}_Y  + x_i'\bar{\eeta}_Y} y^{\bar{\alpha}_Y}\},
	\\
	h_{Y(0)}^{\bS}(y\mid x_i) = \bar{\alpha}_Y y^{\bar{\alpha}_Y-1} \exp\{\bar{\beta}_Y + x'_i\bar{\eeta}_Y \} ,\qquad  G_{Y(0)}^{\bS}(y\mid x_i) =\exp\{-e^{\bar{\beta}_Y   + x'_i\bar{\eeta}_Y}y^{\bar{\alpha}_Y}\}.
\end{array}
$$

\subsection*{Sub-model for $Y_i(0)$ for switchers}
$Y_i(0) \mid S_i(0)=s, s \in \R_+, X_i \sim    s+\mathrm{Weibull}\left(\alpha_Y,\beta_Y +\lambda_0 \log(s)+ X_i'\eeta_Y \right)$, $\alpha_Y>0$, $\beta_Y, \lambda_0 \in \R$, $\eeta_Y \in \R^K$:
\begin{eqnarray*}
	&   f_{Y(0)}(y\!\mid\! s, x_i) \!=\! \alpha_Y (y-s)^{\alpha_Y-1}   \exp\{[\beta_Y \!+\!\lambda_0 \log(s)+x_i'\eeta_Y]\!-\! e^{\beta_Y+\lambda_0 \log(s)\! +\!x_i'\eeta_Y} (y-s)^{\alpha_Y}\} ,&
	\\
	&
	h_{Y(0)}(y\mid s, x_i) = \alpha_Y (y-s)^{\alpha_Y-1}   e^{\beta_Y+\lambda_0 \log(s)+x_i'\eeta_Y}, &\\&
	G_{Y(0)}(y\mid s, x_i) = \exp\{-  e^{\beta_Y+\lambda_0 \log(s)+x_i'\eeta_Y} (y-s)^{\alpha_Y}\}.&
\end{eqnarray*}

\subsection*{Sub-model for $Y_i(1)$ for non-switchers}
$Y_i(1) \mid S_i(0)=\bS, Y_i(0), X_i \sim \kappa Y_i(0)+\mathrm{Weibull}\left(\bar{\nu}_Y,\bar{\gamma}_Y + X_i'\bar{\bzeta}\right)$,  $\kappa \in [0,1]$, $\bar{\nu}_Y>0$, $\bar{\gamma}_Y \in \R$, $\bar{\bzeta} \in \R^K$:
\begin{eqnarray*}
	&   f_{Y(1)}^{\bS}(y\mid y_0,x_i) = \bar{\nu}_Y (y-\kappa y_0)^{\bar{\nu}_Y-1}  \exp\{[\bar{\gamma}_Y+ x_i'\bar{\bzeta}]- e^{\bar{\gamma}_Y+ x_i'\bar{\bzeta}}  (y-\kappa y_0)^{\bar{\nu}_Y}\} , &\\
	&       h_{Y(1)}^{\bS}(y \mid y_0, x_i) = \bar{\nu}_Y (y-\kappa y_0)^{\bar{\nu}_Y-1} e^{\bar{\gamma}_Y + x_i'\bar{\bzeta}}, &\\&
	G_{Y(1)}^{\bS}(y\mid y_0, x_i) = \exp\{- e^{\bar{\gamma}_Y+ x_i'\bar{\bzeta}}  (y-\kappa y_0)^{\bar{\nu}_Y}\}.&
\end{eqnarray*}

\subsection*{Sub-model for $Y_i(1)$ for switchers}
$Y_i(1) \mid  S_i(0)=s, s \in \R_+, Y_i(0), X_i \sim \kappa Y_i(0)+\mathrm{Weibull}\left(\nu_Y,\gamma_Y +\lambda \log(s) + X_i' \bzeta \right)$,  $\kappa \in [0,1]$,  $\nu_Y>0$, $\gamma_Y, \lambda_1 \in \R$, $\bzeta \in \R^K$:
\begin{eqnarray*}
	&   f_{Y(1)}(y\!\mid\! s,  y_0, x_i)\! =\!\nu_Y (y\!-\!\kappa y_0)^{\nu_Y-1}  \exp\{[\gamma_Y\!+\!\lambda_1 \log(s) \!+\! x_i'\bzeta]\!- \! e^{\gamma_Y+\lambda_1 \log(s)+ x_i'\bzeta} (y-\kappa y_0)^{\nu_Y}\} , &\\
	&
	h_{Y(1)}(y\mid s, y_0, x_i) = \nu_Y (y-\kappa y_0)^{\nu_Y-1}  e^{\gamma_Y+\lambda_1 \log(s)+x_i'\bzeta}, &\\&
	G_{Y(1)}(y\mid s, y_0, x_i) = \exp\{-  e^{\gamma_Y+\lambda_1 \log(s)+x_i'\bzeta} (y-\kappa y_0)^{\nu_Y}\}  . &
\end{eqnarray*}

\section{Prior distributions} \label{s:priors}
Under the model specification introduced in the previous Section, we propose to use Normal prior distributions for the parameters of the logistic regression model for the mixing probability $\pi(X_i)$: $(\eta_0, \eeta) \sim  \hbox{N}(\bmu_{\eta}, \sigma^2_{\eta} \, I_{K+1})$, where $I_{r}$ is the $r\times r$ identity matrix.
We use Gamma prior distributions for the shape parameters of the Weibull distributions:
$ \alpha_S \sim \hbox{Gamma}(a_S, b_S)$, $\bar{\alpha}_Y \sim \hbox{Gamma}(\bar{a}_Y, \bar{b}_Y)$, $\alpha_Y \sim \hbox{Gamma}(a_Y, b_Y)$, $\bar{\nu}_Y \sim \hbox{Gamma}(\bar{d}_Y, \bar{s}_Y)$, and $\nu_Y \sim \hbox{Gamma}(d_Y, s_Y)$.
Finally, we use Normal prior distributions for the other parameters of the Weibull distributions:
$\beta_S \sim \hbox{N}(\mu_{\beta_S}, \sigma^2_{\beta_S})$, $\eeta_S \sim \hbox{N}(\bmu_{\eta_S}, \sigma^2_{\eta_S} \, I_{K})$;
$\bar{\beta}_Y \sim \hbox{N}(\mu_{\bar{\beta}_Y},  \sigma^2_{\bar{\beta}_Y})$,
$\bar{\eeta}_Y \sim \hbox{N}(\bmu_{\bar{\eta}_Y},  \sigma^2_{\bar{\eta}_Y} I_{K})$;
$\beta_Y \sim \hbox{N}(\mu_{\beta_Y},  \sigma^2_{\beta_Y})$,
$\eeta_Y \sim \hbox{N}(\bmu_{\eta_Y},  \sigma^2_{\eta_Y} I_{K})$;
$\bar{\gamma}_Y \sim \hbox{N}(\mu_{\bar{\gamma}_Y}, \sigma^2_{\bar{\gamma}_Y})$,
$\bar{\bzeta}_Y \sim \hbox{N}(\mu_{\bar{\zeta}_Y}, \sigma^2_{\bar{\zeta}_Y}I_K)$;
$\gamma_Y \sim \hbox{N}(\mu_{\gamma_Y}, \sigma^2_{\gamma_Y})$,
$\bzeta_Y \sim \hbox{N}(\mu_{\zeta_Y}, \sigma^2_{\zeta_Y}I_K)$;
and $\lambda \sim \hbox{N}(\mu_\lambda, \sigma^2_\lambda)$.

\section{Application: Model and Computational Details} \label{s:app1}
\subsection*{Parametric Assumptions}
\subsubsection*{Sub-model for the Switching Behavior.} 
$\pi  = \bE[\bI\{S_i(0)=\bS\}] =P\left(S_i(0)=\bS\right)$ and 
$S_i(0) \mid S_i(0) \in \R_+ \sim \mathrm{Weibull}\left(\alpha_S,\beta_S\right)$, $\alpha_S>0$, $\beta_S \in \R$:
$$f_{S(0)}(s) = \alpha_S s^{\alpha_S-1}  \exp\{\beta_S - e^{\beta_S}  s^{\alpha_S}\}$$
\begin{eqnarray*}
	h_{S(0)}(s) = \alpha_S s^{\alpha_S-1} e^{\beta_S} &\qquad & G_{S(0)}(s) = \exp\{- e^{\beta_S}  s^{\alpha_S}\}\\
\end{eqnarray*}

\subsubsection*{Sub-model for $Y_i(0)\mid S_i(0)$. \,}
$Y_i(0) \mid   S_i(0) = \bS \sim \mathrm{Weibull}\left(\bar{\alpha}_Y,\bar{\beta}_Y\right)$,
$\bar{\alpha}_Y >0$, $\bar{\beta}_Y \in \R$:
$$
f_{Y(0)}^{\bS}(y) = \bar{\alpha}_Y y^{\bar{\alpha}_Y-1}  \exp\{\bar{\beta}_Y - e^{\bar{\beta}_Y} y^{\bar{\alpha}_Y}\}
$$
\begin{eqnarray*}
	h_{Y(0)}^{\bS}(y)= \bar{\alpha}_Y y^{\bar{\alpha}_Y-1} e^{\bar{\beta}_Y} 
	&\qquad &
	G_{Y(0)}^{\bS}(y) = \exp\{- e^{\bar{\beta}_Y} y^{\bar{\alpha}_Y}\}
\end{eqnarray*}

and  $Y_i(0) \mid  S_i(0) =s, s \in \R_+ \sim s+\mathrm{Weibull}\left(\alpha_Y,\beta_Y +\lambda \log(s) \right)$,
$\alpha_Y>0$, $\beta_Y \in \R$, $\lambda \in \R$:
$$
f_{Y(0)}(y\mid s) =\alpha_Y (y-s)^{\alpha_Y-1}   \exp\{[\beta_Y +\lambda \log(s)]-  e^{\beta_Y+\lambda \log(s)} (y-s)^{\alpha_Y}\}
$$	
\begin{eqnarray*}
	h_{Y(0)}(y|s) = \alpha_Y (y-s)^{\alpha_Y-1}   e^{\beta_Y+\lambda \log(s)}  &\qquad &
	G_{Y(0)}(y|s) = \exp\{-  e^{\beta_Y+\lambda \log(s)} (y-s)^{\alpha_Y} 
\end{eqnarray*}

\subsubsection*{Sub-model for $Y_i(1)\mid Y_i(0),S_i(0)$. \,}
$Y_i(1) \mid  Y_i(0), S_i(0) = \bS \sim \kappa Y_i(0) + \mathrm{Weibull}\left(\bar{\nu}_Y,\bar{\gamma}_Y\right)$ 
$\bar{\nu}_Y >0 $, $\bar{\gamma}_Y \in\R$:
$$
f_{Y(1)}^{\bS}(y\mid y_0) = \bar{\nu}_Y (y-\kappa y_0)^{\bar{\nu}_Y-1}  \exp\{\bar{\gamma}_Y- e^{\bar{\gamma}_Y}  (y-\kappa y_0)^{\bar{\nu}_Y}\}$$
\begin{eqnarray*}
	h_{Y(1)}^{\bS}(y\mid y_0) = \bar{\nu}_Y (y-\kappa y_0)^{\bar{\nu}_Y-1} e^{\bar{\gamma}_Y} 
	&\qquad &	
	G_{Y(1)}^{\bS}(y\mid y_0) = \exp\{- e^{\bar{\gamma}_Y}  (y-\kappa y_0)^{\bar{\nu}_Y}\}
\end{eqnarray*}
and $Y_i(1) \mid   Y_i(0), S_i(0)=s, s \in \R_+ \sim \kappa Y_i(0) +\mathrm{Weibull}\left(\nu_Y,\gamma_Y +\lambda \log(s) \right)$, $\nu_Y>0$, $\gamma_Y \in \R$,  $\lambda \in \R$:
$$
f_{Y(1)}(y\mid s, y_0) = \nu_Y (y-\kappa y_0)^{\nu_Y-1}  \exp\{[\gamma_Y+\lambda \log(s)]-  e^{\gamma_Y+\lambda \log(s)} (y-\kappa y_0)^{\nu_Y}\}
$$
\begin{eqnarray*}
	h_{Y(1)}(y|s, y_0) =\nu_Y (y-\kappa y_0)^{\nu_Y-1}  e^{\gamma_Y+\lambda \log(s)} &\quad&
	G_{Y(1)}(y|s, y_0)= \exp\{-  e^{\gamma_Y+\lambda \log(s)} (y-\kappa y_0)^{\nu_Y}\}\\
\end{eqnarray*}

Therefore, the entire parameter vector is $
\btheta = \big[\pi, \left(\alpha_S, \beta_S\right),
\left(\bar{\alpha}_Y, \bar{\beta}_Y\right), \left(\alpha_Y, \beta_Y\right), 
\left(\bar{\nu}_Y, \bar{\gamma}_Y\right),$ $\left(\nu_Y, \gamma_Y\right), \lambda, \kappa\big].$

\subsection*{Prior distributions}  
Parameters are assumed to be a priori independent, with the following prior distributions. 
We use a conjugate Beta prior distribution for the mixing probability $\pi \sim \hbox{Beta}(a, b)$:
$$
\pi \sim \hbox{Beta}(a, b): p(\pi) = \dfrac{\Gamma(a)\Gamma(b)}{\Gamma(a+b)}\pi^{a-1} (1-\pi)^{b-1}
$$
with $a=b=1$.  Therefore the full conditional distribution of $\pi$ is Beta with parameters $a + \sum_{i=1}^{n}\bI\{S_i(0)=\bS\}$ and  $b +\sum_{i=1}^{n}\bI\{S_i(0)\in \R_+\}$.

We use Gamma priors for the shape parameters of the Weibull distributions, $\alpha_S, \bar{\alpha}_Y, \alpha_Y, \bar{\nu}_Y$ and $\nu_Y $,
$$
\alpha_S \sim \hbox{Gamma}(a_S, b_S): p(\alpha_S) = \dfrac{1}{(b_S)^{a_S} \Gamma(a_S)} \alpha_S^{a_S-1} e^{-\alpha_S/b_S}$$
with $a_S=0.1$ and $b_S=10$,
\begin{eqnarray*}
	&\bar{\alpha}_Y \sim \hbox{Gamma}(\bar{a}_Y, \bar{b}_Y) \qquad 
	\alpha_Y \sim \hbox{Gamma}(a_Y, b_Y)  &\\
	&\bar{\nu}_Y \sim \hbox{Gamma}(\bar{d}_Y, \bar{s}_Y)  
	\qquad \nu_Y \sim \hbox{Gamma}(d_Y, s_Y)&
\end{eqnarray*}
with  $\bar{a}_Y=a_Y=0.1$, $\bar{b}_Y=b_Y =10$,  and    $ \bar{d}_Y= d_Y= 100$, $\bar{s}_Y=s_Y= 0.01$.

We use Normal priors for $\beta_S, \bar{\beta}_Y, \beta_Y, \bar{\gamma}_Y, \gamma_Y$ and $\lambda_Y$:
$$
\beta_S \sim \hbox{N}(\mu_S, \sigma^2_S): p(\beta_S) = \dfrac{1}{\sqrt{2 \pi \sigma^2_S}} \exp\left\{-\dfrac{1}{2\sigma^2_S} (\beta_S  - \mu_S)^2\right\}$$
$$
\bar{\beta}_Y \sim \hbox{N}(\bar{\mu}_Y, \bar{\sigma}^2_Y) \qquad 
\beta_Y \sim \hbox{N}(\mu_Y, \sigma^2_Y) \qquad \bar{\gamma}_Y \sim \hbox{N}(\bar{m}_Y, \bar{\tau}^2_Y)  \qquad  \gamma_Y \sim \hbox{N}(m_Y, \tau^2_Y) 
$$
$$\lambda \sim \hbox{N}(\mu_\lambda, \sigma^2_\lambda)$$
with $\mu_S=\bar{\mu}_Y=\mu_Y=\bar{m}_Y=m_Y=\mu_\lambda=0$ and $\sigma^2_S=\bar{\sigma}^2_Y= \sigma^2_Y= \sigma^2_\lambda=10^4$, and 
$ \bar{\tau}^2_Y=\tau^2_Y = 0.25$.

It is worth noting that we use more informative prior distributions for the parameters $\overline{\nu}_Y$, $\overline{\gamma}_Y$, $\nu_Y$ and $\gamma_Y$ to deal with the difficulty of untying the mixture of switchers and non-switchers under treatment.
Since we never observe the switching behavior for units assigned to the active treatment, there is no unique way to disentangle the mixture of switchers and non-switchers under treatment, and thus we can end up with unrealistic draws for those parameters.
The availability of covariates might, at least partially, address this issue, helping to better disentangle the mixture. 

\subsection*{Complete data posterior distribution}
Let $D^{\mathrm{obs}}_i= \left[ Z_i, C_i, \TS_i,  \mathbb{I}\{S_i^{obs} \leq C_i\},
\TY_i, \mathbb{I}\{\Yobs_i \leq C_i\} \right]$ denote the observed data for unit $i$ and let $\bD^{\mathrm{obs}}= \left[ \bZ, \mathbf{C}, \tilde{\mathbf{S}}^{\mathrm{obs}},  \mathbb{I}\{\mathbf{S}^{\mathrm{obs}} \leq \mathbf{C}\},
\tilde{\bY}^{\mathrm{obs}}, \mathbb{I}\{\mathbf{Y}^{\mathrm{obs}} \leq \mathbf{C}\} \right]$ be the matrix 
stacking observations for all units.
For $\kappa=\kappa_0$, with $\kappa_0 \in (0,1]$, the complete data (w.r.t. the switching status and the survival time under control) posterior distribution for the parameter vector $\btheta = \left[\pi, \left(\alpha_S, \beta_S\right),
\left(\bar{\alpha}_Y, \bar{\beta}_Y\right),\right.$ $\left.\left(\alpha_Y, \beta_Y\right),
\left(\bar{\nu}_Y, \bar{\gamma}_Y\right), \left(\nu_Y, \gamma_Y\right), \lambda, \kappa=\kappa_0\right]$, is

{\small
	\begin{eqnarray*}\label{eq:completekappa0}
		\lefteqn{P\left\{\btheta \mid  \bD^{\mathrm{obs}}, \bbS^\ast(0), \bY^\ast(0)\right\}  \propto }\\&&
		\hspace{-0.5cm}P\{\pi\}  P\{ \alpha_S\} P\{ \beta_S\}
		P\{\bar{\alpha}_Y\} P\{\bar{\beta}_Y\} P\{\alpha_Y\} P\{\beta_Y\}
		P\{\bar{\nu}_Y\} P\{\bar{\gamma}_Y\} P\{\nu_Y\} P\{\gamma_Y\} P\{\lambda\} \delta_{\kappa_0}(\kappa)  \\&&
		\hspace{-0.5cm}
		\times  \!\!\!\!\prod_{i: Z_i=0, S^\ast_i(0)=\bS}  \!\!\! \!\!\! \pi 
		\left[ \bar{\alpha}_Y (\Yobs_i)^{\bar{\alpha}_Y\!-\!1}  \exp\{\bar{\beta}_Y \!-\! e^{\bar{\beta}_Y} (\Yobs_i)^{\bar{\alpha}_Y}\}\right]^{\mathbb{I}\{ \Yobs_i \leq\Cobs_i\}}
		\!\!\exp\{- e^{\bar{\beta}_Y} \Cobs_i^{\bar{\alpha}_Y}\}^{1-\mathbb{I}\{\Yobs_i\leq \Cobs_i\}} \\&&
		\hspace{-0.5cm}
		\times \!\!\!\!\!\!	\prod_{i: Z_i=0, S^\ast_i(0) \in \R_+  }
		\left(1-\pi\right) \, \Bigg\{
		\alpha_S (\Sobs_i)^{\alpha_S-1}  \exp\{\beta_S - e^{\beta_S}  (\Sobs_i)^{\alpha_S}\}  \\&&
		\hspace{-0.2cm}
		\left[\alpha_Y (\Yobs_i\!-\!\Sobs_i)^{\alpha_Y\!-\!1}   \exp\{[\beta_Y \!+\!\lambda \log(\Sobs_i)]-  e^{\beta_Y\!+\!\lambda \log(\Sobs_i)} (\Yobs_i\!-\!\Sobs_i)^{\alpha_Y}\}\right]^{\mathbb{I}\{\Yobs_i\leq\Cobs_i\}} \\&&
		\hspace{-0.2cm}
		\left[\exp\{-  e^{\beta_Y+\lambda \log(\Sobs_i)} (\Cobs_i-\Sobs_i)^{\alpha_Y}\}\right]^{1-{\mathbb{I}\{\Yobs_i\leq\Cobs_i\}}}   \Bigg\}^{\mathbb{I}\{ \Sobs_i\leq\Cobs_i\}}
		\!\! \Bigg\{\! \! \exp\{- e^{\beta_S}  (\Cobs_i)^{\alpha_S}\} \!\!\Bigg \}^{1-\mathbb{I}\{\Sobs_i\leq\Cobs_i\}}
		\\&&
		\hspace{-0.5cm} \times \!\!\!\!\prod_{i: Z_i=1, S^\ast_i(0)=\bS}    \! \!\!\! \pi
		\,	\bar{\alpha}_Y (Y^\ast_i(0))^{\bar{\alpha}_Y\!-\!1}  \exp\{\bar{\beta}_Y \!-\! e^{\bar{\beta}_Y} (Y^\ast_i(0))^{\bar{\alpha}_Y}\}\\ && 
		\left[ \bar{\nu}_Y (\Yobs_i-\kappa Y^\ast_i(0))^{\bar{\nu}_Y\!-\!1}  \exp\{\bar{\gamma}_Y \!-\! e^{\bar{\gamma}_Y} (\Yobs_i-\kappa Y^\ast_i(0))^{\bar{\nu}_Y}\}\right]^{\mathbb{I}\{\Yobs_i\leq\Cobs_i\}}\\&&
		~~~~~~~	\exp\{-e^{\bar{\gamma}_Y} (\Cobs_i-\kappa Y^\ast_i(0))^{\bar{\nu}_Y}\}^{(1-\mathbb{I}\{\Yobs_i\leq\Cobs_i\})\mathbb{I}\{Y^\ast_i(0)\leq \Cobs_i/\kappa\}}
		1^{(1-\mathbb{I}\{\Yobs_i\leq\Cobs_i\})\mathbb{I}\{Y^\ast_i(0)>\Cobs_i/\kappa\}}\\&&
		\hspace{-0.5cm}  \times \!\!\!\!\!\!\prod_{i: Z_i=1, S_i(0) \in \R_+}
		\! \!\!\! \!\!\!\!\left(1-\pi\right) 
		\alpha_S S^\ast_i(0)^{\alpha_S\!-\!1}  \exp\{\beta_S \!-\! e^{\beta_S}  S^\ast_i(0)^{\alpha_S}\}\\&&
		\alpha_Y (Y^\ast_i(0)\!-\! S^\ast_i(0))^{\alpha_Y\!-\!1}   \exp\{[\beta_Y \!+\!\lambda \log(S^\ast_i(0))]-  e^{\beta_Y\!+\!\lambda \log(S^\ast_i(0))} (Y^\ast_i(0)\!-\! S^\ast_i(0))^{\alpha_Y}\}\\&&
		\left[\!\exp\{-e^{\gamma_Y+\lambda \log(S^\ast_i(0))} (\Cobs_i-\kappa Y^\ast_i(0))^{\nu_Y}\}\!\right]^{(1-\mathbb{I}\{\Yobs_i\leq\Cobs_i\})\mathbb{I}\{Y^\ast_i(0)\leq\Cobs_i/\kappa\}	}
		1^{(1-\mathbb{I}\{\Yobs_i\leq\Cobs_i\})\mathbb{I}\{Y^\ast_i(0)>\Cobs_i/\kappa\}}
		\\&&
		~~\left[\nu_Y (\Yobs_i-\kappa Y^\ast_i(0))^{\nu_Y-1}   \exp\{[\gamma_Y +\lambda \log(S^\ast_i(0))]-  e^{\gamma_Y+\lambda \log(S^\ast_i(0))} (\Yobs_i-\kappa Y^\ast_i(0))^{\nu_Y}\}\right]^{\mathbb{I}\{ \Yobs_i\leq \Cobs_i\}} 
	\end{eqnarray*}
}

For $\kappa=0$, the  complete (switching status)  data posterior distribution for the parameter vector $\btheta = \left[\pi, \left(\alpha_S, \beta_S\right),
\left(\bar{\alpha}_Y, \bar{\beta}_Y\right),\right.$ $\left.\left(\alpha_Y, \beta_Y\right),
\left(\bar{\nu}_Y, \bar{\gamma}_Y\right), \left(\nu_Y, \gamma_Y\right), \lambda\right]$,    is

{\small
	\begin{eqnarray*}\label{eq:complete}
		\lefteqn{P\left\{\btheta \mid  \bD^{\mathrm{obs}}, \bbS^\ast(0), \bY^\ast(0)\right\}  \propto }\\&&
		\hspace{-0.5cm}P\{\pi\}  P\{ \alpha_S\} P\{ \beta_S\}
		P\{\bar{\alpha}_Y\} P\{\bar{\beta}_Y\} P\{\alpha_Y\} P\{\beta_Y\}
		P\{\bar{\nu}_Y\} P\{\bar{\gamma}_Y\} P\{\nu_Y\} P\{\gamma_Y\} P\{\lambda\} \delta_{\kappa_0}(\kappa)  \\&&
		\hspace{-0.5cm}
		\times  \!\!\!\!\prod_{i: Z_i=0, S^\ast_i(0)=\bS}  \!\!\! \!\!\! \pi 
		\left[ \bar{\alpha}_Y (\Yobs_i)^{\bar{\alpha}_Y\!-\!1}  \exp\{\bar{\beta}_Y \!-\! e^{\bar{\beta}_Y} (\Yobs_i)^{\bar{\alpha}_Y}\}\right]^{\mathbb{I}\{ \Yobs_i \leq\Cobs_i\}}
		\!\!\exp\{- e^{\bar{\beta}_Y} \Cobs_i^{\bar{\alpha}_Y}\}^{1-\mathbb{I}\{\Yobs_i\leq \Cobs_i\}} \\&&
		\hspace{-0.5cm}
		\times \!\!\!\!\!\!	\prod_{i: Z_i=0, S^\ast_i(0) \in \R_+  }
		\left(1-\pi\right) \, \Bigg\{
		\alpha_S (\Sobs_i)^{\alpha_S-1}  \exp\{\beta_S - e^{\beta_S}  (\Sobs_i)^{\alpha_S}\}  \\&&
		\hspace{-0.2cm}
		\left[\alpha_Y (\Yobs_i\!-\!\Sobs_i)^{\alpha_Y\!-\!1}   \exp\{[\beta_Y \!+\!\lambda \log(\Sobs_i)]-  e^{\beta_Y\!+\!\lambda \log(\Sobs_i)} (\Yobs_i\!-\!\Sobs_i)^{\alpha_Y}\}\right]^{\mathbb{I}\{\Yobs_i\leq\Cobs_i\}} \\&&
		\hspace{-0.2cm}
		\left[\exp\{-  e^{\beta_Y+\lambda \log(\Sobs_i)} (\Cobs_i-\Sobs_i)^{\alpha_Y}\}\right]^{1-{\mathbb{I}\{\Yobs_i\leq\Cobs_i\}}}   \Bigg\}^{\mathbb{I}\{ \Sobs_i\leq\Cobs_i\}}
		\!\! \Bigg\{\! \! \exp\{- e^{\beta_S}  (\Cobs_i)^{\alpha_S}\} \!\!\Bigg \}^{1-\mathbb{I}\{\Sobs_i\leq\Cobs_i\}}
		\\&&
		\hspace{-0.5cm} \times \!\!\!\!\prod_{i: Z_i=1, S^\ast_i(0)=\bS}    \! \!\!\! \pi
		\left[ \bar{\nu}_Y (\Yobs_i)^{\bar{\nu}_Y\!-\!1}  \exp\{\bar{\gamma}_Y \!-\! e^{\bar{\gamma}_Y} (\Yobs_i)^{\bar{\nu}_Y}\}\right]^{\mathbb{I}\{\Yobs_i\leq\Cobs_i\}} 	\exp\{-e^{\bar{\gamma}_Y} (\Cobs_i)^{\bar{\nu}_Y}\}^{(1-\mathbb{I}\{\Yobs_i\leq\Cobs_i\})}
	\\&&
		\hspace{-0.5cm}  \times \!\!\!\!\!\!\prod_{i: Z_i=1, S_i(0) \in \R_+}
		\! \!\!\! \!\!\!\!\left(1-\pi\right) 
		\alpha_S S^\ast_i(0)^{\alpha_S\!-\!1}  \exp\{\beta_S \!-\! e^{\beta_S}  S^\ast_i(0)^{\alpha_S}\}\\&& 
		\left[\!\exp\{-e^{\gamma_Y+\lambda \log(S^\ast_i(0))} (\Cobs_i)^{\nu_Y}\}\!\right]^{(1-\mathbb{I}\{\Yobs_i\leq\Cobs_i\})}
		\\&&
		~~\left[\nu_Y (\Yobs_i)^{\nu_Y-1}   \exp\{[\gamma_Y +\lambda \log(S^\ast_i(0))]-  e^{\gamma_Y+\lambda \log(S^\ast_i(0))} (\Yobs_i)^{\nu_Y}\}\right]^{\mathbb{I}\{ \Yobs_i\leq \Cobs_i\}} 
	\end{eqnarray*}
}


\subsection*{Details of Calculations}
Note that if $\kappa=0$, we only need to impute the missing switching status by drawing from its conditional distribution given $(\mathbf{D}^{\mathrm{obs}}, \btheta)$; we do not need to impute $Y_i(0)$ for treated units. 

The random variables $S^\ast_i(0)$ and $Y^\ast_i(0)$ are independent across units $i=1, \ldots, n$ given $(\mathbf{D}^{\mathrm{obs}}, \btheta)$; therefore, sampling from the distributions of $(\bbS^\ast(0) \mid   \mathbf{D}^{\mathrm{obs}}, \btheta)$ (for $\kappa=0$) and $(\bbS^\ast(0), \bY^\ast(0) \mid \mathbf{D}^{\mathrm{obs}}, \btheta)$ (for $\kappa \in (0,1]$) for data augmentation only involves independent drawing from $(S^\ast_i(0) \mid  D_i^{\mathrm{obs}}, \btheta)$ and $(S^\ast_i(0), Y^\ast_i(0) \mid  D_i^{\mathrm{obs}}, \btheta)$. 

\subsubsection*{Details of Calculations: $\kappa=0$.} 
Let $(\btheta, \bbS^{\ast}(0))$ denote the current state of the chain, with $$\btheta = \big[\pi, \left(\alpha_S, \beta_S\right), \left(\bar{\alpha}_Y, \bar{\beta}_Y\right), \left(\alpha_Y, \beta_Y\right), \left(\bar{\nu}_Y, \bar{\gamma}_Y\right), \left(\nu_Y, \gamma_Y\right), \lambda, \kappa=0\big].$$

\begin{enumerate}
    \item Given the parameter $\btheta$ and observed data, $\bD^{\mathrm{obs}}$, 
    draw the missing data $S^\ast_i(0)$
    \begin{itemize}
	\item[$-$] For control patients, we have
	$$
        S^\ast_i(0)=S_i(0)=\begin{cases}
	\bS & \hbox{if } Z_i=0, \mathbb{I}\{\Sobs_i\leq        C_i\}=0, \mathbb{I}\{\Yobs_i \leq C_i\}=1\\
			\TS_i=\Sobs_i & \hbox{if } Z_i=0,  \mathbb{I}\{\Sobs_i \leq C_i\}=1,  \mathbb{I}\{ \Yobs_i \leq C_i\} \in \{0,1\},\\
		\end{cases}
	$$
	For control patients with $\mathbb{I} \{\Sobs_i\leq C_i\}=0$ and $\mathbb{I}\{\Yobs_i \leq C_i\}=0$, we have
	\begin{align*}
			&\pi_{NS} \equiv\\& P\left(S^\ast_i(0)=\bS \mid  \btheta, Z_i=0, C_i, \tilde{S}^{\mathrm{obs}}_i,\mathbb{I}\{\Sobs_i \leq C_i\}=0,
			\tilde{Y}^{\mathrm{obs}}_i, \mathbb{I}\{\Yobs_i \leq C_i\}=0 \right)=\\&			 	\dfrac{\pi    G^{\bS}_{Y(0)}(C_i)}{\pi  G^{\bS}_{Y(0)}(C_i) + \left(1-\pi\right) G_{S_i(0)}(C_i)    \cdot 1}
	\end{align*}
	Therefore, control patients with $\mathbb{I}\{\Sobs_i\leq C_i\}=0$ and $\mathbb{I}\{\Yobs_i \leq C_i\}=0$ are classified as non-switchers ($S^\ast_i(0)=\bS$) with probability 	$\pi_{NS}$ and as switchers with censored switching time ($S^\ast_i(0)=\tilde{S}_i^{\mathrm{obs}}=C_i$) with probability $1-\pi_{NS}$.
    \item[$-$] For treated patients, we never observe  $S_i(0)$. 
    We use Metropolis-Hasting steps to draw $S_i(0)$ according to $P\left(S_i(0) \mid \btheta,  D_i^{\mathrm{obs}} \right)$. 
    We draw candidate values $S^{\ca}_i(0)$ from a semi-continuous distribution:
    We first draw $n_1$ values from a Bernoulli distribution with probability $\pi$ setting $S^{\ca}_i(0)=\bS$ for treated units for which we obtain a success (a positive value). 
    For treated units for which we obtain a failure, a missing value of $S^{\ca}_i(0)$ is then drawn from the Weibull distribution with parameters $\alpha_S$ and $\beta_S$:  $\mathrm{Weibull}\left(\alpha_S,\beta_S\right)$.
    For each $i$ with $Z_i=1$, we accept $S^{\ca}_i(0)$, setting $S^{\ast}_i(0)=S^{\ca}_i(0)$, with probability $p_i= \min\{p_{S_i(0)}, 1\}$, with
		$$
		p_{S_i(0)}=  
		\begin{cases}
			r_i & \hbox{if } \,   S_i^\ast(0)=\bS,  S_i^{\ca}(0)=\bS\\
			\\
			r_i \cdot\dfrac{\pi}{(1-\pi)  f_{S(0)}(S_i^{\ca}(0))}& \hbox{if } \,    S^\ast_i(0)=\bS,  S_i^{\ca}(0)\in \R_+\\
			\\
			r_i\cdot \dfrac{(1-\pi) f_{S(0)}(S^\ast_i(0))}{\pi}& \hbox{if } \,   S^\ast_i(0)\in \R_+,  S^{\ca}_i(0)=\bS\\
			\\
			r_i \cdot\dfrac{f_{S(0)}(S^\ast_i(0))}{ f_{S(0)}(S^\ca_i(0))}& \hbox{if } \,  S^\ast_i(0)\in \R_+,  S_i^{\ca}(0)\in \R_+\\
		\end{cases}
		$$
    where $ f_{S(0)}(\cdot)$ is the density of the proposal Weibull distribution, $\mathrm{Weibull}\left(\alpha_S,\beta_S\right)$, and 		
	$$
r_i= \dfrac{P\left\{S^{\ca}_i(0)  \mid \btheta, D^{\mathrm{obs}}_i \right\} }{P\left\{S^{\ast}_i(0) \mid \btheta, D^{\mathrm{obs}}_i \right\} }.
$$
	\end{itemize}
	\item Given the imputed complete data, $$\bD= \left[ \bZ, \mathbf{C}, \tilde{\mathbf{S}}^{\mathrm{obs}},  \mathbb{I}\{\mathbf{S}^{\mathrm{obs}} \leq \mathbf{C}\},
	\tilde{\bY}^{\mathrm{obs}}, \mathbb{I}\{\mathbf{Y}^{\mathrm{obs}} \leq \mathbf{C}\},  \bbS^{\ast}(0)\right],$$ we then draw the following sub-vectors of $\btheta$ in sequence, conditional on all others: 
	$\pi$, $\alpha_S$, $\beta_S$, 
	$\bar{\alpha}_Y$, $\bar{\beta}_Y$, $\alpha_Y$, $\beta_Y$,
	$\bar{\nu}_Y$, $\bar{\gamma}_Y$, $\nu_Y$, $\gamma_Y$, $\lambda$.
	We  draw $\pi$ directly from its full conditional distribution, a Beta distribution with parameters $a+\sum_{i=1}^{n} \mathbb{I}\{S_i^\ast(0)=\bS\}$ and 
	$b+\sum_{i=1}^{n} \mathbb{I}\{S_i^\ast(0)\in \R_+\}$.
	We cannot draw directly from the appropriate conditional distributions for the other model parameters, but we use Metropolis--Hasting steps for drawing from their full-conditional distributions.
    For instance, to draw $\alpha_S$, we draw a candidate value $\alpha_S^{\ca}$ from a density $g(\alpha_S \mid \btheta )$. 
    The candidate draw is accepted with probability
    $$
    p_{\alpha_S} = \min \left\{\dfrac{P\left\{[\btheta \setminus \alpha_S],  \alpha_S^{\ca} \mid \bD\right\}}{
		P\left\{[\btheta \setminus \alpha_S], \alpha_S \mid \bD \right\}} \dfrac{g(\alpha_S \mid [\btheta \setminus \alpha_S], \alpha_S^{\ca}   )}{g(\alpha^{\ca}_S \mid [\btheta \setminus \alpha_S], \alpha_S   )}, 1\right\}
    $$
    For the candidate densities, we use Gamma densities for the parameters $\alpha_S$, $\bar{\alpha}_Y$, $\alpha_Y$, $\bar{\nu}_Y$, and $\nu_Y$, and Normal densities for the parameters $\beta_S$, $\bar{\beta}_Y$,   $\beta_Y$, $\bar{\gamma}_Y$, $\gamma_Y$, and $\lambda$, centered at the current values of the parameters. 
    The scaling factors were chosen based on preliminary runs of the chains.
\end{enumerate}

\subsubsection*{Details of Calculations: $\kappa \in (0,1]$.} 
Let $(\btheta, \bbS^{\ast}(0), \bY^{\ast}(0))$ denote the current state of the chain, with $$\btheta = \big[\pi, \left(\alpha_S, \beta_S\right), \left(\bar{\alpha}_Y, \bar{\beta}_Y\right), \left(\alpha_Y, \beta_Y\right), \left(\bar{\nu}_Y, \bar{\gamma}_Y\right), \left(\nu_Y, \gamma_Y\right), \lambda, \kappa=\kappa_0\big] \qquad \kappa_0 \in (0,1].$$
\begin{enumerate}
    \item Given the parameter $\btheta$, observed data, $\bD^{\mathrm{obs}}$, and $\bbS^{\ast}(0)$, draw the missing data $Y^\ast_i(0)$
    \begin{itemize}
        \item[$-$] For control patients, we set $Y^\ast_i(0)=\tilde{Y}_i^{\mathrm{obs}}$
        \item[$-$] For treated patients, we never observe $ Y_i(0)$. We use Metropolis-Hasting steps to draw $Y_i(0)$ according to $P\left(Y_i(0) \mid  \btheta, S_i(0), D_i^{\mathrm{obs}} \right)$. 
        We draw candidate values $Y^{\ca}_i(0)$ from Weibull distributions: 
        $(a)$ For treated patients with $S^\ast_i(0)=\bS$, we draw $Y^{\ca}_i(0)$ from a Weibull distribution with parameters $\left(\bar{\alpha}_Y,\bar{\beta}_Y\right)$; 
        and $(b)$ for treated patients with $S^\ast_i(0) \in \R_+$, we draw $Y^{\ca}_i(0)$ from the following location shifted Weibull distribution: $S^\ast_i(0)+\mathrm{Weibull}\left(\alpha_Y,\beta_Y +\lambda \log(S^\ast_i(0)) \right)$. 
        For each $i$ with $Z_i=1$, we accept $Y^{\ca}_i(0)$, setting $Y^{\ast}_i(0)=Y^{\ca}_i(0)$, with probability $p_i= \min\{	p_{Y_i(0)}, 1\}$, with
        $$ 
        p_{Y_i(0)} =  \begin{cases} 
        r_i \cdot\dfrac{f^{\bS}_{Y(0)}(Y^\ast_i(0))}{f^{\bS}_{Y(0)}(Y_i^{\ca}(0))} & \hbox{if } \,     S^\ast_i(0)=\bS, \mathbb{I}\{\Yobs_i \leq C_i \}=0\\ 
        \\
        r_i \cdot\dfrac{f^{\bS}_{Y(0)}(Y^\ast_i(0))}{f^{\bS}_{Y(0)}(Y_i^{\ca}(0))} & \hbox{if } \,     S^\ast_i(0)=\bS, \mathbb{I}\{ \Yobs_i \leq C_i \}=1, Y_i^{\ca}(0) \leq  \Yobs_i/\kappa  \\
        \\
        r_i \cdot\dfrac{f_{Y(0)}(Y_i(0))}{f_{Y(0)}(Y_i^{\ca}(0))} & \hbox{if } \,     S^\ast_i(0)\in \R_+, \mathbb{I}\{\Yobs_i \leq C_i \}=0\\
        \\
        r_i \cdot\dfrac{f_{Y(0)}(Y^\ast_i(0))}{f_{Y(0)}(Y_i^{\ca}(0))} & \hbox{if } \,     S^\ast_i(0)\in \R_+, \mathbb{I}\{ \Yobs_i \leq C_i \}=1, Y_i^{\ca}(0) \leq  \Yobs_i/\kappa  \\ 
        \\
        0 & \hbox{if } \, \mathbb{I}\{\Yobs_i \leq C_i \}=1, Y_i^{\ca}(0) > \Yobs_i/\kappa \\
        \end{cases}
        $$
        where $ f^{\bS}_{Y(0)}(\cdot)$ and $ f_{Y(0)}(\cdot)$ are the densities of the proposal  Weibull distributions, and 
        $$
        r_i= \dfrac{P\left\{Y^{\ca}_i(0) \mid \btheta, D^{\mathrm{obs}}_i,S^{\ast}_i(0)  \right\} }{P\left\{Y^{\ast}_i(0) \mid \btheta, D^{\mathrm{obs}}_i, S^{\ast}_i(0) \right\} }. 
        $$
        Note that we do not set $p_{Y_i(0)}=0$ for $\mathbb{I}\{\Yobs_i \leq C_i \}=0$ and  $Y_i^{\ca}(0) > C_i/\kappa$, because, in principle, the survival outcome under control, $Y_i(0)$, can be greater than $C_i/\kappa$: 
        For some units, we can have $Y_i(1)/\kappa \geq Y_i(0) > C_i/\kappa$. 
        For this type of units, the probability that $Y_i(1)>C_i$ is one.
        \end{itemize}
        \item Given the parameter $\btheta$, the observed data, $\bD^{\mathrm{obs}}$, and $\bY^\ast(0)$ draw, the missing data $S^\ast_i(0)$. 
        \begin{itemize}
        \item[$-$] For  control patients, we have
        $$
        S^\ast_i(0)=S_i(0)=\begin{cases}
        \bS & \hbox{if } Z_i=0, \mathbb{I}\{\Sobs_i \leq C_i\}=0, \mathbb{I}\{\Yobs_i \leq C_i\}=1\\
        \TS_i=\Sobs_i & \hbox{if } Z_i=0,  \mathbb{I}\{\Sobs_i \leq C_i\}=1,  \mathbb{I}\{\Yobs_i \leq C_i\} \in \{0,1\}.
        \end{cases}
        $$
For control patients with $\mathbb{I}\{\Sobs_i \leq C_i\}=0$ and $\mathbb{I}\{\Yobs_i \leq C_i\}=0$, we have
\begin{eqnarray*}
&\pi_{NS} \equiv P\left(S^\ast_i(0)=\bS \mid  \btheta, Z_i=0, C_i, \tilde{S}^{\mathrm{obs}}_i,\mathbb{I}\{\Sobs_i \leq C_i\}=0,
\tilde{Y}^{\mathrm{obs}}_i, \mathbb{I}\{\Yobs_i \leq C_i\}=0 \right)=&\\
&	\dfrac{\pi    G^{\bS}_{Y(0)}(C_i)}{\pi  G^{\bS}_{Y(0)}(C_i) + \left(1-\pi\right) G_{S_i(0)}(C_i)    \cdot 1}&
\end{eqnarray*}
Therefore, control patients with $\mathbb{I}\{\Sobs_i \leq C_i\}=0$ and $\mathbb{I}\{\Yobs_i \leq C_i\}=0$ are classified as non-switchers
($S^\ast_i(0)=\bS$) with probability $\pi_{NS}$ and as switchers with censored switching time ($S^\ast_i(0)=\tilde{S}_i^{\mathrm{obs}}=C_i$)
with probability $1-\pi_{NS}$.	

\item[$-$] For treated patients, we never observe $S_i^\ast(0)=S_i(0)$. 
We use Metropolis-Hasting steps to draw  $S_i(0)$ according to $P\left(S_i(0) \mid  \btheta, D_i^{\mathrm{obs}} \right)$. 
We draw candidate values $S^{\ca}_i(0)$ from a semi-continuous distribution:
We first draw $n_1$ values from a Bernoulli distribution with probability $\pi$ setting $S^{\ca}_i(0)=\bS$ for treated units for which we obtain a success (a positive value).
For treated units for which we obtain a failure, a missing value of $S^{\ca}_i(0)$ is then drawn from the Weibull distribution with parameters $\alpha_S$ and $\beta_S$:  $\mathrm{Weibull}\left(\alpha_S,\beta_S\right)$.
For each $i$ with $Z_i=1$, we accept $S^{\ca}_i(0)$, setting $S^{\ast}_i(0)=S^{\ca}_i(0)$,  with probability $p_i= \min\{	p_{S_i(0)}, 1\}$, with
$$
p_{S_i(0)}=  
\begin{cases}
r_i & \hbox{if } \,   S^\ast_i(0)=\bS,  S_i^{\ca}(0)=\bS\\
\\
r_i \cdot\dfrac{\pi}{(1-\pi)  f_{S(0)}(S_i^{\ca}(0))}& \hbox{if } \,    S^\ast_i(0)=\bS,  S_i^{\ca}(0)\in \R_+,
S_i^{\ca}(0) \leq Y_i^\ast(0)\\
\\
r_i\cdot \dfrac{(1-\pi) f_{S(0)}(S^\ast_i(0))}{\pi}& \hbox{if } \,   S^\ast_i(0)\in \R_+,  S^{\ca}_i(0)=\bS\\
\\
r_i \cdot\dfrac{f_{S(0)}(S^\ast_i(0))}{ f_{S(0)}(S^\ca_i(0))}& \hbox{if } \,  S^\ast_i(0)\in \R_+,  S_i^{\ca}(0)\in \R_+,
S_i^{\ca}(0) \leq Y_i^\ast(0)\\
\\
0 & \hbox{if } \,    S_i^{\ca}(0)\in \R_+,  S_i^{\ca}(0) > Y_i^\ast(0)\\
\end{cases}
$$
where $ f_{S(0)}(\cdot)$ is the density of the proposal  Weibull distribution, $\mathrm{Weibull}\left(\alpha_S,\beta_S\right)$, and 
$$
r_i= \dfrac{P\left\{S^{\ca}_i(0), \mid \btheta, D^{\mathrm{obs}}_i,  Y_i^\ast(0) \right\} }{P\left\{S^{\ast}_i(0), \mid \btheta, D^{\mathrm{obs}}_i,  Y_i^\ast(0) \right\} }.
$$
\end{itemize}
\item Given the imputed complete data, $$\bD= \left[ \bZ, \mathbf{C}, \tilde{\mathbf{S}}^{\mathrm{obs}},  \mathbb{I}\{\mathbf{S}^{\mathrm{obs}} \leq \mathbf{C}\},
\tilde{\bY}^{\mathrm{obs}}, \mathbb{I}\{\mathbf{Y}^{\mathrm{obs}} \leq \mathbf{C}\},  \bbS^{\ast}(0),
\bY^{\ast}(0)\right],$$ we then draw for the following sub-vectors of $\btheta$ in sequence, conditional on all others: 
$\pi$, $\alpha_S$, $\beta_S$, 
$\bar{\alpha}_Y$, $\bar{\beta}_Y$, $\alpha_Y$, $\beta_Y$,
$\bar{\nu}_Y$, $\bar{\gamma}_Y$, $\nu_Y$, $\gamma_Y$, $\lambda$, using the procedure described in step 2. in Section ``Details of Calculations: $\kappa=0$.''
\end{enumerate}

\section{Application: Additional Results} \label{s:app2}
\subsection*{Convergence Checks}
\begin{figure}
\begin{center}
\begin{tabular}{ccc}
\includegraphics[width=4.0cm]{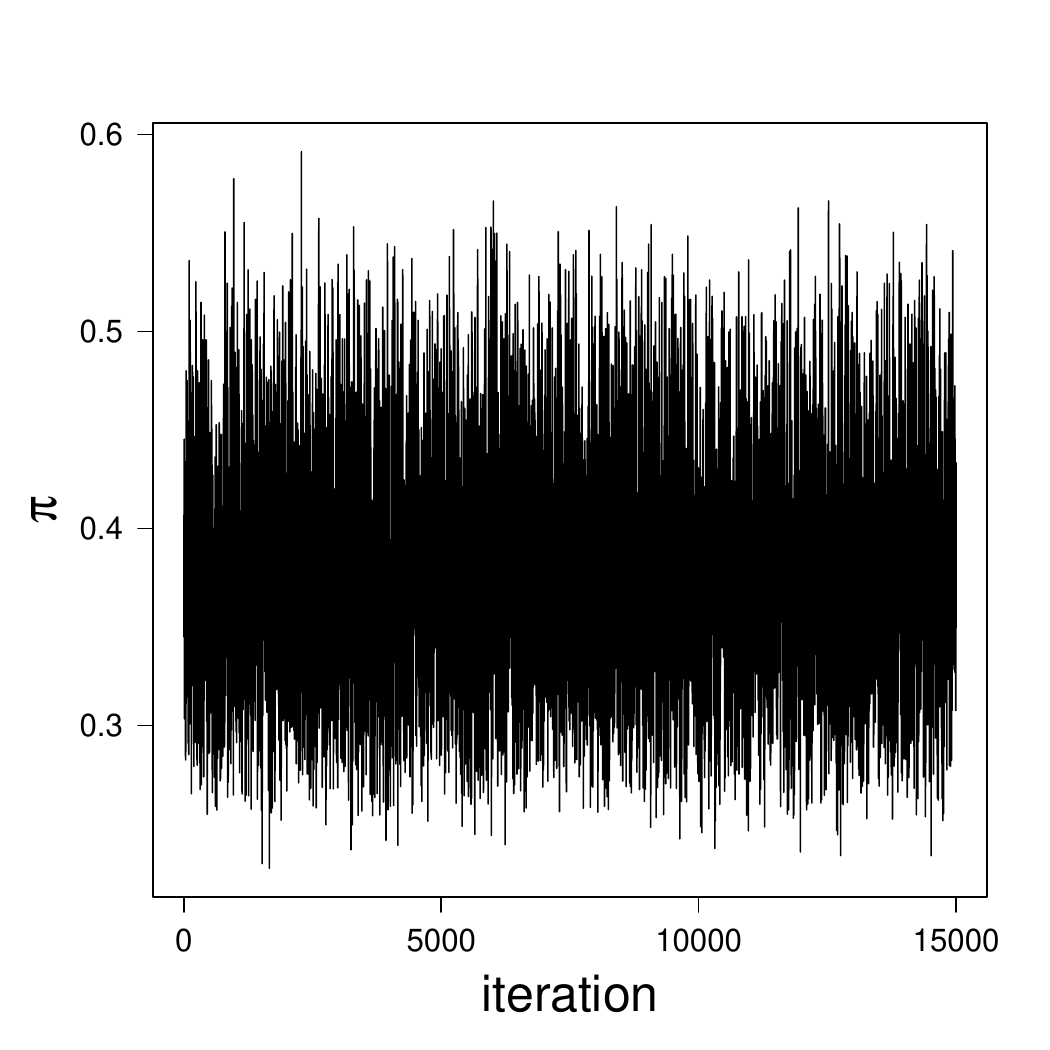} & 
\includegraphics[width=4.0cm]{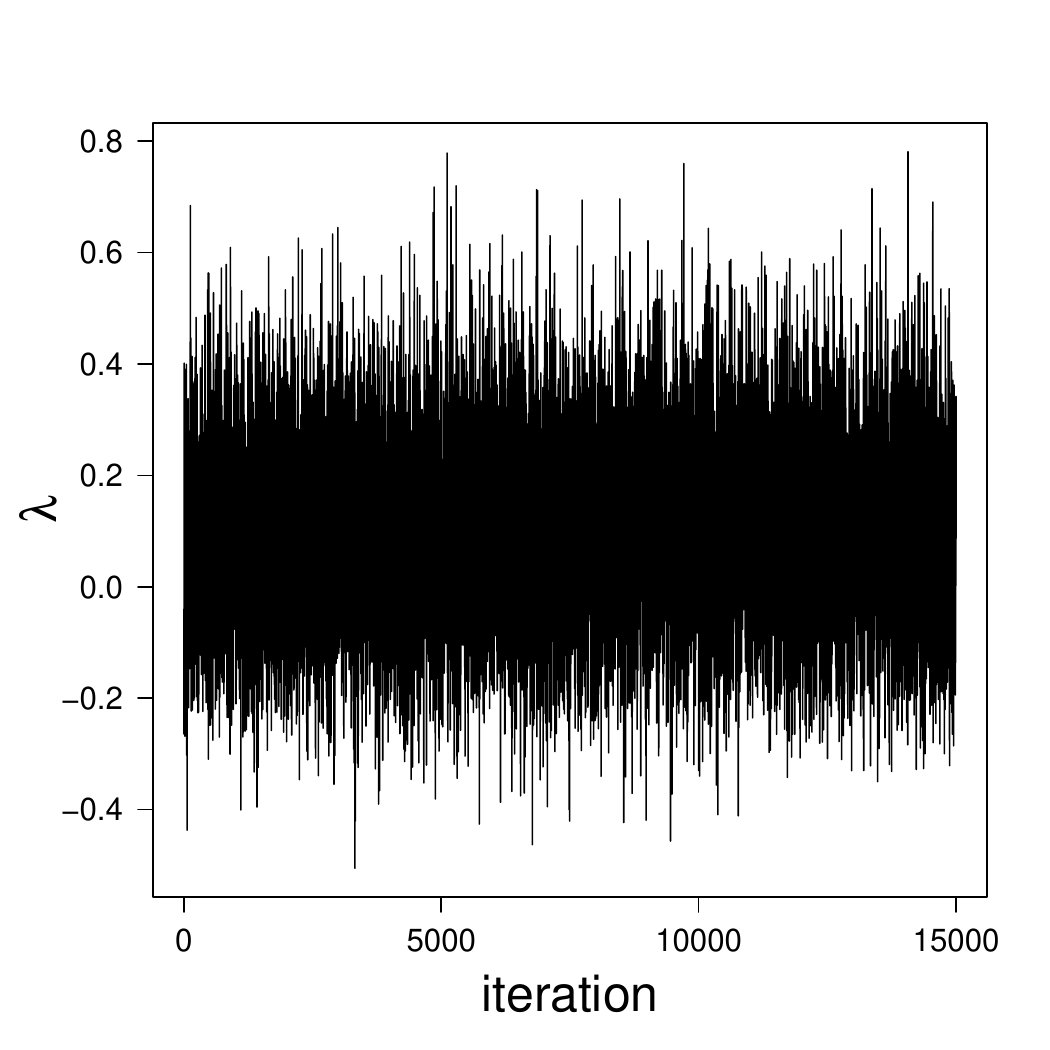}&
\includegraphics[width=4.0cm]{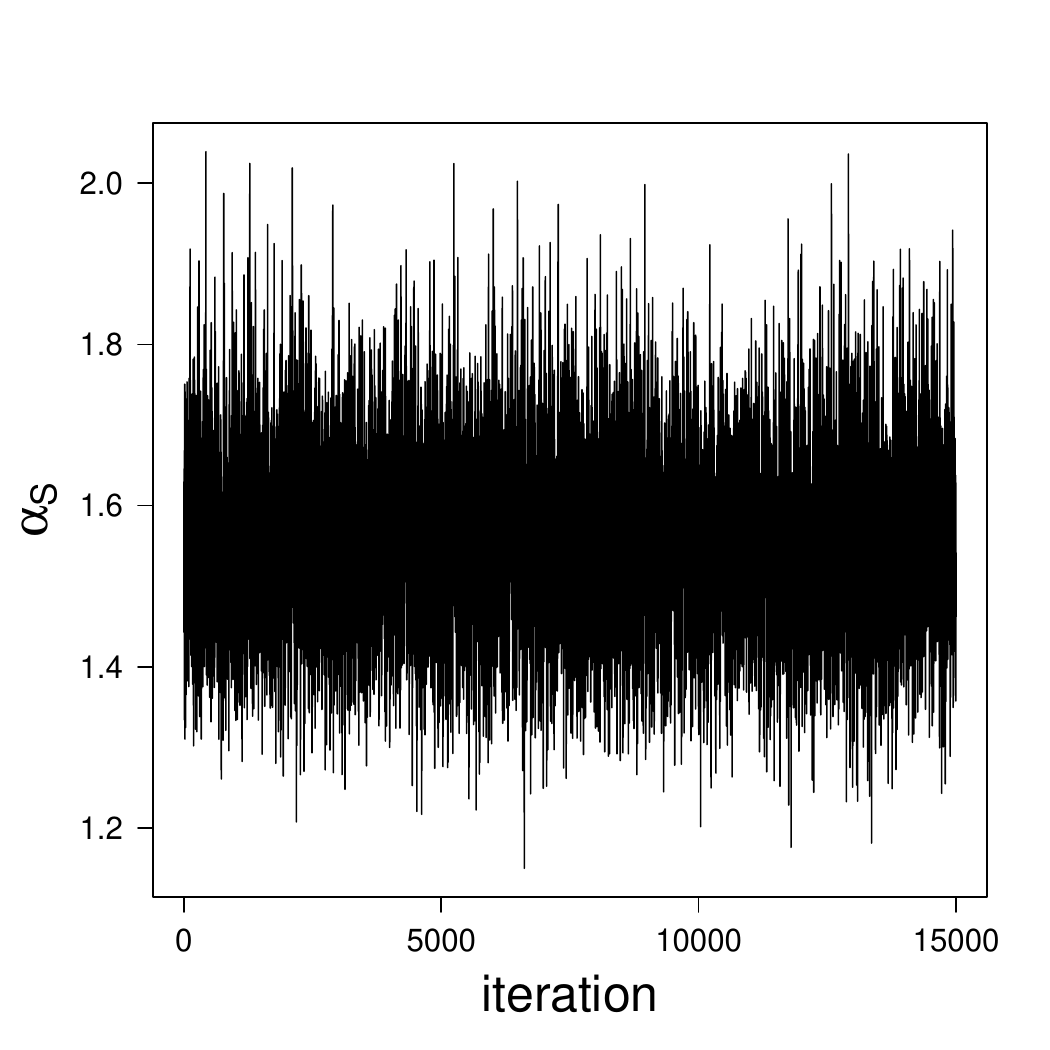}\\
\includegraphics[width=4.0cm]{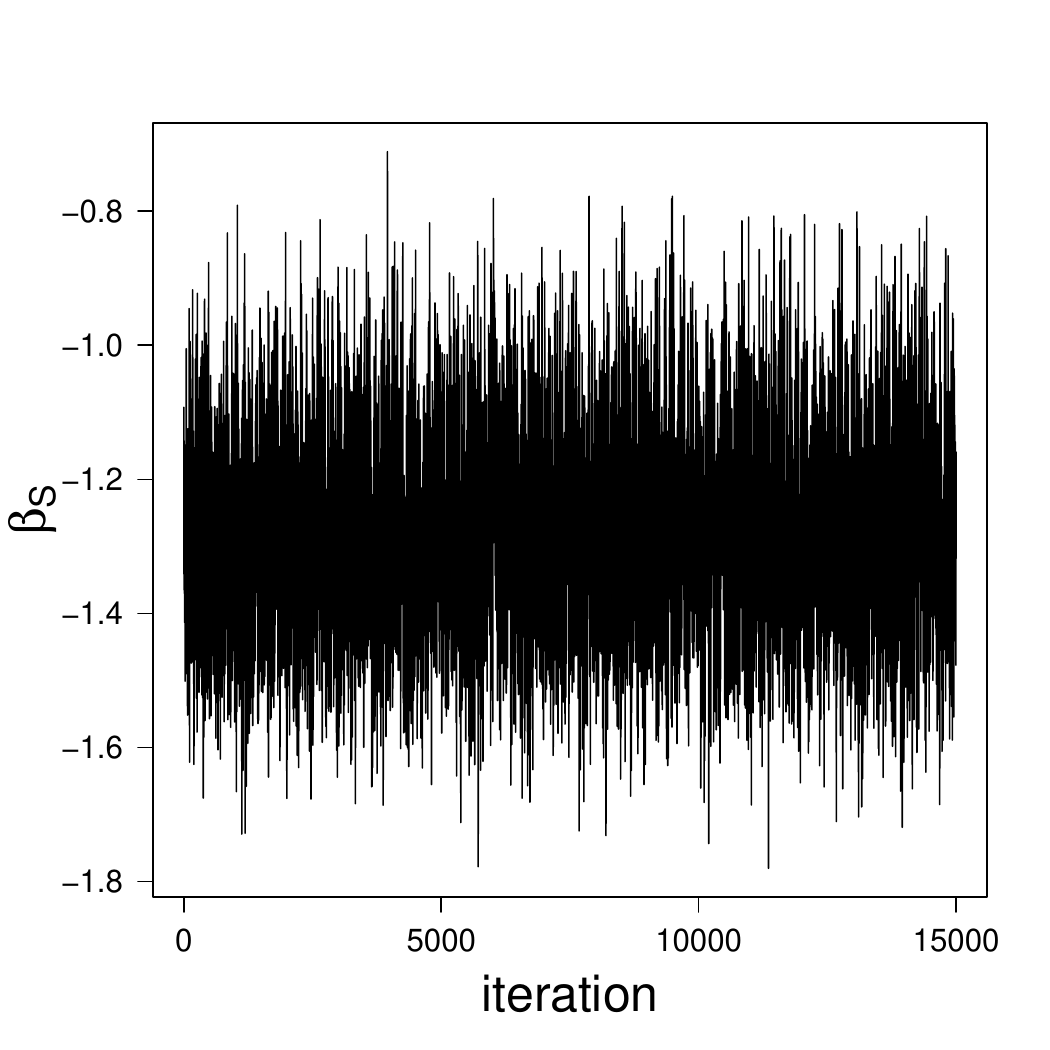} &
\includegraphics[width=4.0cm]{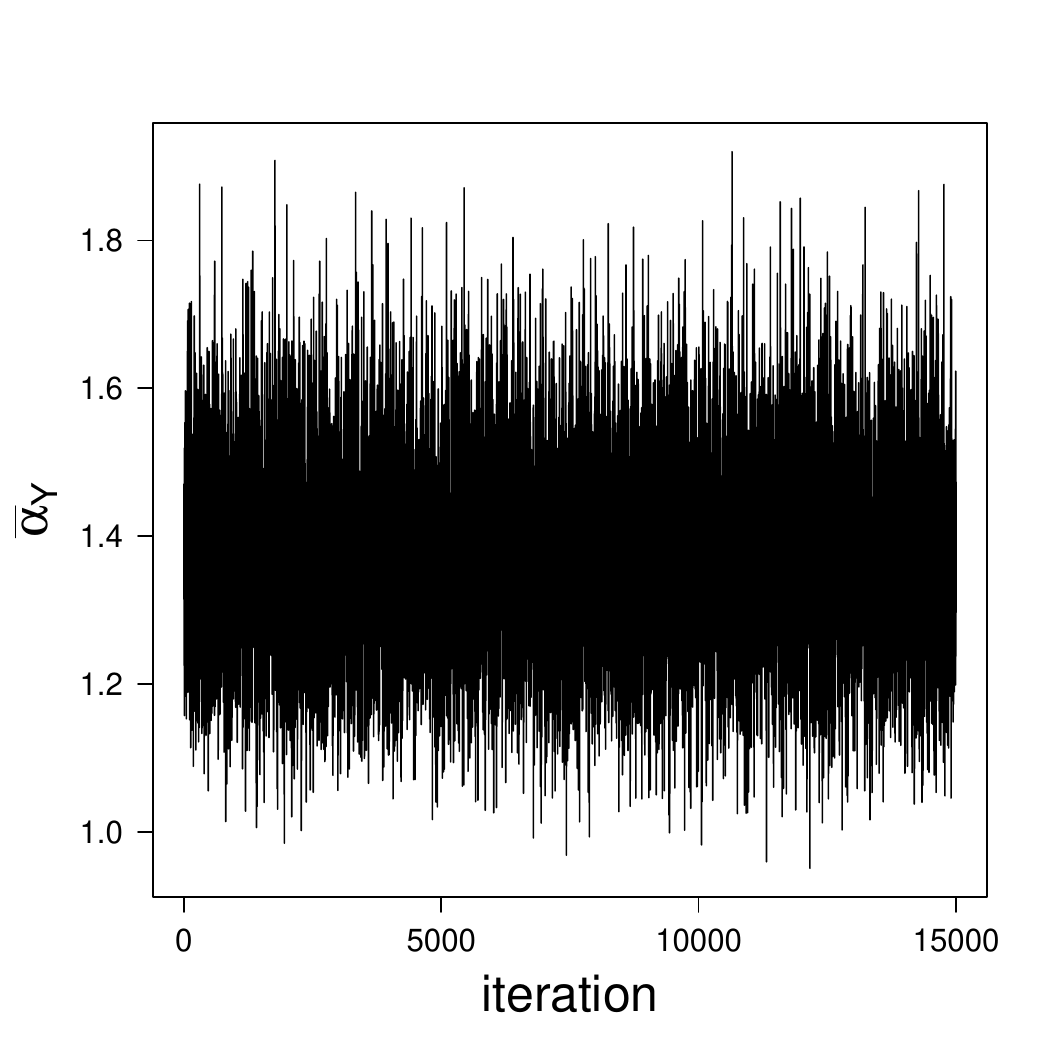}& \includegraphics[width=4.0cm]{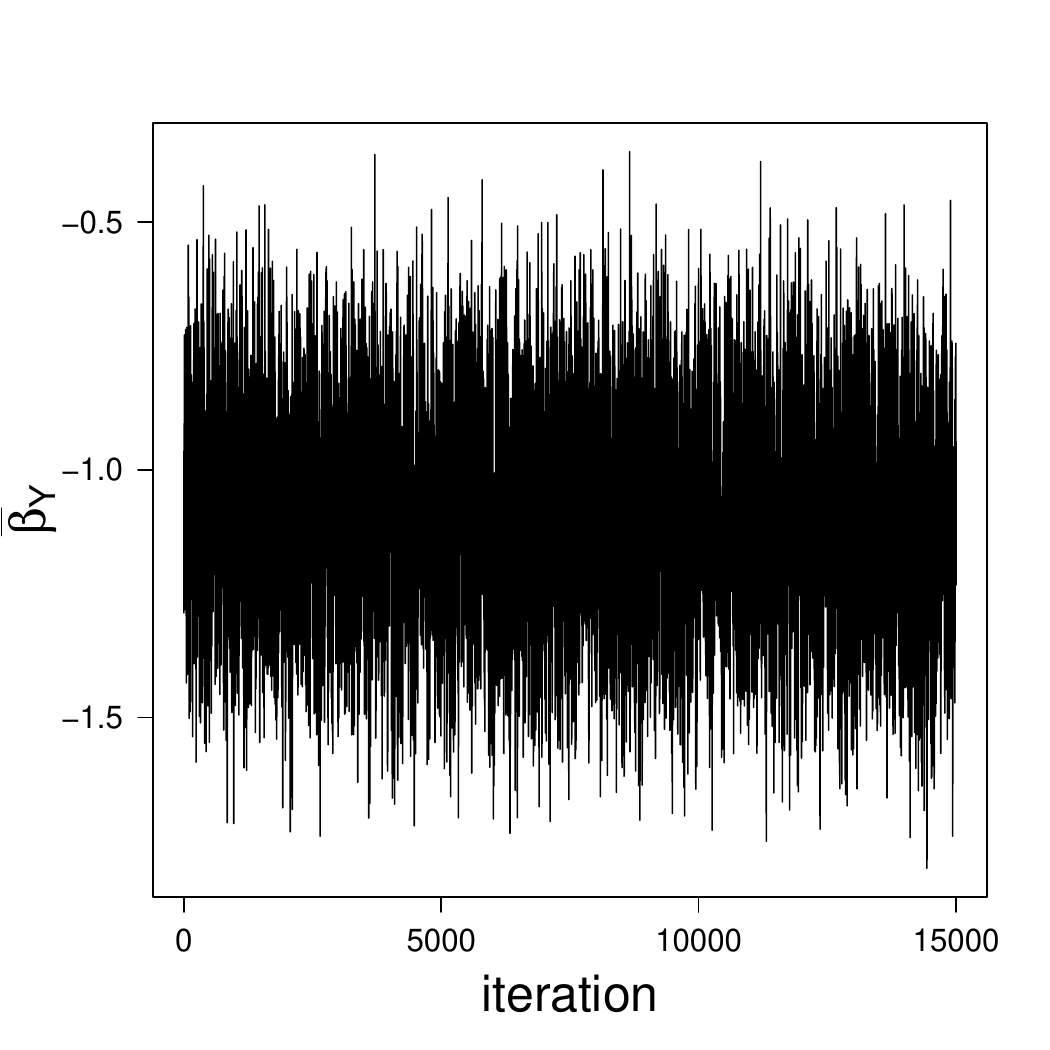}\\
\includegraphics[width=4.0cm]{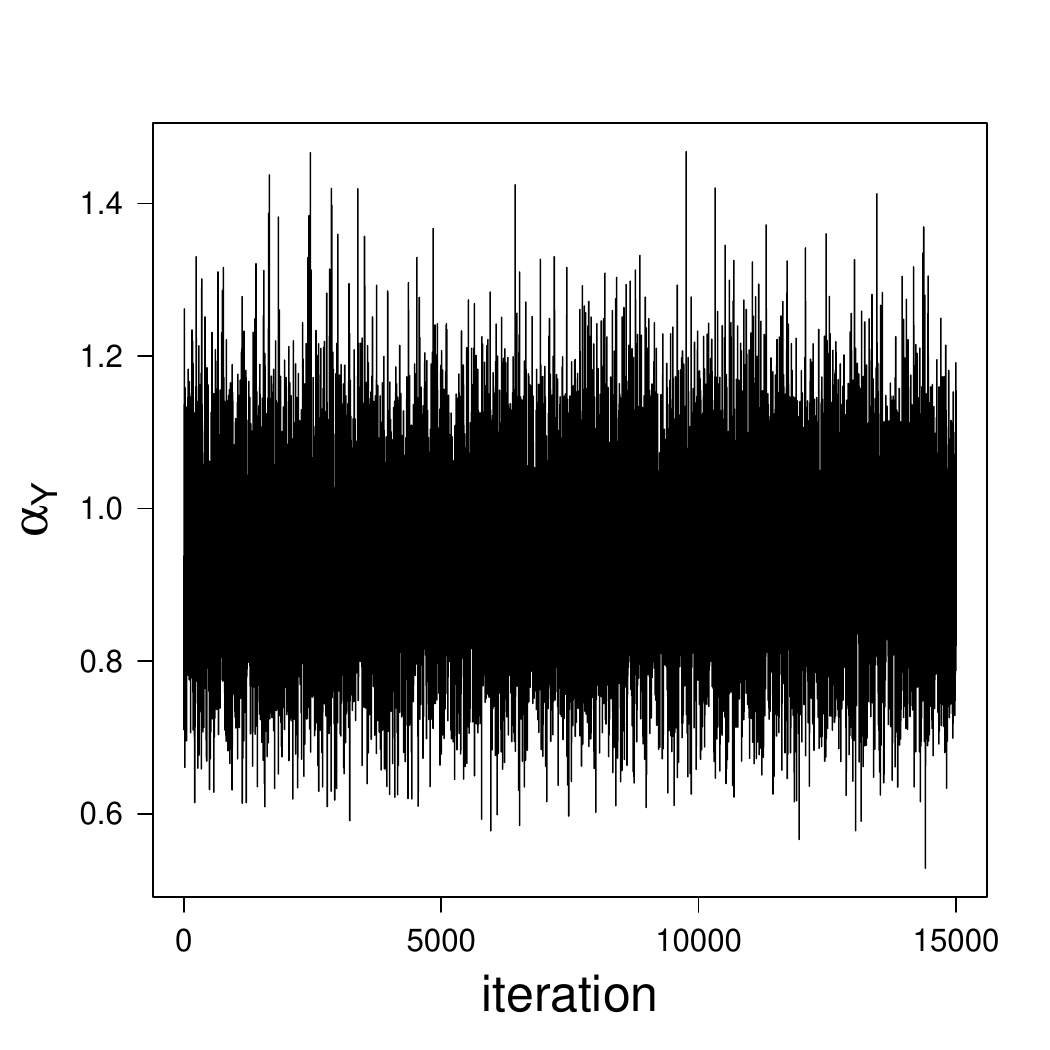}&
\includegraphics[width=4.0cm]{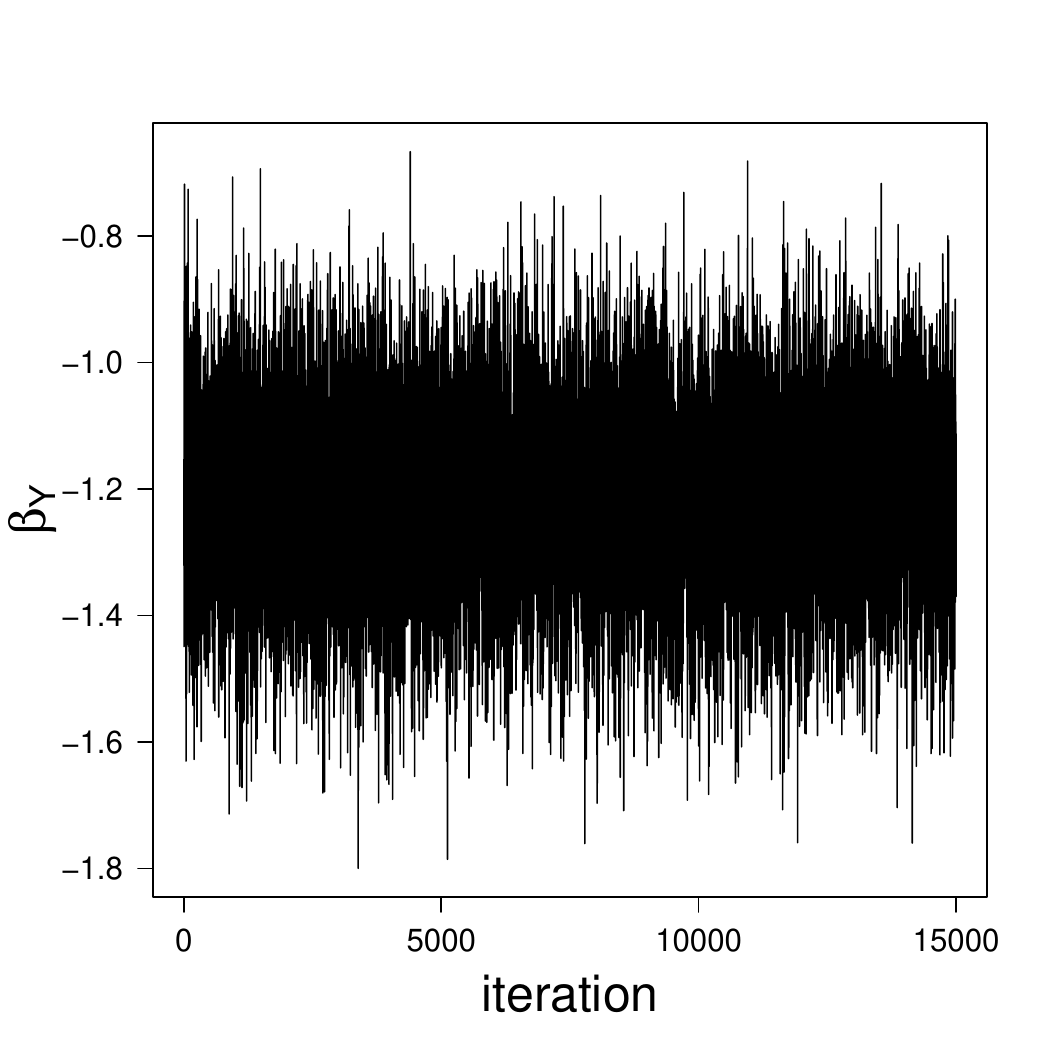}&
\includegraphics[width=4.0cm]{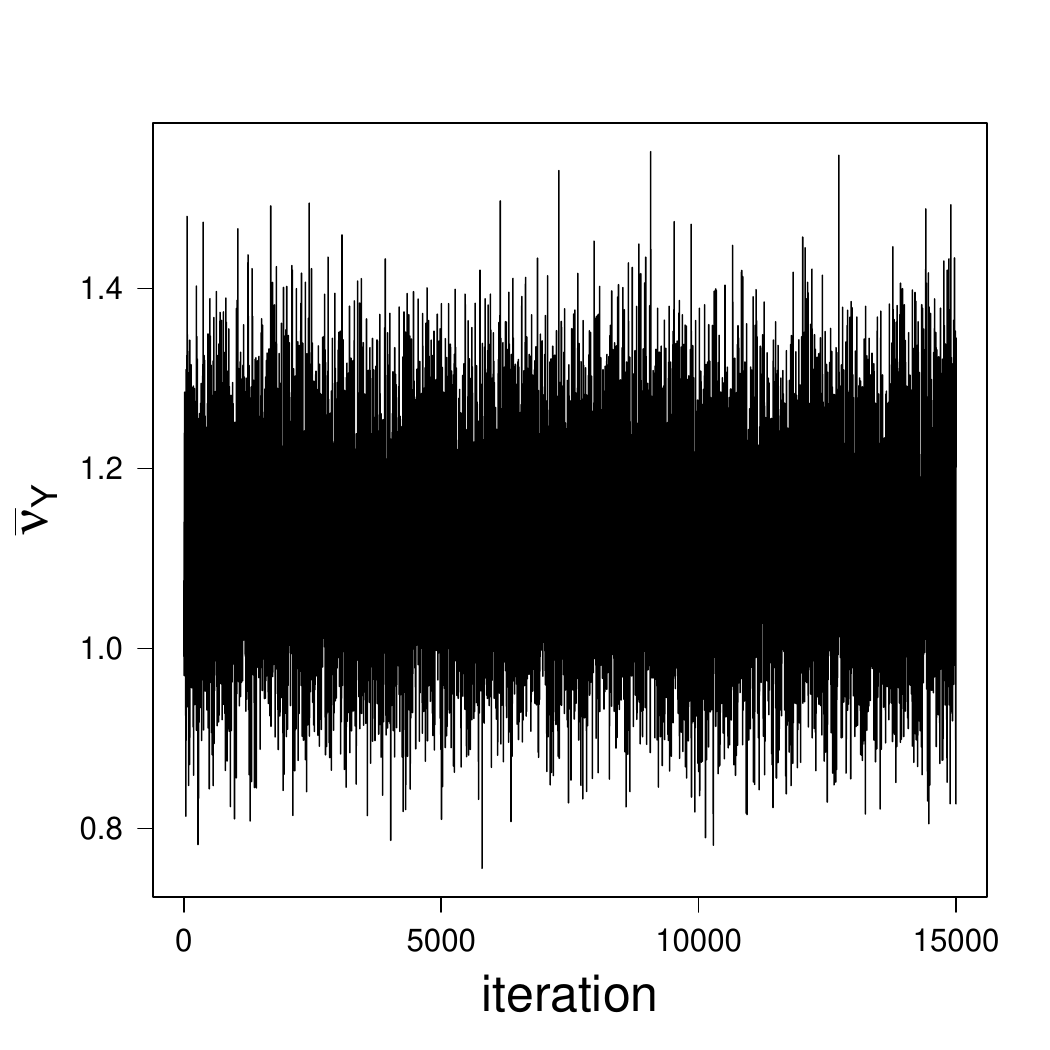}\\
\includegraphics[width=4.0cm]{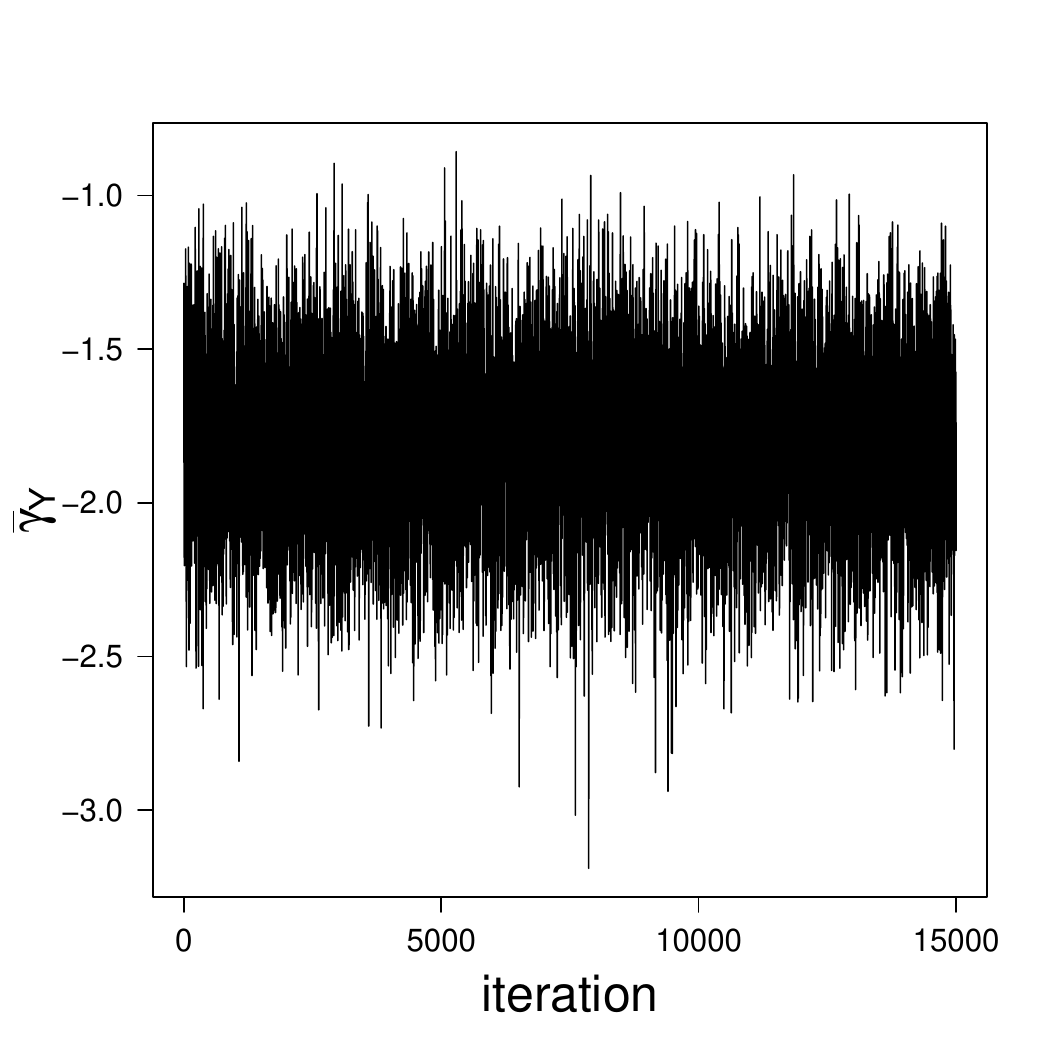}&
\includegraphics[width=4.0cm]{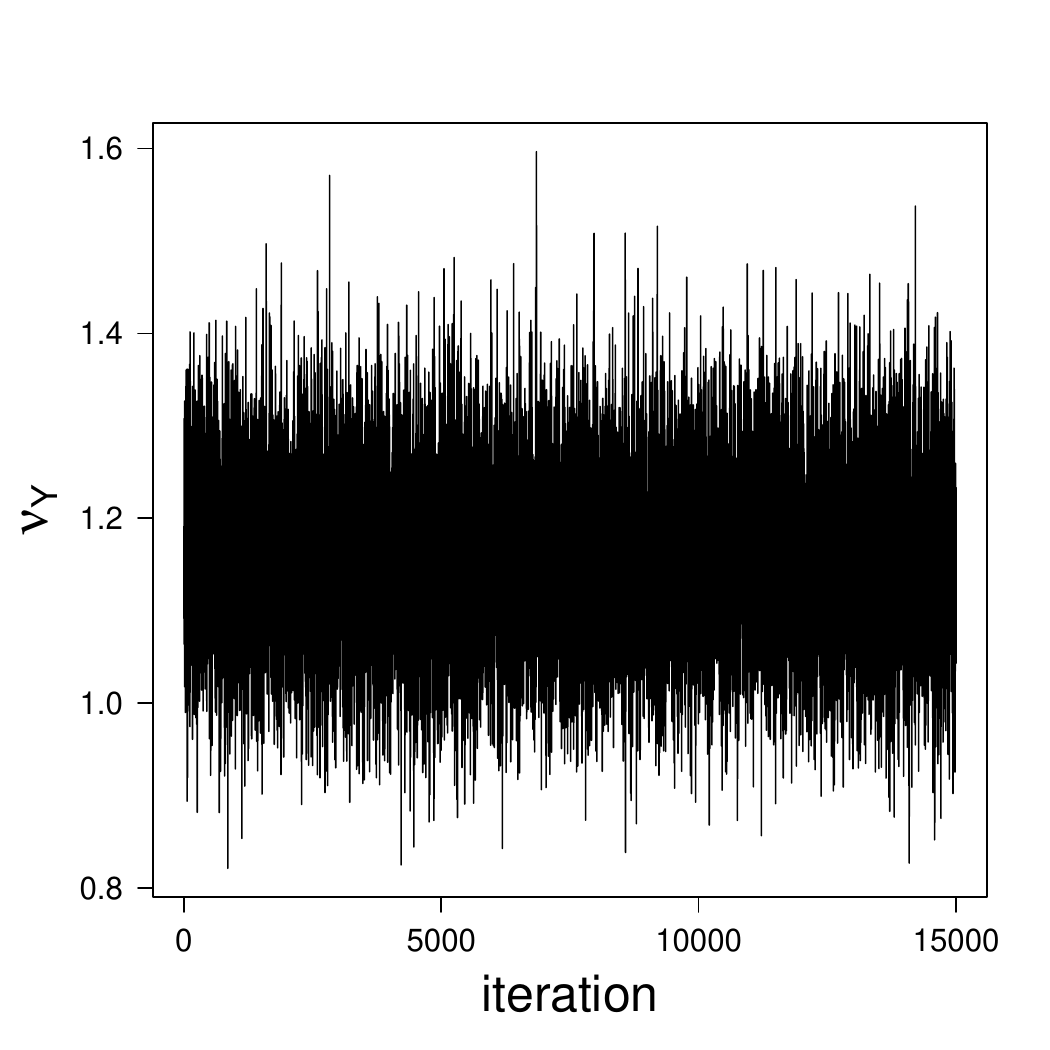}&
\includegraphics[width=4.0cm]{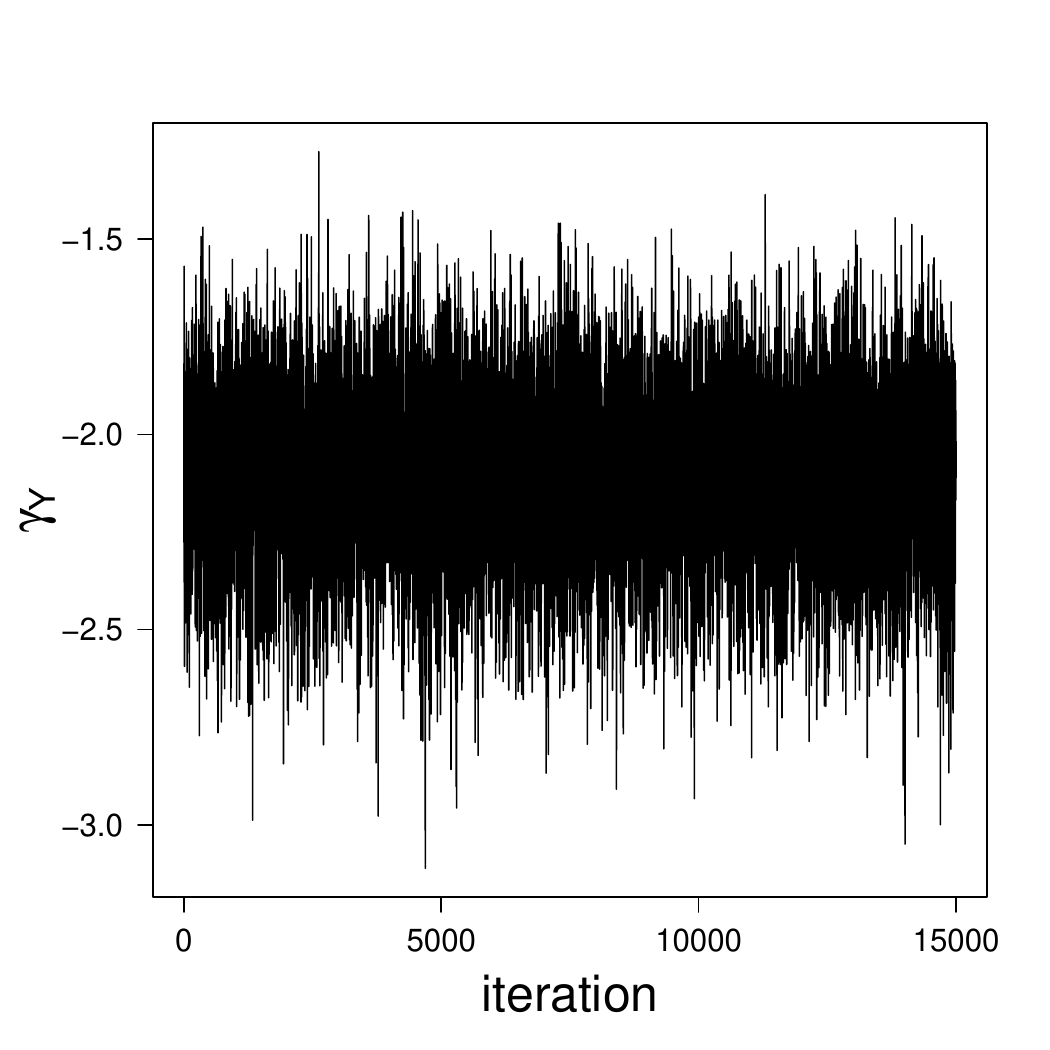}\\
\\
\end{tabular}
\end{center}
\caption{Trace plots of the model parameters \label{fig:TP}}
\end{figure}

\begin{figure}
\begin{center}
\begin{tabular}{cccc}
\includegraphics[width=4.0cm]{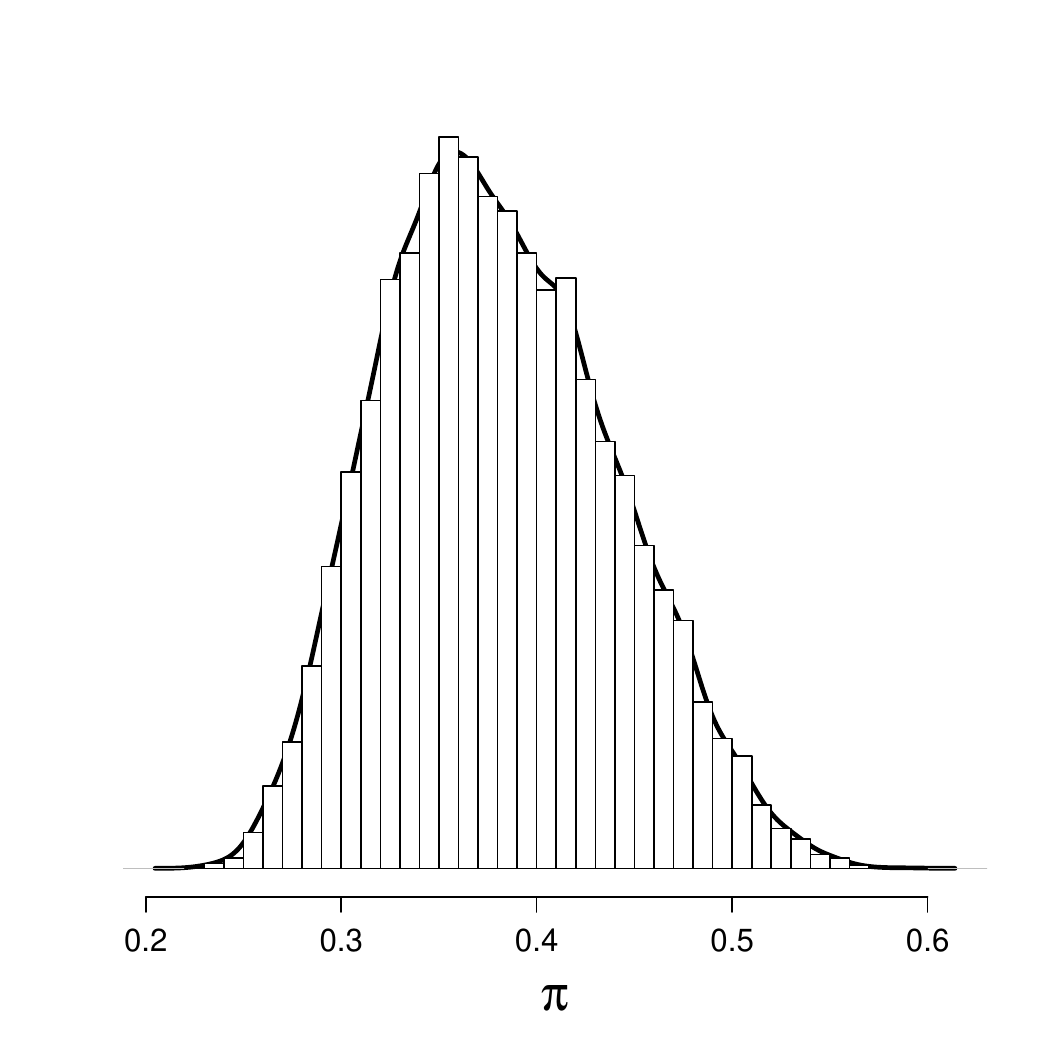} & 
\includegraphics[width=4.0cm]{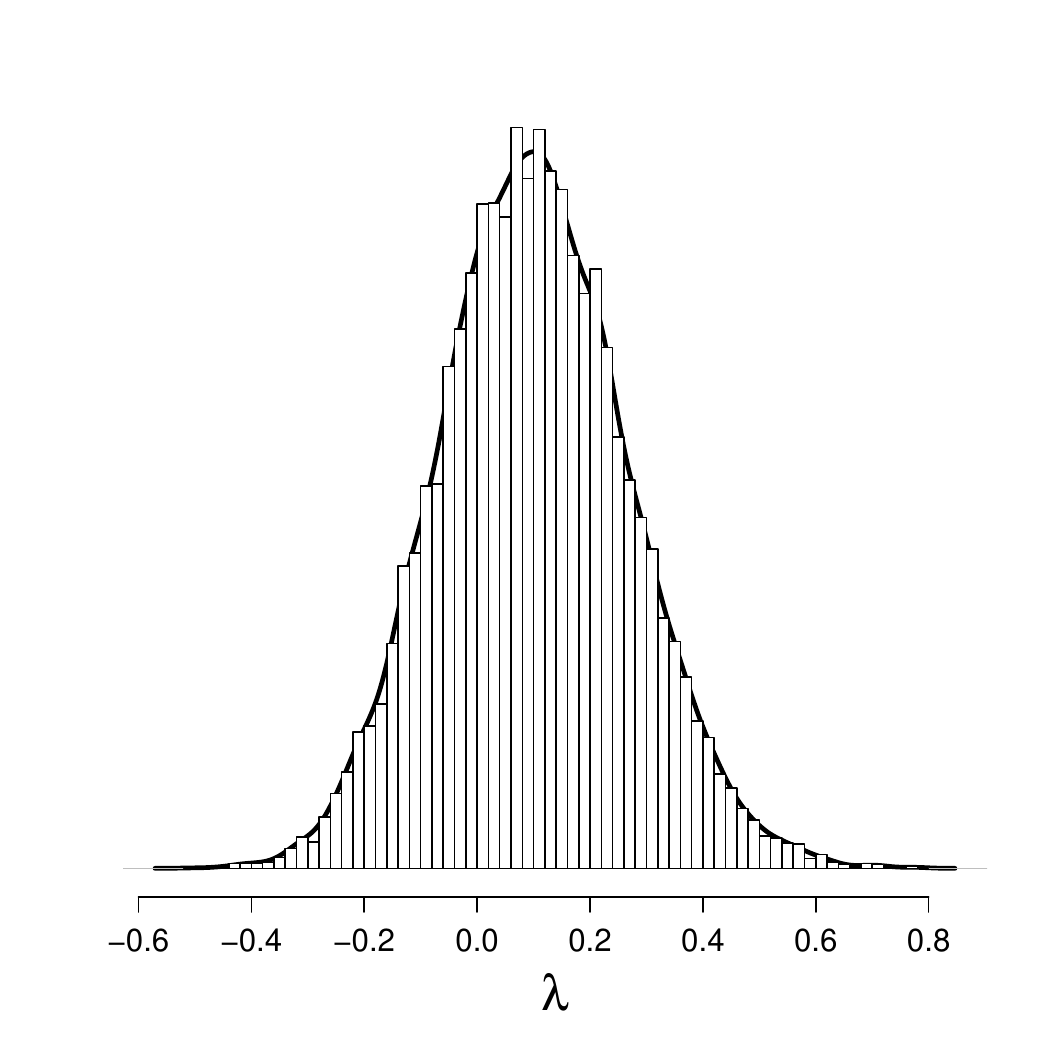}&
\includegraphics[width=4.0cm]{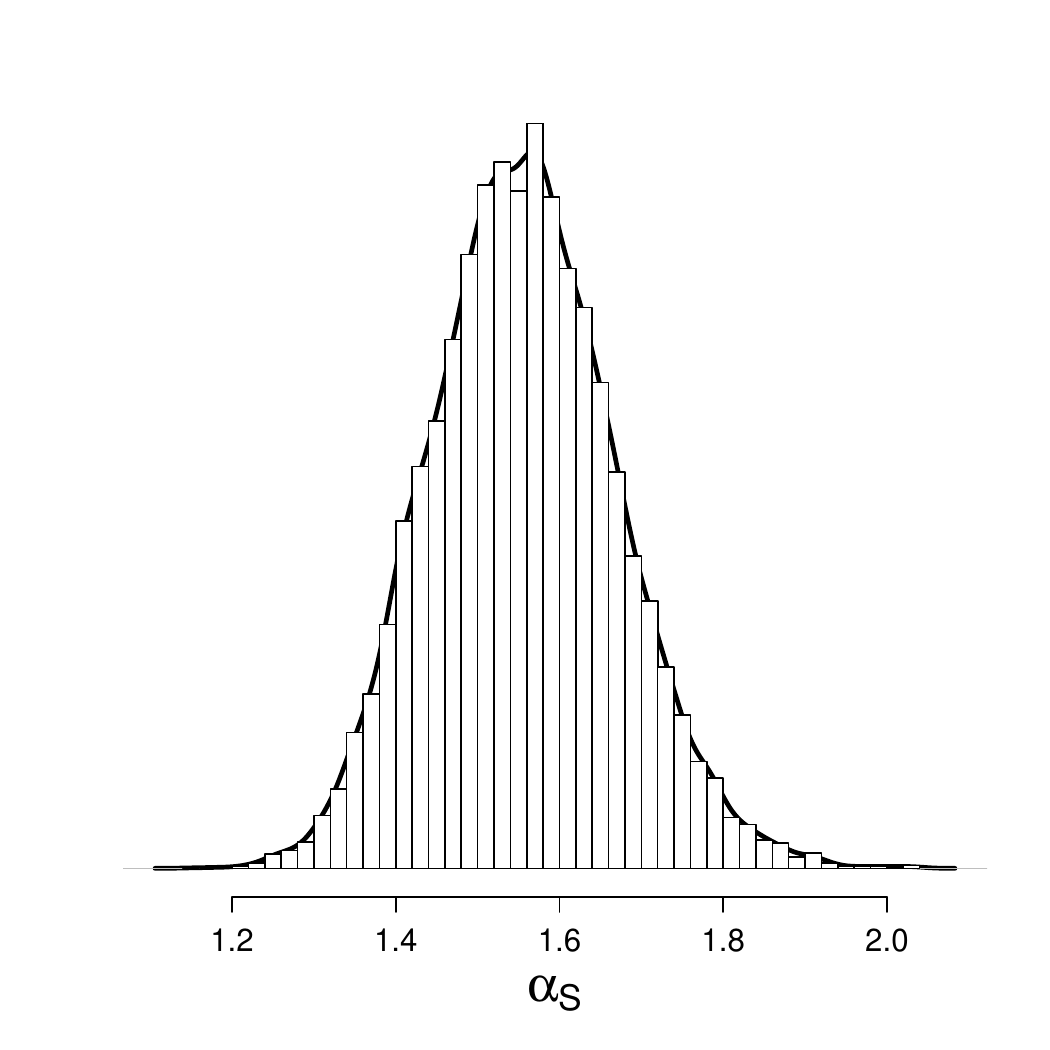}\\
\includegraphics[width=4.0cm]{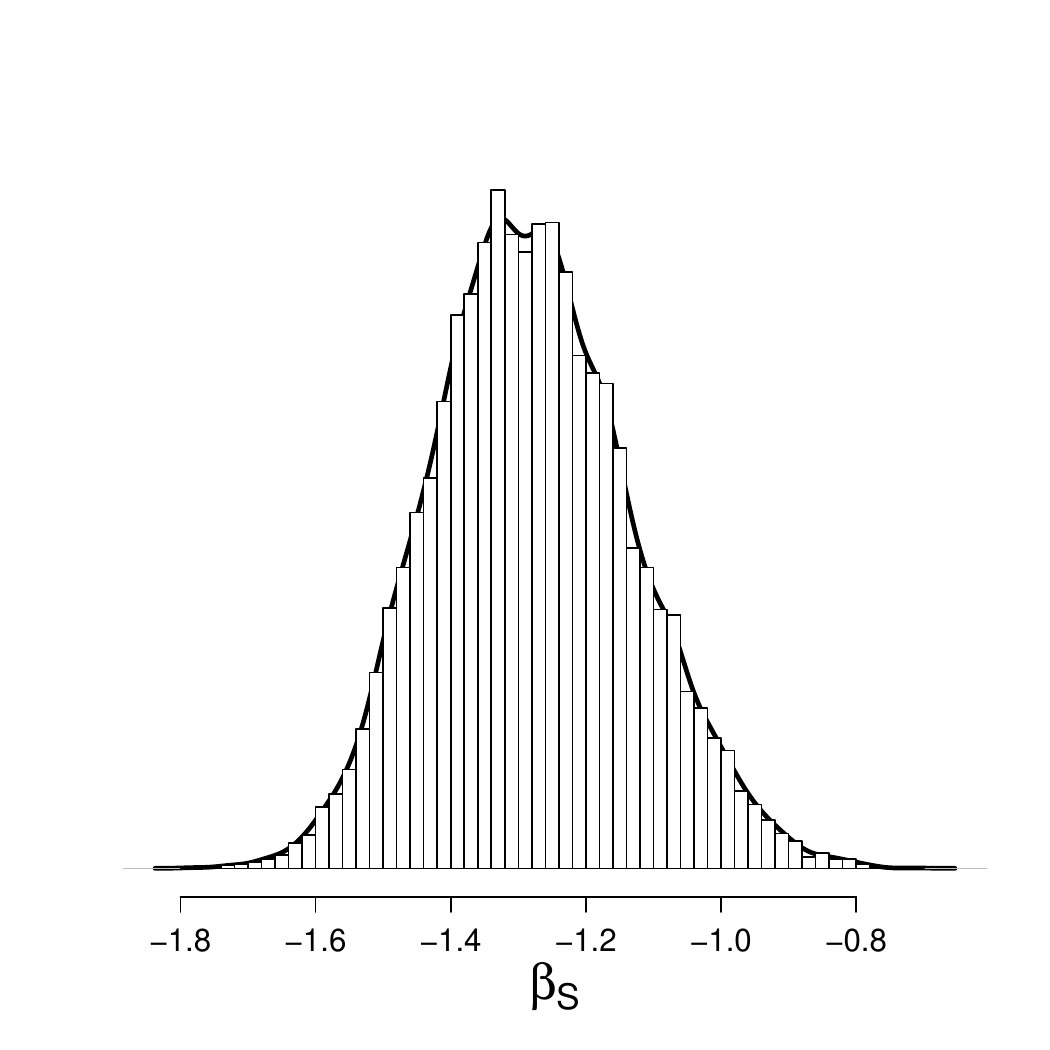}&
\includegraphics[width=4.0cm]{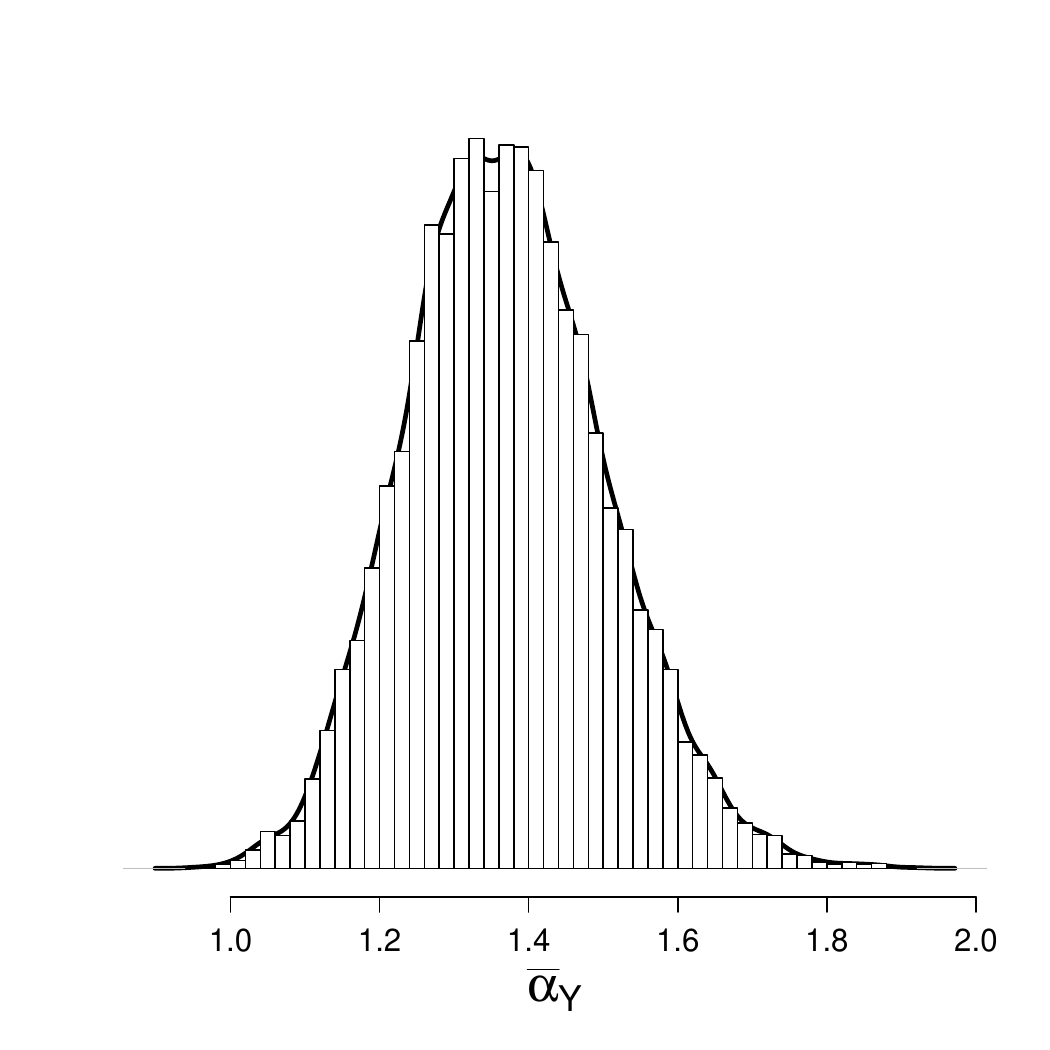}&
\includegraphics[width=4.0cm]{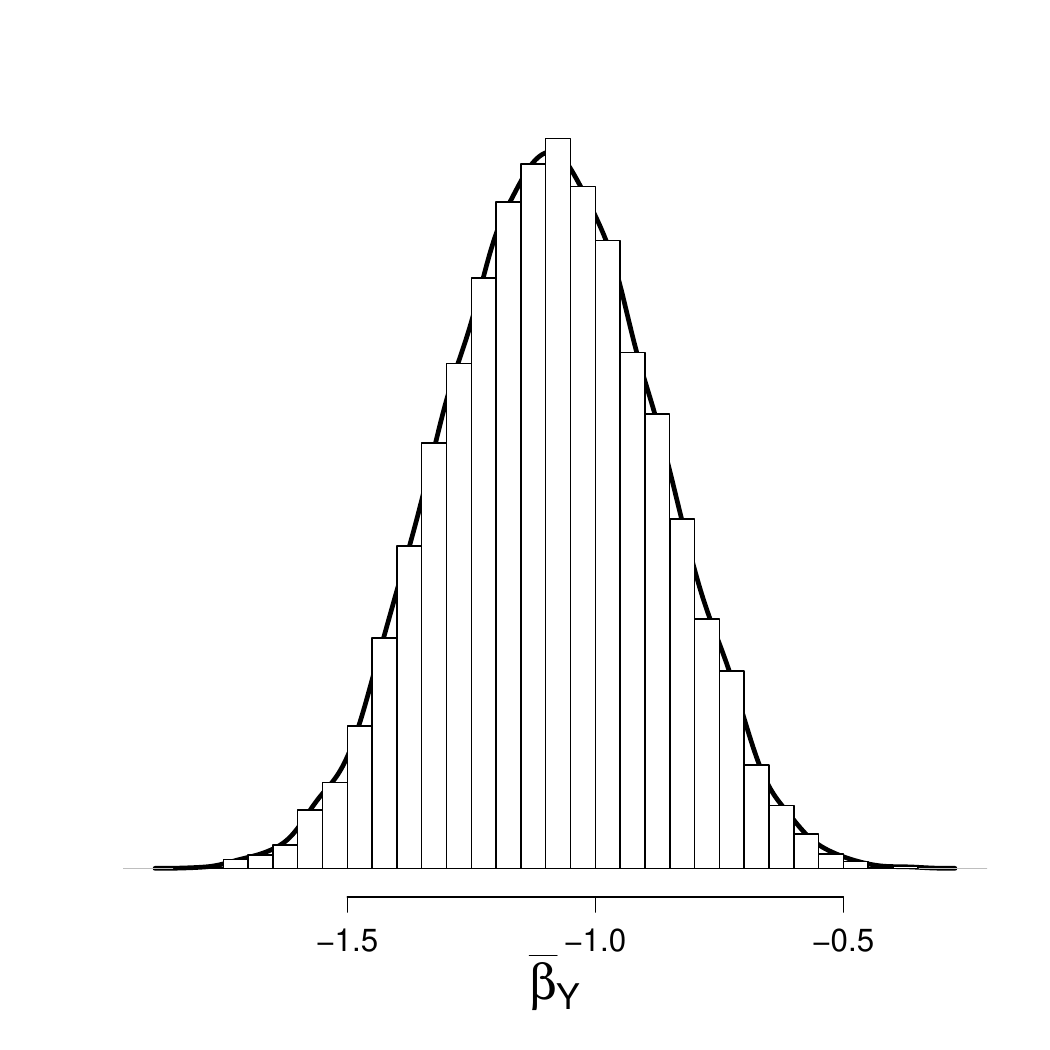}\\
\includegraphics[width=4.0cm]{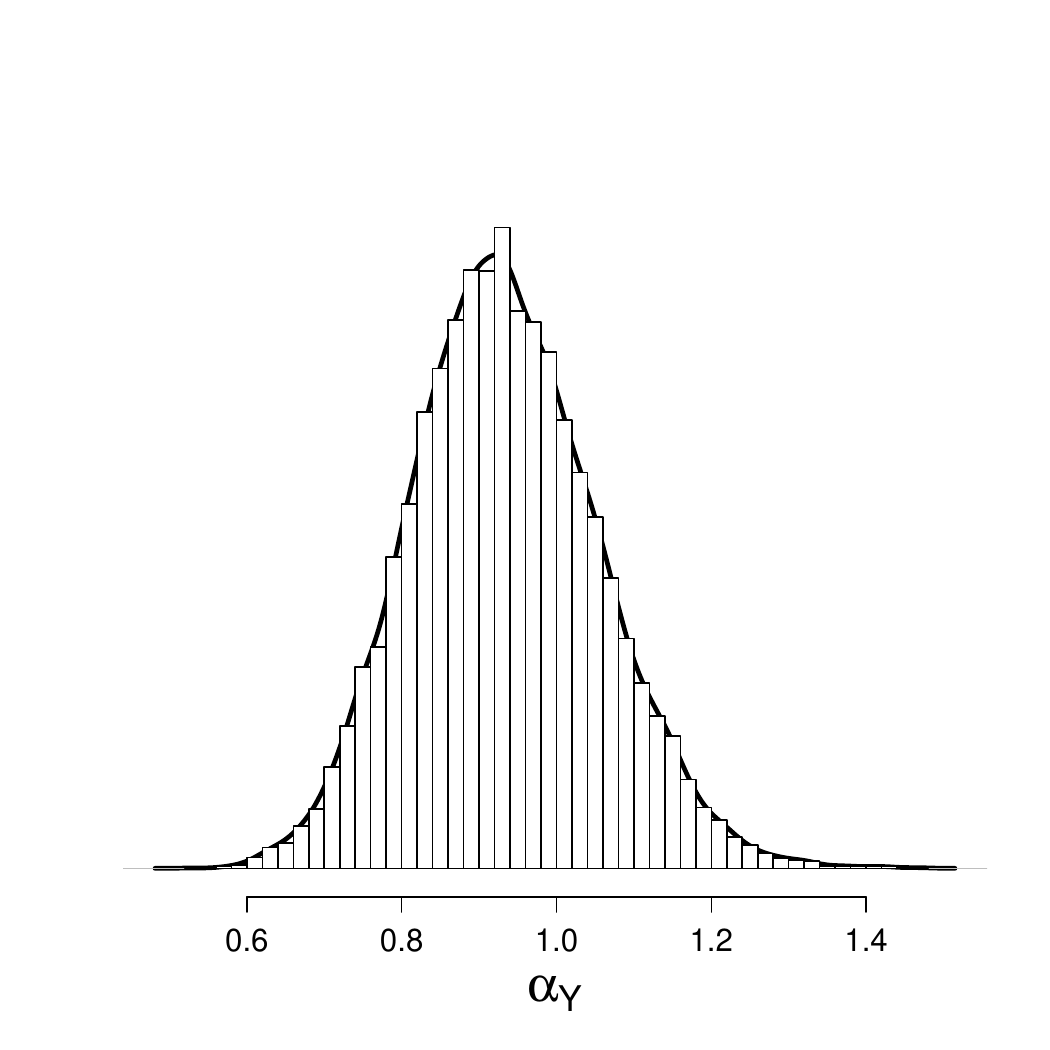}&
\includegraphics[width=4.0cm]{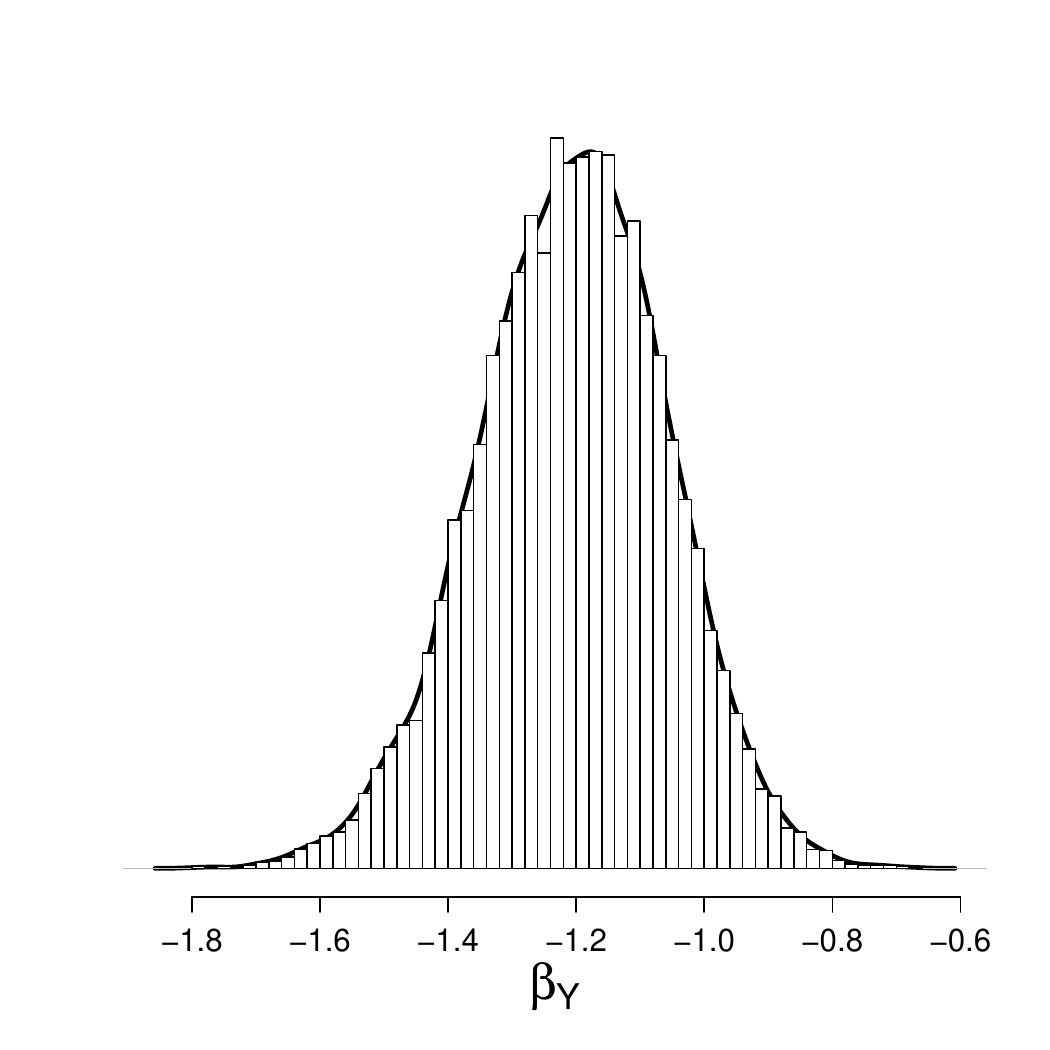}&
\includegraphics[width=4.0cm]{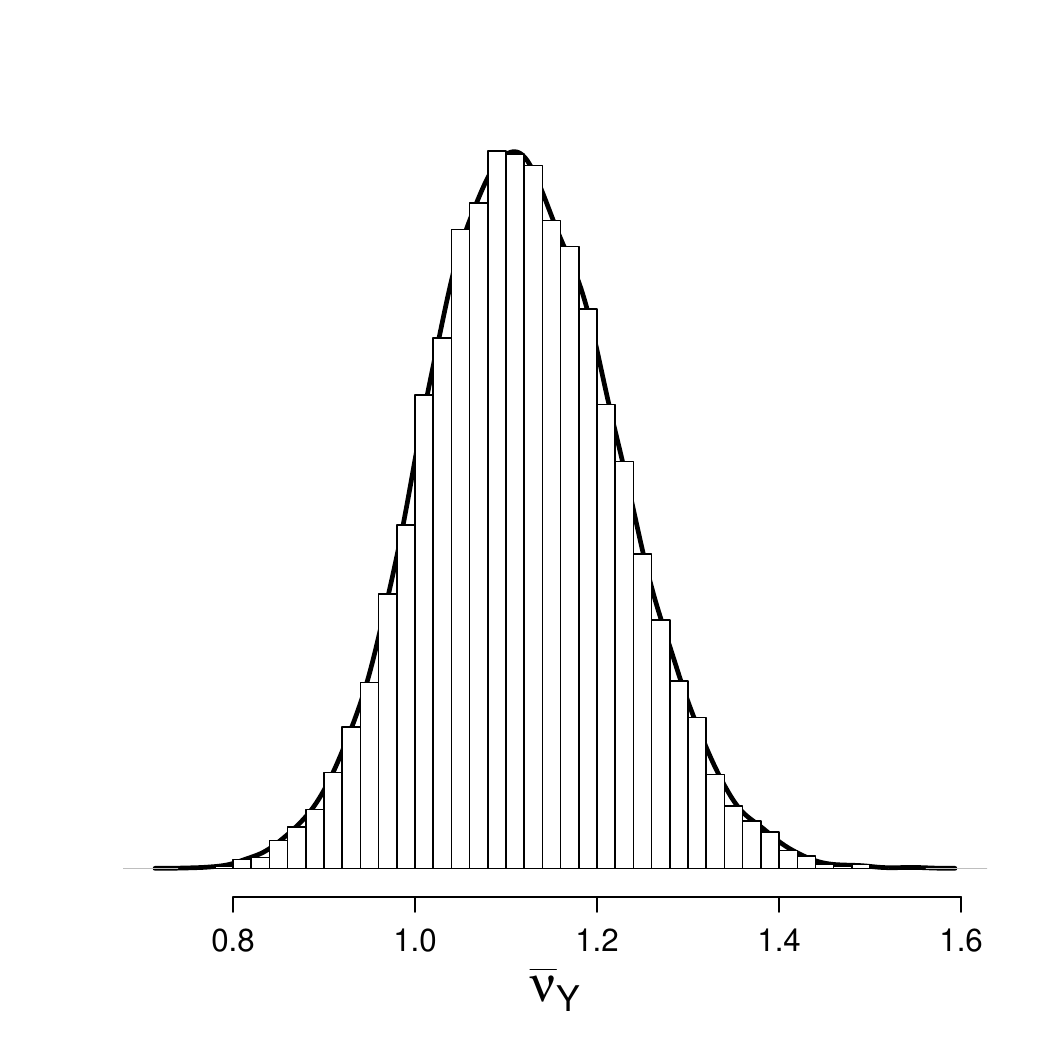}\\
\includegraphics[width=4.0cm]{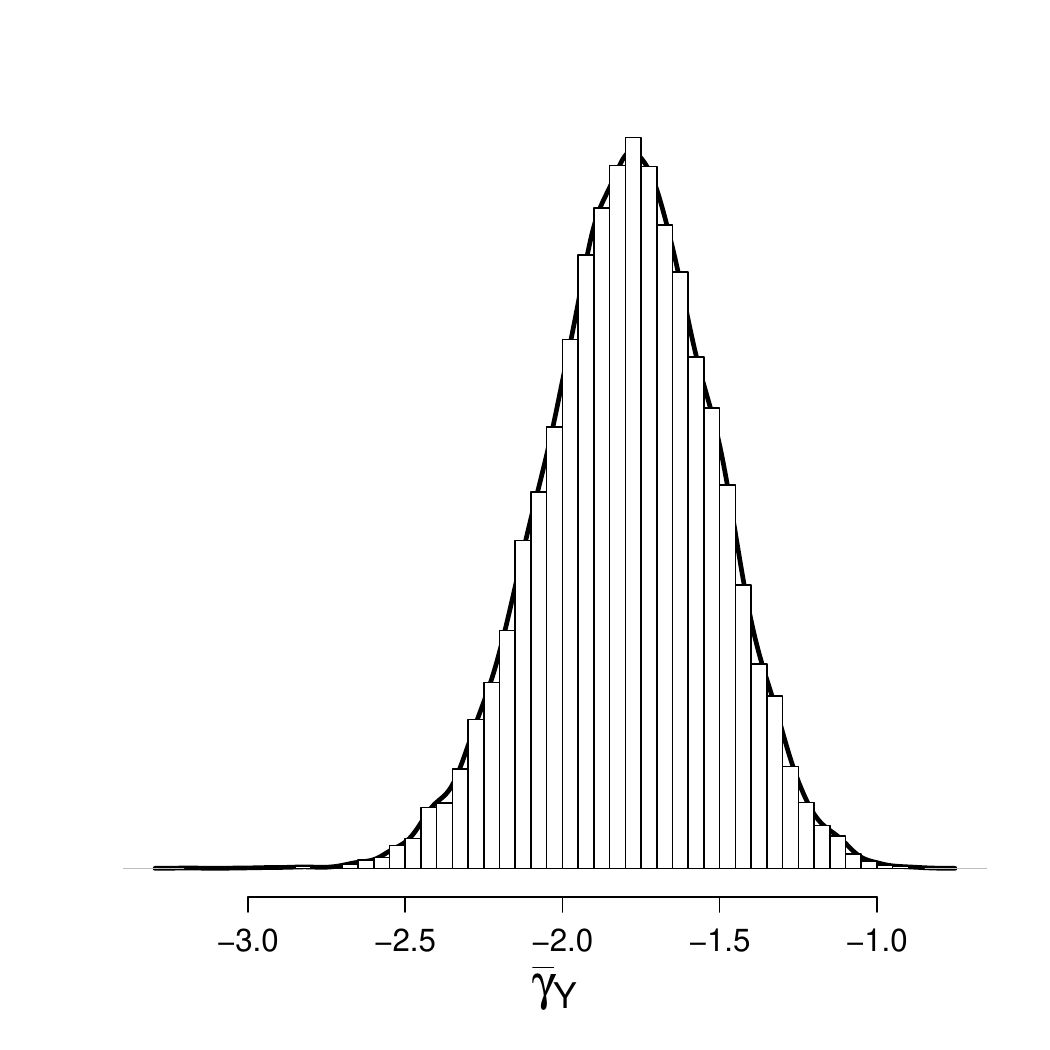}&
\includegraphics[width=4.0cm]{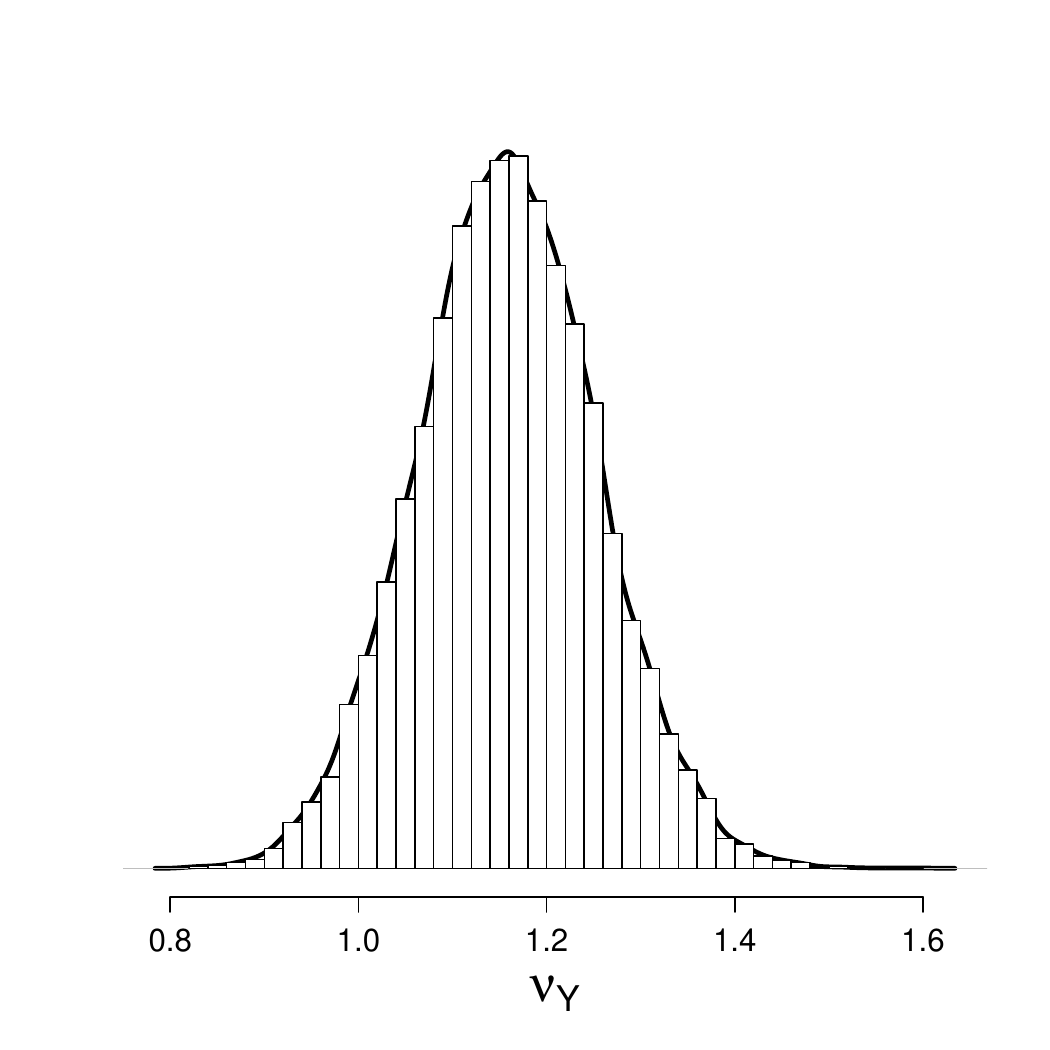}&
\includegraphics[width=4.0cm]{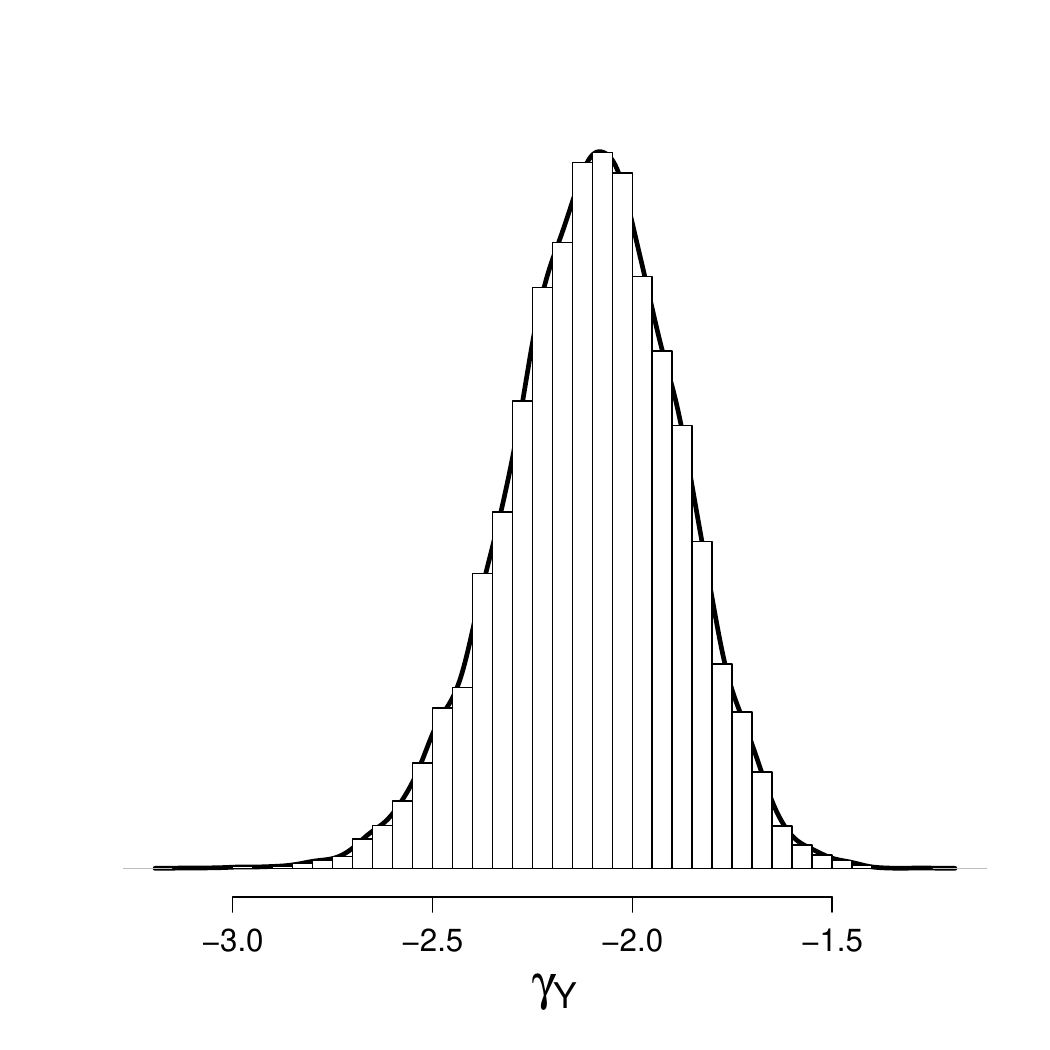}\\
\\
\end{tabular}
\end{center}
\caption{Posterior density of the model parameters\label{fig:PD}}
\end{figure}

We use the potential scale-reduction statistic \cite[][]{GelmanRubin1992} to assess convergence of the MCMC algorithm; the potential scale reduction statistic takes on values around 1 for all the model parameters, showing no evidence against convergence (see Table \ref{tab:pd}).
Figure \ref{fig:TP} shows the trace plots, which exhibit up-and-down variation with no long-term trends or drift, showing further evidence that convergence has been reached. 
Finally, Figure \ref{fig:PD} shows the posterior distributions of the model parameters, which are generally well-shaped.

\begin{table}\caption{Summary statistics of the posterior distributions \label{tab:pd}}
$$\begin{array}{crcrrrrrc}
&&\multicolumn{5}{c}{\qquad \qquad Percentiles}\\
\cline{4-8}
\hbox{Parameter}& Mean & sd & 2.5\% & 25\% & 50\% & 75\% & 97.5\% & \widehat{R}\\ 
\hline
\pi & 0.38 & 0.06 & 0.28 & 0.34 & 0.38 & 0.42 & 0.50 & 1.000 \\ 
\alpha_Y & 1.56 & 0.11 & 1.35 & 1.48 & 1.55 & 1.63 & 1.79 & 1.000 \\ 
\beta_S & -1.28 & 0.15 & -1.55 & -1.38 & -1.28 & -1.18 & -0.98 & 1.001 \\ 
\overline{\alpha}_Y & 1.37 & 0.13 & 1.13 & 1.28 & 1.37 & 1.46 & 1.64 & 1.000 \\ 
\overline{\beta}_Y & -1.09 & 0.21 & -1.50 & -1.24 & -1.09 & -0.95 & -0.69 & 1.001 \\ 
\alpha_Y & 0.93 & 0.12 & 0.71 & 0.85 & 0.93 & 1.01 & 1.18 & 1.000 \\ 
\beta_Y & -1.21 & 0.15 & -1.51 & -1.30 & -1.20 & -1.10 & -0.93 & 1.000 \\ 
\overline{\nu}_Y & 1.12 & 0.10 & 0.92 & 1.05 & 1.12 & 1.19 & 1.33 & 1.000 \\ 
\overline{\gamma}_Y & -1.79 & 0.27 & -2.33 & -1.97 & -1.79 & -1.61 & -1.29 & 1.000 \\ 
\nu_Y & 1.16 & 0.09 & 0.97 & 1.09 & 1.16 & 1.22 & 1.35 & 1.000 \\ 
\gamma_Y & -2.10 & 0.21 & -2.54 & -2.24 & -2.09 & -1.95 & -1.70 & 1.000 \\ 
\lambda & 0.10 & 0.17 & -0.21 & -0.01 & 0.10 & 0.21 & 0.43 & 1.001 \\ 
\hline
\end{array}
$$
\end{table}

\subsection*{Distributional Causal Effects for Switchers}
\begin{figure}
	\begin{center}
		\begin{tabular}{c}
			$\DCE(y \mid s)$  \vspace{-0.4cm}\\
			\includegraphics[width=7cm]{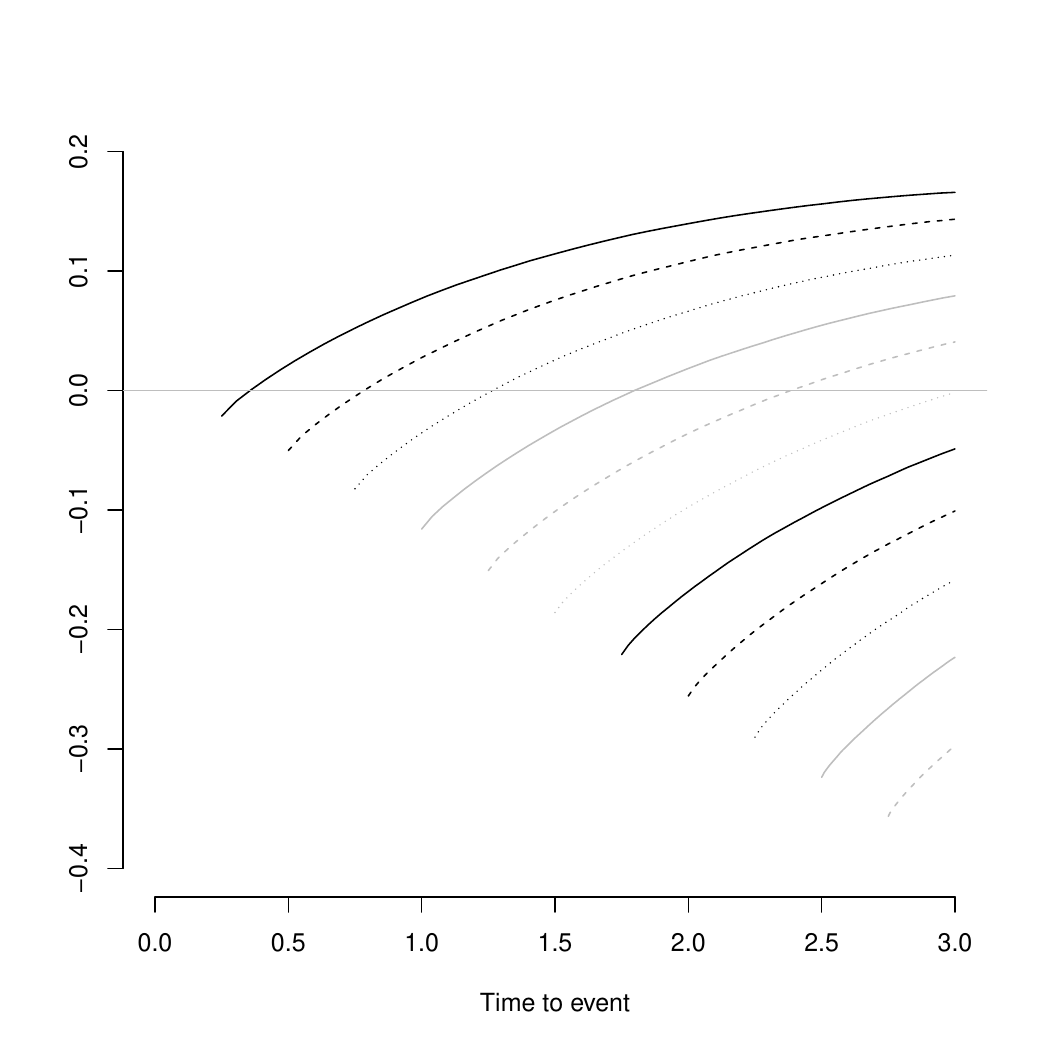}\\
			\\
		\end{tabular}
	\end{center}
	\caption{Principal stratification analysis: Posterior median of the distributional causal effects for switchers, $\DCE(y \mid s)$, for $s=0.25, 0.50, \ldots, 2.50, 2.75$} \label{Fig4app} 
\end{figure}

Figure~\ref{Fig4app} shows the posterior median of the distributional causal effects for switchers. 
The distributional causal effects are almost always positive, with an increasing trend over time for switchers who would switch to zidovudine early after the assignment. 
For switchers who would switch to zidovudine between 0.25 and 1.25 years after the assignment, the distributional causal effects are negative for early durations greater than the switching time, and become positive for later durations. 
The later the switching time, the longer the durations until which the distributional causal effects are negative. 
For instance, the distributional causal effects for patients who would switch to zidovudine $0.25$ years after the assignment are negative, ranging between $-0.021$ and $-0.009$, for few durations longer than 0.25 years (between 0.25 and about 0.3 years). 
The distributional causal effects for patients who would switch to zidovudine $1.25$ years after the assignment are negative, ranging between $-0.151$ and $-0.001$, for durations between 1.25 and about 2.37 years.
These results are, at least partially, driven by the natural constraint $S_i(0) < Y_i(0)$. 
Therefore, we need to interpret distributional causal effects for patients who would switch at a given time $s$, bearing in mind how such effects are defined. 
Indeed, a DCE is a comparison between the probability under assignment to immediate treatment with zidovudine that switchers at time $s$ will survive beyond any specified time, $y$, 
and the probability under assignment to deferred treatment with zidovudine that those switchers will survive beyond $y$, 
given that they have survived beyond the time-to-switching, $s$. 
It is then sensible that some distributional causal effects are negative also for $y\geq s$, especially for long switching durations; in fact, immediate versus deferred treatment with zidovudine should have a very strong effect for making these distributional effects positive. 

\subsection*{Sensitivity Analysis to $\kappa$} \label{sens_kappa}
 
We conduct a sensitivity analysis to $\kappa$, the partial association between $Y_i (1)$ and $Y_i(0)$ given the switching status, $S_i(0)$.
We derive the posterior distribution of the causal estimands for $\kappa=0, 0.25, 0.5, 0.75, 1$, using the same priors for other parameters as in Section~5.2 (see Web-Appendix~\ref{s:app1} for details).
Table \ref{tab4_Sens} and Figures~\ref{Fig_SensACE_S_kappa}-\ref{Fig_SensDCE_S_kappa} present the results.
Results display some sensitivity to $\kappa$. 

\begin{figure}
\begin{center}
\vspace{-0.3cm}
\begin{tabular}{cc}
\multicolumn{2}{c}{	$\kappa=0$ }  \vspace{-0.5cm}\\
  \multicolumn{2}{c}{\includegraphics[width=4.5cm]{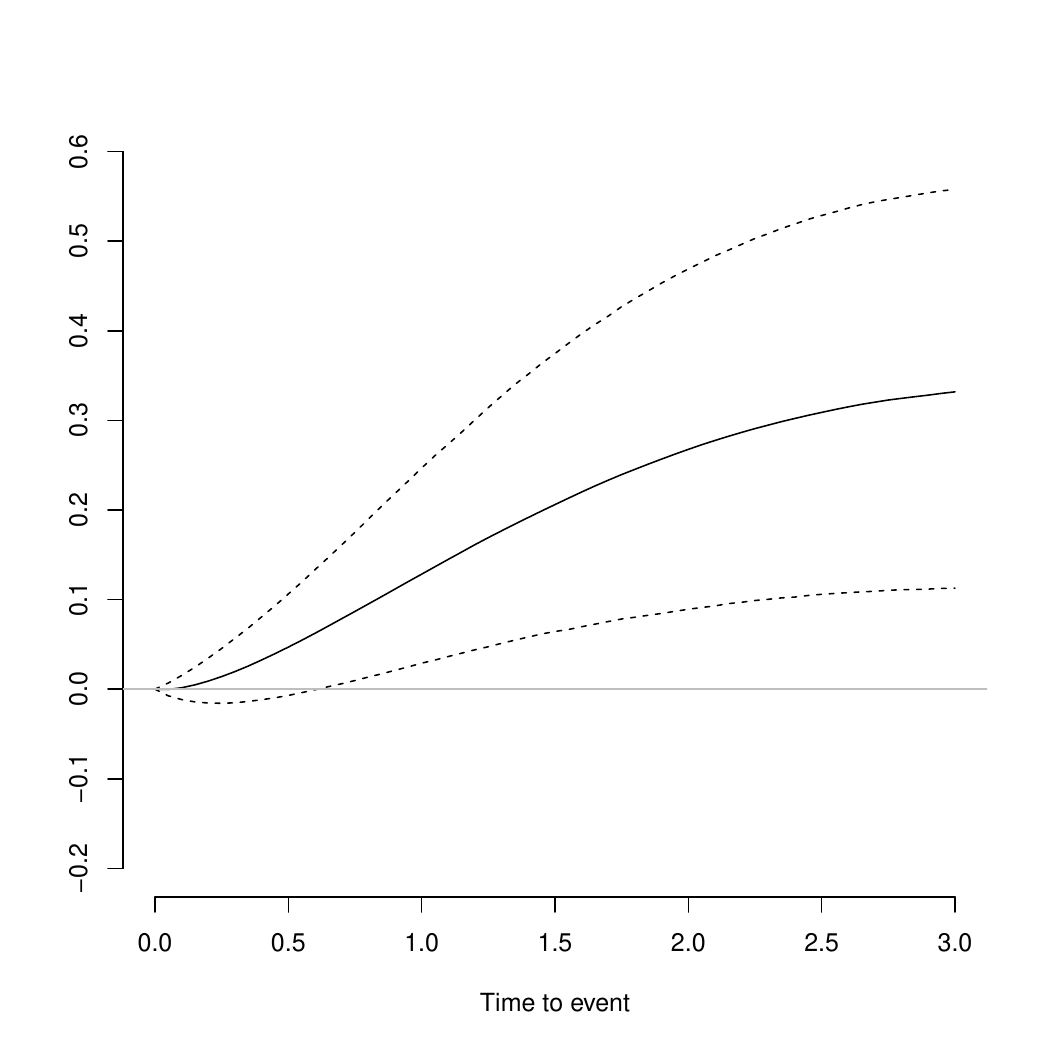}}\\
$\kappa=0.25$ & $\kappa=0.5$ \vspace{-0.5cm}\\
\includegraphics[width=4.5cm]{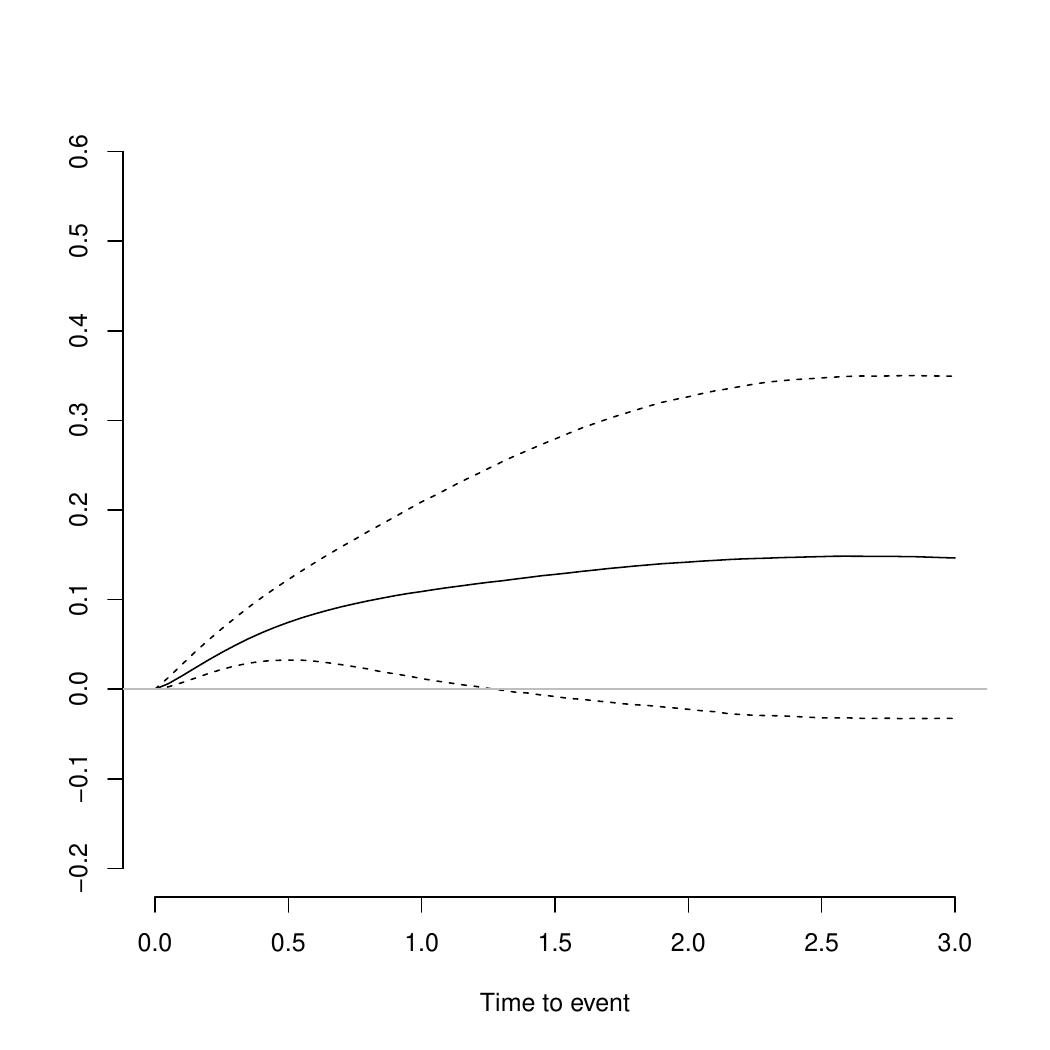}&
\includegraphics[width=4.5cm]{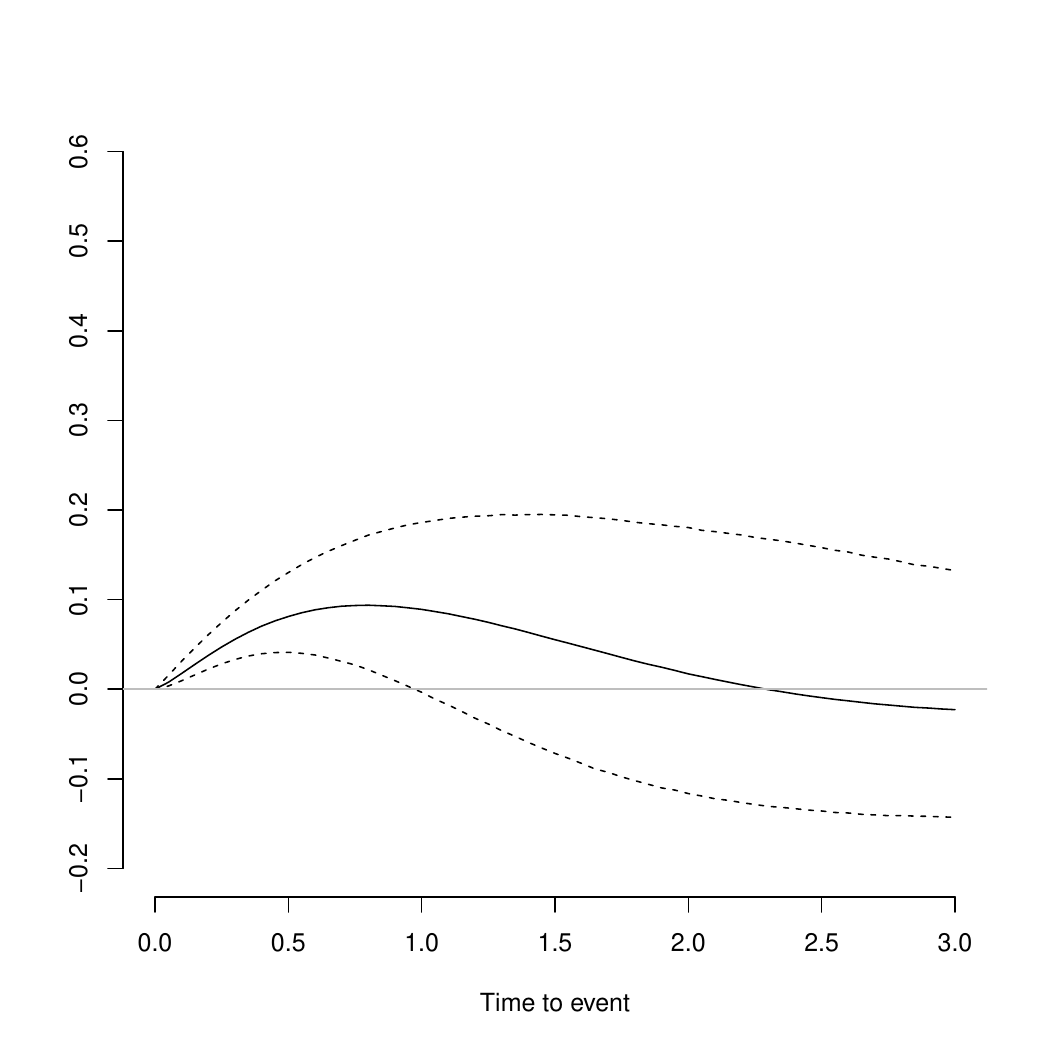}\\
$\kappa=0.75$ & $\kappa=1$\vspace{-0.5cm}\\
\includegraphics[width=4.5cm]{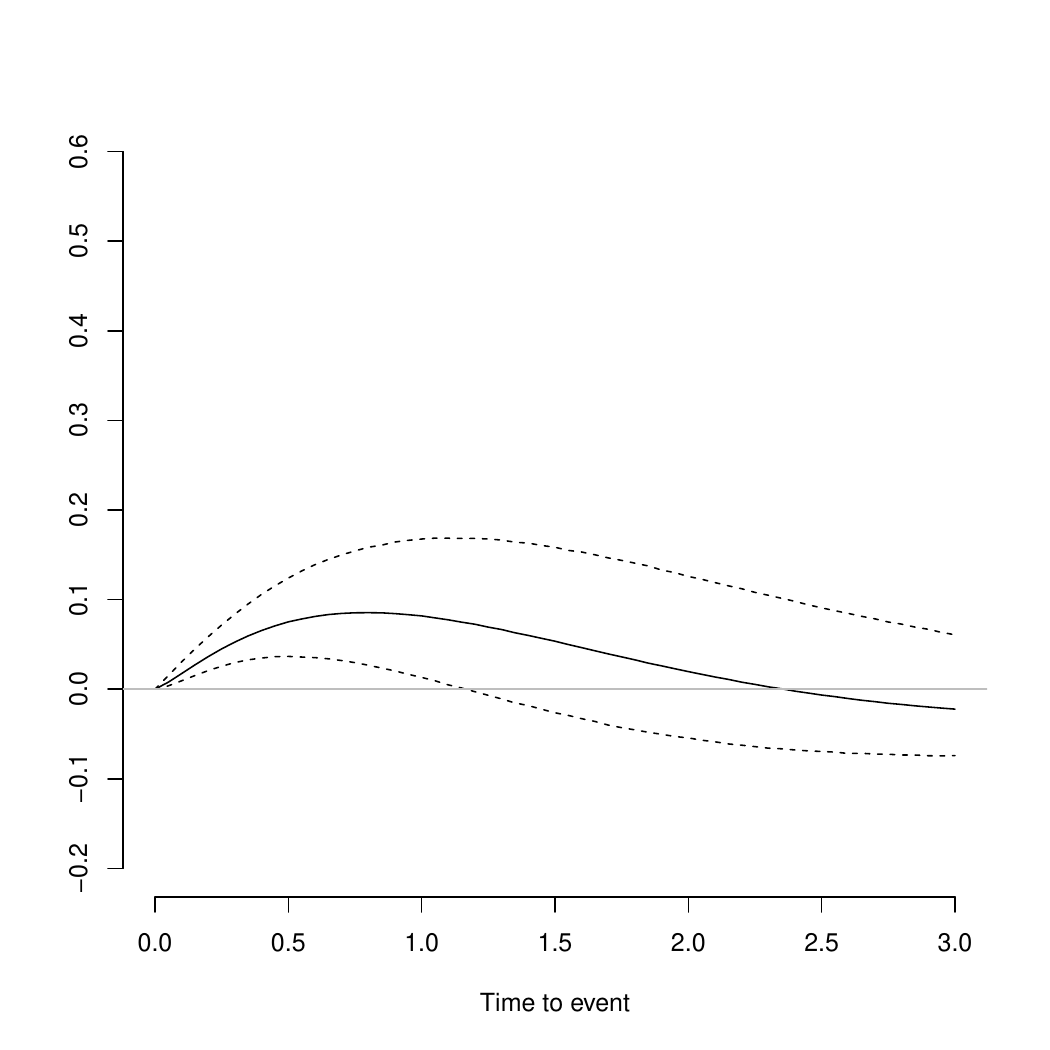} & 
\includegraphics[width=4.5cm]{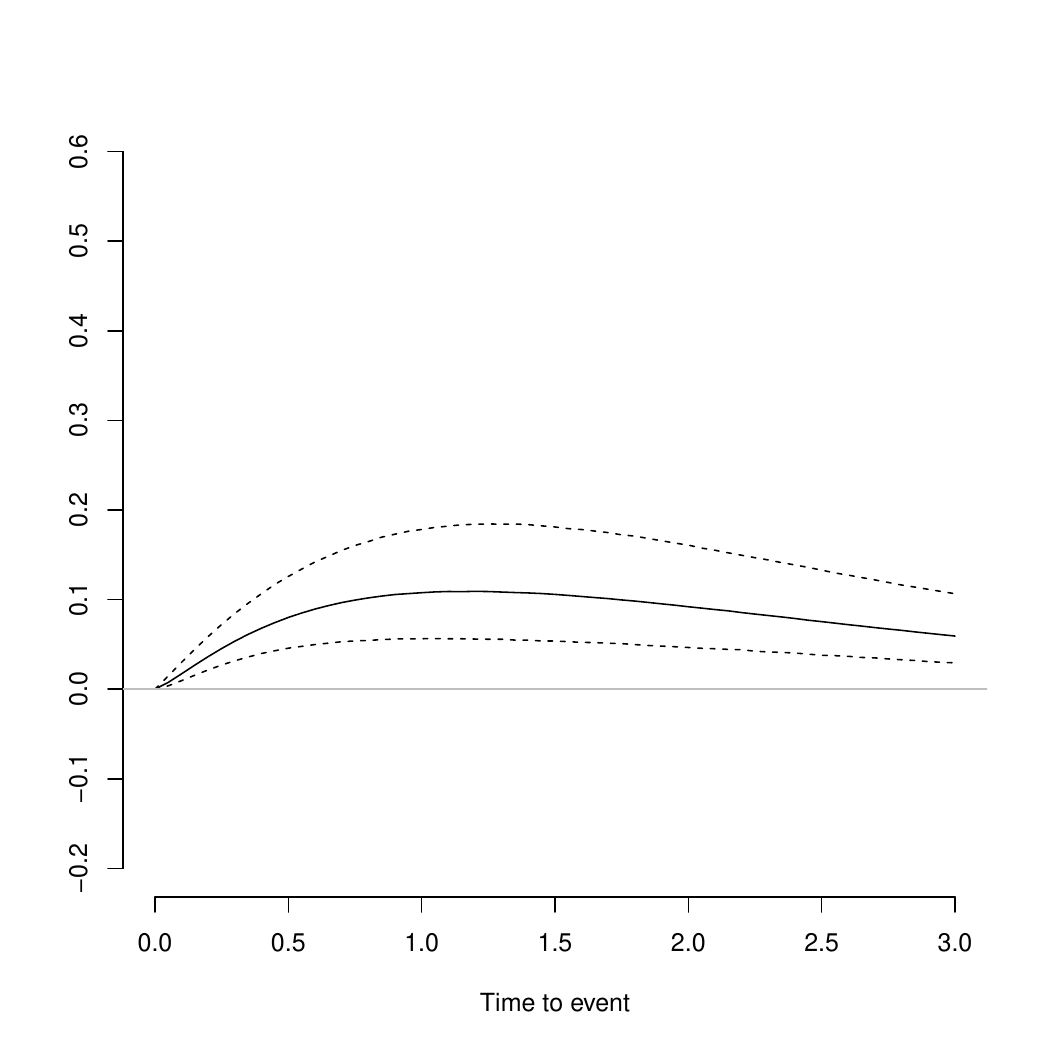}  \\       %
\end{tabular}
\end{center}
\caption{Principal stratification analysis:  Posterior median (solid line) and 95\% posterior credible interval (dashed lines) of distributional causal effects for never switchers, $\DCE(y \mid \bS)$, for different values of $\kappa$} \label{Fig_SensDCE_NS_kappa}
\end{figure}

\begin{figure}
	\begin{center}
		\vspace{-0.3cm}
		\begin{tabular}{cc}
			\multicolumn{2}{c}{	$\kappa=0$ }  \vspace{-0.5cm}\\
			\multicolumn{2}{c}{\includegraphics[width=4.5cm]{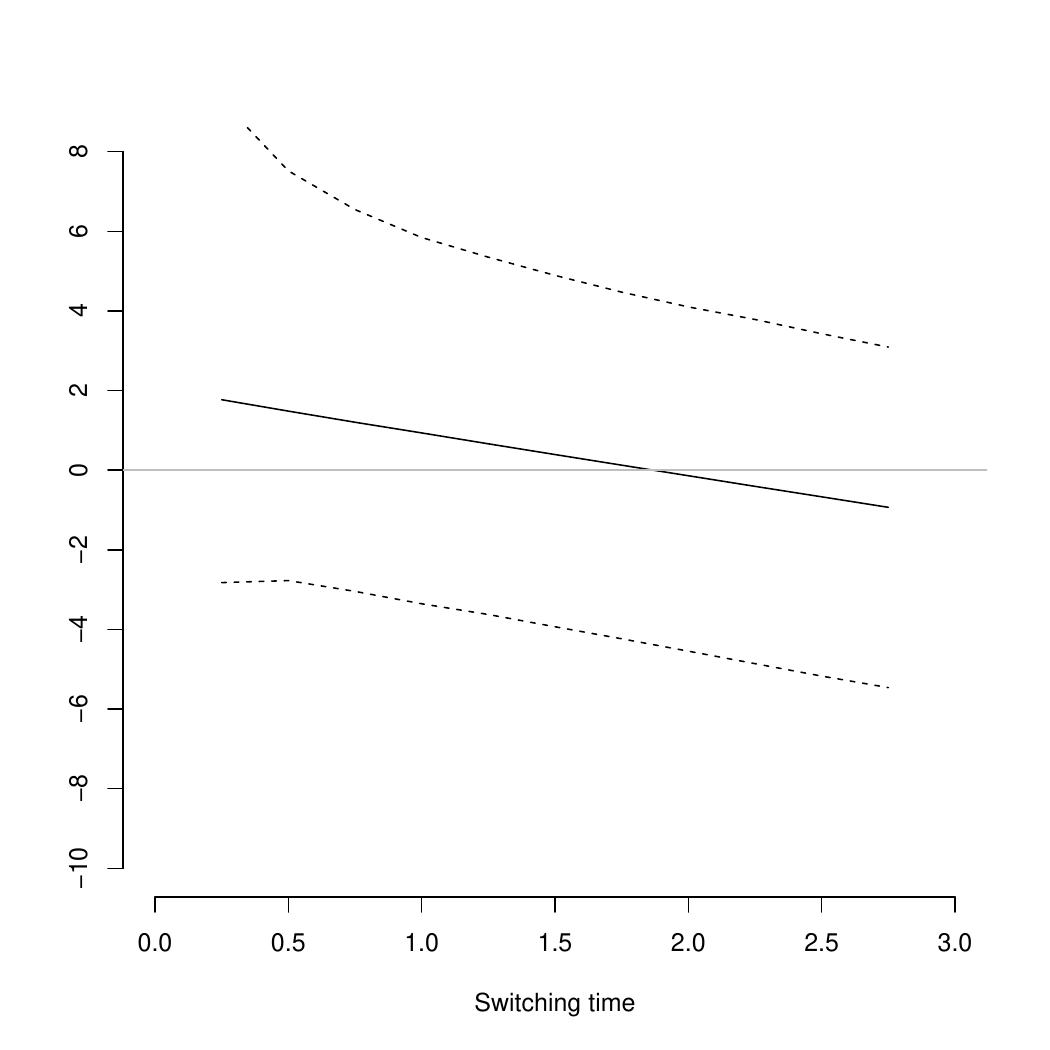}}\\
			$\kappa=0.25$ & $\kappa=0.5$ \vspace{-0.5cm}\\
			\includegraphics[width=4.5cm]{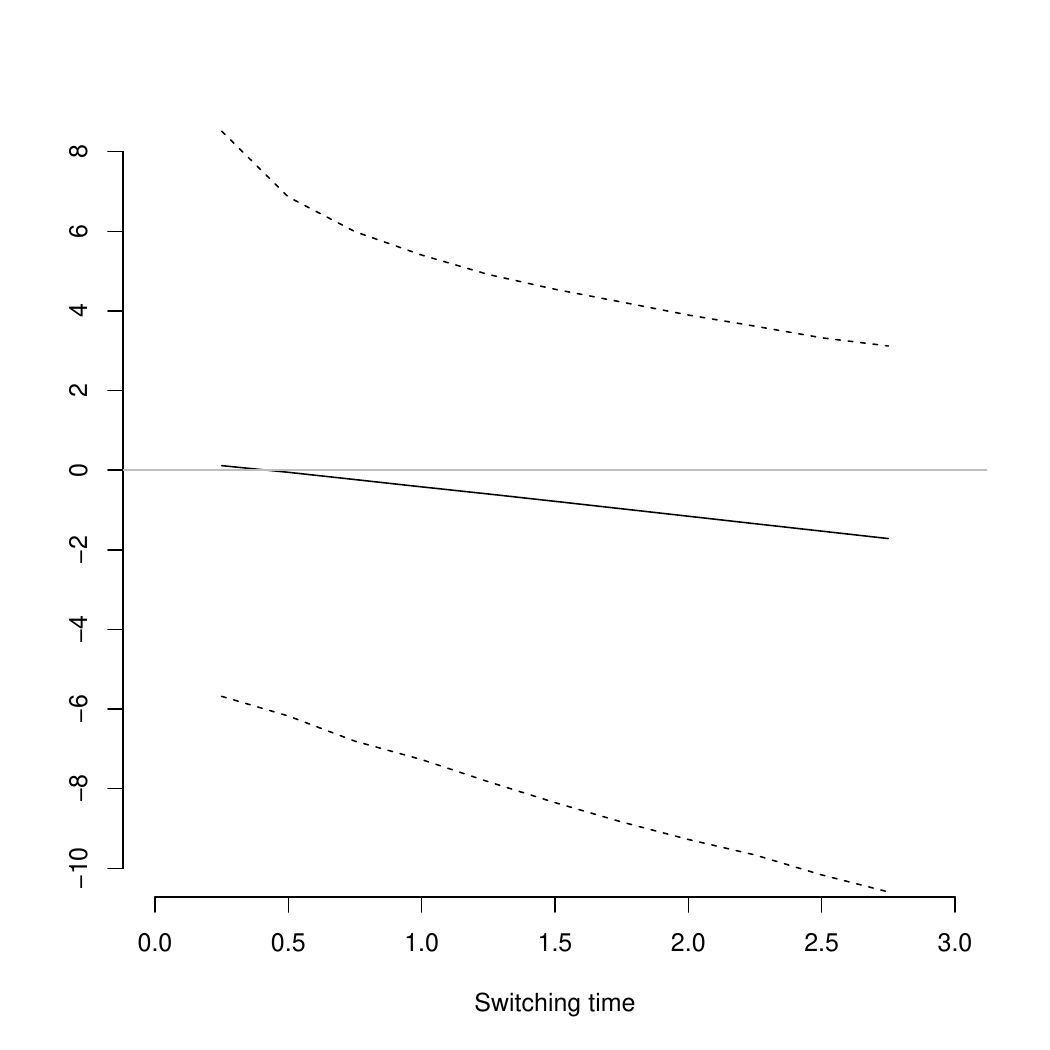}&
			\includegraphics[width=4.5cm]{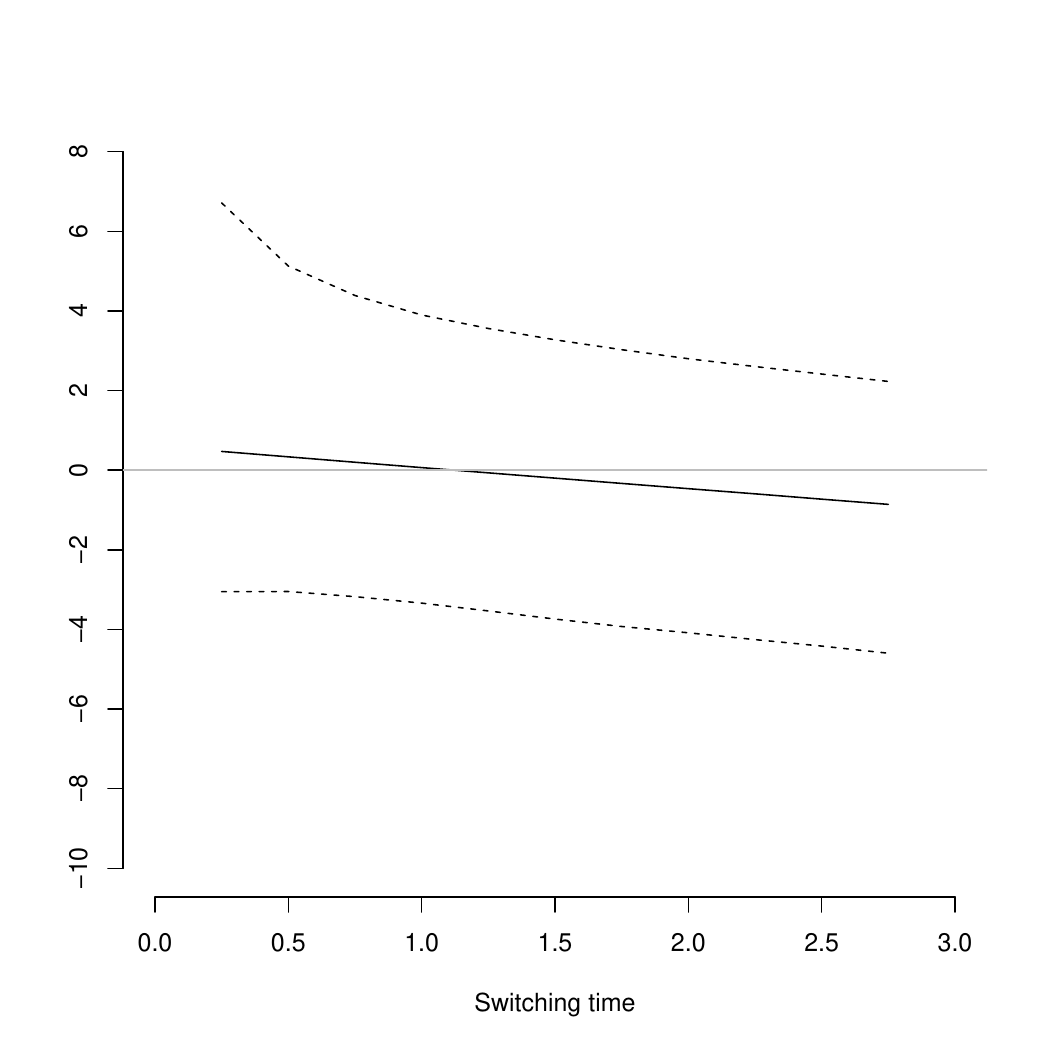}\\
			$\kappa=0.75$ & $\kappa=1$\vspace{-0.5cm}\\
			\includegraphics[width=4.5cm]{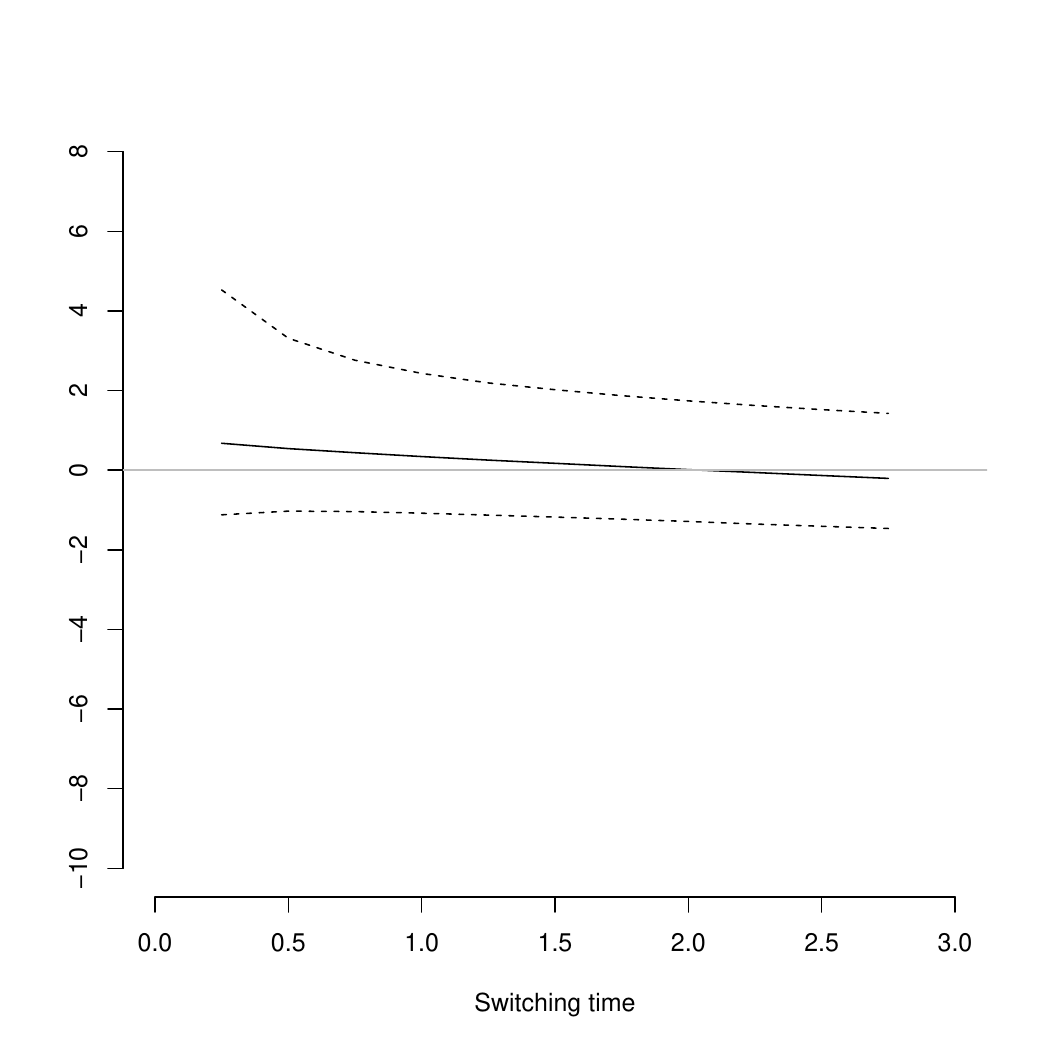} & 
			\includegraphics[width=4.5cm]{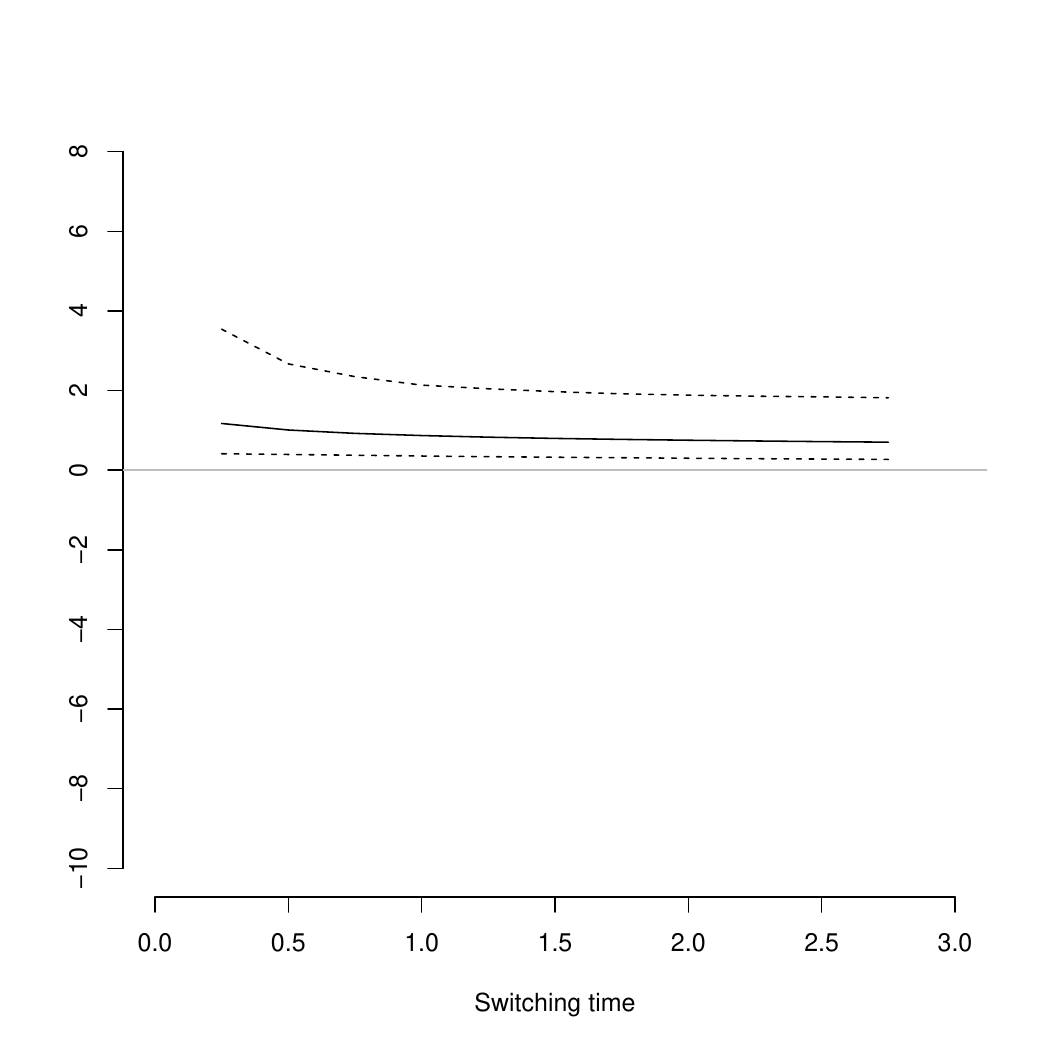}  \\       %
		\end{tabular}
	\end{center}
	\caption{Principal stratification analysis: Posterior median (solid line) and 95\% posterior credible interval (dashed lines) of average causal effects for switchers, $\ACE(s)$, $s \in\mathbb{R}_+$ for different values of $\kappa$} \label{Fig_SensACE_S_kappa}
\end{figure}

\begin{figure}
\begin{center}
\vspace{-0.3cm}
	\begin{tabular}{cc}
	\multicolumn{2}{c}{	$\kappa=0$ }  \vspace{-0.5cm}\\
\multicolumn{2}{c}{\includegraphics[width=4.5cm]{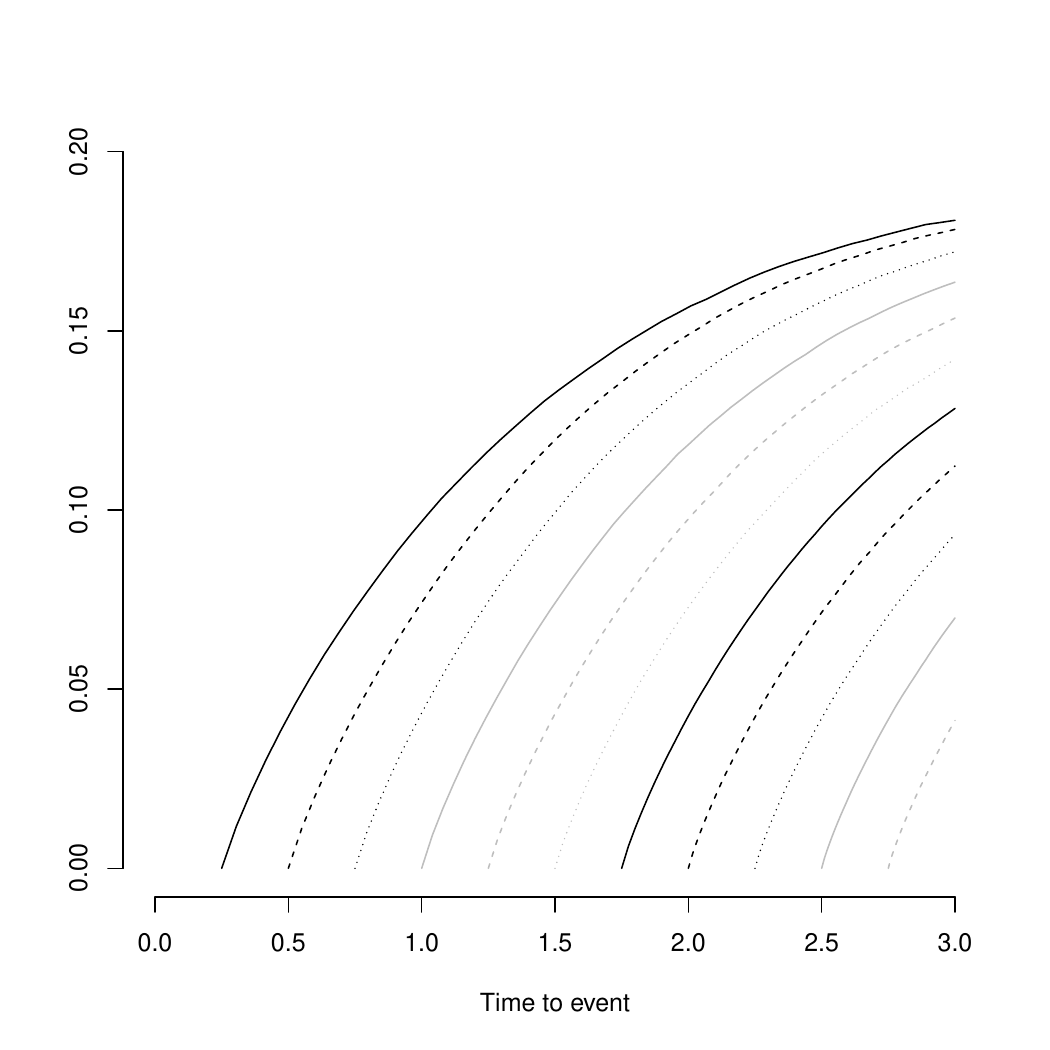}}\\
	$\kappa=0.25$ & $\kappa=0.5$ \vspace{-0.5cm}\\
	\includegraphics[width=4.5cm]{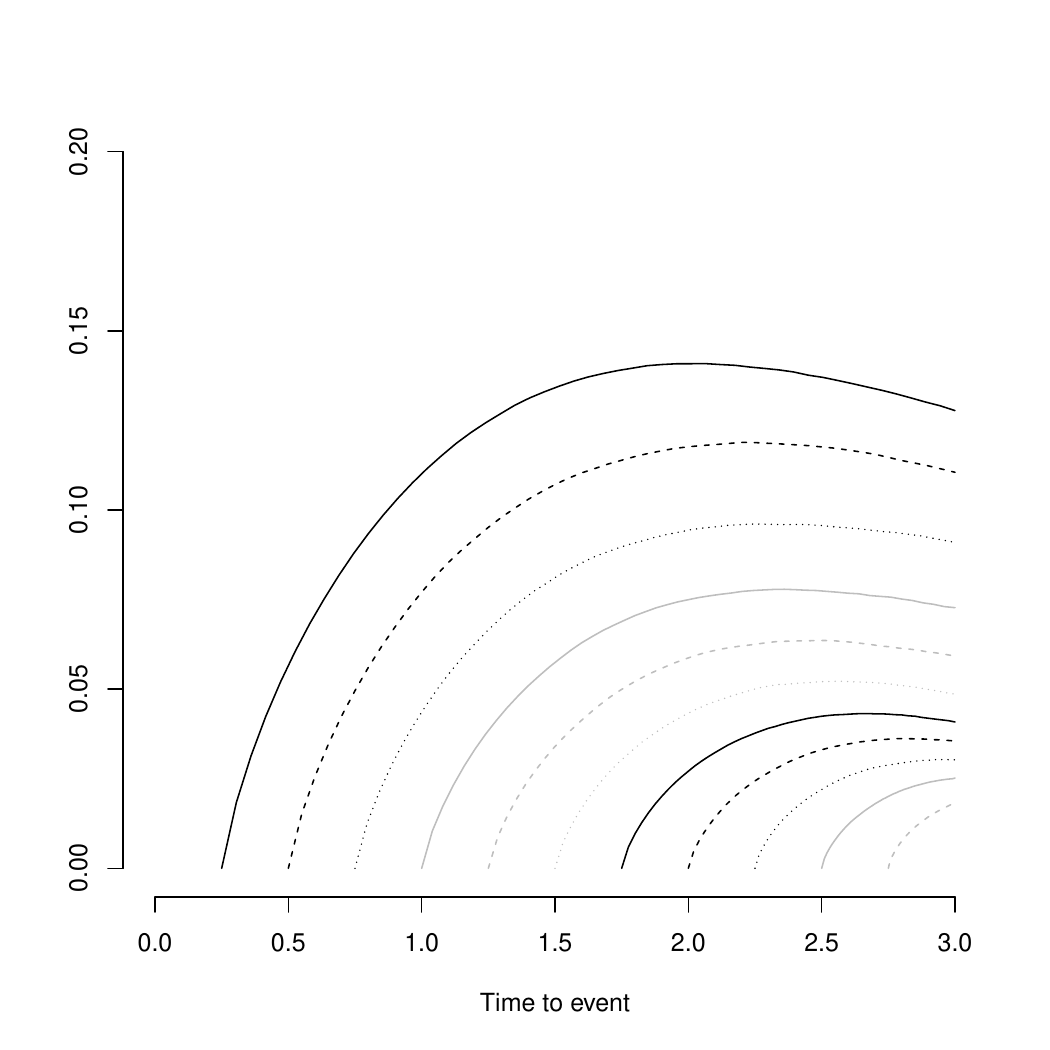}&
	\includegraphics[width=4.5cm]{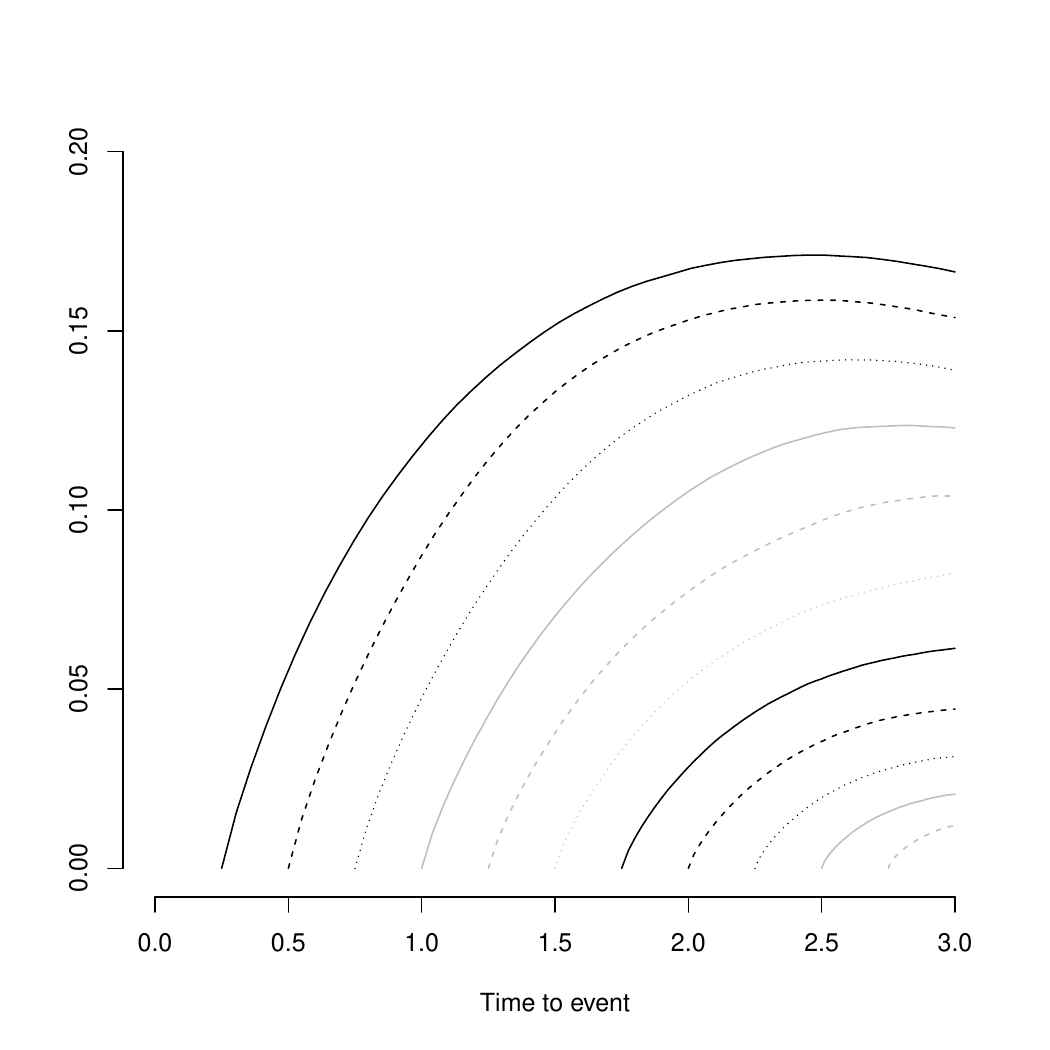}\\
	$\kappa=0.75$ & $\kappa=1$\vspace{-0.5cm}\\
	\includegraphics[width=4.5cm]{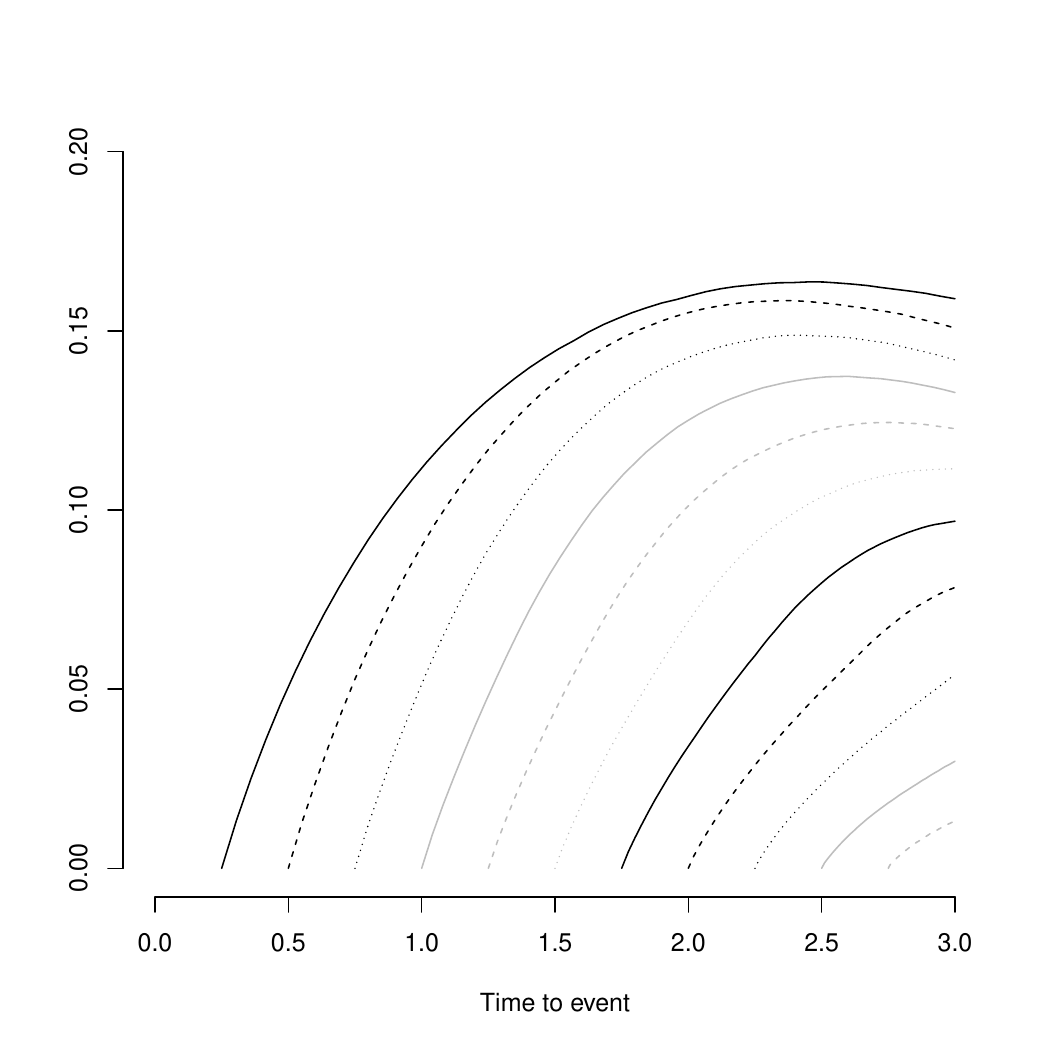} & 
	\includegraphics[width=4.5cm]{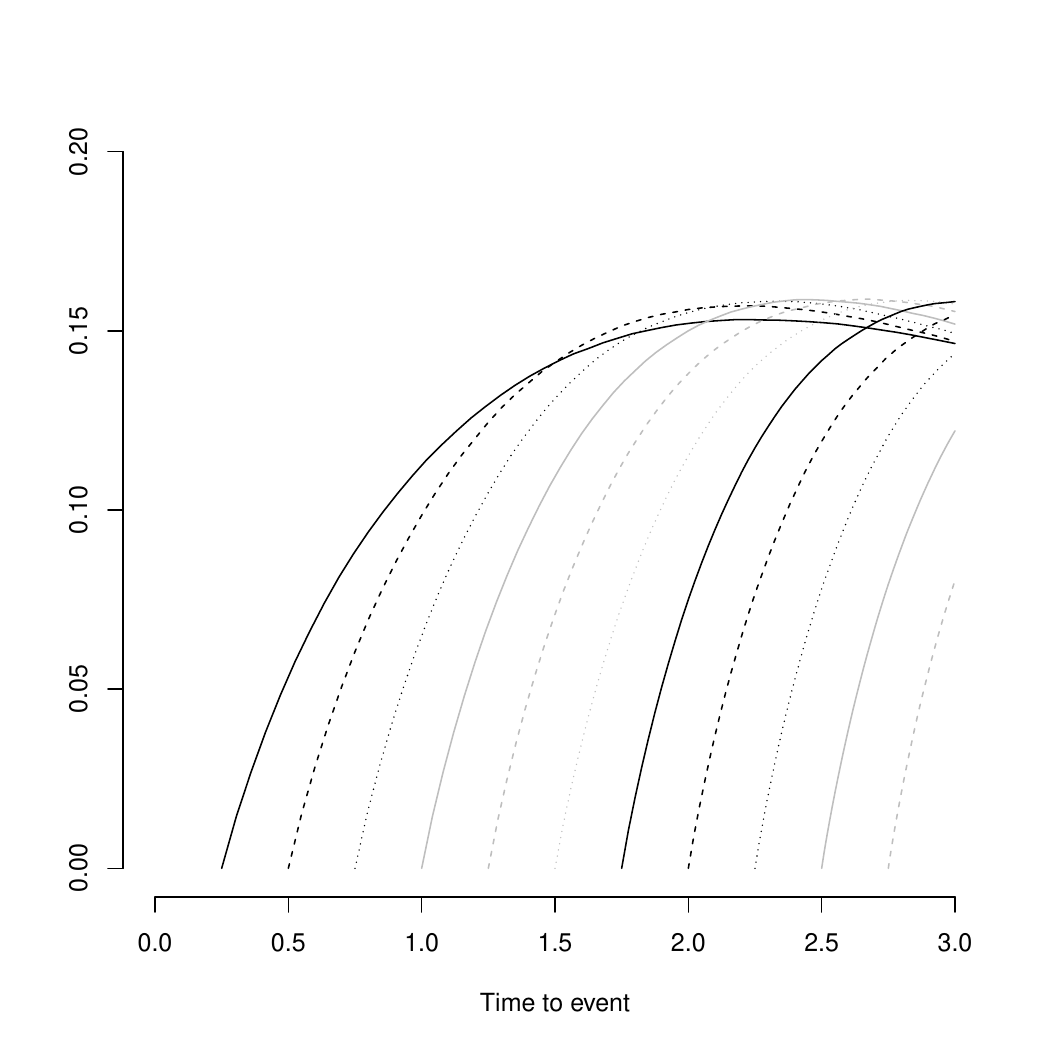}  \\       %
\end{tabular}
\end{center}
\caption{Principal stratification analysis: Posterior median of conditional distributional causal effects for switchers, $\cDCE(y \mid s)$, at time $s=0.25, 0.50, \ldots, 2.50, 2.75$ for different values of $\kappa$} \label{Fig_SensCDCE_S_kappa}
\end{figure}

\begin{figure}
	\begin{center}
		\vspace{-0.3cm}
		\begin{tabular}{cc}
			\multicolumn{2}{c}{	$\kappa=0$ }  \vspace{-0.5cm}\\
			\multicolumn{2}{c}{\includegraphics[width=4.5cm]{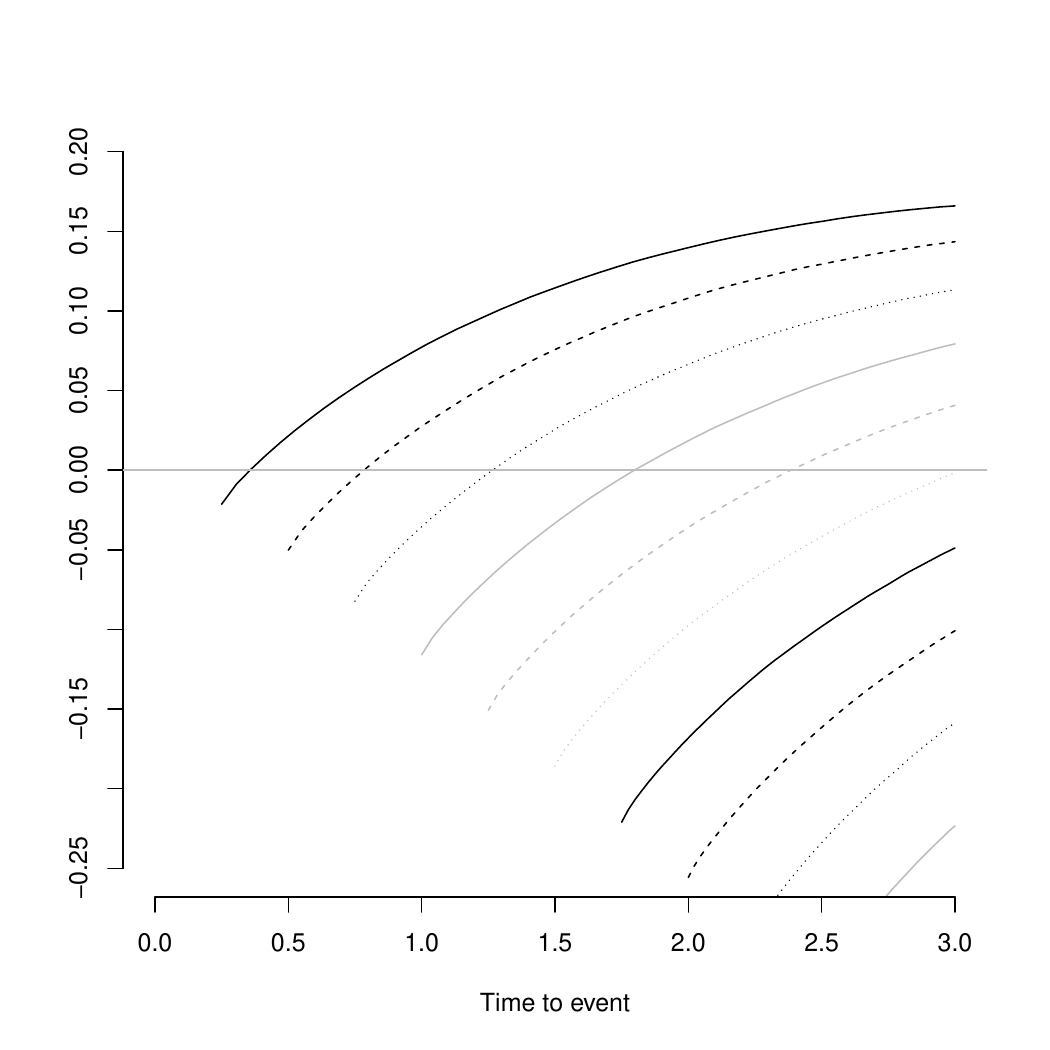}}\\
			$\kappa=0.25$ & $\kappa=0.5$ \vspace{-0.5cm}\\
			\includegraphics[width=4.5cm]{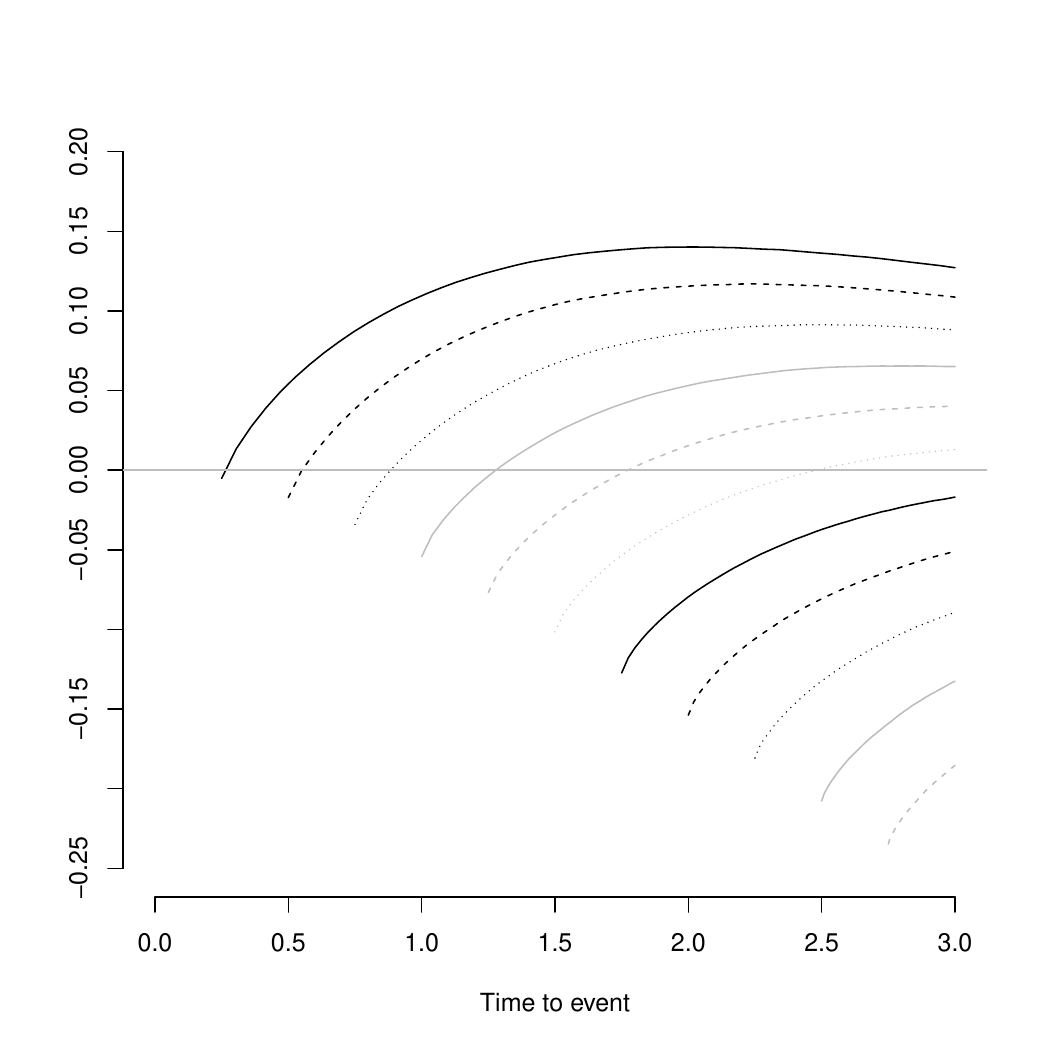}&
			\includegraphics[width=4.5cm]{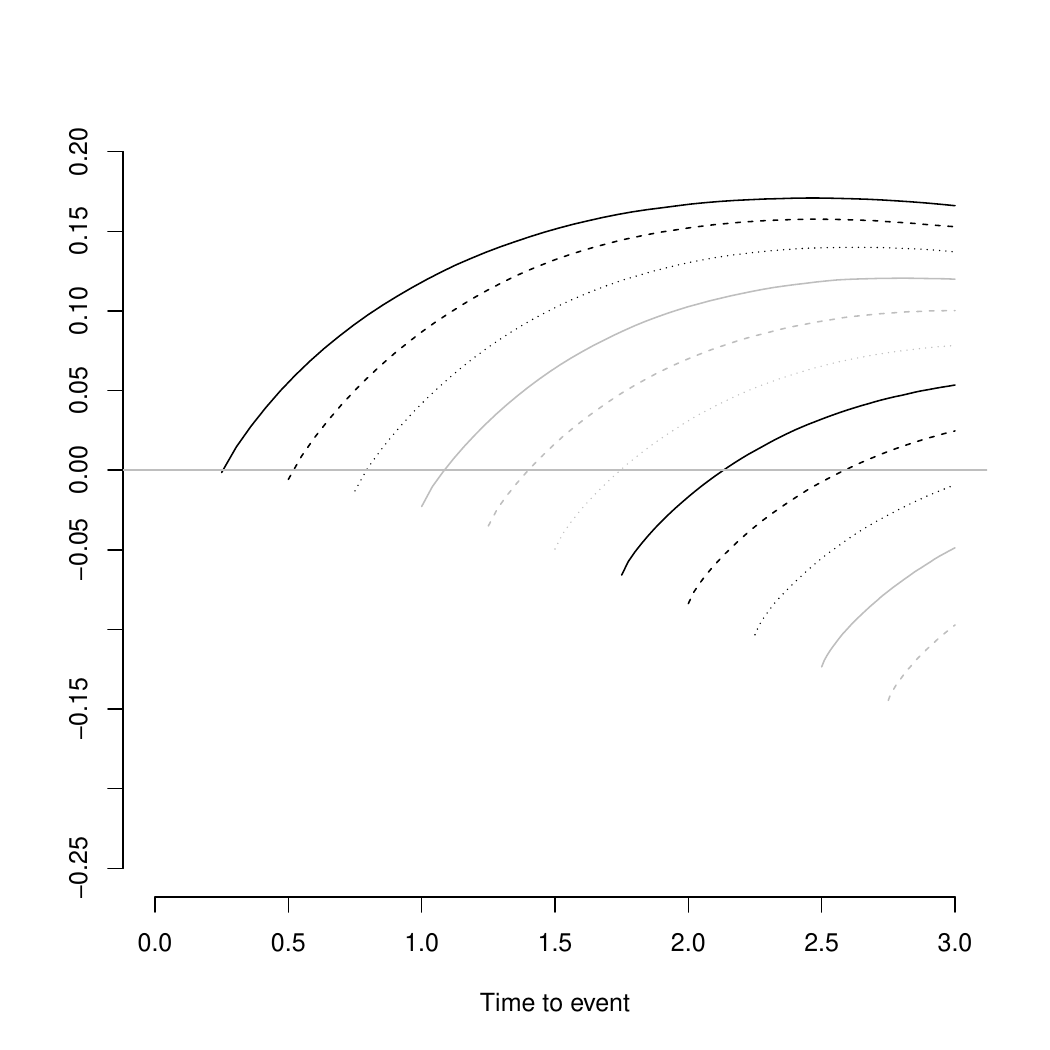}\\
			$\kappa=0.75$ & $\kappa=1$\vspace{-0.5cm}\\
			\includegraphics[width=4.5cm]{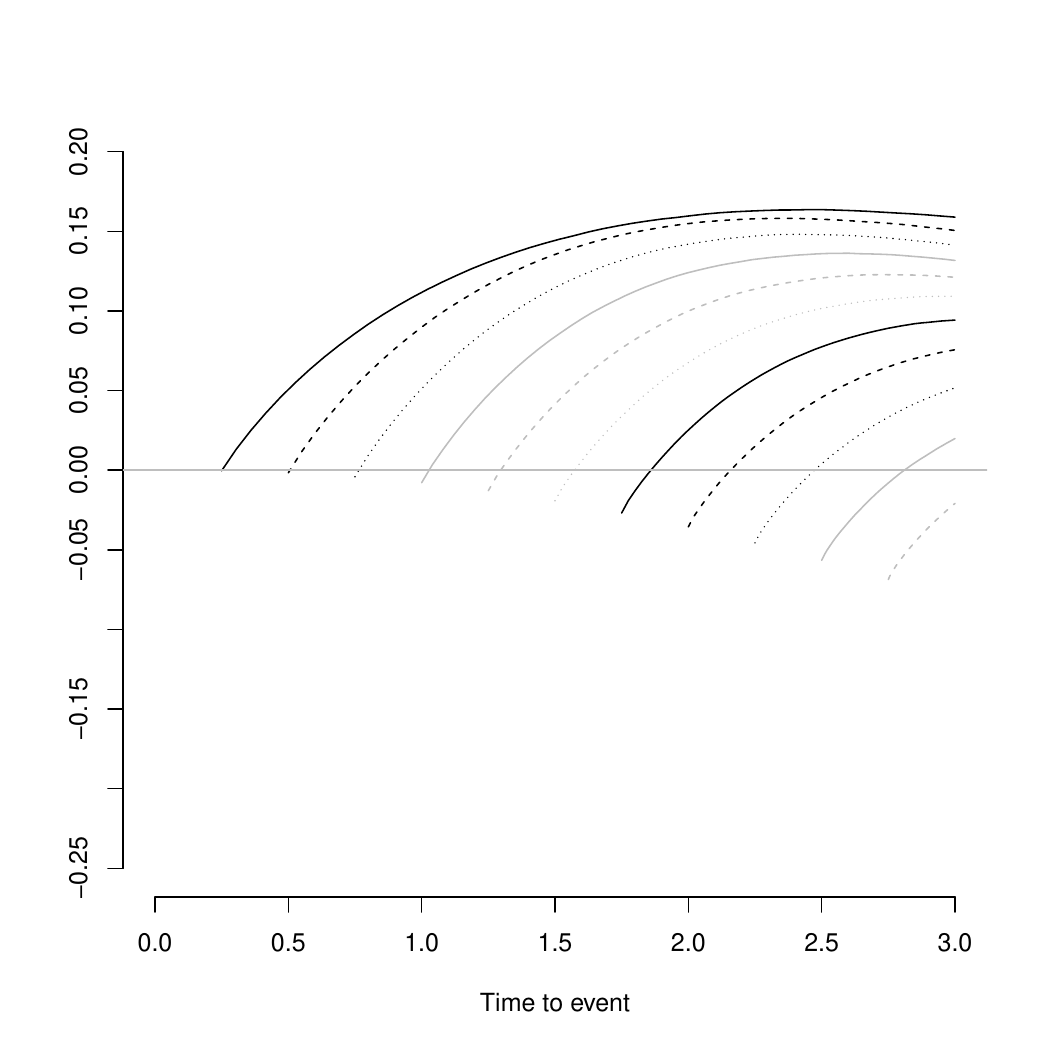} & 
			\includegraphics[width=4.5cm]{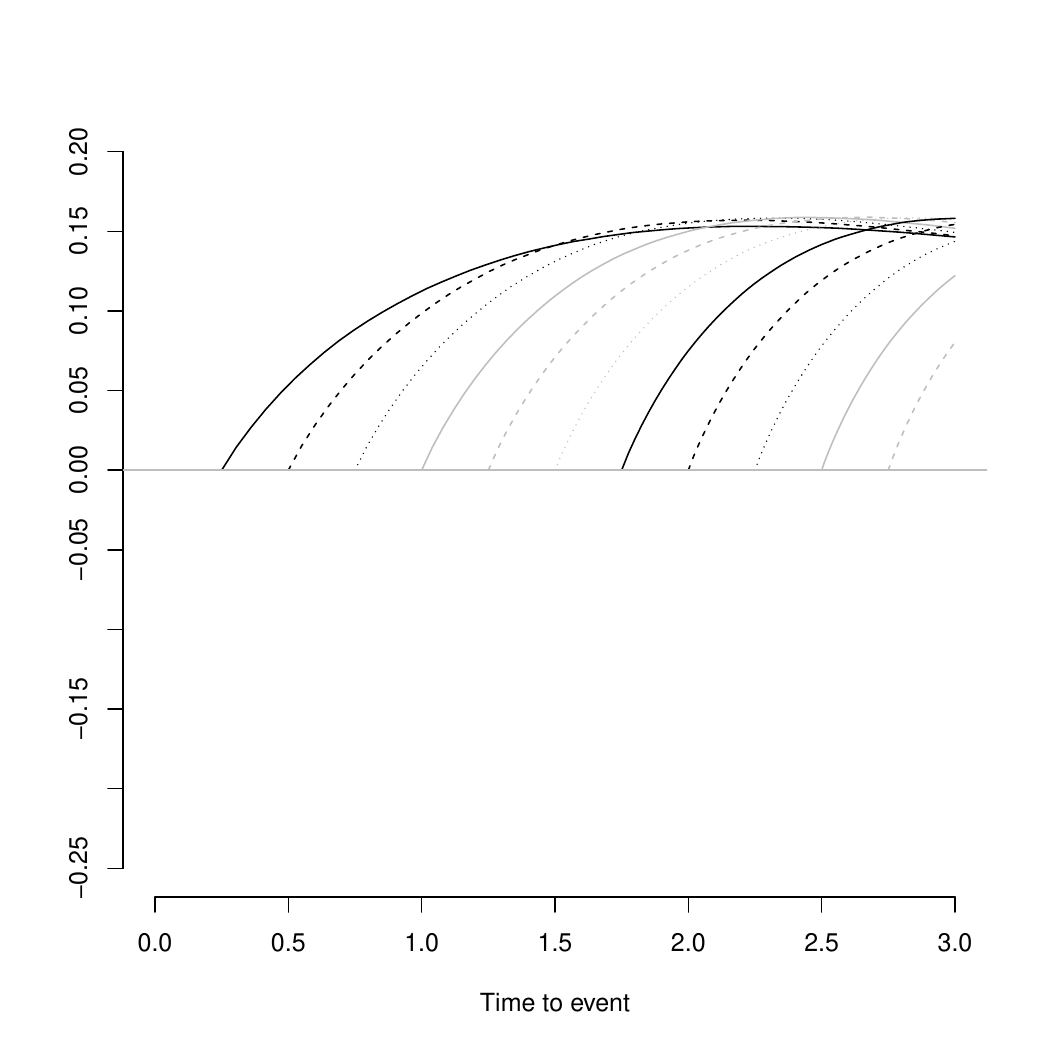}  \\       %
		\end{tabular}
	\end{center}
	\caption{Principal stratification analysis: Posterior median of distributional causal effects for switchers, $\DCE(y \mid s)$, at time $s=0.25, 0.50, \ldots, 2.50, 2.75$ for different values of $\kappa$} \label{Fig_SensDCE_S_kappa}
\end{figure}

In Table \ref{tab4_Sens} and Figure~\ref{Fig_SensDCE_NS_kappa}, the posterior distributions of the average causal effects and the distributional causal effects for non-switchers suggest that the evidence in favor of beneficial effects of immediate versus deferred treatment with zidovudine on survival time for never-switchers weakens when the assumption of independence between the potential survival outcomes, $Y_i(0)$ and $Y_i(1)$ (i.e., $\kappa=0$) is relaxed allowing for values of $\kappa$ greater than zero.
For $\kappa=1$ (which implies monotonicity, i.e., $Y_i(1) \geq Y_i(0) $), we still find evidence that immediate versus deferred treatment with zidovudine increases survival time for non-switchers, both on average and over time. However, the posterior distributions of the average causal effect and the distributional causal effects for non-switchers are centered on smaller values and have smaller posterior variances than those obtained for $\kappa=0$, leading to tighter 95\% posterior credible intervals.
It is also worth noting that the distributional causal effects, $DCE(y \mid \bS)$, show a different time trend for $\kappa=1$ than for $\kappa=0$: 
For $\kappa=1$ they increase over time from 0 to 0.109 up to $y=1.25$ years and then start to decrease, although they are always positive with rather tight 95\% posterior credible intervals including only positive values.
For $\kappa =0.25$, the posterior distributions of the average causal effect and the distributional causal effects for non-switchers are still centered on positive values. 
Still, they have a rather large posterior variability, leading to 95\% posterior credible intervals that cover zero except for the 95\% posterior credible intervals for distributional causal effects, $DCE(y \mid \bS)$ for $y\leq 1.25$.
For $\kappa =0.5, 0.75$, the posterior medians of the average causal effects for non-switchers are very close to zero, and the 95\% posterior credible intervals cover zero. 
Therefore, there is no evidence that immediate versus deferred treatment with zidovudine increases survival time for non-switchers on average. 
For non-switchers, we find positive and statistically significant, even if small, distributional causal effects for times to event $y\leq 0.95$ and $y\leq1.15$, respectively, for $\kappa =0.5$ and $\kappa=0.75$.
Then, distributional causal effects start to decrease, also reaching negative values for $y\geq 2.30$ ($\kappa =0.5$) and $y\geq 2.40$
($\kappa=0.75$); however, they are statistically negligible with 95\% posterior credible intervals always covering zero.

Figure~\ref{Fig_SensACE_S_kappa} shows that the estimates of the average causal effects for switchers are statistically negligible as those we obtained for $\kappa=0$ for $\kappa=0.25,0.5, 0.75$.
Instead, we find evidence that immediate versus deferred treatment with zidovudine increases the average survival time for switchers irrespective of the time to switching for $\kappa=1$, under which monotonicity $Y_i(1) \geq Y_i(0) $ holds.

\begin{table}\caption{Principal stratification analysis: Posterior median and 95\% posterior credible interval for causal estimands for non-switchers for different values of $\kappa$ } \label{tab4_Sens}  
$$
\begin{array}{l rl rl rc}
\hline
\vspace{-0.25cm}\\
\kappa &\multicolumn{2}{c}{\bE[Y_i(0)\mid S_i(0)=\bS]}& \multicolumn{2}{c}{\bE[Y_i(1)\mid S_i(0)=\bS]} &
\multicolumn{2}{c}{\ACE(\bS)}\\
\vspace{-0.3cm}\\
\hline \vspace{-0.3cm}\\
\kappa =0    & 2.05 & (1.44; 2.99)   & 4.76 & (2.80; 9.80)    & 2.66& (0.71; \ \ 7.73) \\
\kappa =0.25 & 1.87 & (1.38; 2.88)  & 2.78 & (1.87; 4.58)     & 0.86&  (-0.09;  2.49) \\
\kappa =0.50 & 1.91 & (1.37; 3.09)  & 1.98 & (1.46; 2.78)     & 0.02&  (-0.70; 0.72) \\
\kappa =0.75 & 2.10 & (1.48; 3.15)  & 2.11 & (1.59; 2.94)     & -0.01&  (-0.36; 0.40) \\
\kappa =1    & 2.06 & (1.47; 2.96)  & 2.43 & (1.81; 3.35)     & 0.36&  (0.19; \ \ 0.59) \\
\hline
\end{array}
$$
\end{table}

Figure~\ref{Fig_SensCDCE_S_kappa} compares the posterior medians of $\cDCE(y \mid s)$ for $s=0.25, 0.50, \ldots,$ $2.50, 2.75$. 
From Figure~\ref{Fig_SensCDCE_S_kappa}, the posterior medians of $\cDCE(y \mid s)$ show a tread increase throughout the years at $\kappa=0$, but have an asymmetrical inverted U-shape skewed to the right at $\kappa\in \{ 0.25, 0.5, 0.75, 1\}$, at least for  switchers who would switch relatively soon.

Figure~\ref{Fig_SensDCE_S_kappa} compares the posterior medians of $\DCE(y \mid s)$  for$s=0.25, 0.50, \ldots, 2.50,$ $2.75$.
Note that at $\kappa=1$, $\DCE(y \mid s)=\cDCE(y \mid s)$.
Two major patterns appear in the posterior medians of $\DCE(y \mid s)$. 
First, the posterior medians are negative for some durations greater than the switching time both at $\kappa=0$ and $\kappa\in \{ 0.25, 0.5 , 0.75\} $, but the posterior medians of $\DCE(y \mid s)$ at $\kappa \in \{ 0.25, 0.5, 0.75\}$ turn to be positive at earlier durations. 
Second, the posterior medians for switchers who would switch to zidovudine early after the assignment show an increasing trend over time at $\kappa=0$, whereas those derived at $\kappa\in \{ 0.25, 0.5, 0.75,1\}$ follow an asymmetrical inverted U-shape skewed to the right.

\subsection*{Sensitivity Analysis to the Prior Distribution for $\lambda$}
\begin{figure}
\begin{center}
\begin{tabular}{cc}
	$\lambda \sim N(0, 1)$ & $\lambda \sim N(0, 10)$     \vspace{-0.35cm}\\
	\includegraphics[width=4.75cm]{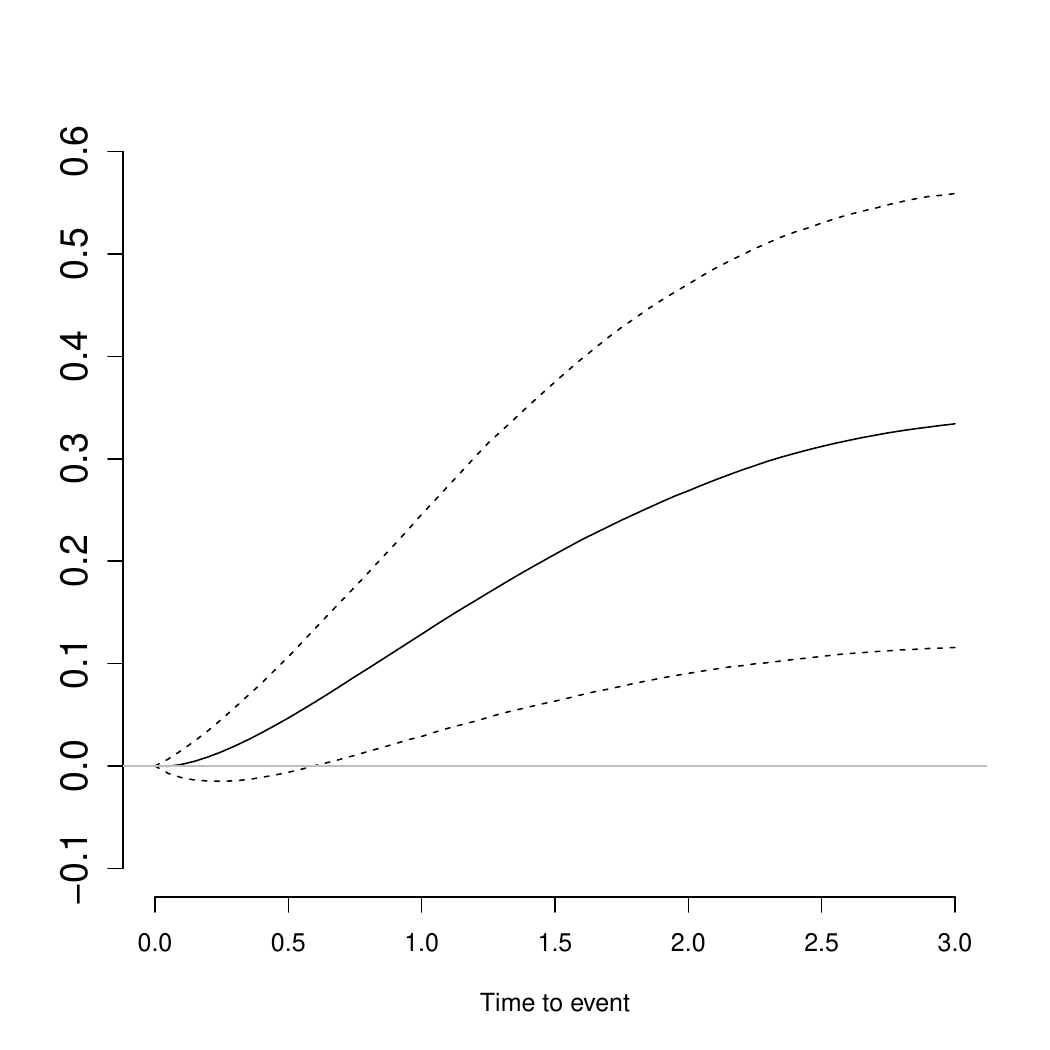}&
	\includegraphics[width=4.75cm]{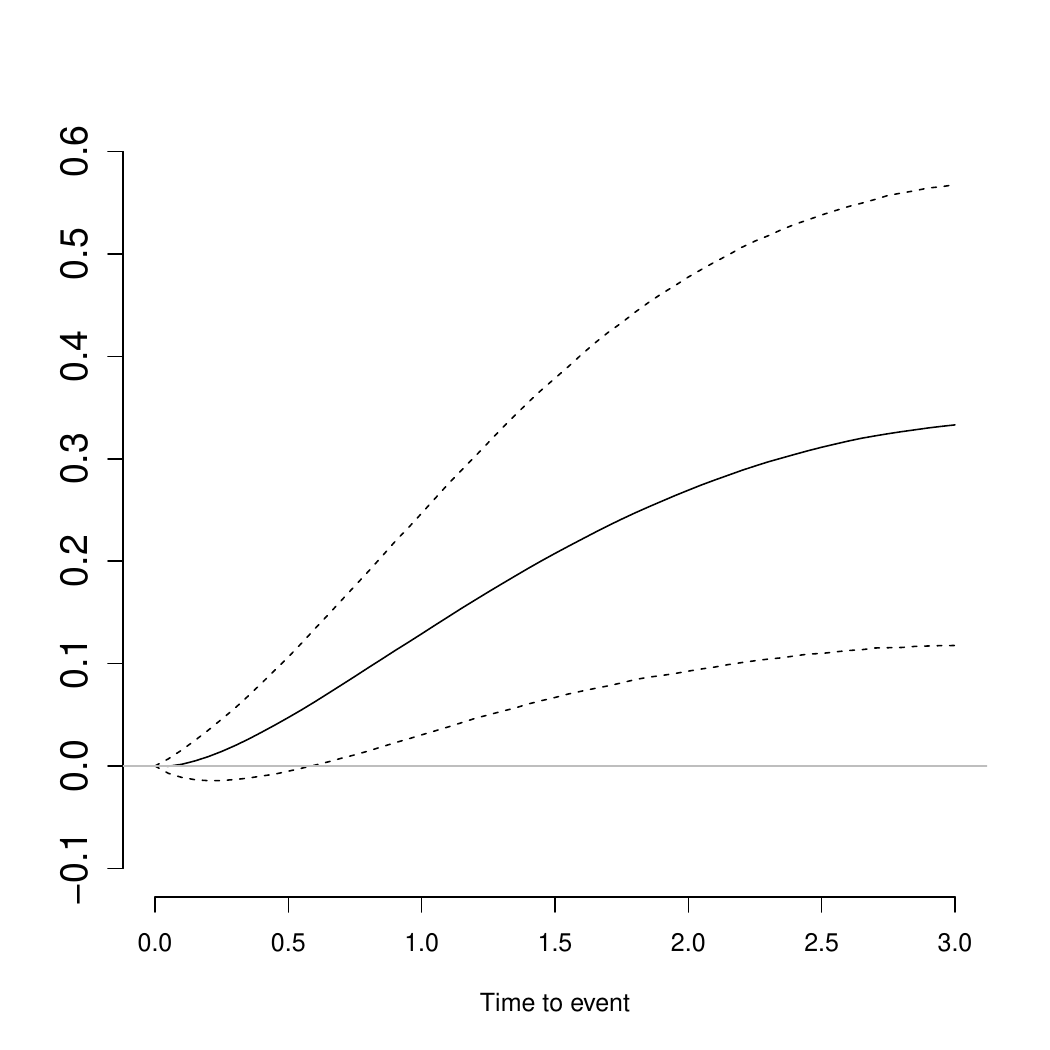}\\
	  $\lambda \sim N(0, 10\,000)$ &$\lambda \sim \hbox{Uniform}(\R)$     \vspace{-0.35cm}\\
	\includegraphics[width=4.75cm]{Figures/PS_Analysis/Fig3_DCE_NS.pdf}&  
	\includegraphics[width=4.75cm]{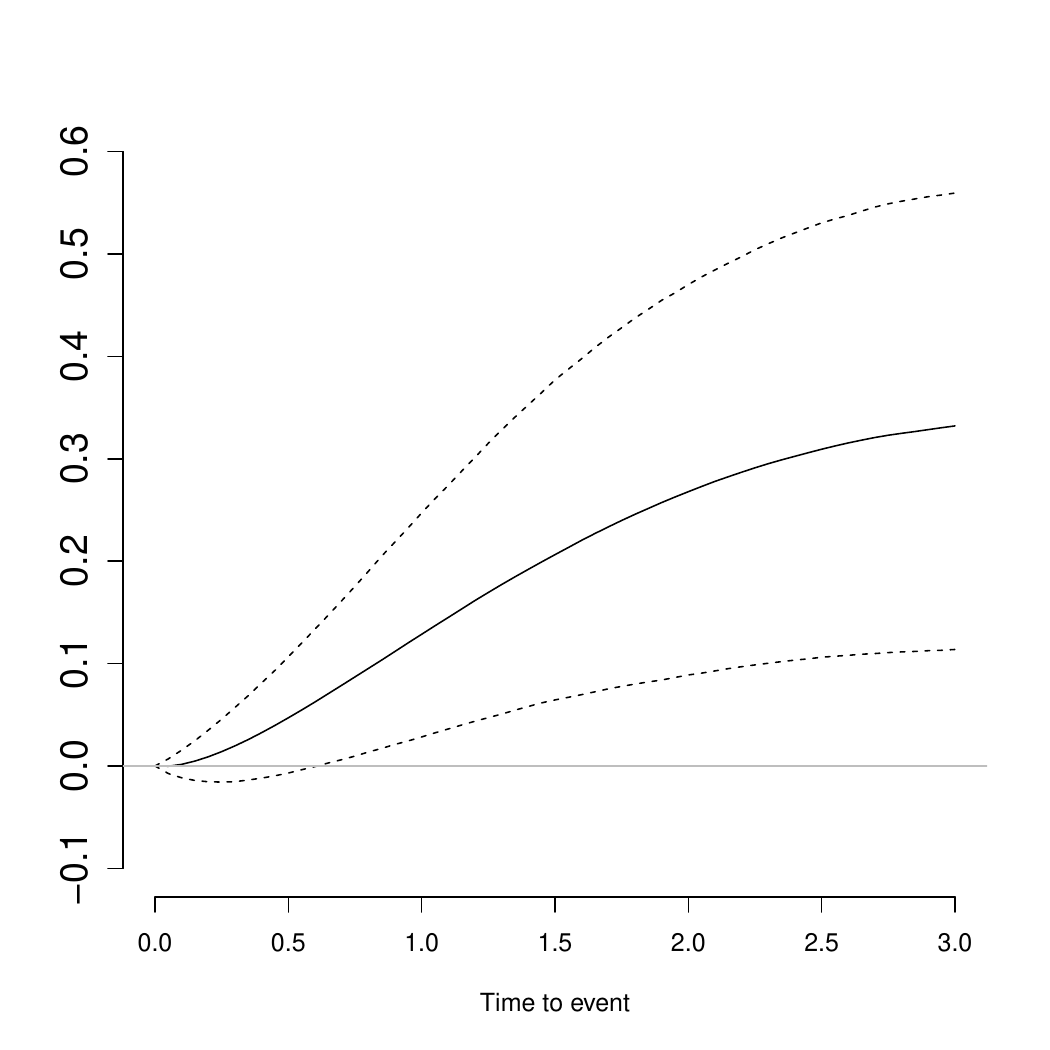} \\
\end{tabular}
\end{center}
\caption{Principal stratification analysis: Posterior median (solid line) and 95\% posterior credible interval (dashed lines) of distributional causal effects for never switchers, $\DCE(y \mid \bS)$, for different prior distributions for $\lambda$ with $\kappa=0$} \label{Sens_Lambda_DCE_NS}
\end{figure}

\begin{figure}
\begin{center}
\begin{tabular}{cc}
$\lambda \sim N(0, 1)$ & $\lambda \sim N(0, 10)$  \vspace{-0.35cm}\\
\includegraphics[width=4.75cm]{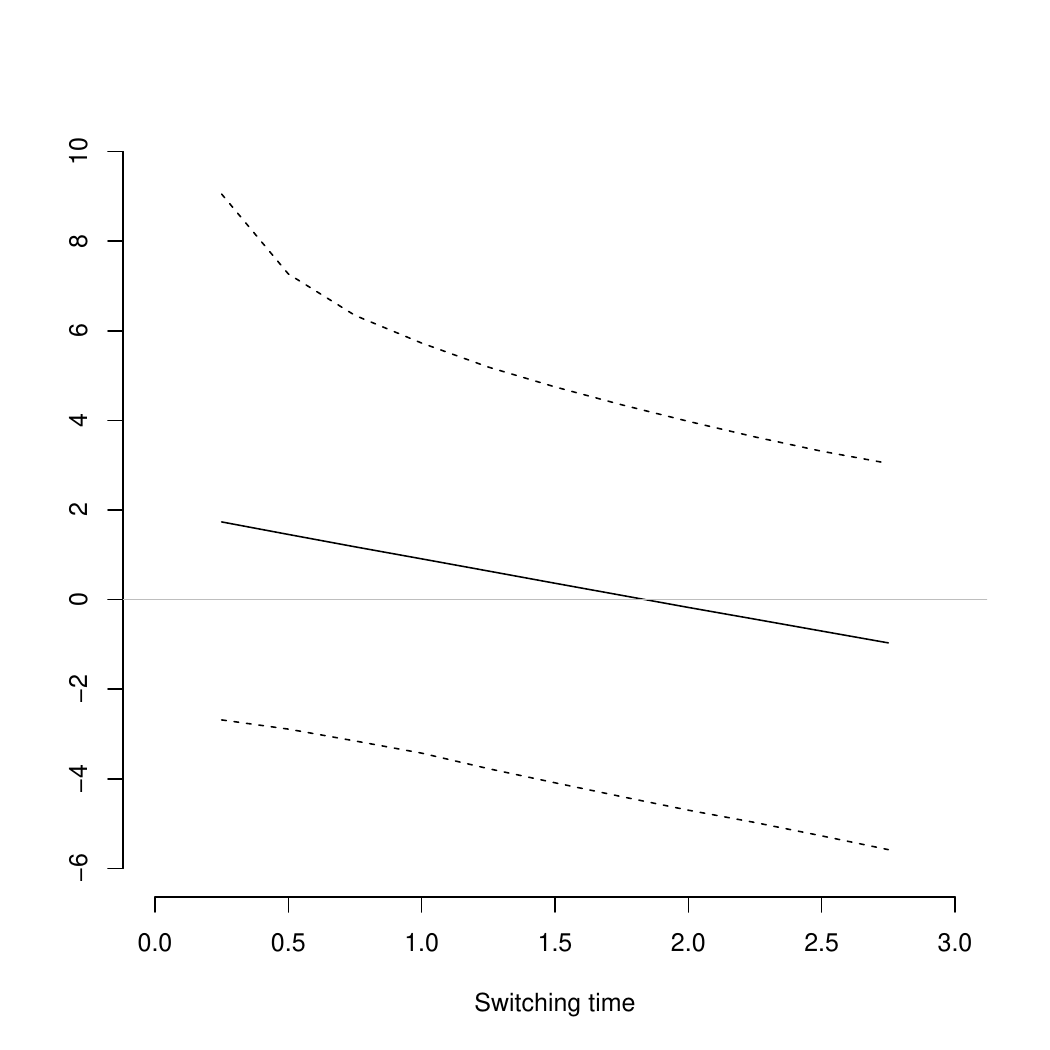}&
\includegraphics[width=4.75cm]{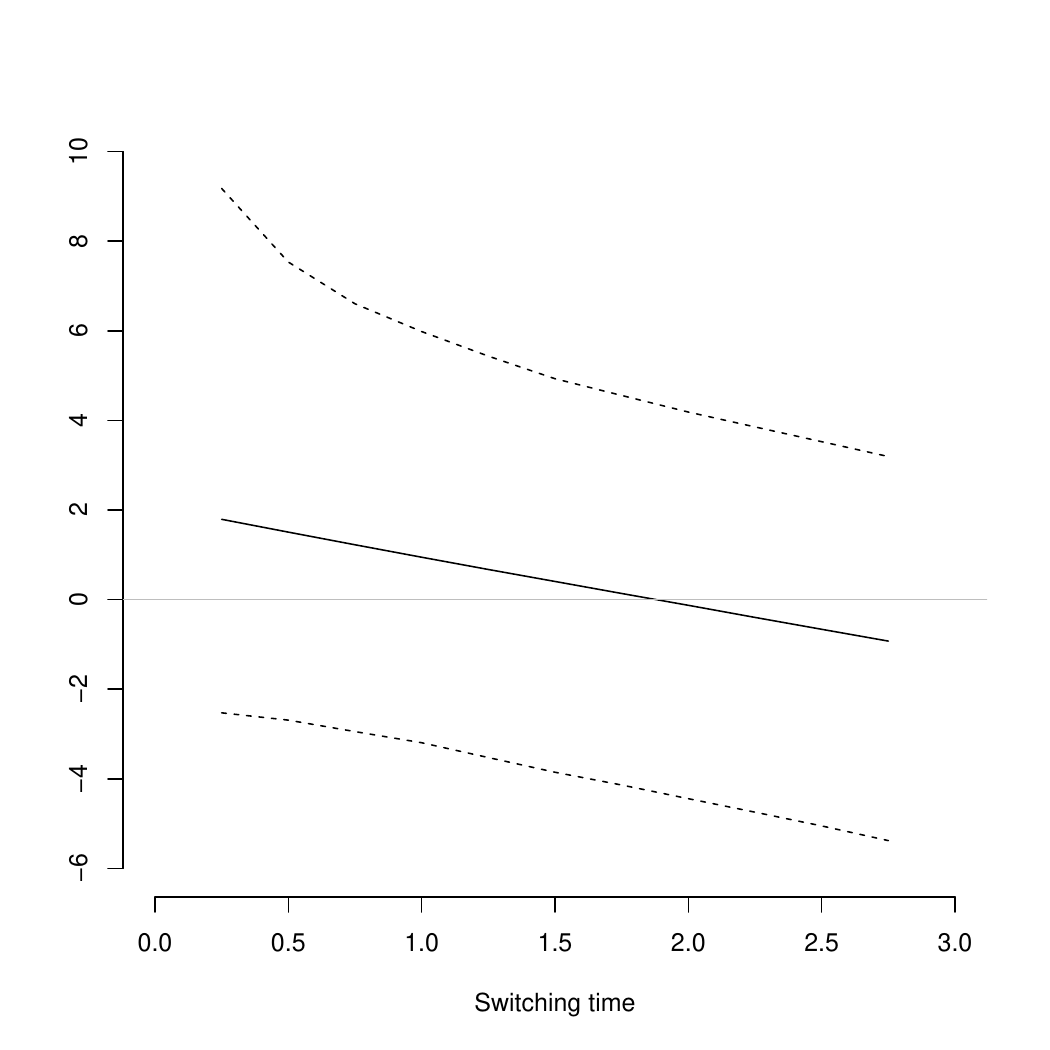}\\
$\lambda \sim N(0, 10\,000)$ &$\lambda \sim \hbox{Uniform}(\R)$     \vspace{-0.35cm}\\
\includegraphics[width=4.75cm]{Figures/PS_Analysis/Fig2_ACECeS.pdf}&
\includegraphics[width=4.75cm]{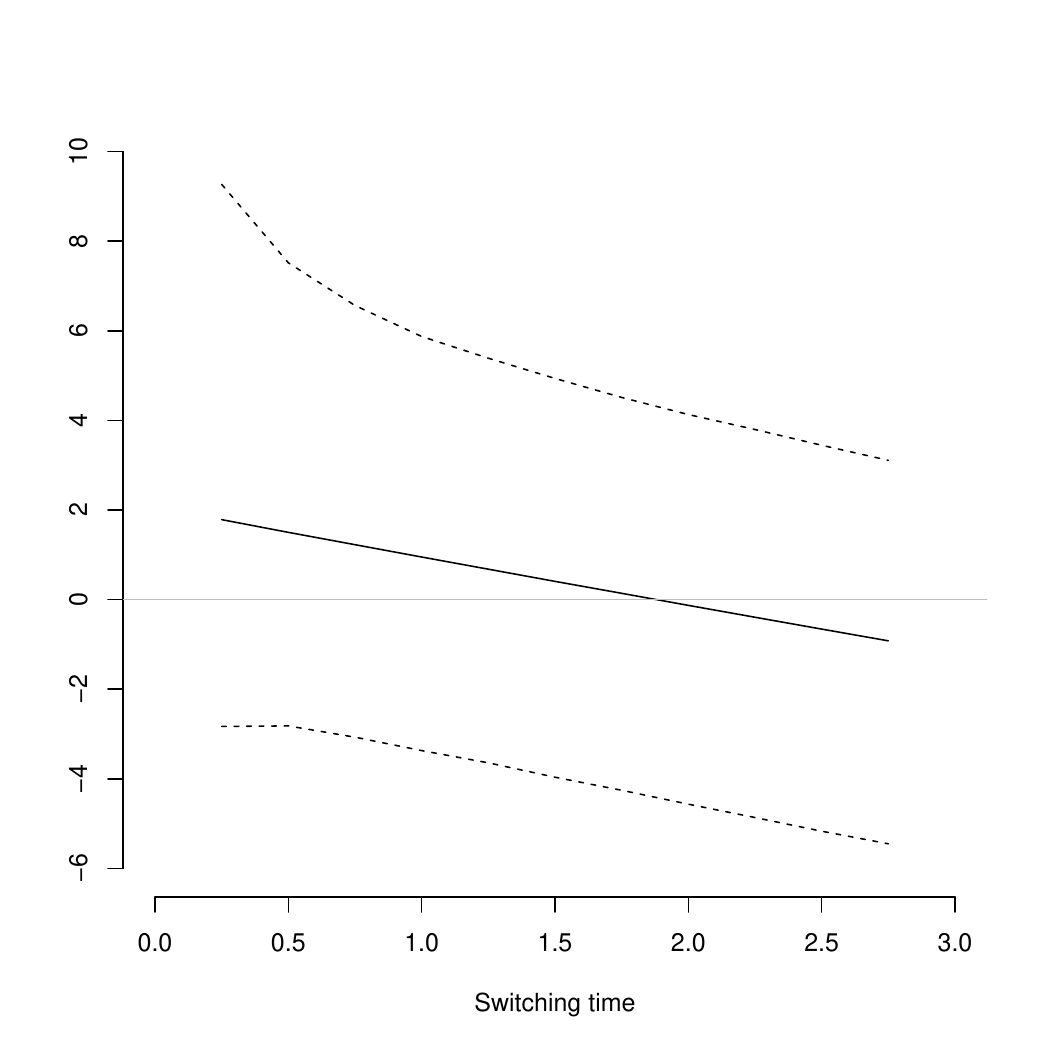}\\
\end{tabular}
\end{center}
\caption{Principal stratification analysis: Posterior median (solid line) and 95\% posterior credible interval (dashed lines) of average causal effects for switchers, $\ACE(s)$, $s \in \R_+$, for different prior distributions for $\lambda$ with $\kappa=0$} \label{Sens_Lambda_ACE_S}
\end{figure}

\begin{figure}
	\begin{center}
		\begin{tabular}{cc}
			$\lambda \sim N(0, 1)$ & $\lambda \sim N(0, 10)$  \vspace{-0.35cm}\\
			\includegraphics[width=4.75cm]{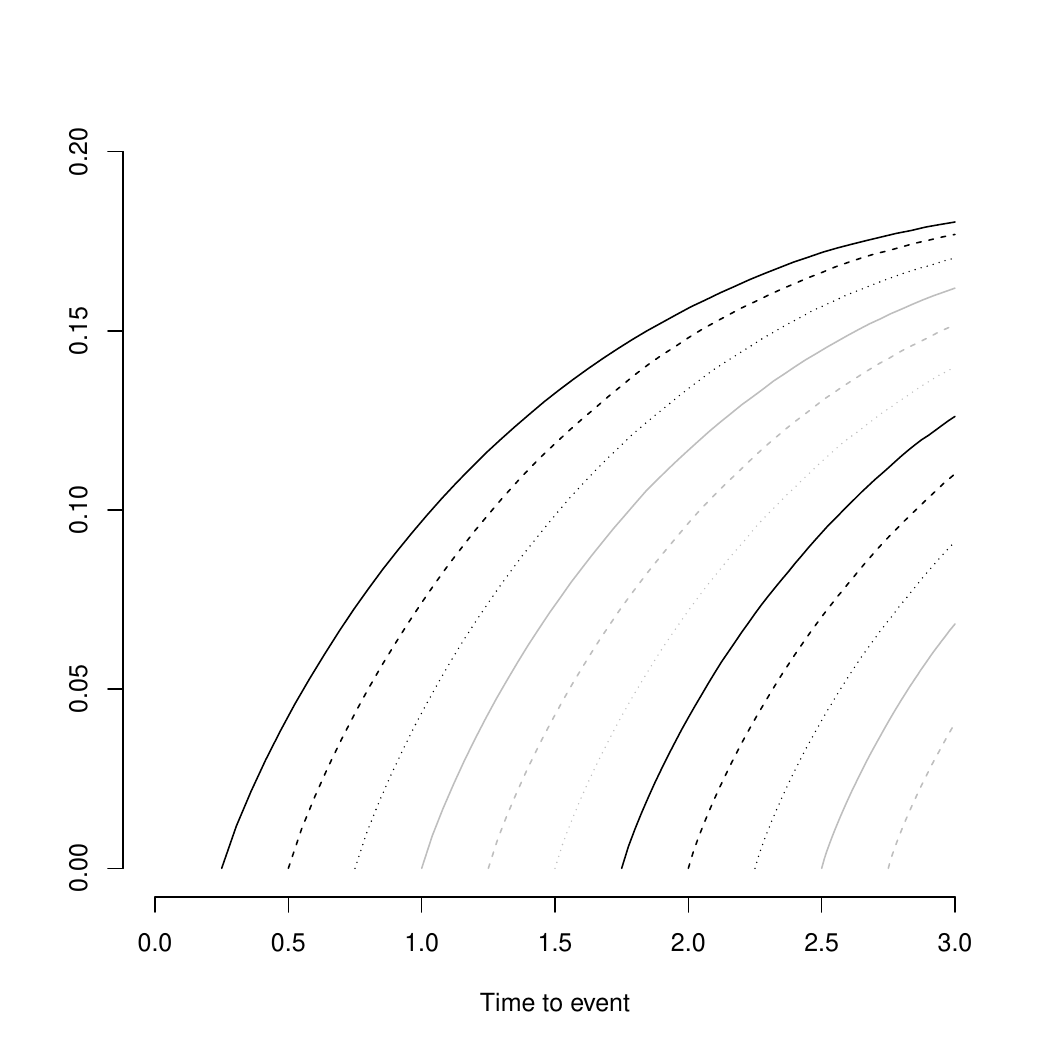}& 
			\includegraphics[width=4.75cm]{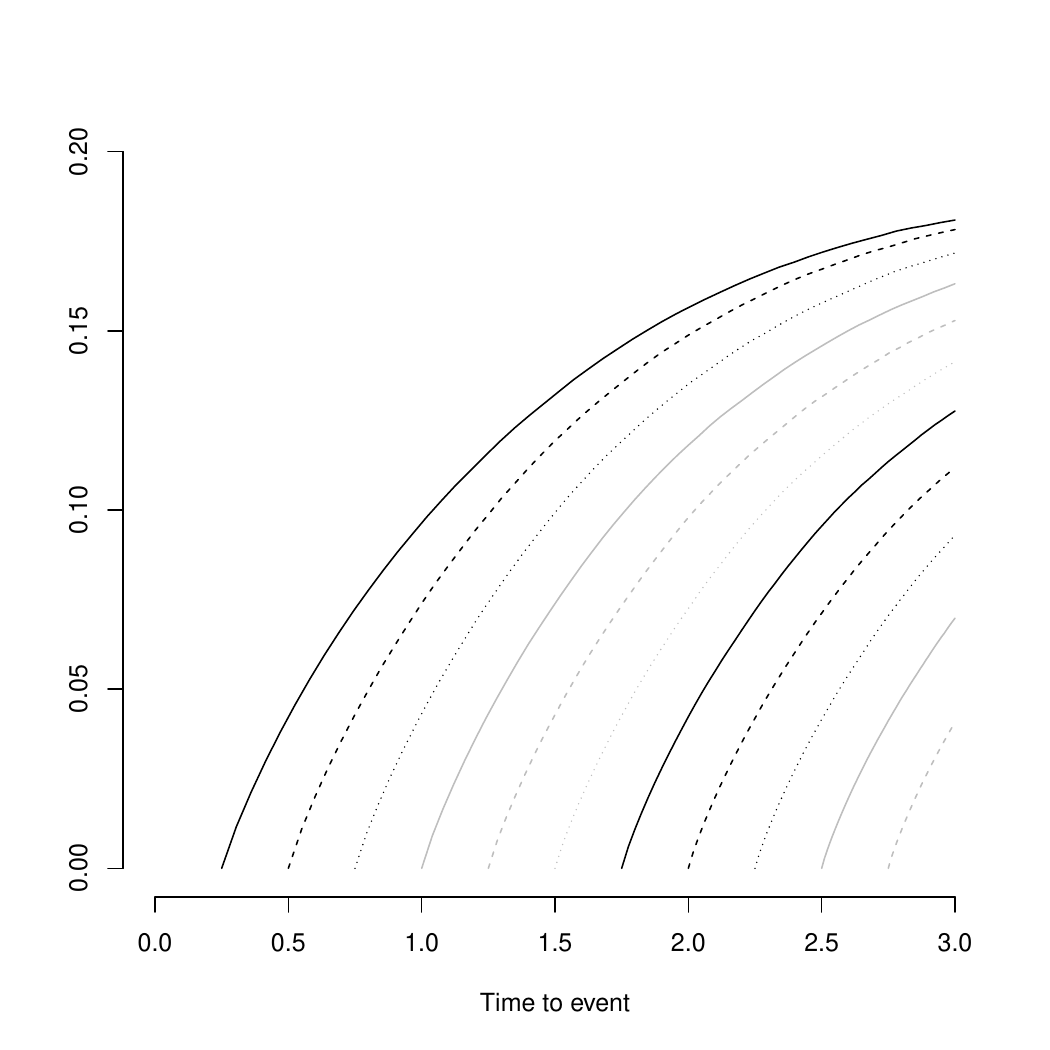}\\
			$\lambda \sim N(0, 10\,000)$ &$\lambda \sim \hbox{Uniform}(\R)$     \vspace{-0.35cm}\\
			\includegraphics[width=4.75cm]{Figures/PS_Analysis/Fig4b_cDCE.pdf}&
			\includegraphics[width=4.75cm]{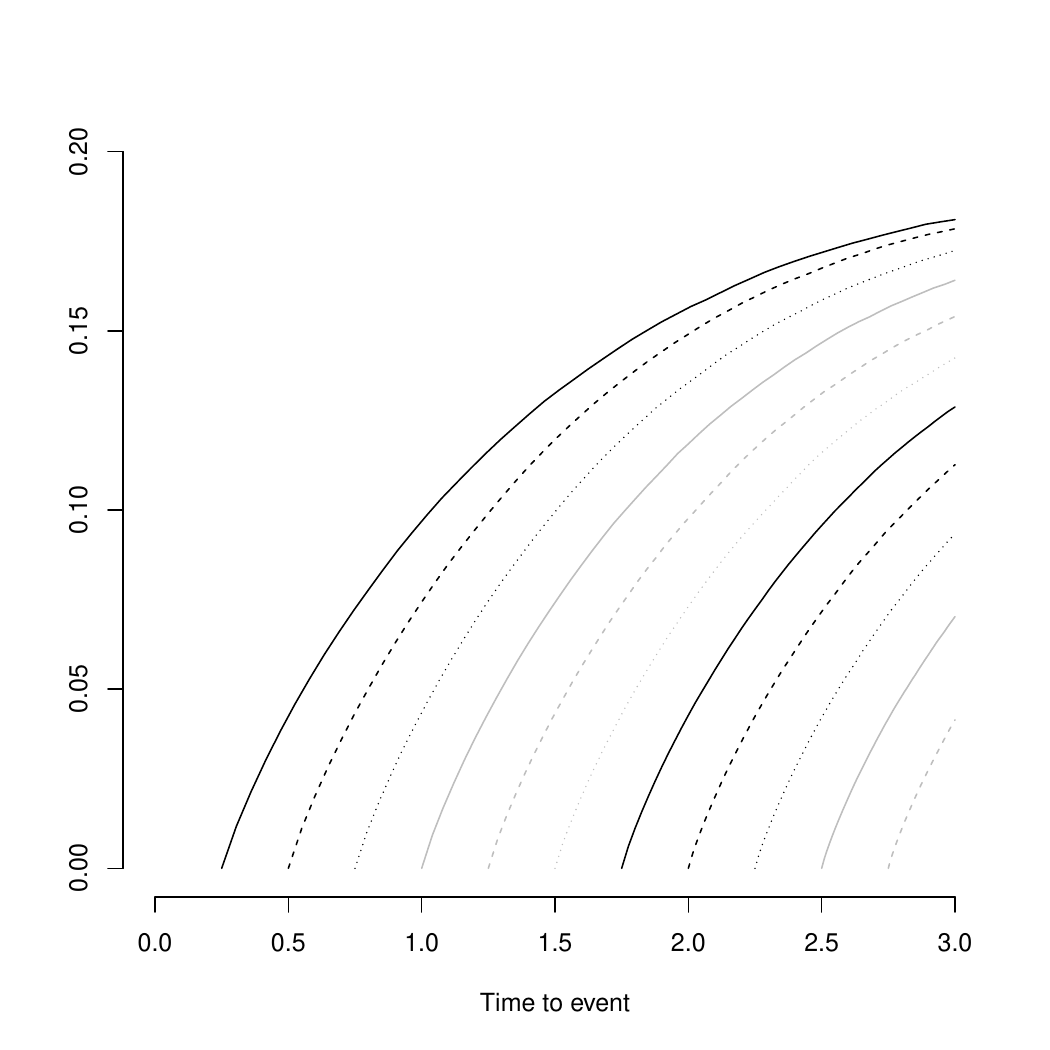} \\
 
		\end{tabular}
	\end{center}
	\caption{Principal stratification analysis: Posterior median (solid line) and 95\% posterior credible interval (dashed lines) of conditional distributional causal effects for switchers at time $s=0.25, 0.50, \ldots, 2.50, 2.75$, $\cDCE(y \mid s)$, for different prior distributions for $\lambda$ with $\kappa=0$} \label{Sens_Lambda_cDCE_S}
\end{figure}

\begin{figure}[h]
	\begin{center}
		\begin{tabular}{cc}
				$\lambda \sim N(0, 1)$ & $\lambda \sim N(0, 10)$  \vspace{-0.35cm}\\
			\includegraphics[width=4.75cm]{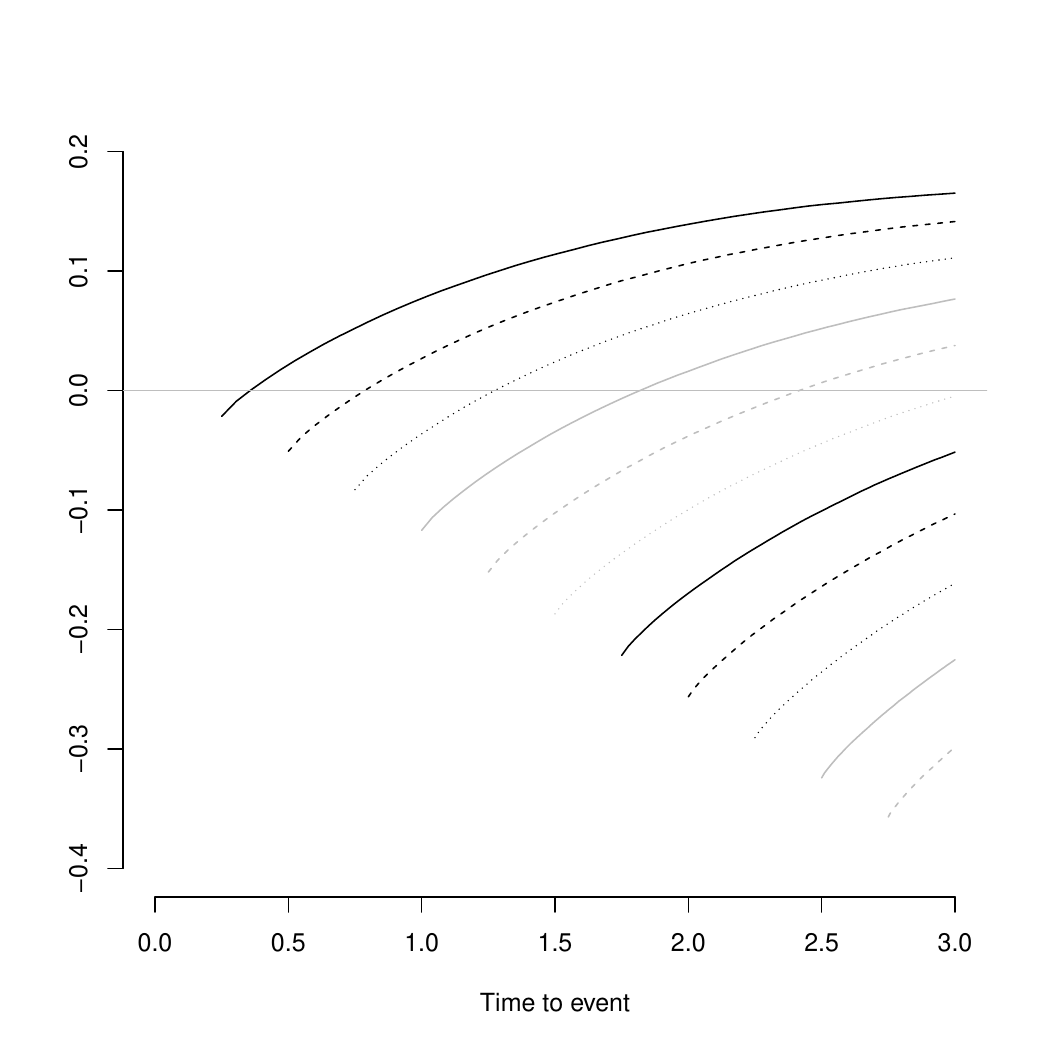}&\hspace{-0.25cm}
			\includegraphics[width=4.75cm]{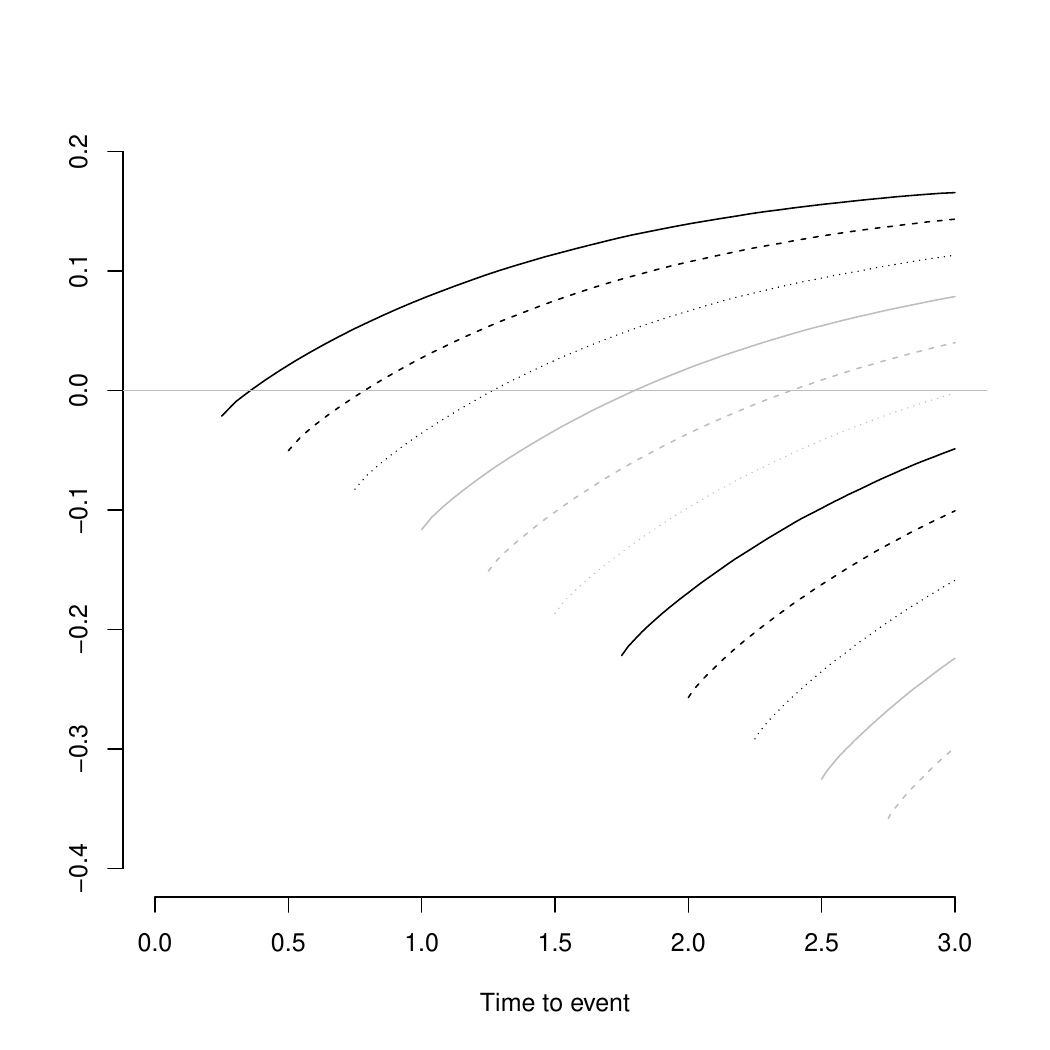}\\
			  $\lambda \sim N(0, 10\,000)$ & $\lambda \sim \hbox{Uniform}(\R)$  \vspace{-0.35cm}\\
			\includegraphics[width=4.75cm]{Figures/PS_Analysis/Appendix/Fig4a_DCE.pdf}& 
			\includegraphics[width=4.75cm]{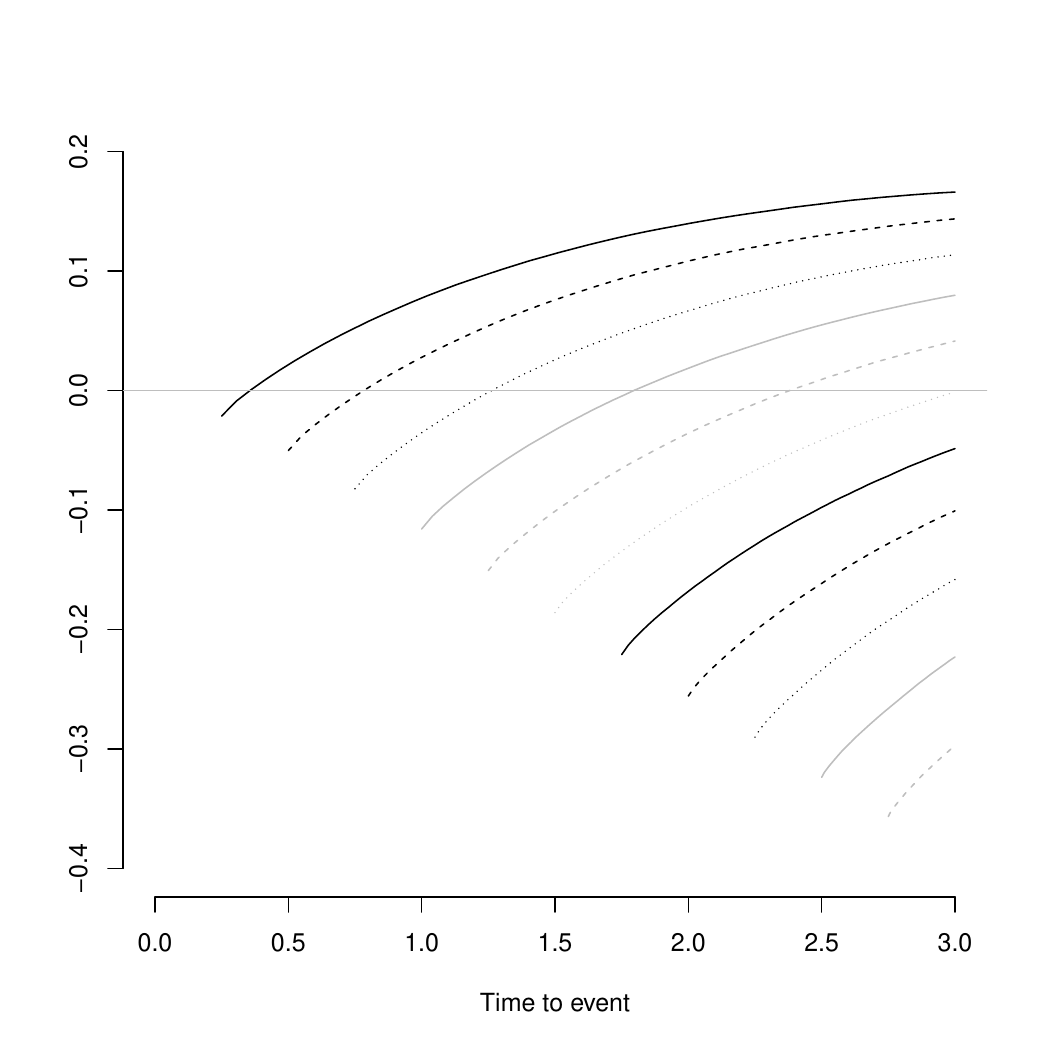}\\
			\\
		\end{tabular}
	\end{center}
	\caption{Principal stratification analysis: Posterior median of the distributional causal effects for switchers, at time $s=0.25, 0.50, \ldots, 2.50, 2.75$,  $\DCE(y \mid s)$, for different prior distributions for $\lambda$ ($\kappa=0$)} \label{Sens_Lambda_DCE_S}
\end{figure}

Previous results are obtained using a weakly informative prior distribution for  $\lambda$, namely, N$(0,10^4)$.
We assess the sensitivity of the results to the prior specification for $\lambda$ by specifying three alternative priors. 
We consider two normal priors with smaller variances, N$(0,1)$ and N$(0,10)$, and an improper prior uniformly over the whole real line. 
The hyperparameters of the prior distributions for the other model parameters are set to the same values as in Section~\ref{s:app1}.
We focus on the scenario with $\kappa=0$.
Table \ref{tab4_Lambda} and Figures \ref{Sens_Lambda_DCE_NS}-\ref{Sens_Lambda_DCE_S} present the results, showing that inference is robust with respect to the prior specification for $\lambda$. 
We see that the posterior distribution of the causal estimands changes only slightly using different prior distributions for $\lambda$.
Moreover, the posterior distribution of $\lambda$ is robust to different prior specifications.
The posterior mean of $\lambda$ remains approximately $0.10$, with a standard deviation of $0.17$, irrespective of the prior specification. 
Although the 95\% posterior credible intervals cover $0$, the posterior probability that the parameter $\lambda$  is positive ranges between $71.5\%$ and $72.7\%$ using different priors. 
Thus, there appears to be some evidence that the death hazard increases as the time of switching increases, suggesting that the residual lifetime after switching is shorter for patients who would switch later than for patients who would switch earlier.

\begin{table}\caption{Principal stratification analysis: Summaries of posterior distributions of causal estimands for non-switchers for different prior distributions for $\lambda$ ($\kappa=0$)} \label{tab4_Lambda} 
{\normalsize
$$
\begin{array}{l ccc c ccc }
\hline
\vspace{-0.3cm}\\
& \multicolumn{3}{c}{\lambda \sim N(0, 1)}   & & \multicolumn{3}{c}{\lambda \sim N(0, 10)}
\\
\cline{2-4}     \cline{6-8}  
\vspace{-0.3cm}\\
&&\multicolumn{2}{c}{95\% \hbox{ PCI}} & & & \multicolumn{2}{c}{95\% \hbox{ PCI}}  \\
\hbox{Estimand}& 0.50 & 0.025 & 0.975& & 0.50 & 0.025 & 0.975 \\
\hline
\vspace{-0.3cm}\\
\bE[Y_i(0)\mid S_i(0)=\bS]  &  2.06 & 1.45 & 3.01  & & 2.04 & 1.43 & \ \ 2.98\\
\bE[Y_i(1)\mid S_i(0)=\bS]  &  4.78 & 2.83 & 9.92  & & 4.76 & 2.81 & 10.12 \\
\ACE(\bS)                   &  2.68 & 0.72 & 7.79  & & 2.66 & 0.72 &  \ \ 8.02  \\
\\
\vspace{-0.3cm}\\
& \multicolumn{3}{c}{\lambda \sim N(0, 10\,000)}  & & \multicolumn{3}{c}{\lambda \sim \hbox{Uniform}(\R)} \\
\cline{2-4}     \cline{6-8}  
\vspace{-0.3cm}\\
&&\multicolumn{2}{c}{95\% \hbox{ PCI}} & & & \multicolumn{2}{c}{95\% \hbox{ PCI}} \\
\hbox{Estimand}& 0.50 & 0.025 & 0.975& & 0.50 & 0.025 & 0.975 \\
\hline
\vspace{-0.3cm}\\
\bE[Y_i(0)\mid S_i(0)=\bS] &    2.05 & 1.44 & 2.99  & & 2.04 & 1.44 &3.00\\
\bE[Y_i(1)\mid S_i(0)=\bS] &    4.76 & 2.80 & 9.80  & & 4.75 & 2.81 & 9.80 \\
\ACE(\bS)                   &   2.66 & 0.71 & 7.73  & & 2.65 & 0.71 & 7.74 \\
\hline
\end{array}
$$
}
\end{table}

\subsection*{Sensitivity Analysis to the Parametric Assumption $\lambda_1=\lambda_0$}
We assess the sensitivity of the results to the parametric assumption $\lambda_1=\lambda_0$ by deriving the posterior distributions of the causal estimands of interest when we relax it.

\begin{figure}
\begin{center}
\begin{tabular}{cc}
\multicolumn{2}{c}{$\DCE(y \mid \bS)$}\\
Under the model with $\lambda_0= \lambda_1$ & Under the model with $\lambda_0\neq \lambda_1$ \\
\vspace{-0.75cm}\\
\includegraphics[width=0.45\textwidth]{Figures/PS_Analysis/Fig3_DCE_NS.pdf}  & \includegraphics[width=0.45\textwidth]{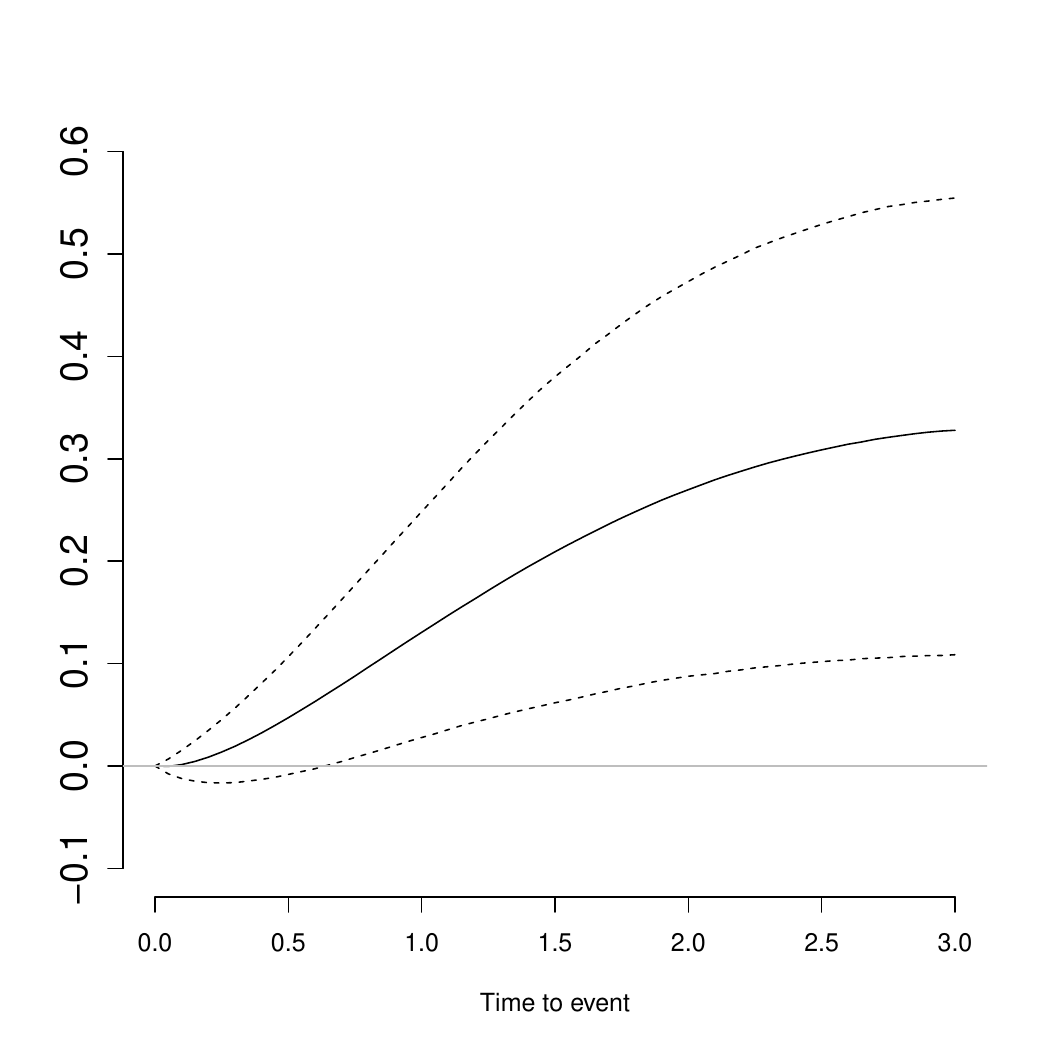} 
\end{tabular}
\end{center}
\caption{Principal stratification analysis: Posterior median (solid line) and 95\% posterior credible interval (dashed lines) of the distributional causal effects for non-switchers  under the model with $\lambda_0= \lambda_1$ (left) and under the model with $\lambda_0\neq \lambda_1$ (right)} \label{Fig:la0la1_ns} 
\end{figure}

\begin{figure}
\begin{center}
\begin{tabular}{cc}
	\multicolumn{2}{c}{ACE$(s)$, $s \in\mathbb{R}_+$}\\
Under the model with $\lambda_0= \lambda_1$ & Under the model with $\lambda_0\neq \lambda_1$ \\
\includegraphics[width=0.45\textwidth]{Figures/PS_Analysis/Fig2_ACECeS.pdf} &\includegraphics[width=0.45\textwidth]{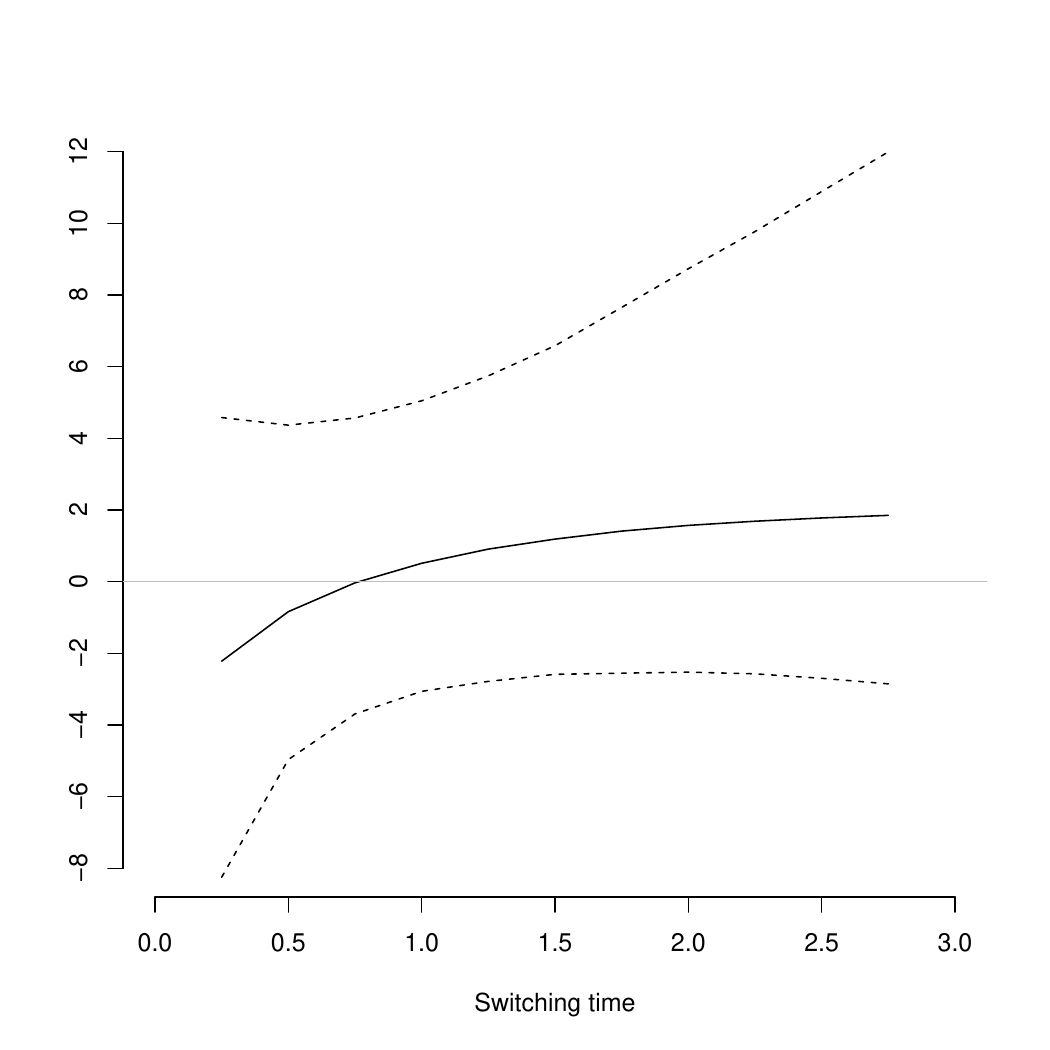}  
\end{tabular}
\end{center}
\caption{Principal stratification analysis: Posterior medians (solid lines) and 95\% posterior credible intervals (dashed lines) of average causal effects for switchers, under the model with $\lambda_0= \lambda_1$ (left) and under the model with $\lambda_0\neq \lambda_1$ (right)} \label{Fig:la0la1_aces} 
\end{figure}

\begin{figure}
\begin{center}
\begin{tabular}{cc }
	\multicolumn{2}{c}{	$cDCE(y \mid s)$ for $s=0.25, 0.5, \ldots, 2.5, 2.75$}\\
	Under the model with  $\lambda_0=\lambda_1$ & 	Under the model with  $\lambda_0\neq \lambda_1$ \vspace{-0.5cm}\\
\vspace{-0.5cm}
\includegraphics[width=5cm]{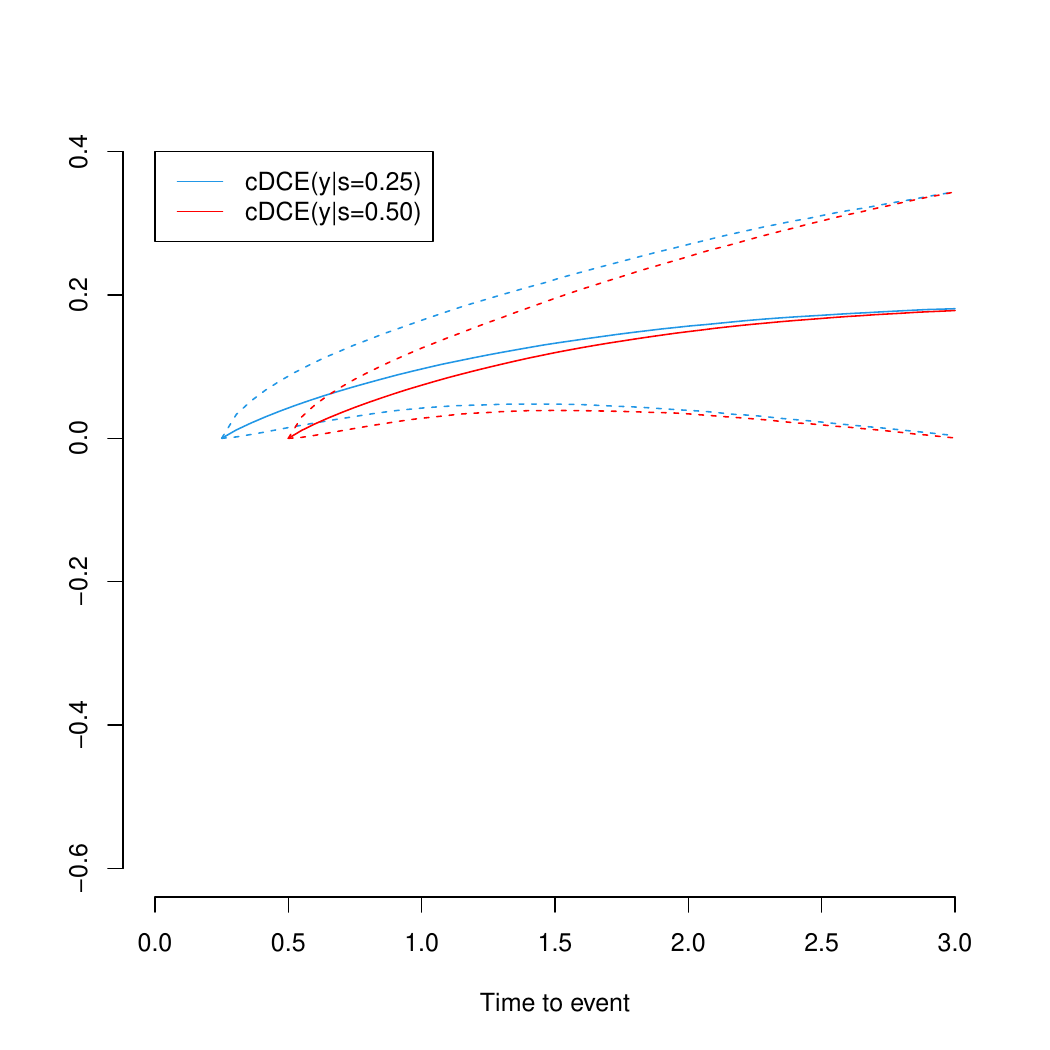} &
\includegraphics[width=5cm]{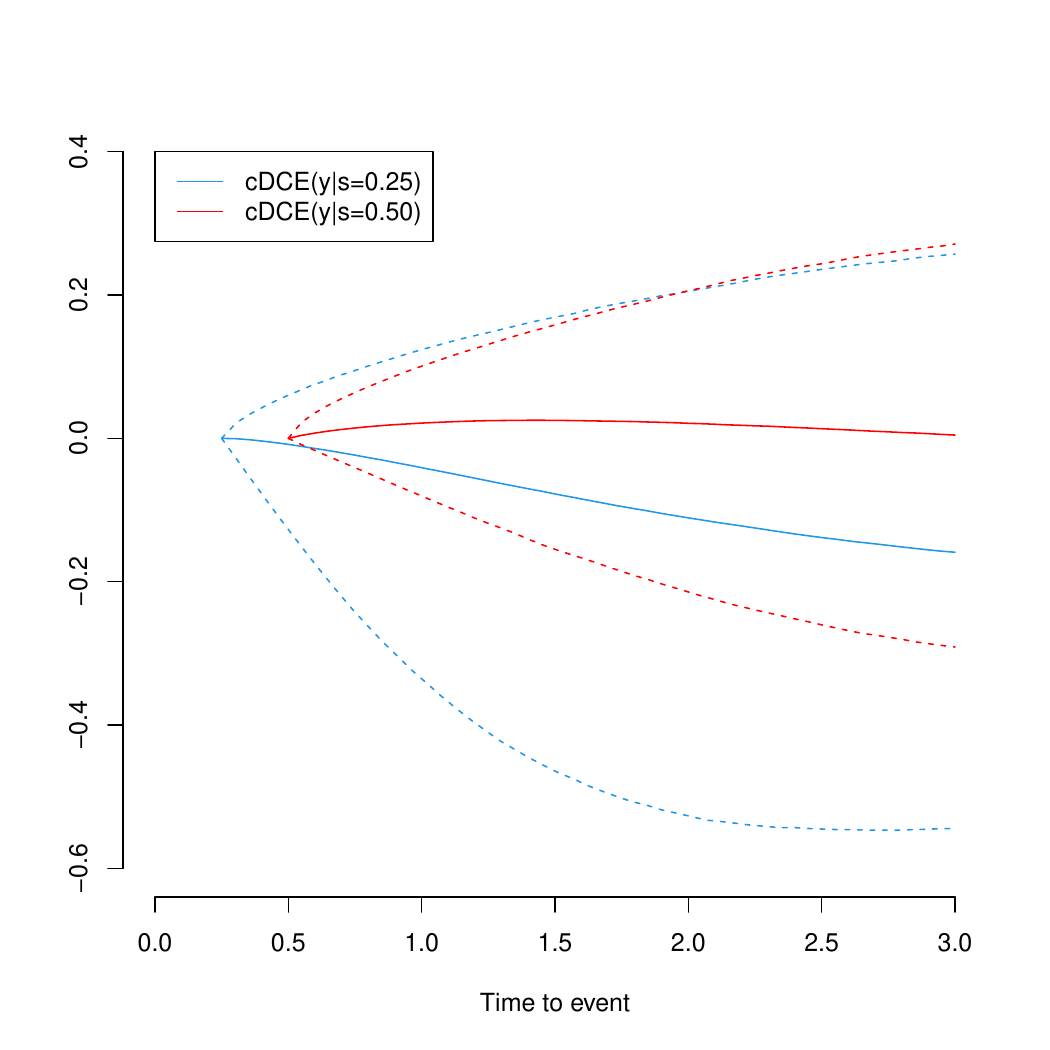}  \\
\vspace{-0.5cm}
\includegraphics[width=5cm]{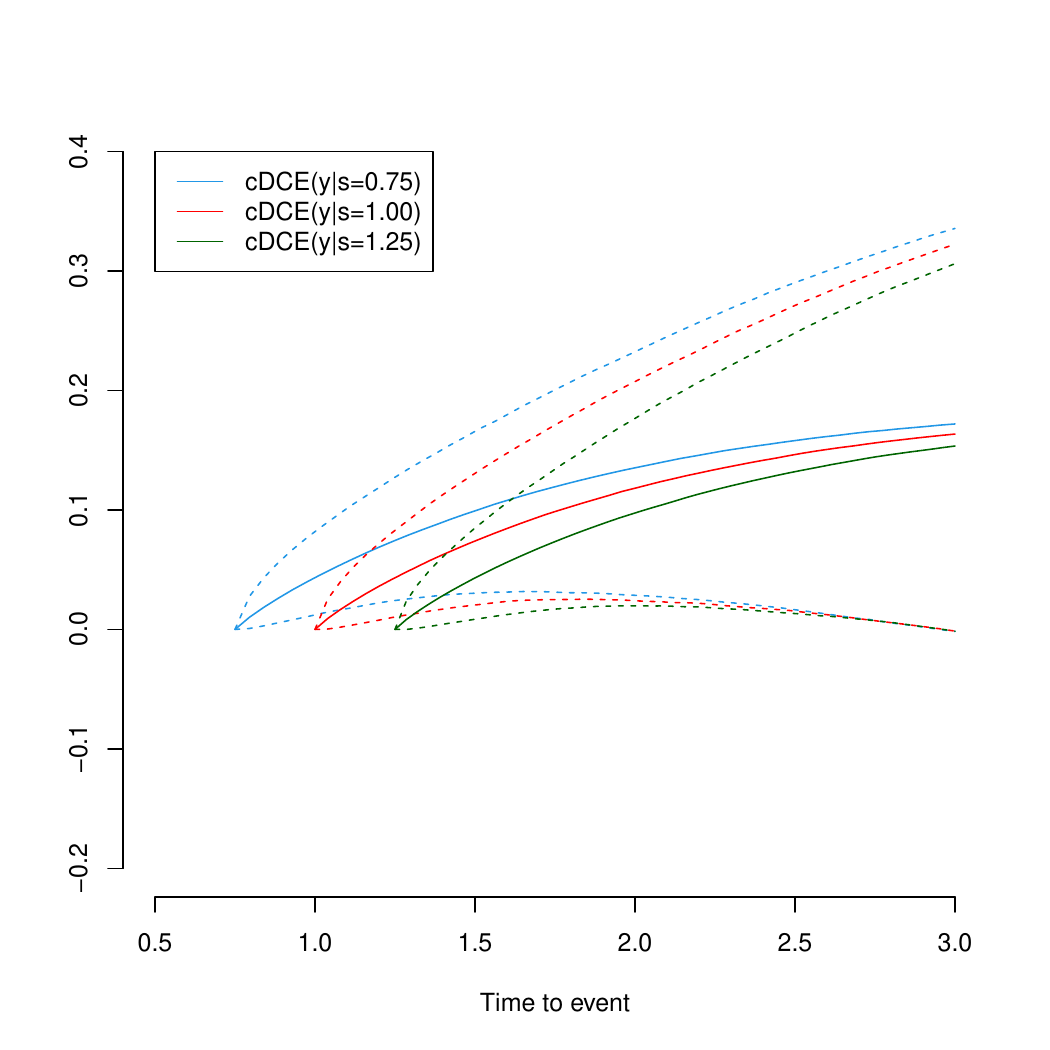} & \includegraphics[width=5cm]{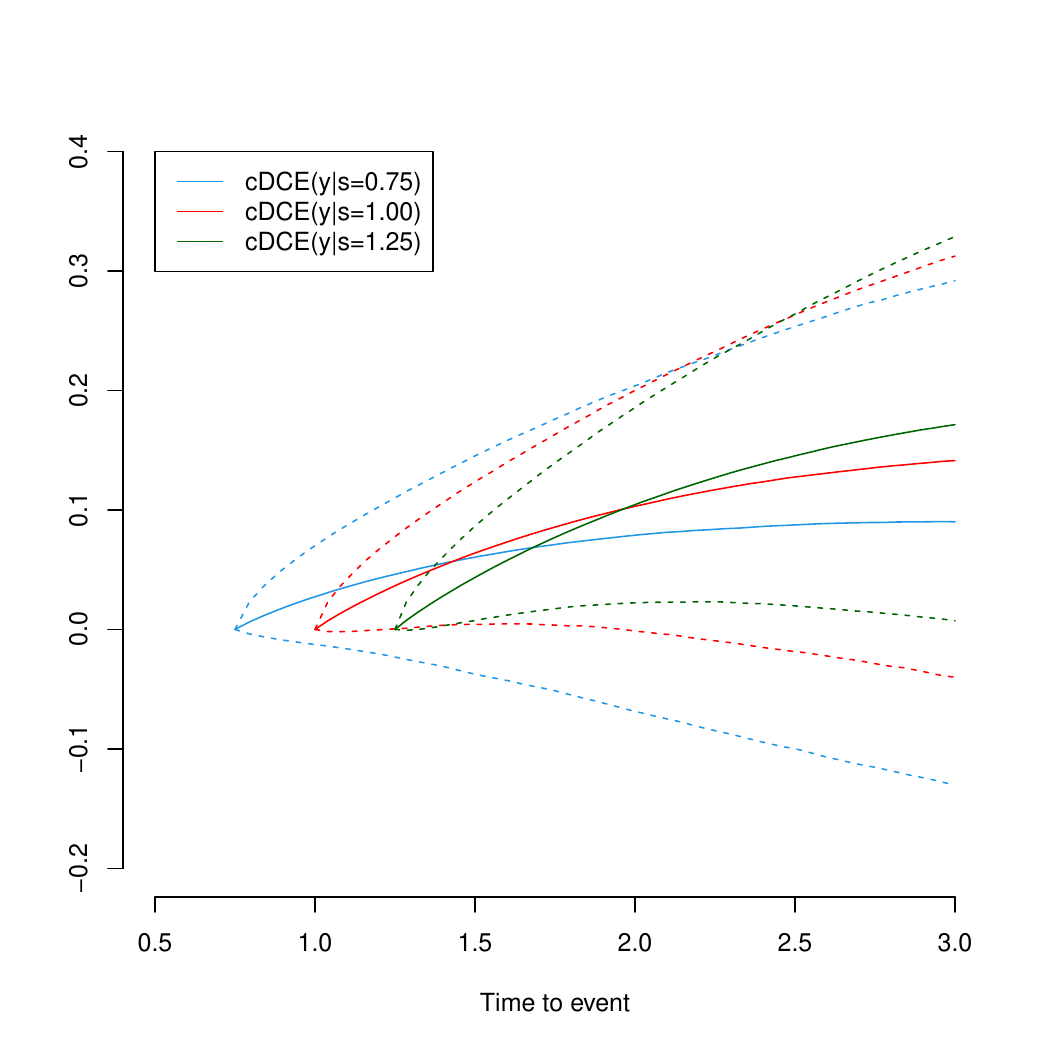}\\
\vspace{-0.5cm}
\includegraphics[width=5cm]{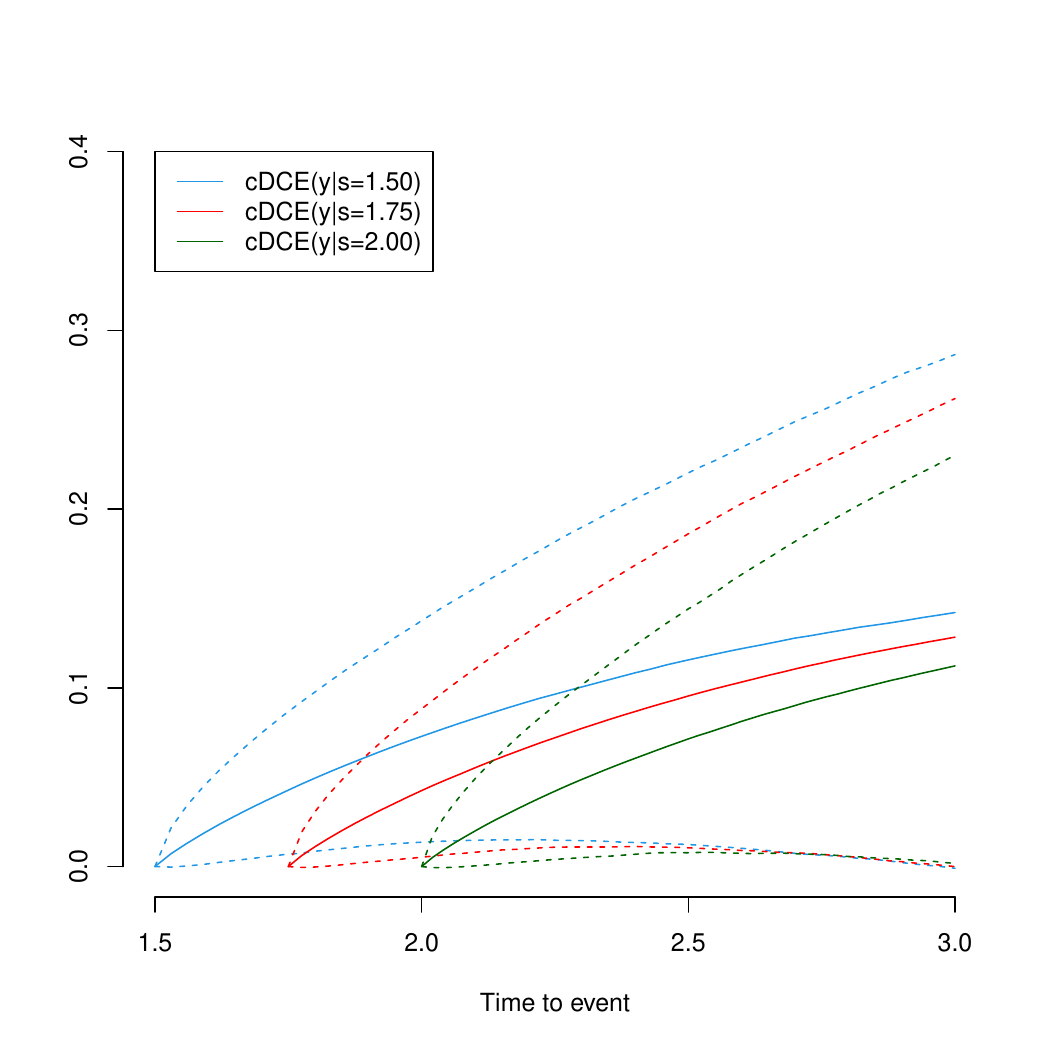} & \includegraphics[width=5cm]{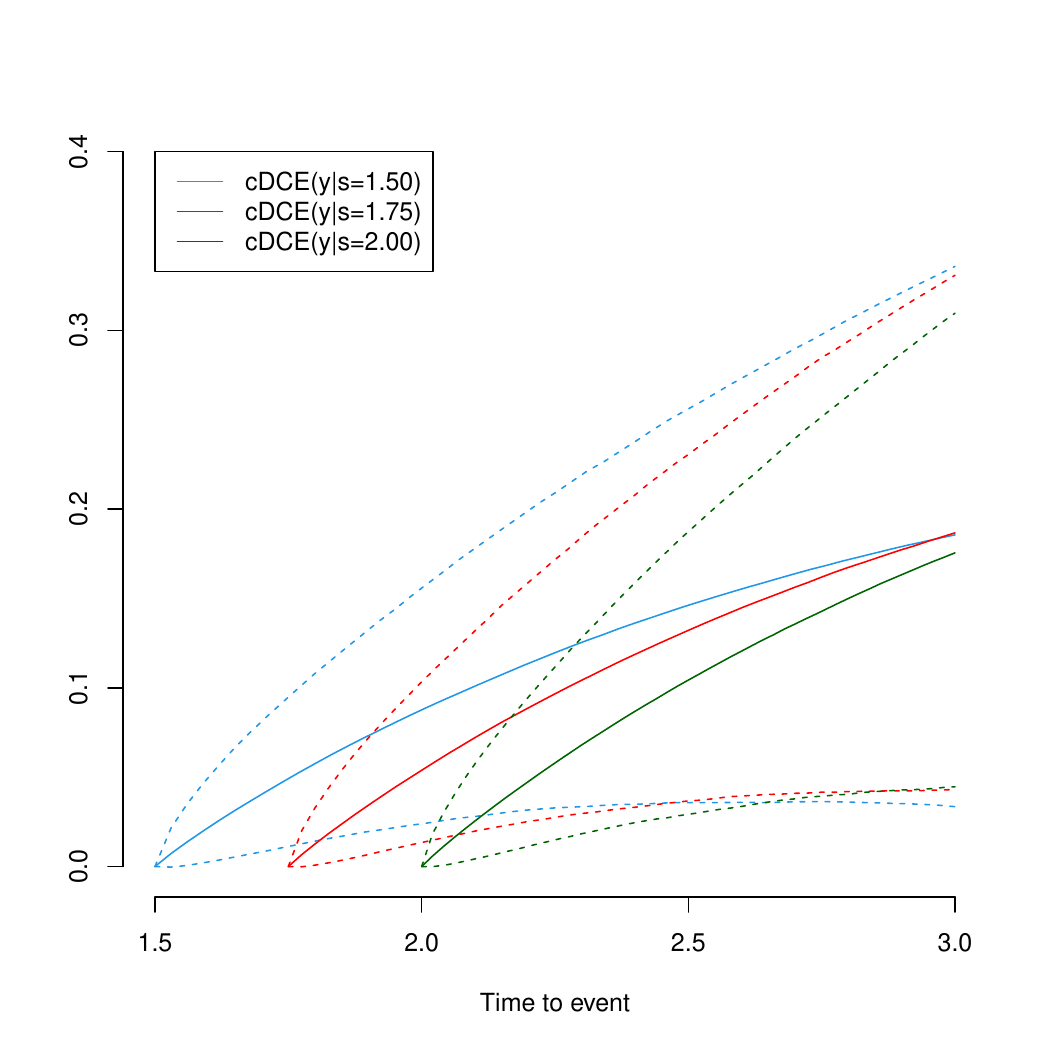} \\
\vspace{-0.5cm}
\includegraphics[width=5cm]{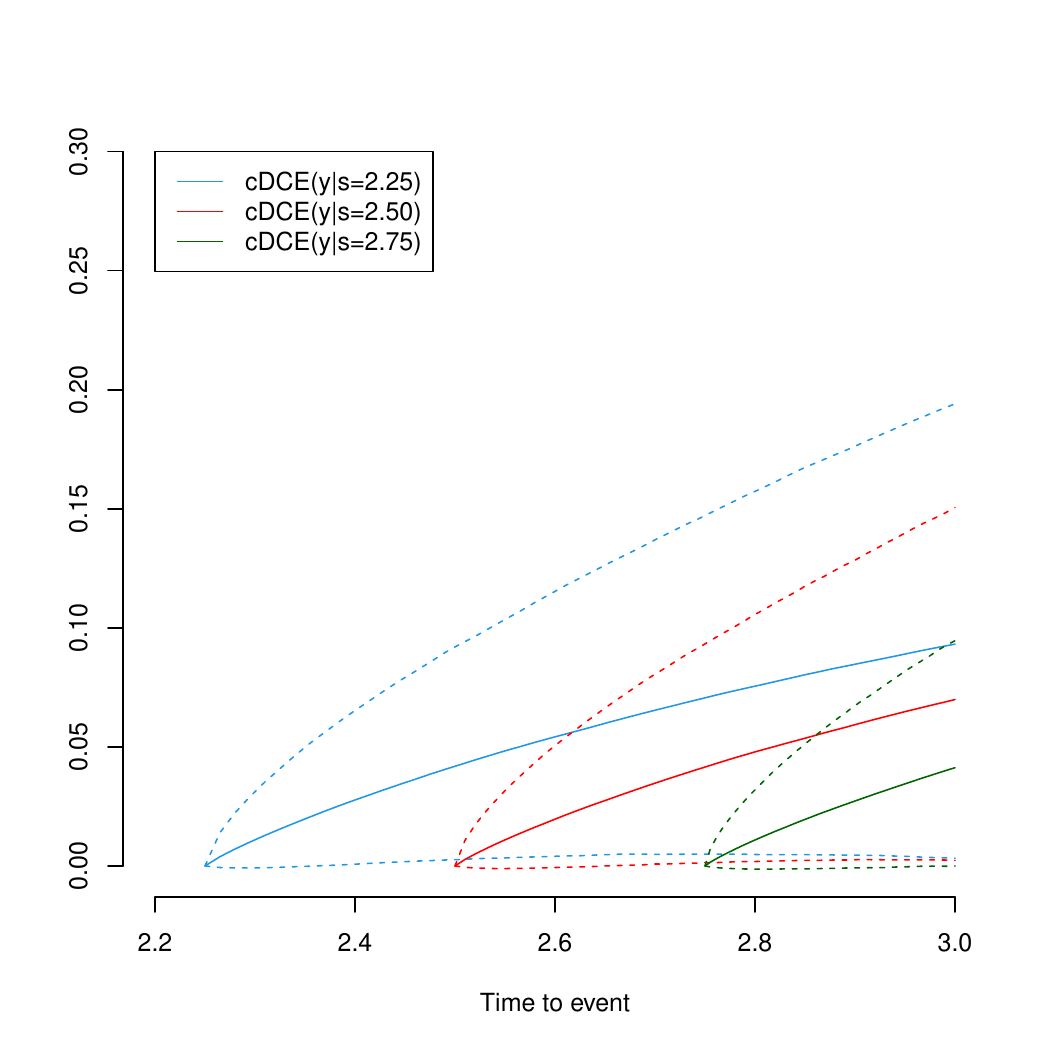}&
\includegraphics[width=5cm]{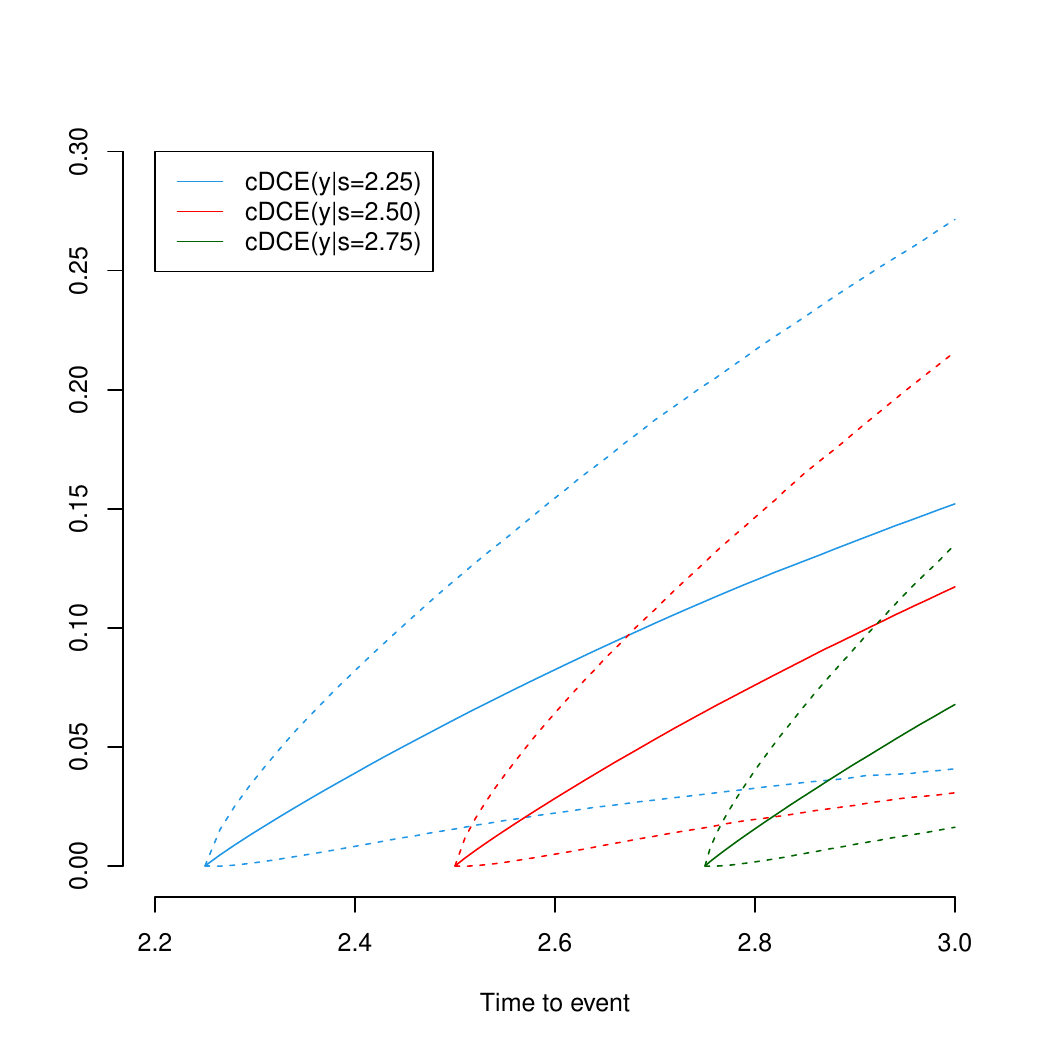}
\end{tabular}
\end{center}
\caption{Principal stratification analysis: Posterior medians (solid lines) and 95\% posterior credible intervals (dashed lines) of the conditional distributional causal effects for switchers under the model with $\lambda_0=\lambda_1$ (first column) and under the model with $\lambda_0\neq \lambda_1$ (second column).} \label{Fig:la0la1_dces}
\end{figure}

Table~\ref{tab:la0la1_ns} and Figure~\ref{Fig:la0la1_ns} show the results for never-switchers and Figures~\ref{Fig:la0la1_aces} and \ref{Fig:la0la1_dces} show the results for switchers.

Relaxing the parametric assumption $\lambda_0=\lambda_1$  does not affect the results for never-switchers (see Table~\ref{tab:la0la1_ns} and Figure~\ref{Fig:la0la1_ns}) and slightly changes the results for switchers, by leading to posterior distributions of the causal effects for switchers with a larger posterior variability (see Figures~\ref{Fig:la0la1_aces} and \ref{Fig:la0la1_dces}). 
The increased uncertainty in the causal estimands for switchers makes it more difficult to draw firm causal conclusions for them, especially for early switchers. 
For instance, for early switchers who would switch earlier than 1 year, we find positive and statistically significant distributional causal effects under the model with $\lambda_0=\lambda_1$ and statistically negligible distributional causal effects under the model with $\lambda_0\neq \lambda_1$ (see the graphs in the first row of Figure~\ref{Fig:la0la1_dces}).

\begin{table}\caption{Principal stratification analysis: Summaries of posterior distributions of causal estimands for non-switchers  under the model with   $\lambda_0=\lambda_1$ and under the model with   $\lambda_0\neq\lambda_1$} \label{tab:la0la1_ns} 
{\normalsize
$$
\begin{array}{l ccc c ccc }
\hline
\vspace{-0.3cm}\\
&\multicolumn{7}{c}{\hbox{Under the model with}}\\
& \multicolumn{3}{c}{\lambda_0=\lambda_1}  & & \multicolumn{3}{c}{  \lambda_0\neq \lambda_1} \\
\cline{2-4}     \cline{6-8}  
\vspace{-0.3cm}\\
&&\multicolumn{2}{c}{95\% \hbox{ PCI}} & & & \multicolumn{2}{c}{95\% \hbox{ PCI}} \\
\hbox{Estimand}& 0.50 & 0.025 & 0.975& & 0.50 & 0.025 & 0.975 \\
\hline
\vspace{-0.3cm}\\
\bE[Y_i(0)\mid S_i(0)=\bS] &    2.05 & 1.44 & 2.99  & & 1.95 & 1.41 & 2.91\\
\bE[Y_i(1)\mid S_i(0)=\bS] &    4.76 & 2.80 & 9.80  & & 4.39 & 2.63 & 9.18 \\
\ACE(\bS)                   &   2.66 & 0.71 & 7.73  & & 2.39 & 0.63 & 7.08 \\
\hline
\end{array}
$$
}
\end{table}

\subsection*{Posterior Predictive Checks}
We evaluate the influence of the parametric assumptions using posterior predictive checks \cite[e.g.,][]{Guttman1967, Rubin1984}, by computing a Bayesian posterior predictive $p$-value ($PPPV$) for various discrepancy measures \cite[][]{Meng1994, Gelman_et_al:96, ForastiereMealliMiratrix2018}. 
A discrepancy measure is a known, real-valued function of the nuisance parameters, the imputed switching status, and the observed data. 
The corresponding Bayesian $PPPV$ is defined as the integral average over the joint posterior distribution of the missing switching statuses and model parameters of the probability that the discrepancy measure calculated for replicated data is more extreme than the value for observed data. Replicated data are drawn from the posterior predictive distribution of the hypothesized model.
A $PPPV$ is a measure of model misfit, with the model including both the prior distribution and the likelihood. Extreme values (close to 0 or 1) of a $PPPV$ would indicate that the model cannot adequately preserve features of the data reflected in the discrepancy measure.

We conduct model checking under conditionally independent potential survival outcomes with $\kappa=0$.
Let $r$ be the study type indicator: $r=\mathrm{obs}$ for the observed study and $r=\mathrm{rep}$ for a  replicated study. 
We generate the replicated data using the observed value of the assignment variable, entry, and censoring time, that is, we set $Z_i^{\mathrm{rep}}=Z^{\mathrm{obs}}_i=Z_i$ and $C_i^{\mathrm{rep}}=\Cobs_i$ for all $i=1, \ldots, n$.

We measure the goodness-of-fit of the posited model using three types of posterior predictive discrepancy measures:
\begin{enumerate}
    \item \textit{BIC posterior predictive discrepancy measure.}
    $$
    BIC^r= -2\left( {\mathscr{L}}\{\btheta \mid \kappa=0,\bD^r \} + \sharp   \{\btheta \setminus \kappa\} \cdot \log(n) \right),
    $$ 
    where $\bD^r=[\bZ, \bC,\tilde{\bbS}^{r}, \mathbb{I}\{\mathbf{S}^{r} \leq \mathbf{C} \},	\tilde{\bY}^{r},\mathbb{I}\{\mathbf{Y}^{r} \leq \mathbf{C} \}, \bbS^{\ast,r}(0)]$ is the $n \times 7$ matrix of the complete switching status data, $ {\mathscr{L}}\{\btheta \mid \kappa=0, \bD^r\}$ is the complete switching status-data likelihood function for $\kappa=0$, and $\sharp \{\btheta \setminus \kappa\}$ is the number of parameters excluding $\kappa$ ($\sharp \{\btheta \setminus \kappa\}=12$ in our study). 
    \item  \textit{Deviance posterior predictive discrepancy.} 
    The deviance is defined as the sum of the deviance residuals for the Weibull model. 
    We calculate the deviance posterior predictive discrepancy measure separately for the survival time and the switching time under control for switchers.
    For $r=$ obs, rep,
	{\footnotesize
		\begin{eqnarray*}
			\lefteqn{Deviance^r_Y(\bD,\btheta) = }\\&&
		\hspace{-0.65cm}	-2 \sum_{i: Z_i=0,1} \left\{ \sum_{i: Z_i=z, S^{\ast,r}_i(0)=\bS}  
		\!\!\!	\left[M^{\bS}_{Y(z)}(\tilde{Y}_i^r) + \mathbb{I}\{Y^r_i \leq \Cobs_i\} \log \left(\mathbb{I}\{Y^r_i \leq \Cobs_i\}-M^{\bS}_{Y(z)}(\tilde{Y}_i^r)\right) \right]+\right.\\&&  	\hspace{-0.65cm} \left.
			\sum_{i: Z_i=z, S^{\ast,r}_i(0) \in \R_+}  
		\!\!\!\!\!	\left[M_{Y(z)}(\tilde{Y}_i^r \mid S^{\ast,r}_i(0)) + \mathbb{I}\{Y^r_i \leq \Cobs_i\} \log \left(\mathbb{I}\{Y^r_i \leq \Cobs_i\}-M_{Y(z)}(\tilde{Y}_i^r\mid S^{\ast,r}_i(0))\right) \right] \right\}\;,
		\end{eqnarray*}
	}
	where $M^{\bS}_{Y(z)}(\cdot)$ and $M_{Y(z)}(\cdot \mid S_i(0))$ are the martingale residuals for non-switchers and switchers, respectively:
	$M^{\bS}_{Y(z)}(\tilde{Y}^r_i) =  \mathbb{I}\{Y^r_i \leq \Cobs_i\}-\Lambda^{\bS}_{Y(z)}(\tilde{Y}_i^r)$
	if $S_i(0)=\bS$, and $M_{Y(z)}(\tilde{Y}^r_i  \mid S_i(0)) =  \mathbb{I}\{Y^r_i \leq \Cobs_i\}-\Lambda_{Y(z)}(\tilde{Y}_i^r\mid S_i(0))$
	if $ S_i(0)\in \R_+$, with $\Lambda^{\bS}_{Y(z)}(\cdot)$ and $\Lambda_{Y(z)}(\cdot \mid S_i(0))$ denoting the   cumulative hazards for the Weibull model \citep{TherneauGrambschFleming:1990}. Similarly, we define	
{\small	\begin{eqnarray*}
		\lefteqn{	Deviance^r_S(\bD, \btheta) = }\\&&
	\hspace{-0.5cm}	-2   	\sum_{i: Z_i=0, S^{\ast,r}_i(0)\in \R_+}  M_{S(0)}(\tilde{S}_i^r)
		+ \mathbb{I}\{S^{\ast,r}_i(0) \leq \Cobs_i\} \log \left(\mathbb{I}\{S^{\ast,r}_i(0) \leq \Cobs_i\}-M_{S(0)}(\tilde{S}_i^r)\right).
	\end{eqnarray*}
}
    \item \textit{Kaplan--Meier posterior predictive discrepancy.}  
    We calculate Kaplan--Meier estimates of the survival curves for the time-to-death/disease progression, separately for non-switchers and switchers, and for the time-to-switching under control for switchers.
	
    For data from study type $r$, $r= $ obs, rep, for each time-point $t$, let  $d_Y^r(t \mid S_i^{\ast,r}(0) \in \cA)$ be the number of events (deaths or disease progressions) at time $t$ among patients with $S_i^{\ast,r}(0) \in \cA$, $\cA=\{\bS\}$ (non-switchers) and $\cA=\R_+$ (switchers), and let $d_S^r(t \mid Z_i=0, S_i^{\ast,r}(0) \in \R_+)$ be the number of switchers assigned to the control treatment who switch at time $t$. 
    Let $R_Y^r(t \mid S_i^{\ast,r}(0) \in \cA)$  denote the number of subjects with $S_i^{\ast,r}(0) \in \cA$, $\cA=\{\bS\},\R_+$,  at risk of death or disease progression at time $t$, and let $R_S^r(t \mid Z_i=0, S_i^{\ast,r}(0) \in \R_+)$ denote the number of switchers assigned to the control treatment at risk of switching at time $t$. 
    We define
	$$
	KM^r_{\cA}(t; \bD,\btheta) = \prod_{i: S_i^{\ast,r}(0)\in \cA} \left[1- \dfrac{d_Y^r(t \mid S_i^{\ast,r}(0)\in \cA)}{R_Y^r(t \mid S_i^{\ast,r}(0) \in \cA)}\right] \qquad \cA=\{\bS\},\R_+;
	$$
	and
	$$
	KM^r(t; \bD,\btheta) = \prod_{i: Z_i=0, S_i^{\ast,r}(0)\in \R_+} \left[1- \dfrac{d_S^r(t \mid Z_i=0, S_i^{\ast,r}(0) \in\R_+)}{R_S^r(t \mid Z_i=0, S_i^{\ast,r}(0) \in \R_+)}\right].
	$$
\end{enumerate}

Following \citet{BFHR2003}, we then consider posterior predictive discrepancy measures aimed to assess the ability of the model to preserve features in the outcome distributions of non-switchers and switchers that we think can be very influential in estimating the average and distributional causal effects. 

Define the following subsets of units in the study of type $r$ ($r=$ obs, rep):
$$
\Dr_{\cA,z} =\{i: \mathbb{I}\{Y^r_i \leq \Cobs_i\} \mathbb{I}\{S^{\ast,r}_i(0) \in \cA\}  \mathbb{I}\{Z_i=z\}=1\} 
$$
for $\cA= \{\bS\}$ and $\cA=\R_+$, and $z=0,1$; and
$$
\Dr =\{i: \mathbb{I}\{S^{\ast,r}_i(0) \leq \Cobs_i\} \mathbb{I}\{S^{\ast,r}_i(0) \in \R_+\}  \mathbb{I}\{Z_i=0\}=1\} 
$$

Let $\overline{Y}_{\cA,z}^r$ and $s^{2,r}_{Y,\cA, z}$ be the mean and the variance of the survival outcome, $Y_i$, for units belonging to $\Dr_{\cA,z}$, for which we observe $Y_i(z)$ in study $r$. 
Similarly, let $\overline{S}^r$ and $s^{2,r}_S$ the mean and the variance of the switching time, $S_i^{\ast}(0)$, for units belonging to $\Dr$, for which $S_i^{\ast}(0)$ is observed in study $r$. 
Then,
$$
\begin{array}{ccc}
	Signal^r_{\cA}(\bD, \btheta) = \left| \overline{Y}_{\cA,1}^r-\overline{Y}_{\cA,0}^r \right| &\qquad & Signal^r(\bD, \btheta) =  \overline{S}^r\\
	\\
	Noise^r_{\cA}(\bD, \btheta) = \sqrt{\dfrac{s^{2,r}_{Y,\cA, 0}}{\sharp \Dr_{\cA,0}}+ \dfrac{s^{2,r}_{Y,\cA, 1}}{\sharp \Dr_{\cA,1}}} &\qquad & Noise^r(\bD, \btheta) = \sqrt{\dfrac{s^{2,r}_{S}}{\sharp \Dr}} \\
	\\
	Ratio^r_{\cA}(\bD, \btheta)=\dfrac{Signal^r_{\cA}(\bD, \btheta)}{Noise^r_{\cA}(\bD, \btheta)} &\qquad & Ratio^r(\bD, \btheta)=\dfrac{Signal^r(\bD, \btheta)}{Noise^r(\bD, \btheta)}\\
	\\
\end{array}
$$
where $\sharp \Dr_{\cA,z} = \sum_{i=1}^{n} \mathbb{I}\{i \in \Dr_{\cA,z}\}$ 
and 
$\sharp \Dr = \sum_{i=1}^{n} \mathbb{I}\{i \in \Dr\}$ are the number of units in the $r$ data belonging to the $\Dr_{\cA,z}$ and $\Dr$ group, respectively.

It is worth noting that these measures are not treatment effects, but they provide information on whether the model can preserve broad features of signal, noise, and signal-to-noise ratio in the survival time distributions for non-switchers and switchers and in the switching time distribution for switchers assigned to the control arm. 

\begin{table}\caption{Bayesian Posterior Predictive $p-$values} \label{tab5}
	$$
	\begin{array}{llcccc}
		\hline
		\multicolumn{2}{l}{\hbox{Variable}} &\hbox{Deviance} &  \hbox{Signal} & \hbox{Noise} & \hbox{Signal to noise}\\
		\hline
		\multicolumn{2}{l}{\hbox{Survival time}}& 0.917\\
		&\hbox{\textit{Non-Switchers}} && 0.258  & 0.595  & 0.251 \\
		&\hbox{\textit{Switchers}}     && 0.621  & 0.773  & 0.542 \\
		\vspace{-0.15cm}\\
		\multicolumn{2}{l}{\hbox{Switching time}} & 0.485  & 0.378  & 0.343  & 0.549 \\
		\hline
		\vspace{-0.2cm}\\
		\multicolumn{6}{c}{PPPV \hbox{ for BIC}: 0.497} 
	\end{array}
	$$
\end{table}
\begin{figure}
	\begin{center}
		\begin{tabular}{cc}
	\hspace{-1cm}		\includegraphics[width=7.5cm]{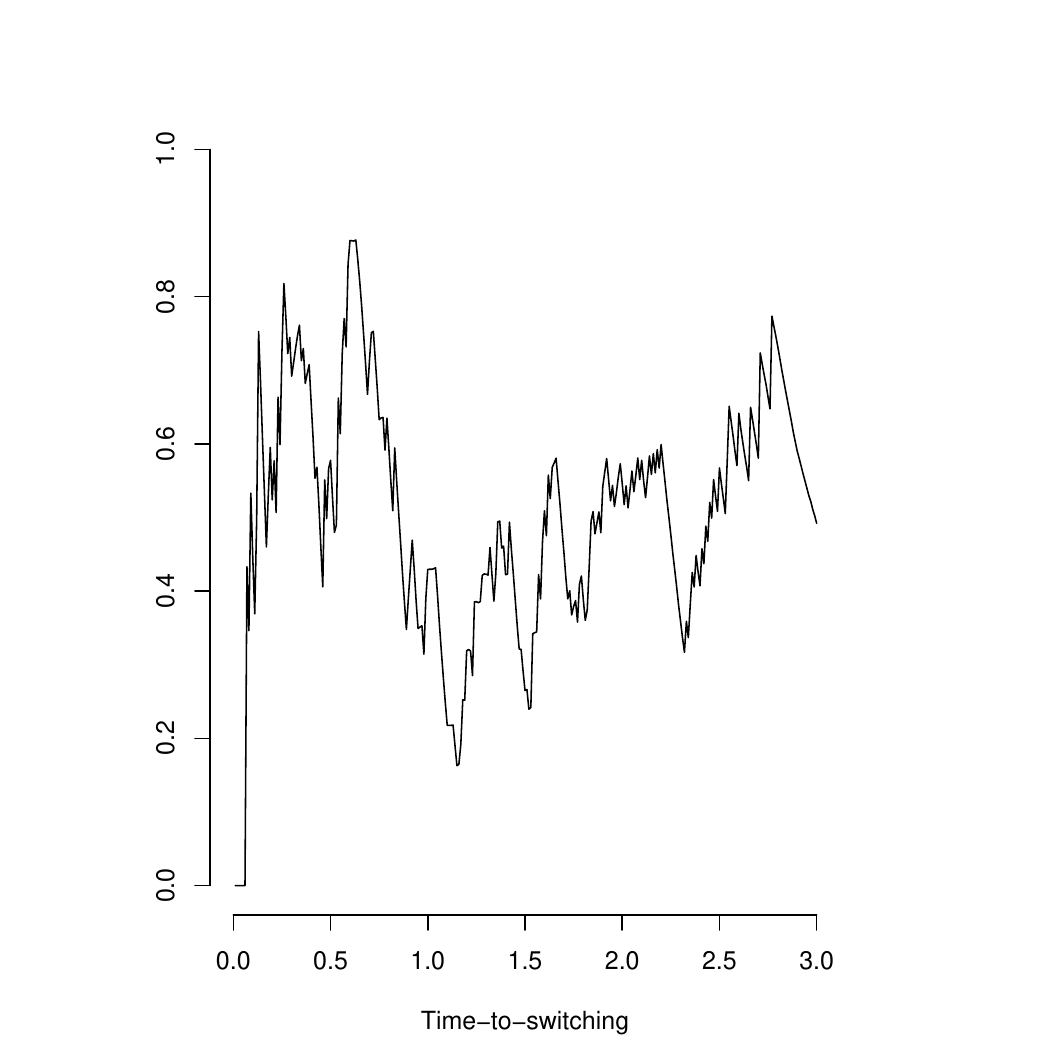}&\hspace{-1cm}
			\includegraphics[width=7.5cm]{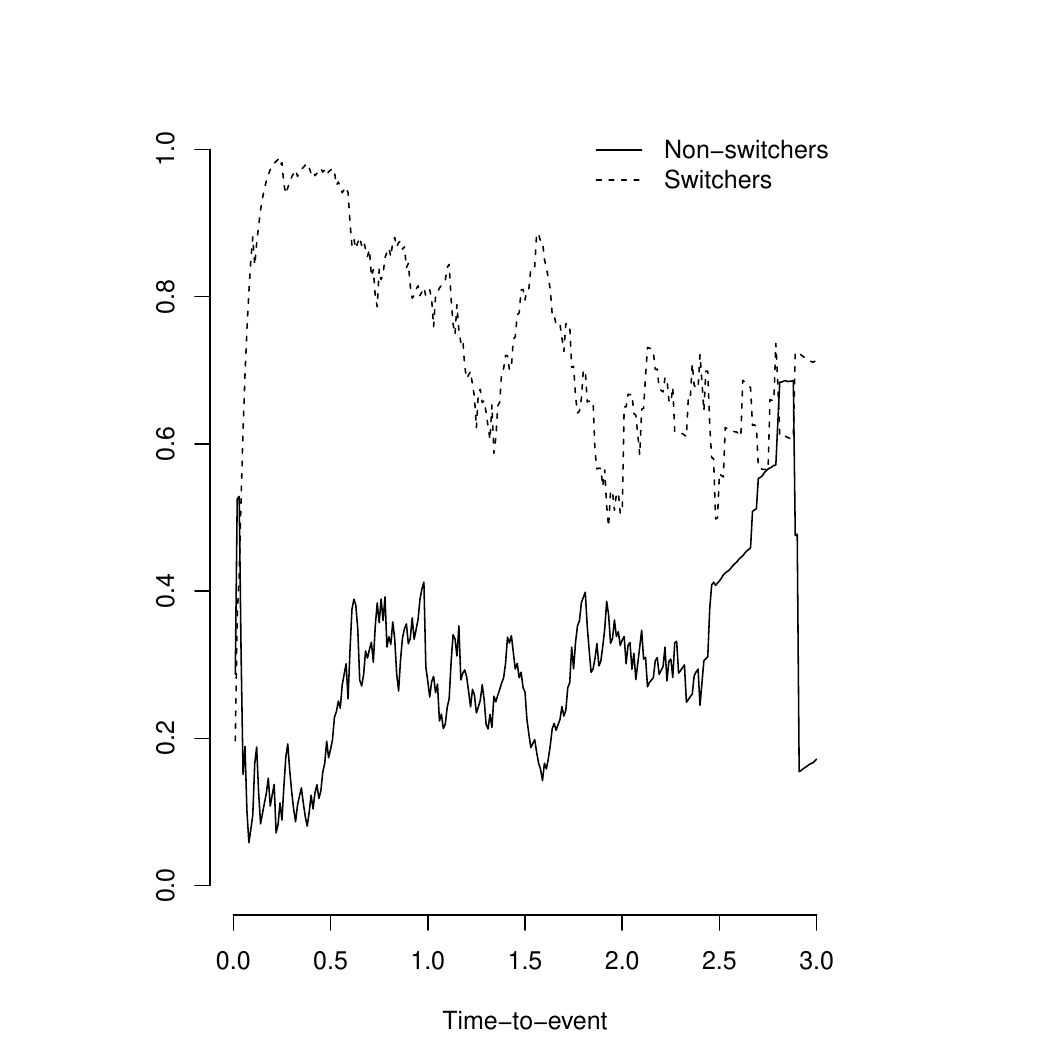} \\
			\\
		\end{tabular}
	\end{center}
	\caption{Bayesian posterior predictive $p-$values for Kaplan-Meier posterior predictive discrepancy measures ($t=0.01, 0.02, \ldots, 2.00, 3.00$)} \label{Fig5}  
\end{figure}

Table \ref{tab5} shows the Bayesian $PPPV$s. 
The $PPPV$ for the BIC is $0.497$, and the $PPPV$s for the deviance posterior predictive discrepancy measures are $0.485$ for the switching time and $0.917$ for the survival time, suggesting that our model fits the data pretty well. 
The $PPPV$s for the Kaplan-Meier posterior predictive discrepancy measures are also sufficiently far away from 0 and 1 for all time points $t$; the only exceptions are the Kaplan-Meier posterior predictive discrepancy for the time-to-switching under control for switchers for times shorter than $0.06$ (approximately 22 days) and times between $0.17$ and $0.54$ (approximately between 2 and 6.5 months) for the time-to-death/disease progression for switchers. 

It is worth noting that, in the observed data, no patient assigned to the control treatment is observed to switch to the active treatment within 22 days, and only 28 patients are observed to either die or experience a progression of the disease between 2 and 6.5 months. 
Results provide no special evidence for specific influences of the model too.
The estimated Bayesian $PPPV$s for the signal, noise, and signal-to-noise ratio posterior predictive discrepancy measures range between $0.251$ and $0.773$, suggesting that our model successfully replicates the corresponding measure of location, dispersion, and their relative magnitude. 







\bibliographystyle{ba}
\bibliography{TreatmentSwitchingR1.bib}


\end{document}